\documentclass[11pt,twoside]{ucthesis}
\def\dsp{\def\baselinestretch{2.0}\large\normalsize}
\dsp

\usepackage{graphicx}
\usepackage{subfig}
\usepackage{amsmath}
\usepackage{amsfonts}
\usepackage{amssymb}
\usepackage{mathrsfs}

\newcommand{\ud}{\mathrm{d}}
\newcommand{\bsy}{\mathbf}
\newcommand{\de}{\partial}
\newcommand{\bnabla}{\boldsymbol{\nabla}}

\newcommand{\be}{\begin{equation}}
\newcommand{\bes}{\begin{equation*}}
\newcommand{\ee}{\end{equation}}
\newcommand{\ees}{\end{equation*}}

\begin{document}

% Declarations for Front Matter

\title{Nonlinear Processes in Coronal Heating and Slow Solar Wind
         Acceleration}
\author{Antonio Franco Rappazzo}
\degreeyear{2006}
\degreesemester{}
\degree{Doctor of Philosophy}
\chair{Professor Ignatius Arrogant}
\othermembers{Professor Ivory Insular\\
Professor General Reference}
\numberofmembers{3}
%\prevdegrees{B.A. (University of Northern South Dakota at Hoople) 1978\\
%M.S. (Ed's School of Quantum Mechanics and Muffler Repair Shop) 1989}
\field{Aboriginal Basketry}
\campus{Berkeley}

\maketitle
\approvalpage
\copyrightpage
\blankpage

%\begin{abstract}
%\include{abstract}
%\end{abstract}

\begin{frontmatter}

\begin{dedication}
\null\vfil
\begin{flushright}
``...highly nonlinear problems of magnetohydrodynamics...are too difficult for exact analytical treatment and must be left to the mastication of computer wallahs.''\\\vspace{8pt}
Donald Lynden-Bell, Mon.\ Not.\ R.\ Astron.\ Soc.\ 267, 146 (1994)
\end{flushright}
\vfil\null
\end{dedication}
%{\large
%``That's enough, now this book is yours."\\\vspace{4pt}
%}
%\texttt{Charles Bukowski}\hspace*{10pt}   \\\vspace{12pt}
%(\emph{from the introduction to}\hspace*{10pt} \\\vspace{4pt}
%\emph{``Ask the Dust'' by John Fante})\hspace*{10pt}
%\end{flushright}

\blankpage
\tableofcontents
%\listoffigures
%\listoftables
\begin{acknowledgements}
This work has been supported in part by a fellowship of the
Universit\`a degli Studi di Pisa.

I am grateful to the many people who helped me throughout
this work. Many thanks to my official and unofficial supervisors, 
Russell Dahlburg, Giorgio Einaudi, Steve Shore and Marco Velli.
In particular I would like to thank Russ Dahlburg, who made 
possible a one year visit at Naval Research Laboratory.
I would also like to thank the IPAM program 
``Grand Challenge Problems in Computational Astrophysics'' 
at UCLA, where part of this work has been performed.

The 3D visualization of the numerical simulations  has been made
easy (and possible) thanks to Viggo Hansteen, who has introduce 
me to OpenDX, a user-friendly program with which all the 3D figures 
and movies have been done.
I would also like to thank Francesco Malara for the many stimulating 
and useful discussions we have had while I was writing the manuscript,
and Zoran Mikic, Jon Linker and Roberto Lionello for their suggestions.
Finally, I would like to thank the referees who have carefully read the
manuscript: Claudio Chiuderi, Francesco Pegoraro and  Dalton Schnack.

I would also like to thank my friends and colleagues who have made
less painful my stay in Pisa, in particular Andrea Dotti, Daniele Ippolito,
Gianluca Lamanna and, last but not least, my cousin Massimo Romano who 
has patiently and repeatedly given me hospitality during the last few years.
\end{acknowledgements}
\blankpage
\end{frontmatter}

\part{Coronal Heating}

\chapter{The Coronal Heating Problem} \label{ch:chp}

The Sun, our nearest star, is both the source of energy for life on Earth
and a unique physics laboratory. Energy produced in the core of the Sun
is transported to its surface and atmosphere. From here, through radiation and 
a flow of particles (the solar wind), it affects the Earth and forms the so-called 
heliosphere, a cavity inside the local interstellar medium which extends beyond the 
solar system boundaries. 

The visible surface of the Sun, the photosphere, has a temperature of about 
$6000\ K$ and emits mostly visible light. High energy radiation 
(EUV, FUV, up to X-rays) is produced in the upper layers of the atmosphere: the 
chromosphere, the transition region and the Corona. On the other hand the 
steady production of this kind of radiation requires high temperatures, and the
highest temperatures are localized in the Corona. 
Hence this gives rise to the so-called \emph{Coronal heating problem}: 
Corona is counterintuitively much hotter (millions of degrees $K$) than the 
underlying photosphere.
While there is general agreement that the magnetic field plays a fundamental role, 
and that the source of this energy derives from convective motions in the 
photosphere (which have more than enough energy), the debate currently focuses 
on what physical mechanisms can transfer, store, and dissipate this energy between 
the photosphere and the corona.

\section{The Sun: its Interior and Atmosphere}

The solar interior is conveniently separated into zones by the different physical processes at work.
The pressure, density, and temperature decrease going outward, from the center up to 
the surface, located at 1 solar radius ($1\ R_{\odot} \sim 7 \times 10^5\, km$).
Energy is generated by nuclear reactions in the \emph{core}, the innermost 25\% of the radius. 
This energy is transported outward by radiation through the \emph{radiative zone} and by 
convective flows  through the \emph{convection region}, the outermost 30\%.

The convection zone is the outermost layer of the solar interior. It extends from a depth of
about $2 \times 10^5\, km$ up to the visible surface. At the base of the convection zone the 
temperature is about $2\times10^6\, K$. This temperature is low enough that the heavier ions
are not totally ionized. This increases the cross section for the radiation, ultimately trapping
more heat, which in turns makes the plasma unstable to convection. 
Convection occurs when the temperature gradient required to radiatively transport the energy
is larger than the adiabatic gradient, i.e. the rate at which the temperature would fall if a 
volume of material expanded adiabatically while rising toward the surface.
Where this occurs a parcel of plasma displaced upward will be warmer than its surroundings 
and will continue to rise, i.e. convection instability sets in. 
Convective motions are quite efficient in carrying heat to the solar surface, and
while the plasma rises it expands and cools. At the visible surface the 
temperature drops to $5800 \ K$ and the density to $2 \times 10^{-7} \ g\, cm^{-3}$.

The photosphere is the visible surface of the Sun. Since the Sun is made of ionized gas, 
this surface is not sharply defined, and it is actually  a layer about $100\ km$ thick 
(very thin compared to the $7 \times 10^5\, km$ radius of the Sun).
A number of features, which are relevant for coronal heating models,  are observed in the 
photosphere, including the dark sunspots, the bright faculae, and granules. 
\begin{figure}[t]
\begin{center}
\includegraphics[width=0.5\textwidth]{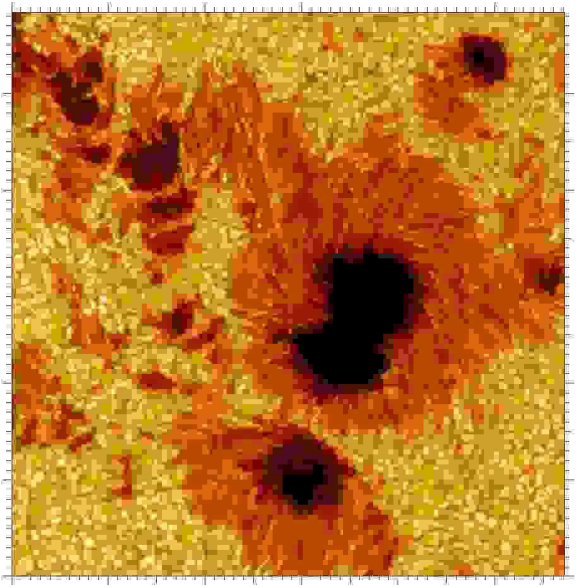}
\caption{Large field-of-view image, in visible light, of sunspots in Active Region 10030 
observed on 15 July 2002 by the Swedish 1-m Solar Telescope on the island of La Palma, 
Spain. It shows part of a sunspot group near disk center. Distance between tick marks is 
1000 km. Credits: the Royal Swedish Academy of Sciences. \label{fig:spot} }
\end{center}
\end{figure}

Sunspots (Fig.~\ref{fig:spot}) appear as comparatively isolated dark regions on the surface.
They typically last for several days, although very large ones may live for 
several weeks. Sunspots are magnetic regions on the Sun characterized by a magnetic field 
of the order of about $100-1000$ gauss, noticeably stronger that the surrounding field 
(a few gauss). Sunspots usually come in 
groups with pairs of spots, with opposite field polarity. The field is strongest in the umbra,
the darkest part of the sunspots, while it is weaker and more horizontal in the penumbra,
the brighter part surrounding the umbra.
Temperatures in the dark centers of sunspots drop to about $3700 \ K$, compared to 
$5800\ K$ for the surrounding photosphere, due to the presence of the strong magnetic
field which partially inhibits convective motions.

Faculae are bright areas that are most easily seen near the limb, or edge, of the 
solar disk. These are also magnetic regions but the field is concentrated in much 
smaller bundles than in sunspots. While the sunspots tend to make the Sun look darker, the 
faculae make it look brighter. During a sunspot cycle the faculae actually win out over the 
sunspots and make the Sun appear slightly (about 0.1\%) brighter at sunspot maximum that 
at sunspot minimum.
\begin{figure}[t]
\begin{center}
\includegraphics[width=0.6\textwidth]{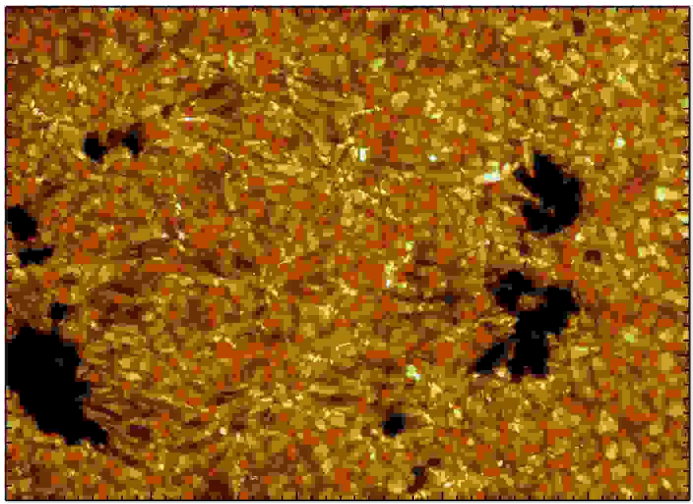}
\caption{Image of a solar active region taken on 13 May 2003 near the center of the solar disk 
at heliographic coordinates N10 E15 degrees. The tick marks are 1000 km apart. The image is 
a filtergram taken in 430 nm ``G-band'' light at the Swedish 1-meter Solar Telescope. The image 
was taken by Dr.\ Tom Berger of the Lockheed Martin Solar and Astrophysics Lab, Palo Alto, 
California.  \label{fig:gr}  }
\end{center}
\end{figure}

Granules (Fig.~\ref{fig:gr} and also Fig.~\ref{fig:spot}) are small (their typical size is about 
$1000\ km$) cellular features that cover the entire Sun 
except for those areas covered by sunspots. These are the tops of convection cells 
where hot plasma rises up from the interior in the bright areas, spreads out across the surface, 
cools and then sinks inward along the dark lanes. Individual granules last for only about 8 
minutes. The granulation pattern is continually changing as old granules are pushed aside 
by newly emerging ones (see movie from the Swedish 1-meter Solar Telescope). 
The average speed of the flow within the granules is $1\ km\, s^{-1}$, but it can reach 
supersonic speeds  of more than $7\ km\, s^{-1}$ . This flow, and its interaction with the
magnetic field, generate waves which then propagates toward the upper layers of the
atmosphere.

Supergranules are another flow pattern present in the photosphere, at a larger scale 
(about $35.000\ km$ across) than granules. These features also cover the entire Sun and are 
continually evolving. Individual supergranules last for a day or two and have flow
speeds of about $0.5\ km\, s^{-1}$. 

The chromosphere is an irregular layer, $2000-3000\ km$ thick, above the photosphere  
where the temperature rises from $4400\ K$ (the temperature minimum) at its base to 
about $2 \times 10^4\, K$ at the top.  
At these higher temperatures hydrogen emits light that gives off a reddish color 
(H$\alpha$ emission). This colorful (chromo = color) emission can be 
seen in prominences that project above the limb of the sun during total solar eclipses. 
The chromosphere is the site of activity as well. Solar flares, prominence and 
filament eruptions, and the flow of material in post-flare loops can all be observed over the 
course of just a few minutes.

The transition region is a thin and very irregular layer of the Sun's atmosphere that 
separates the chromosphere from the much hotter corona. The temperature changes 
rapidly from $2\times 10^4\ K$ up to $1\times 10^6\ K$ over a $30\ km$ distance.
Hydrogen is ionized at these temperatures and is therefore difficult to detect. Instead 
the light emitted by the transition region is dominated by such ions as 
C IV, O IV, and Si IV that emit light in the far ultraviolet region ($< 2000$ \AA) of the solar 
spectrum that is only accessible from space.

\section{The Solar Corona}

\begin{figure}[t]
\begin{center}
\includegraphics[width=0.6\textwidth]{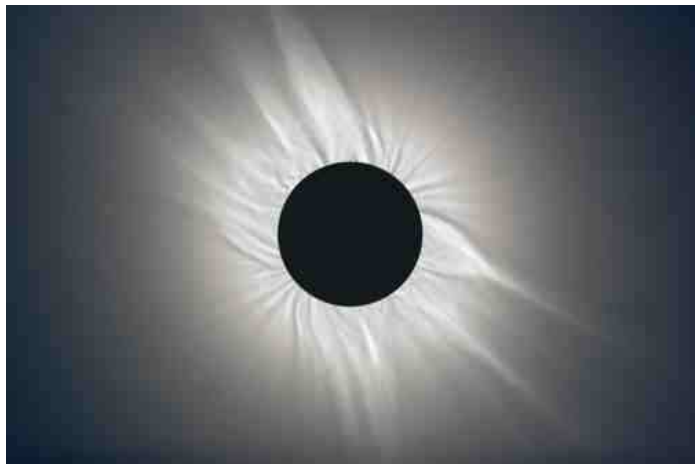}
\caption{Image of the solar corona in visible light taken on 29 March 2006 during total
solar eclipse in Side, Turkey. Light from the photosphere is diffused through Thompson 
scattering. The image was taken by Koenraad van Gorp.  \label{fig:ecl}  }
\end{center}
\end{figure}

The Corona is the Sun's outer atmosphere. Most of the visible light coming from the 
corona is light coming from the photosphere diffused trough Thompson scattering.
This results in a faint emission compared to the photosphere, and it only becomes visible 
during total eclipses of the Sun, as shown in Fig.~\ref{fig:ecl}, or using a special
instrument called coronagraph. 
The corona displays a variety of features including 
streamers, plumes, and loops. These features change from eclipse to eclipse and the 
overall shape of the corona changes with the sunspot cycle. However, during the few 
fleeting minutes of totality few, if any, changes are seen in these coronal features.

Early observations by Harkness and Young in 1869 of the visible spectrum of the corona 
revealed bright emission lines at 
wavelengths that did not correspond to any known materials. This led astronomers to 
propose the existence of ``coronium'' as the principal gas in the corona. The true nature of 
the corona remained a mystery until 1939, when Grotrian and Edlen identified the Fe XIV and
Ni XVI lines in the corona. The corona is in fact heated to temperatures greater than 
$1\times 10^6\ K$. At these high temperatures 
both hydrogen and helium (the two dominant elements), and even
minor elements like carbon, nitrogen, and oxygen are totally ionized. 
Only the heavier trace elements like iron and calcium are able to retain a few of their 
electrons. It is emission from these highly ionized elements that produces the 
spectral emission lines that were so mysterious to early astronomers.

The corona shines brightly in x-rays because of its high temperature. On the other hand, the 
cooler solar photosphere emits very few x-rays. This allows us to view the corona across the 
disk of the Sun when we observe the Sun in X-rays. To do this we must first design optics 
that can image x-rays and then we must get above the Earth's atmosphere, which shields
the Earth from this high energy radiations. In the early 1970s 
Skylab carried an x-ray telescope that revealed coronal holes and coronal bright points for 
the first time (these features were actually visible in earlier sounding rocket data, but they
were not recognized). In the 1990s Yohkoh provided a wealth of information and images 
on the solar corona. Today  SOHO  and TRACE satellites are obtaining new and 
exciting observations of the Sun's corona, its features, and its dynamic character.

Coronal loops are found above sunspots and active regions. These structures are 
associated with the closed field lines that connect magnetic regions on the solar 
surface. Many coronal loops last for days or weeks but most change quite rapidly. 

Coronal holes, where the corona is dark,  were discovered when 
X-ray telescopes were first flown above the earth's atmosphere to reveal the structure of the 
corona across the solar disc. They are associated with ``open'' magnetic field lines 
and are often found at the Sun's poles (a region that is currently studied by the Ulysses 
mission). The high-speed solar wind is known to originate in coronal holes.

Different layers of the solar atmosphere are characterized by different features,
but most of these features are connected one another by the magnetic field.
At solar minimum the solar magnetic field is approximately a dipole field, with the 
magnetic axis aligned to the rotational axis, so that we have ``open'' magnetic
field around polar regions and closed magnetic structures
around the equator. This picture is, of course, an approximation and the major 
departures are at the large scales the presence of an heliospheric current sheet
(a region where the magnetic field changes rapidly its polarity),
and at smaller scales a more disordered structure, which becomes more important
during solar maxima.

The origin of the solar magnetic field is an active field of research.
Not being a solid body the surface of the Sun is characterized by differential 
rotation, i.e. the poles have a lower rotation rate than the equator. In fact, the rotation period
is of 35 days at the poles and 25 days at the equator.  Helioseismology has
demonstrated that differential rotation extends down to the convection zone. 
The deeper zones, the radiative zone and the core, appear to rotate as a solid body. 
The mechanism, called \emph{magnetic dynamo}, which leads to the formation of the 
solar magnetic field seems to be connected with convection. This intense magnetic 
field rises toward the solar surface due to magnetic buoyancy and emerges at the photosphere.

\begin{figure}[t]
   \centering
   \begin{minipage}[c]{0.5\linewidth}
       \centering \includegraphics[width=1\textwidth]{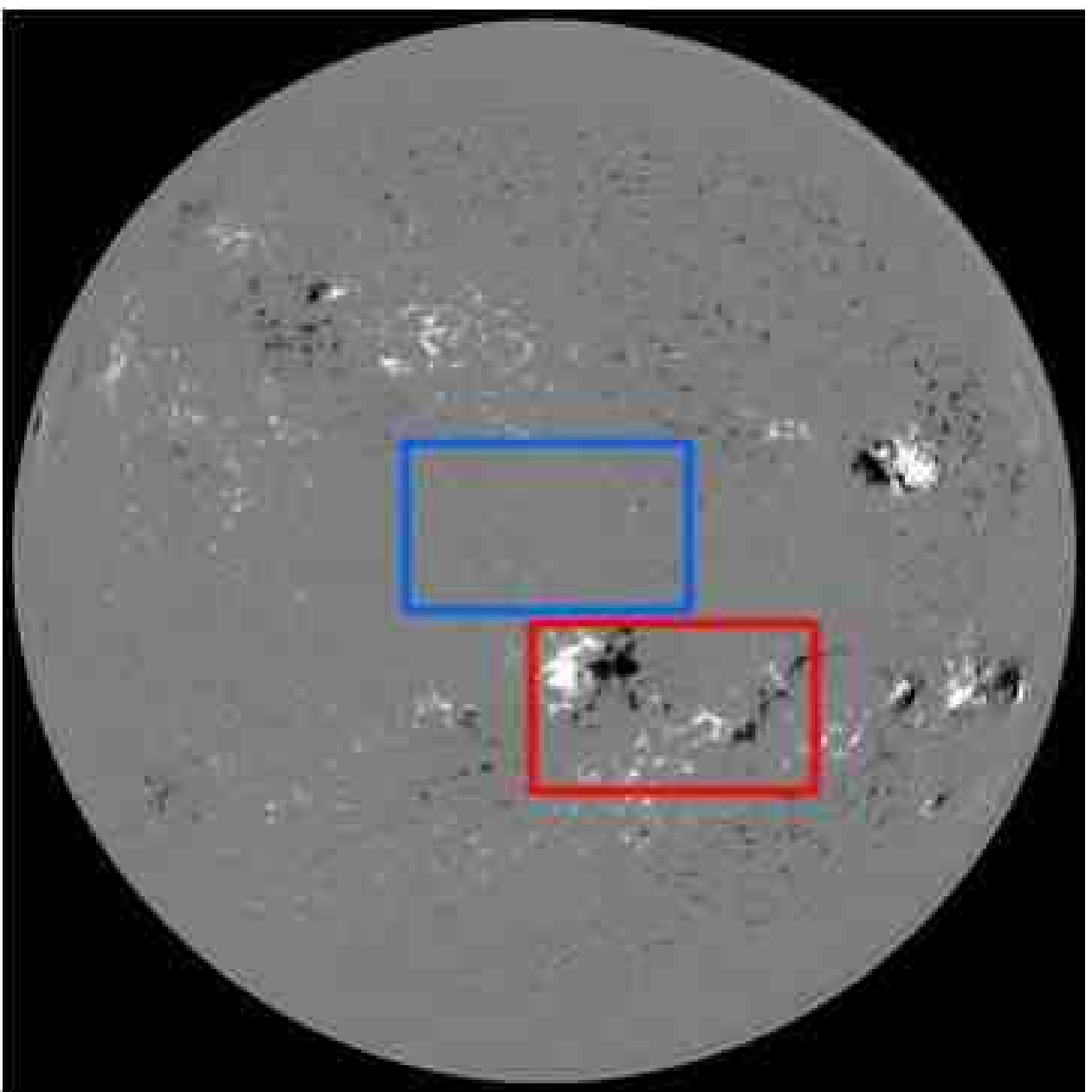}
   \end{minipage}%
   \begin{minipage}[c]{0.5\linewidth}
       \centering \includegraphics[width=0.8\textwidth]{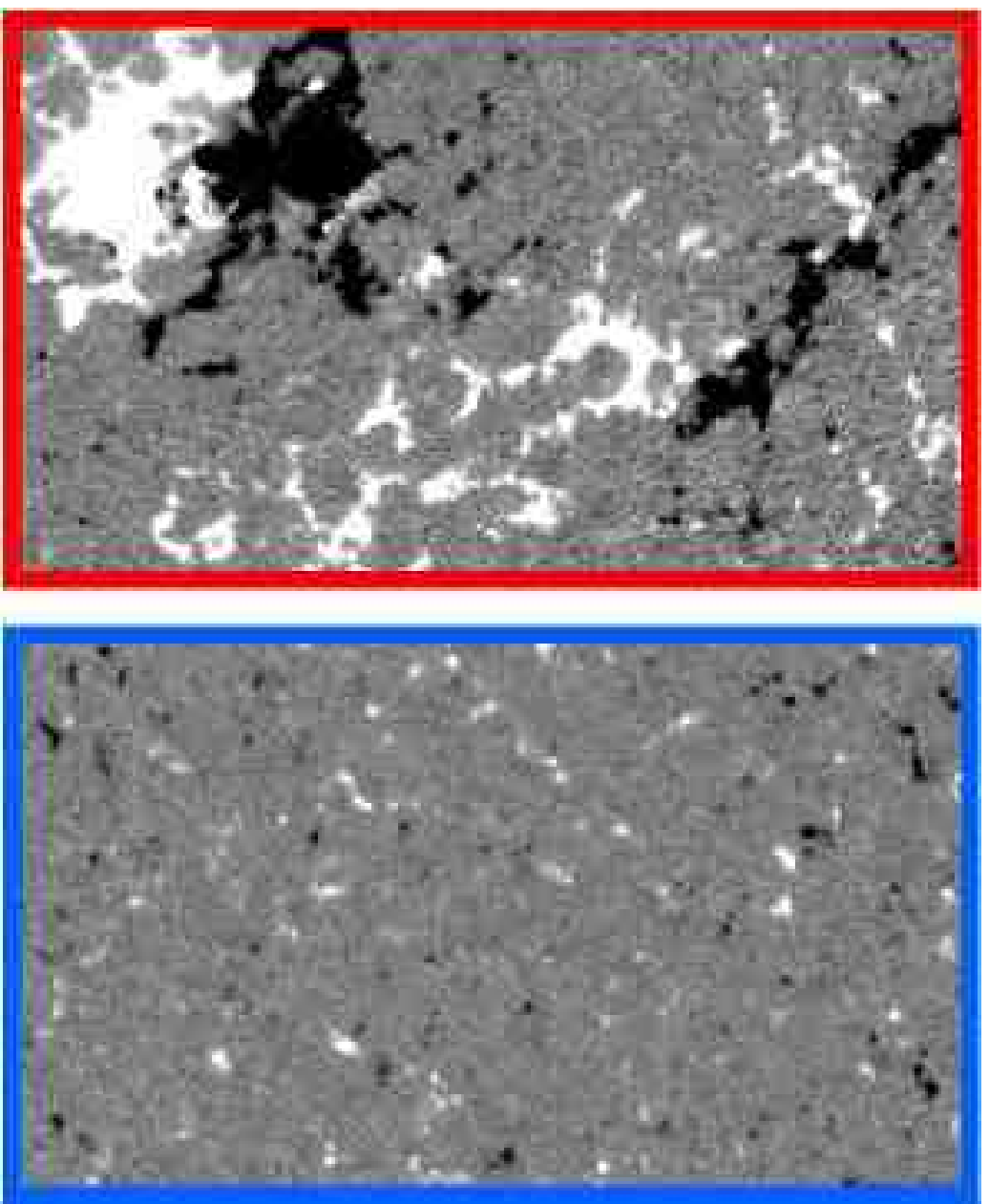}
   \end{minipage}
   \caption{\emph{Left}: Full disk magnetogram showing the strenght and polarity
                   of the line-of-sight component of the photospheric magnetic field. The
                   red rectangle highlights an active region, while the blue one highlights a
                   quiet-Sun region. \emph{Right}: Close-ups of the active and quiet-Sun
                   regions. These magnetograms have been obtained with the Michelson
                   Doppler Imager (MDI) onboard SOHO.
     \label{fig:actr}  }
\end{figure}

\begin{figure}[t]
   \centering
   \begin{minipage}[c]{0.5\linewidth}
       \centering \includegraphics[width=0.9\textwidth]{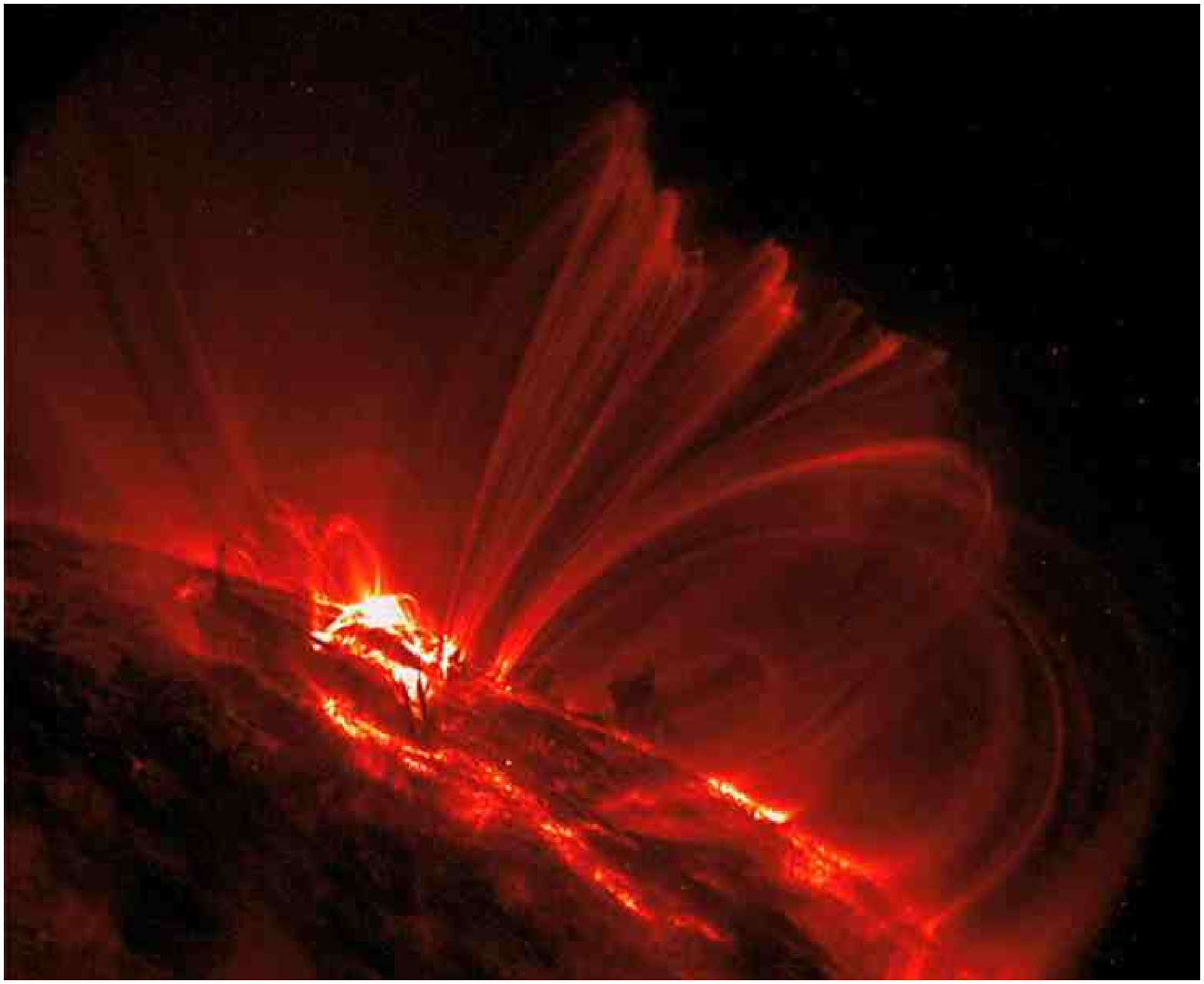}
   \end{minipage}%
   \begin{minipage}[c]{0.5\linewidth}
       \centering \includegraphics[width=0.9\textwidth]{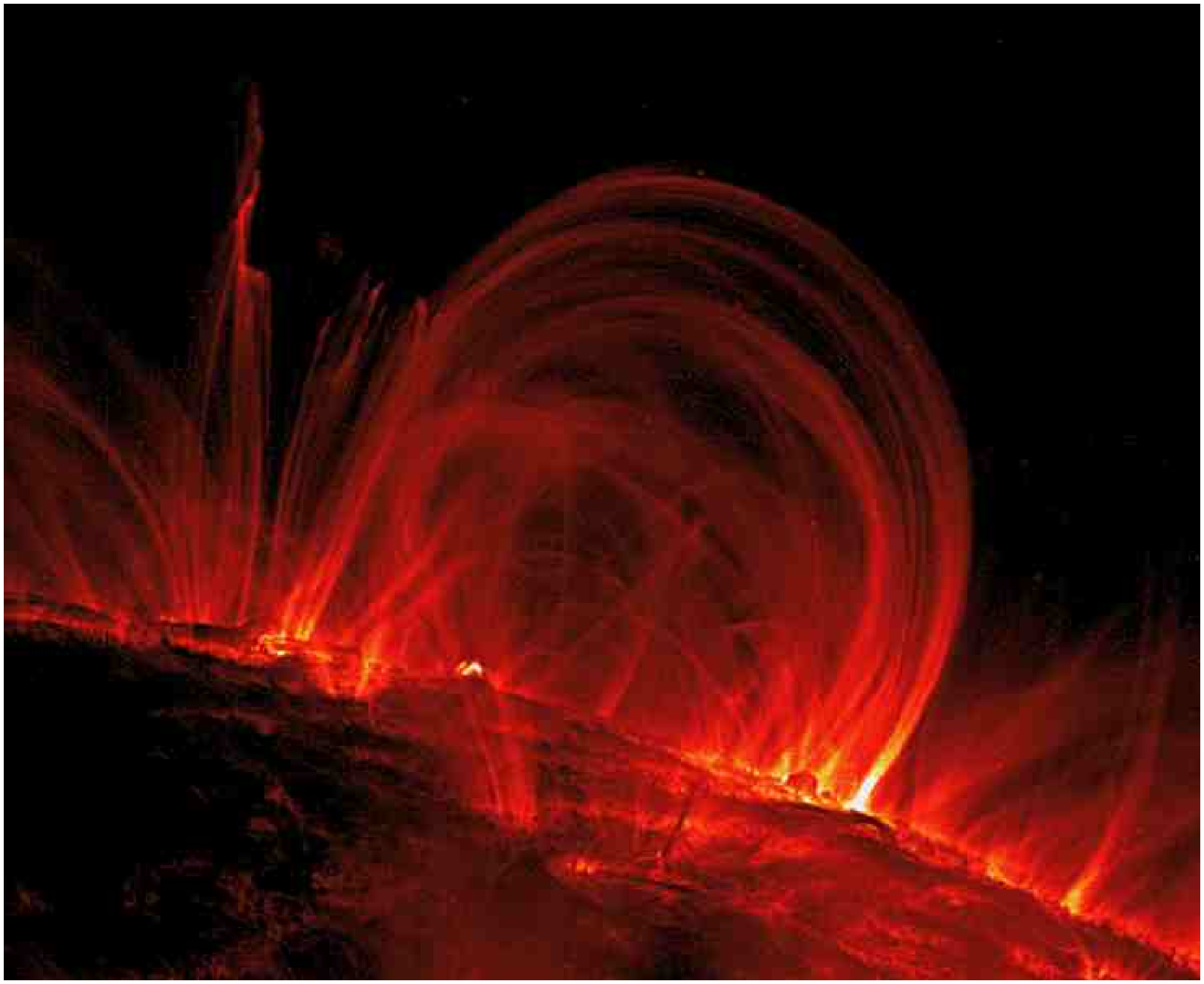}
   \end{minipage}
   \caption{\emph{Left}: Active regions 9628 and 9632 are rotating over the southwest limb of 
   the Sun in this image, showing the loops arching above them. This image 
   was taken in the Fe IX/X 171 \AA \ channel at 23:59 UT on January 10, 2001. North is to the 
   left, west to the top. The field of view is 512 arc sec N/S and 384 arc sec E/W.  
   \emph{Right}: This image  of coronal loops over the eastern limb of the Sun was taken in 
   the 171 \AA \ pass band, characteristic of plasma at 1 MK, on November 6, 1999, 
   at 02:30 UT. 
   Both images have been obtained by the Transition Region And Coronal Explorer
   (TRACE) satellite.
   \label{fig:loop}  }
\end{figure}

The organization of the  magnetic field at the photospheric level gives rise 
to two different kind of regions, so-called  ``active regions'' and  ``quiet-Sun regions'', 
as shown in Fig.~\ref{fig:actr}. 
All the solar surface is characterized by a magnetic field with a mixed polarity, 
but in active regions the unipolar areas are bigger and the magnetic field stronger.
This structure is commonly called a \emph{magnetic carpet}.
There is general agreement that the active region field is the direct result of the 
large-scale solar dynamo. The continuous presence of quiet-sun regions also during
solar minima, when the number of active regions considerably decreases, suggests 
that this flux might be generated by local dynamo action just below the sun's surface, 
driven by granular and supergranular flows \cite{men89,dur93,pet93,lin95}
The opposite polarity regions of the magnetic carpet are connected by closed magnetic field lines,
extending up to the upper layers of the solar atmosphere, called loops. 
In correspondence of active regions, where the magnetic field is stronger, these
loops shine bright in EUV and x-rays as shown in Fig.~\ref{fig:loop}

\begin{figure}[p]
      \centering
      %%---- start ----
      \subfloat[Blue Continuum]{
               \label{fig:sp:a}             %% label for subfigure
               \includegraphics[width=0.3\linewidth]{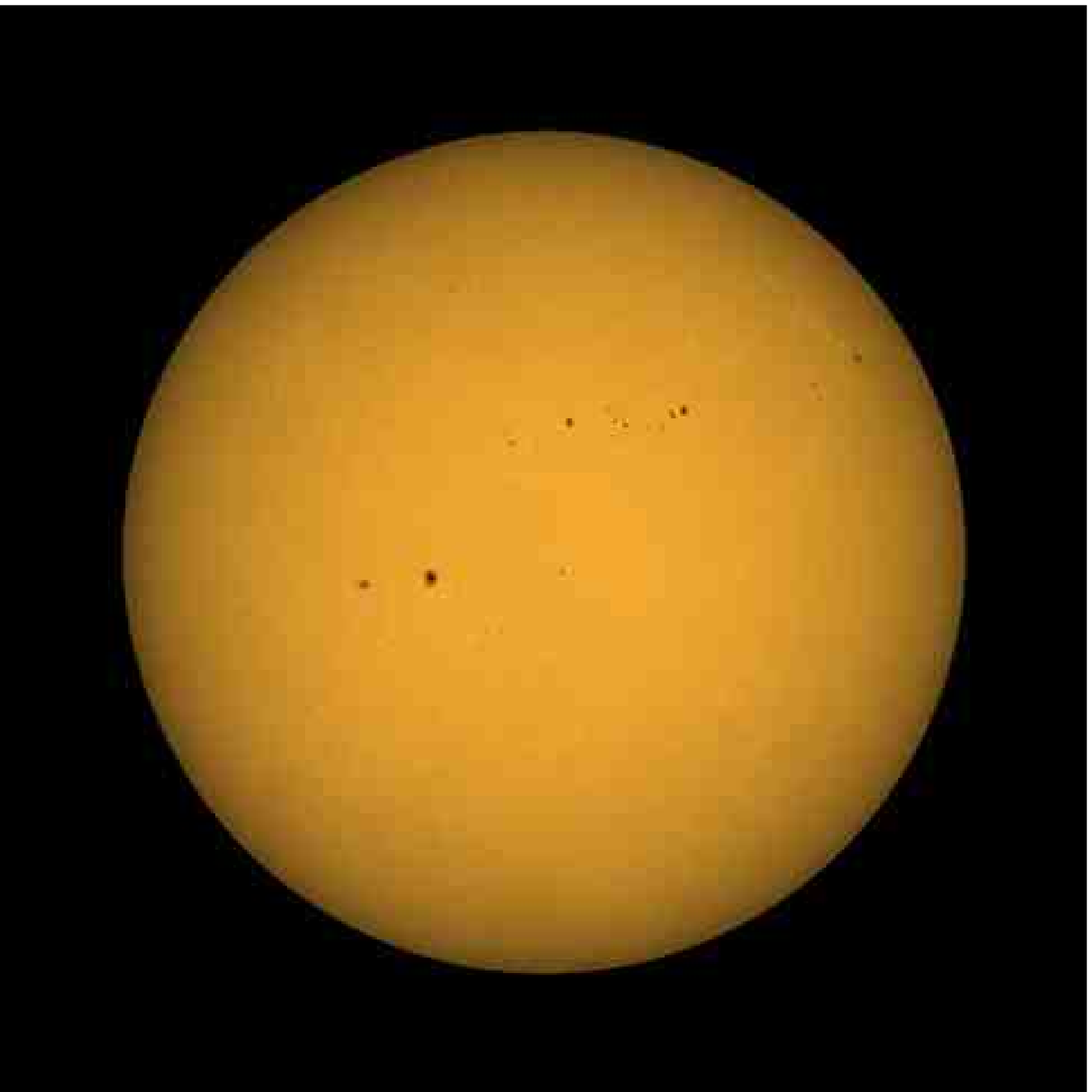}}
      \hspace{0.01\linewidth}
     %%---- start ----
      \subfloat[Magnetogram]{
               \label{fig:sp:b}             %% label for subfigure
               \includegraphics[width=0.3\linewidth]{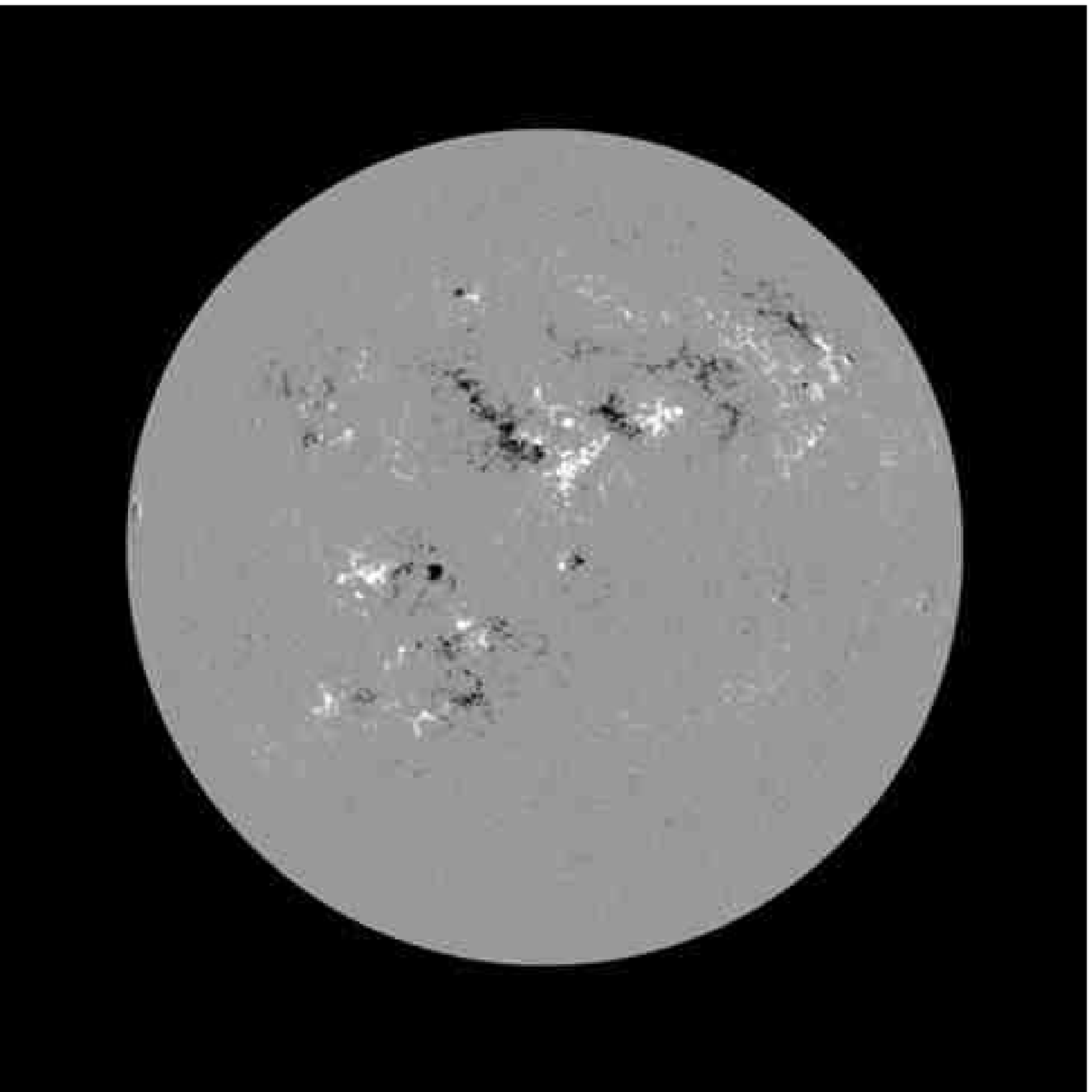}}
      \hspace{0.01\linewidth}
      %%---- start ----
      \subfloat[Ca II K]{
               \label{fig:sp:c}             %% label for subfigure
               \includegraphics[width=0.3\linewidth]{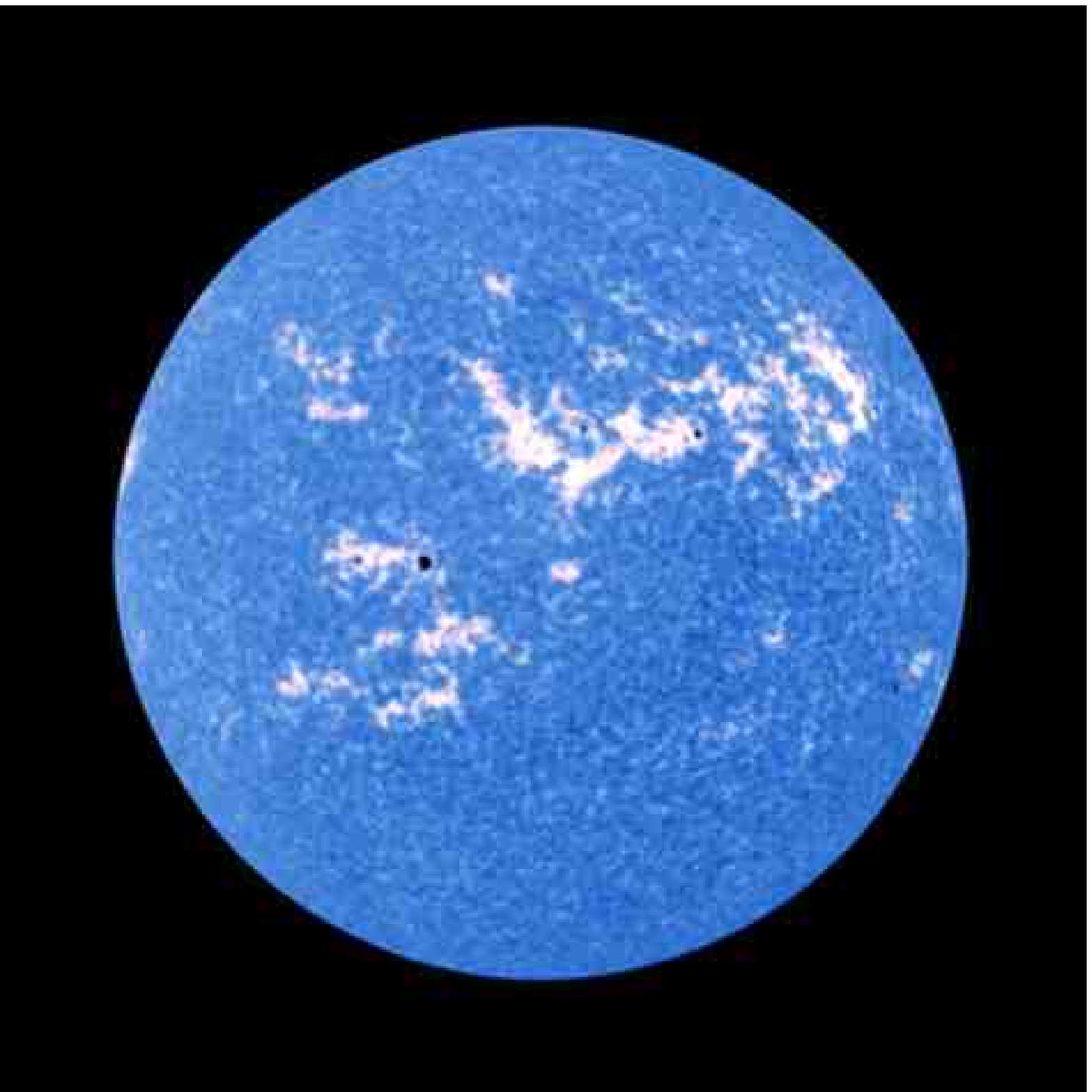}}\\[20pt]
      %%---- start ----
      \subfloat[$H_{\alpha}$]{
               \label{fig:sp:d}             %% label for subfigure
               \includegraphics[width=0.3\linewidth]{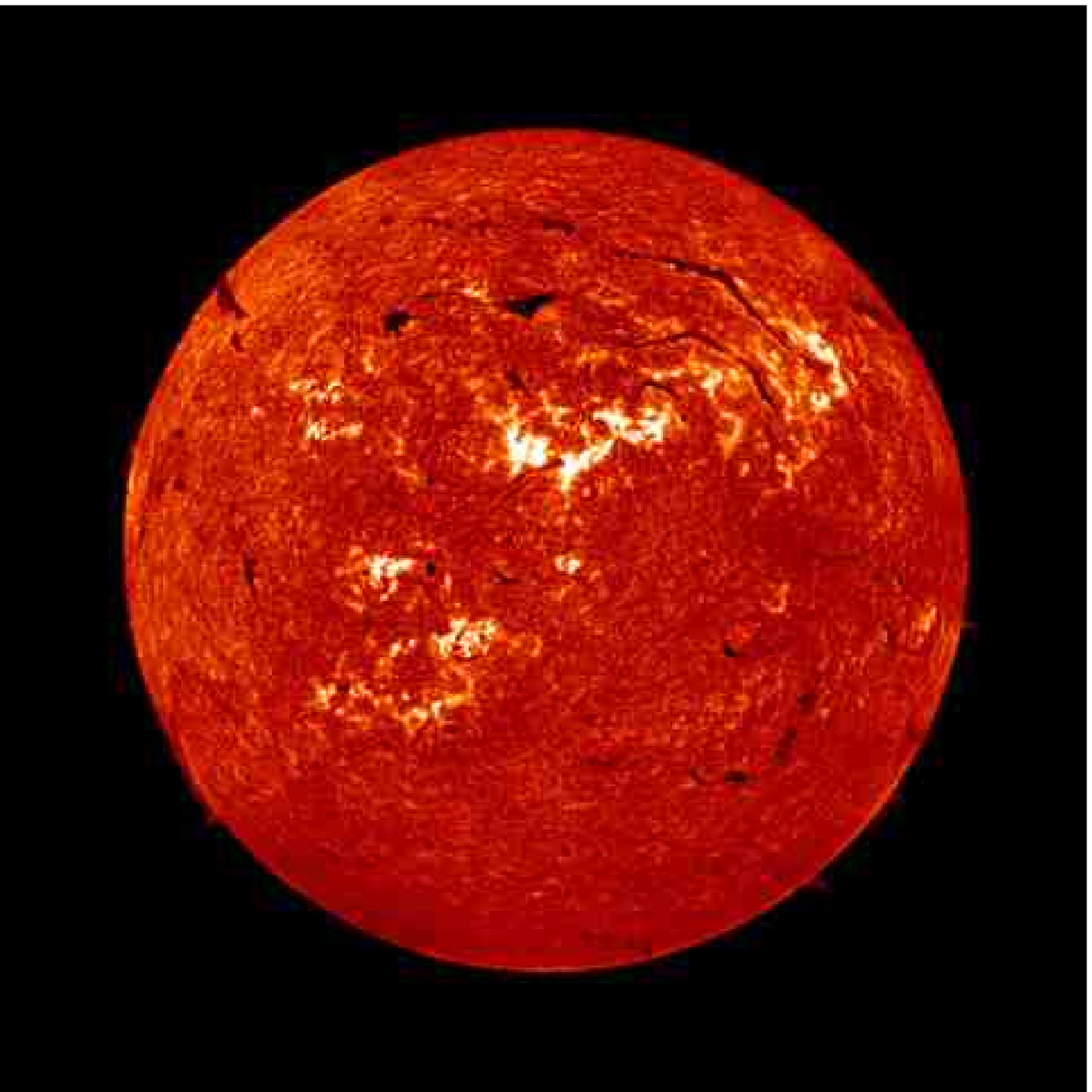}}
      \hspace{0.01\linewidth}
      %%---- start ----
      \subfloat[He II 304 \AA]{
               \label{fig:sp:e}             %% label for subfigure
               \includegraphics[width=0.3\linewidth]{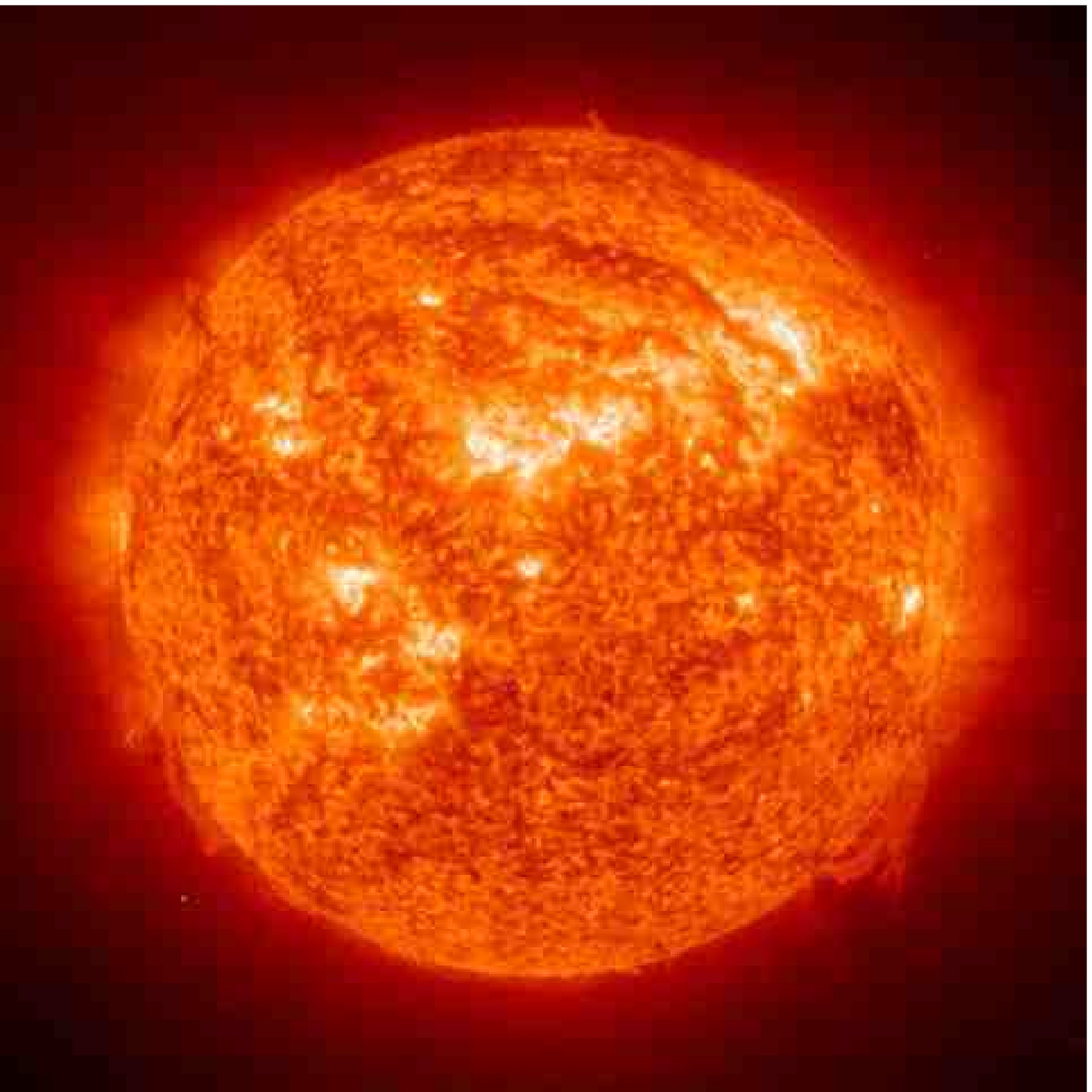}}
      \hspace{0.01\linewidth}
      %%---- start ----
      \subfloat[Fe IX 171 \AA]{
               \label{fig:sp:f}             %% label for subfigure
               \includegraphics[width=0.3\linewidth]{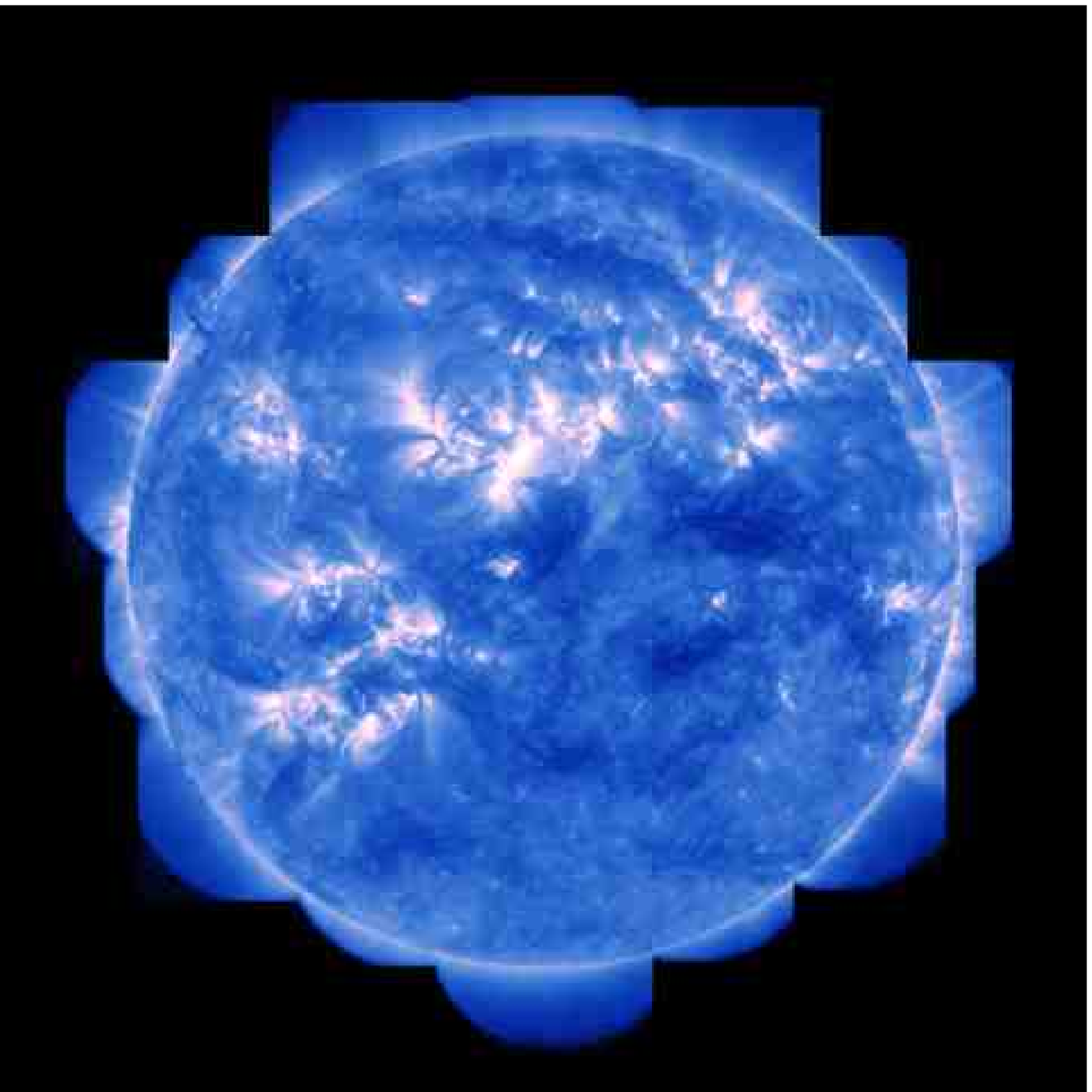}}\\[20pt]
      %%---- start ----
      \subfloat[Fe XII 195 \AA]{
               \label{fig:sp:g}             %% label for subfigure
               \includegraphics[width=0.3\linewidth]{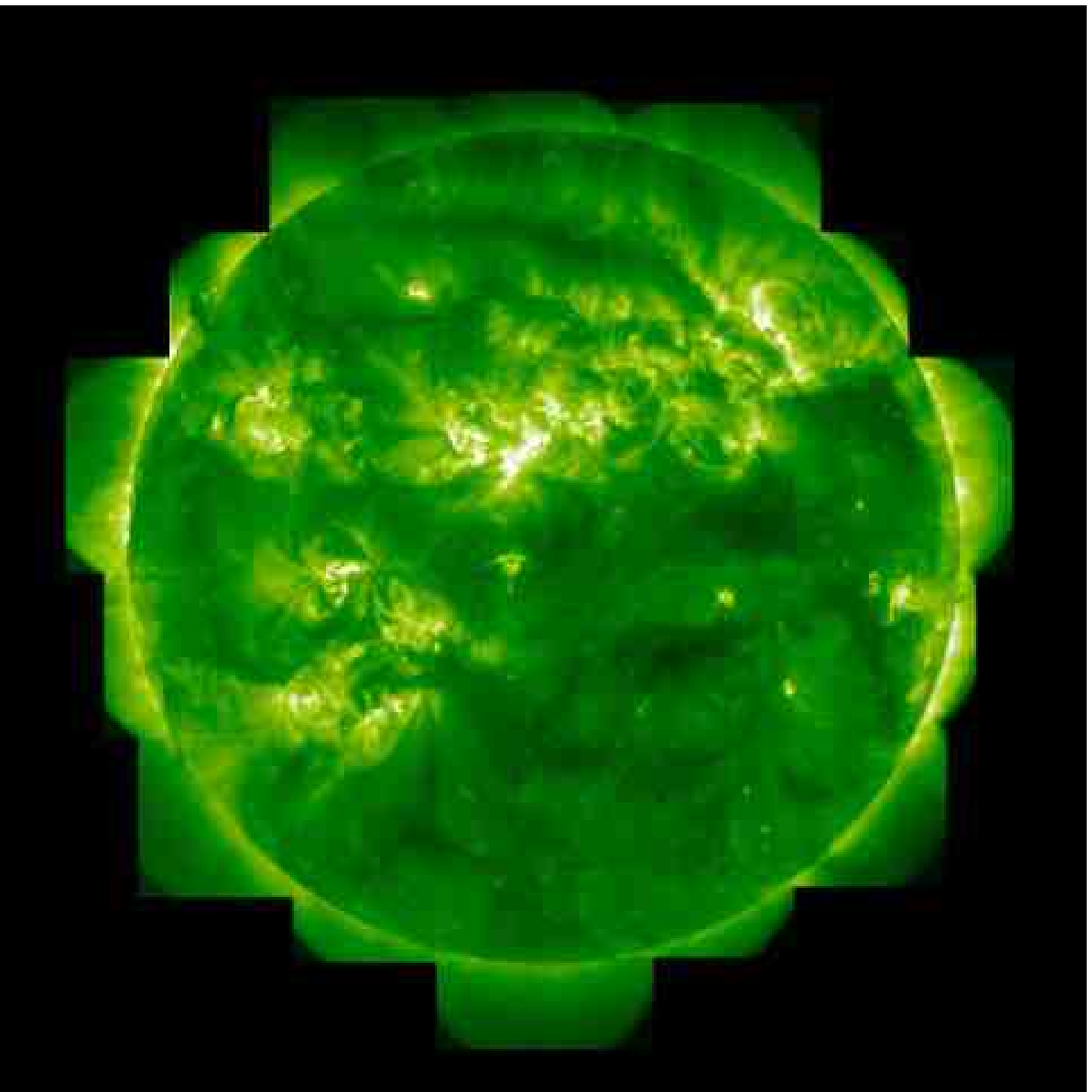}}
      \hspace{0.01\linewidth}
      %%---- start ----
      \subfloat[Fe XV 284 \AA]{
               \label{fig:sp:h}             %% label for subfigure
               \includegraphics[width=0.3\linewidth]{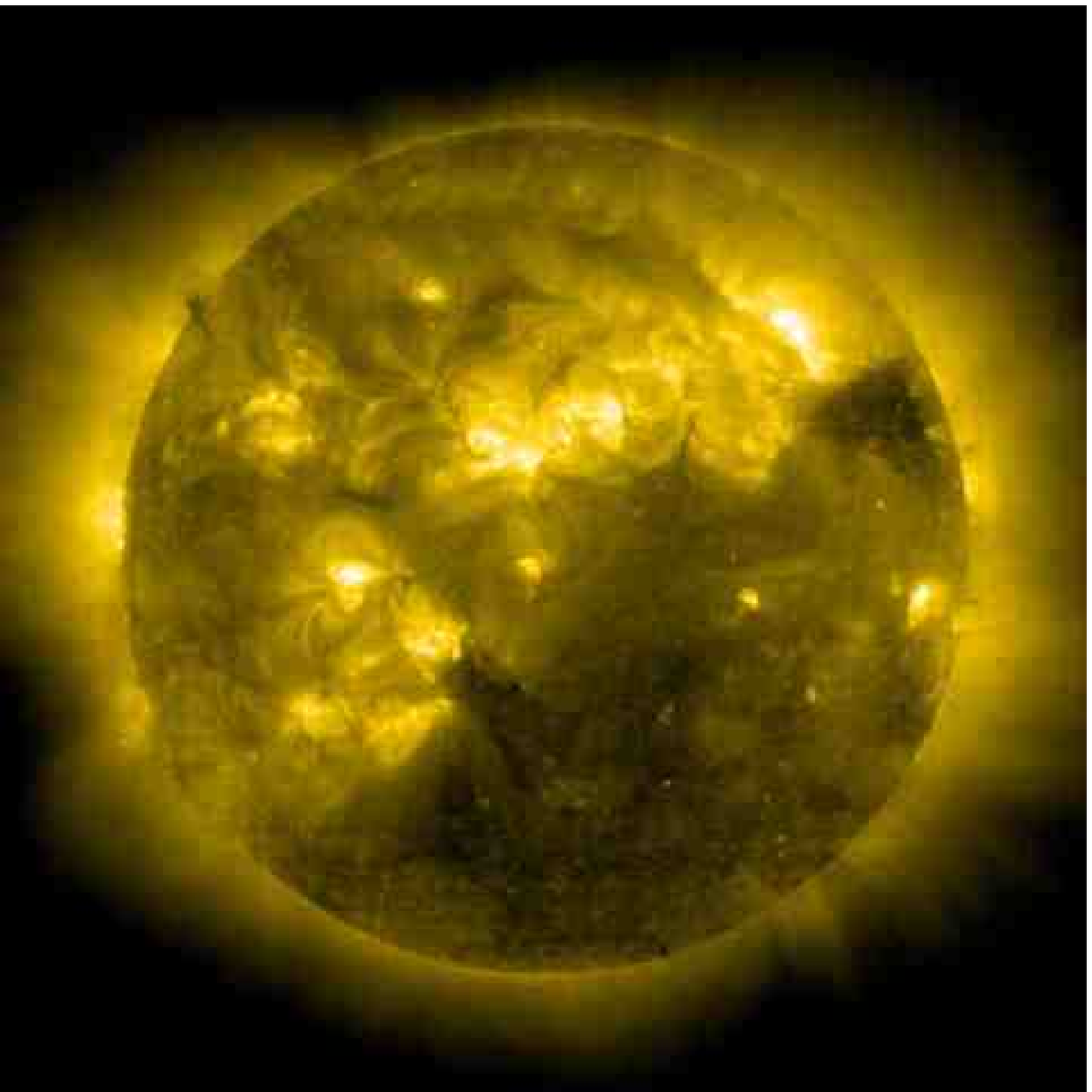}}
      \hspace{0.01\linewidth}
      %%---- start ----
      \subfloat[Soft X-rays]{
               \label{fig:sp:i}              %% label for subfigure
               \includegraphics[width=0.3\linewidth]{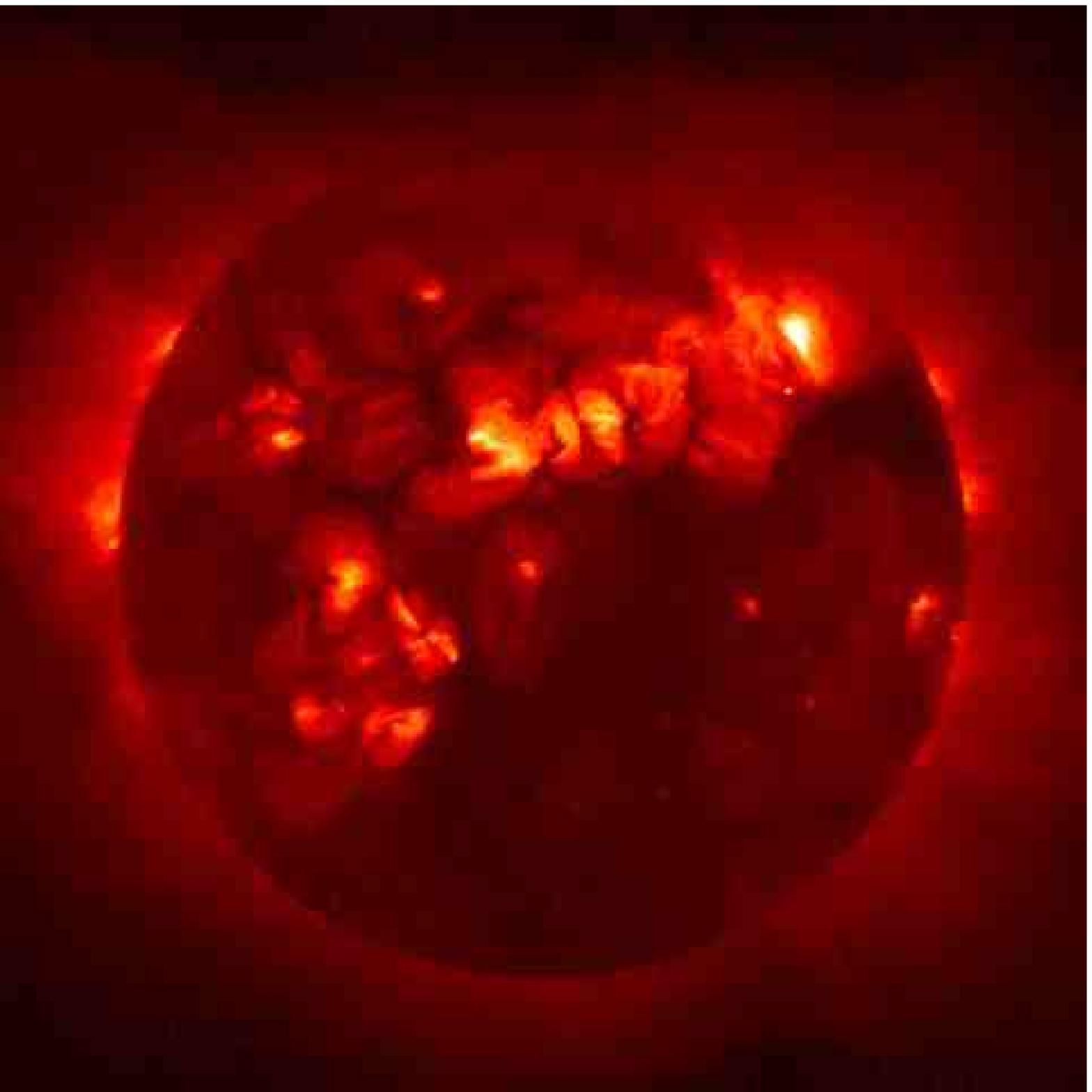}}
      \caption{Multi-wavelength images of the Sun taken on 8 February 2001, all within hours
      of each other.  Credits: ESA/NASA SOHO (a, b, e, h),  TRACE (f, g), the 
      Japanese-American YOHKOH satellite (i), and the Big Bear Solar Observatory (c, d). }
      \label{fig:sp}                           %% label for entire figure
\end{figure}

The intimate relationship of the magnetic field and the structures of the outer atmosphere is 
illustrated in Figure~\ref{fig:sp} that shows images of the Sun at different wavelengths,
corresponding to different temperatures and layers.
In particular the blue continuum image~\ref{fig:sp:a} shows the photosphere, 
image~\ref{fig:sp:b} is a magnetogram
showing the line-of-sight component of the magnetic field (white for out-going directed field, black 
for in-going field), the upper photosphere shown in the light of Ca II K in image~\ref{fig:sp:c},  the
chromosphere in $H\alpha$ line in image~\ref{fig:sp:d}, the transition region in the light of
He II at 304 \AA \ in image~\ref{fig:sp:e}, and then the corona seen in lines Fe IX 171 \AA \ formed at 
$\sim 7.5\times 10^5\, K$ in~\ref{fig:sp:f}, Fe XII 195 \AA \ at $\sim 1.5 \times 10^6\ K$  in~\ref{fig:sp:g},
Fe XV 284 \AA \ at $\sim 2.5 \times 10^6\ K$ in~\ref{fig:sp:h}, and $3-5 \times 10^6\ K$ in the
soft x-ray region in~\ref{fig:sp:i}. The main information we have from these pictures is the 
correspondence between the regions of strong magnetic field at the photospheric level 
(\ref{fig:sp:b}) and the activity in all the upper layers of the solar atmosphere up to the X-ray 
corona. The main structures seen in Figure~\ref{fig:sp:a} are the sunspots. Emergence of the 
magnetic field at the photosphere~\ref{fig:sp:b} occurs in and near the spots and elsewhere. 
The upper photosphere~\ref{fig:sp:c},
chromosphere~\ref{fig:sp:d}, and transition region~\ref{fig:sp:e} show local brightening, heating, 
at the locations of strong magnetic field. The coronal images~\ref{fig:sp:f}-\ref{fig:sp:i} show a complex
of loops. Note that all space is not covered by loops and that hotter temperature loops tend to be 
nested inside loops of cooler temperatures.

\section{Overview of Coronal Heating Models}

In the 1930s Edlen, Grotrian, and Lyot established that the solar corona has a temperature
of the order of $10^6 \ K$. After this, the fundamental step was done in the 1940s by 
Biermann~\cite{bier46}, Alfv\'en~\cite{alf47}, and Schwarzschild~\cite{sch48}, who
pointed out that the high temperatures of the corona are a direct consequence of the
convective motions at the photospheric level. The convection does a mechanical
work, which subsequently is transported and dissipated in the chromosphere and the 
corona.

As pointed out in the previous paragraphs, the solar corona consists of magnetically
confined regions characterized by closed structures (loops) roughly located around
the equator, and regions of ``open'' magnetic field (coronal holes) roughly located
around the northern and southern poles. Observations of the solar corona in EUV 
and X-rays show that great part of this emission takes place in the closed structures.
This is the reason for which  the ``open'' magnetic regions, which look faint in this
high-energy range of the spectrum, are called \emph{coronal holes}. In fact they look
like ``holes'' in an otherwise bright corona.

Whether the mechanism which heats open and closed magnetic regions may be similar
or not, in this work we focus our attention on the magnetically confined regions of the
solar corona.

The active X-ray corona has a temperature of $1-5 \times 10^6\ K$, an electron and
proton numerical density of about $10^{10}\ cm^{-3}$, and is magnetically confined
in the $\sim 10^2-10^3$ gauss magnetic field of active regions. These high temperatures are
maintained by a heat input of about $10^7\ erg\ cm^{-2}\ s^{-1}$ \cite{wit77}.

In the 1940s Biermann, Alfv\'en, and Schwarzschild supposed that waves, 
generated at the photospheric level by convective motions, and then
propagating upward into the corona, would have dissipated their energy leading 
to the high temperatures observed. Although the mechanisms which are able to transfer,
store and dissipate the energy are still a matter of debate, the basic idea remains 
unchanged. 
The waves which are generated at the photospheric level include sound waves, 
gravitational waves, and magnetohydrodynamics waves. More recently it has been
shown \cite{ost61,ste74,su81,pri82} that all but Alfv\'en waves are dissipated and/or 
refracted before reaching the corona. Then, while these other waves contribute to chromospheric
heating, it is only Alfv\'en waves which are able to reach the corona. 

The Alfv\'en wave is a purely magnetohydrodynamic phenomenon, and it is essentially an
oscillation due to magnetic field line tension. In fact transverse motions of the magnetic field
lines cause a force that tries to restore them to straight-line form. Linearizing the equations of 
magnetohydrodynamics (hereafter MHD) in the simple case of a homogeneous plasma 
embedded in a 
homogeneous magnetic field $\boldsymbol{B}_0$, Alfv\`en waves are found as a transverse 
incompressible wave, propagating in the direction of the wave vector $\boldsymbol{k}$ 
with the dispersion relation:
\begin{equation} \label{eq:rd}
\omega = v_{\mathcal A}\, k\, \cos \theta
\end{equation}
where  $\omega$ is the wave frequency,
$v_{\mathcal A} = B_0 / \sqrt{4\pi \rho}$ the Alfv\'en velocity,  
$k$ the wave vector modulus and $\theta$ is the angle between the magnetic field $\bsy{B}_0$
and the wave vector $\bsy{k}$.
Alfv\'en waves due to photospheric motions are expected to have periods 
comparable to the 300 seconds characteristic time of granules, whose characteristic dimension
$l$ is of the order of $1000\ km$ and their velocity is of the order of $1\ km\ s^{-1}$. Shorter periods
have been detected, but at noticeably reduced power levels.

Coronal loop length is of the order of $10^4-10^5$ km, with a typical sound speed of the order of
$2\times 10^7\, cm\, s^{-1}$, and Alfv\'en speed roughly 10 times larger, $2\times 10^8\, cm\, s^{-1}$.
An important parameter in plasma physics is the ratio between kinetic and magnetic pressure
$\beta = 8\pi p / B^2$, that in the case of a coronal loop is of the order of $\beta \sim 2\times 10^{-2}$,
i.e. a coronal loop is a magnetic dominated system. 

The basic problem with wave heating is that the wavelengths of an Alfv\'en wave with a period 
typical of photospheric motions are too large to match coronal loop lengths. 
In fact from the dispersion relation~(\ref{eq:rd}) (with $v_{\mathcal A} = 2\times 10^3\, km\, s^{-1}$),
for a period $T \sim 10^2\, s$ it follows a wavelength $\lambda \sim 2\times 10^5\, km$, which is
of the same order as the length of the longest loops. Then they are quasi-static displacements
of the magnetic fields rather than waves. 

Traditionally the mechanisms responsible for coronal heating have been divided into two main
groups: AC (alternate current) and DC (direct current).

\subsection{High Frequency Models}

AC heating models propose two different mechanisms, Phase mixing (Heyvaerts \& 
Priest~\cite{hp83}) and resonant absorption (Davila~\cite{dav87}), in order to facilitate
dissipation of these waves within about one wave-period. They both occur when
Alfv\'en velocity is not uniform. 

Phase mixing occurs when Alfv\'en waves propagate at different phase velocities along
nearby fieldlines, making them come out of phase. 
 
Resonant absorption occurs whenever the Alfv\'en velocity is nonuniform (e.g. if density
is not uniform)  in a loop cross section. The wave amplitude is enhanced in a narrow layer 
where the local Alfv\'en resonance frequency matches the frequency of the global loop
oscillations.  Gradients in the magnetic and velocity fields are very large in this layer, and the
wave energy is easily dissipated by Ohmic and viscous processes.

\subsection{Low Frequency Models} \label{sec:lfm}

Alternatively it has been proposed (Parker~\cite{park79}, \cite{park83}, \cite{park86}, 
\cite{park88}) that the X-ray corona is heated by dissipation at the many small current 
sheets forming in a coronal loop as  a consequence of the continuous shuffling and 
intermixing of the footpoints of the field in the photospheric convection. 
The formation of these current sheets is conjectured by Parker~\cite{park91} in the 
following way. 
Consider a region $0 \le x, y \le l, \  0 \le z \le L$, embedded in a uniform magnetic field
aligned along the z-direction $\boldsymbol{B_0} = B_0\, \bsy{e}_z$ . Top and
bottom boundaries are located at the planes $z=0$ and  $z=L$. Supposing that in the plane
$z=0$ we have a zero velocity field, and that in $z=L$ an incompressible 2-D, i.e. $v_z = 0$,
velocity pattern. A continuous mapping of the footpoints velocity pattern in the perpendicular magnetic field ($b_x, b_y$) is produced. This field spontaneously produces tangential discontinuities: the 
discontinuities appear in the initially continuous field at the boundaries between local regions of different winding patterns. The tangential discontinuities (current sheets) become 
increasingly severe with the continuing winding and interweaving, eventually producing intense 
magnetic dissipation in association with magnetic reconnection. Parker suggested that
this dissipation is largely in the form of bursts of rapid reconnection. It is this sporadic explosive
dissipation at the tangential discontinuities in the bipolar fields on the sun that creates the active X-ray corona. The heating occurs in bursts, which are estimated to involve individually 
$10^{23}-10^{25}\, ergs$. Such a burst is too small to be observable and he refers to the individual burst as a ``nanoflare'', because it is 9 orders of magnitude smaller than a large flare of 
$10^{32}\, ergs\, cm{-2}\, s^{-1}$.

He eventually computes the energy input. Magnetic field lines connect the fixed footpoints
at $z=0$ with the moving ones in $z=L$, which move about with the velocity pattern imposed.
The field lines have more or less a uniform deviation $\Theta(t)$ to the vertical, where
\begin{equation}
\tan \Theta(t) \sim \frac{vt}{L}
\end{equation}
supposing $\Theta(t) < 1$. The vertical component of the field is $B_0$, indicating with $b_{\perp}$
the orthogonal component we have 
\begin{equation}
b_{\perp} = B_0 \tan \Theta(t) \sim \frac{B_0 vt}{L}
\end{equation}
the field line tension opposes this movement with a stress of the order of $b_{\perp} B_0 /4\pi$,
so that the work for unitary time and surface (the power) done by the photosphere is
\begin{equation} \label{eq:park}
W \sim \frac{v b_{\perp} B_0}{4\pi} = \frac{B_0^2}{4\pi} \frac{v^2 t}{L} \, ergs\, cm^{-2}\, s^{-1}.
\end{equation}
The input flux then increases linearly with time. We know from observations that the input
flux is of the order of $W\sim 10^7  \, ergs\, cm^{-2}\, s^{-1}$. For a magnetic field $B_0 = 10^2\, G$,
a velocity field $v = 1\, km\, s^{-1}$, and a loop length $L = 10^5\, km$, it follows from 
equation~(\ref{eq:park}) that the observed input flux is reached at $t \sim 5 \times 10^4\, s$,
when $b_{\perp} \sim B_0/4$ and $\Theta \sim 14^{\circ}$ (the so-called ``Parker angle'').
At this point it is conjectured that bursty rapid reconnection dissipates $b_{\perp}$  as
rapidly as it is produced by the velocity forcing at the photosphere. In this way the input energy
flux is on the average always of the order of $W\sim 10^7  \, ergs\, cm^{-2}\, s^{-1}$.

 \chapter{The Reduced MHD Model for Coronal Heating} \label{ch:model}

\begin{figure}[t]
   \centering
   \begin{minipage}[c]{0.45\linewidth}
       \centering \includegraphics[width=1\textwidth]{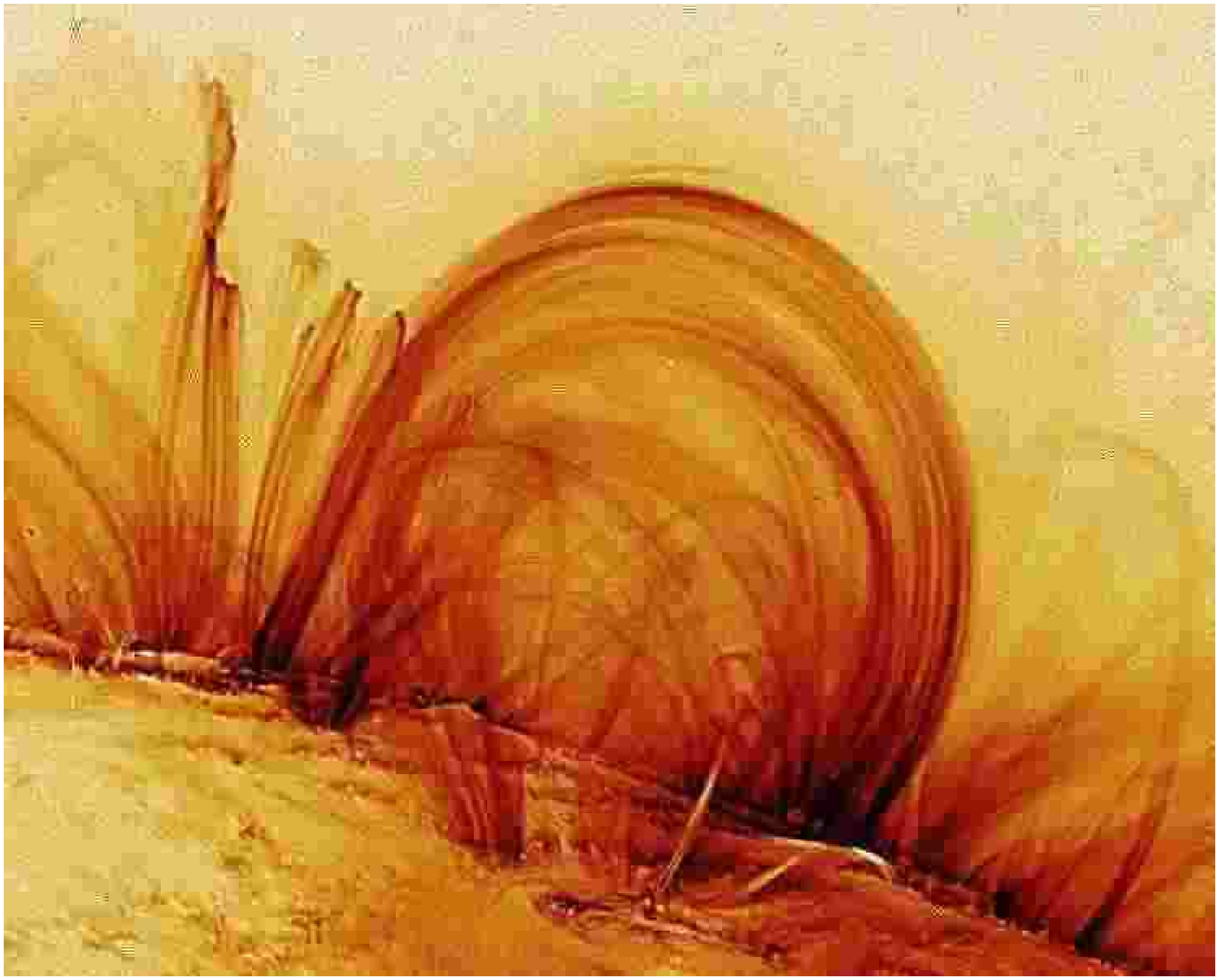}
   \end{minipage}%
   \begin{minipage}[c]{0.55\linewidth}
       \centering \includegraphics[width=0.8\textwidth]{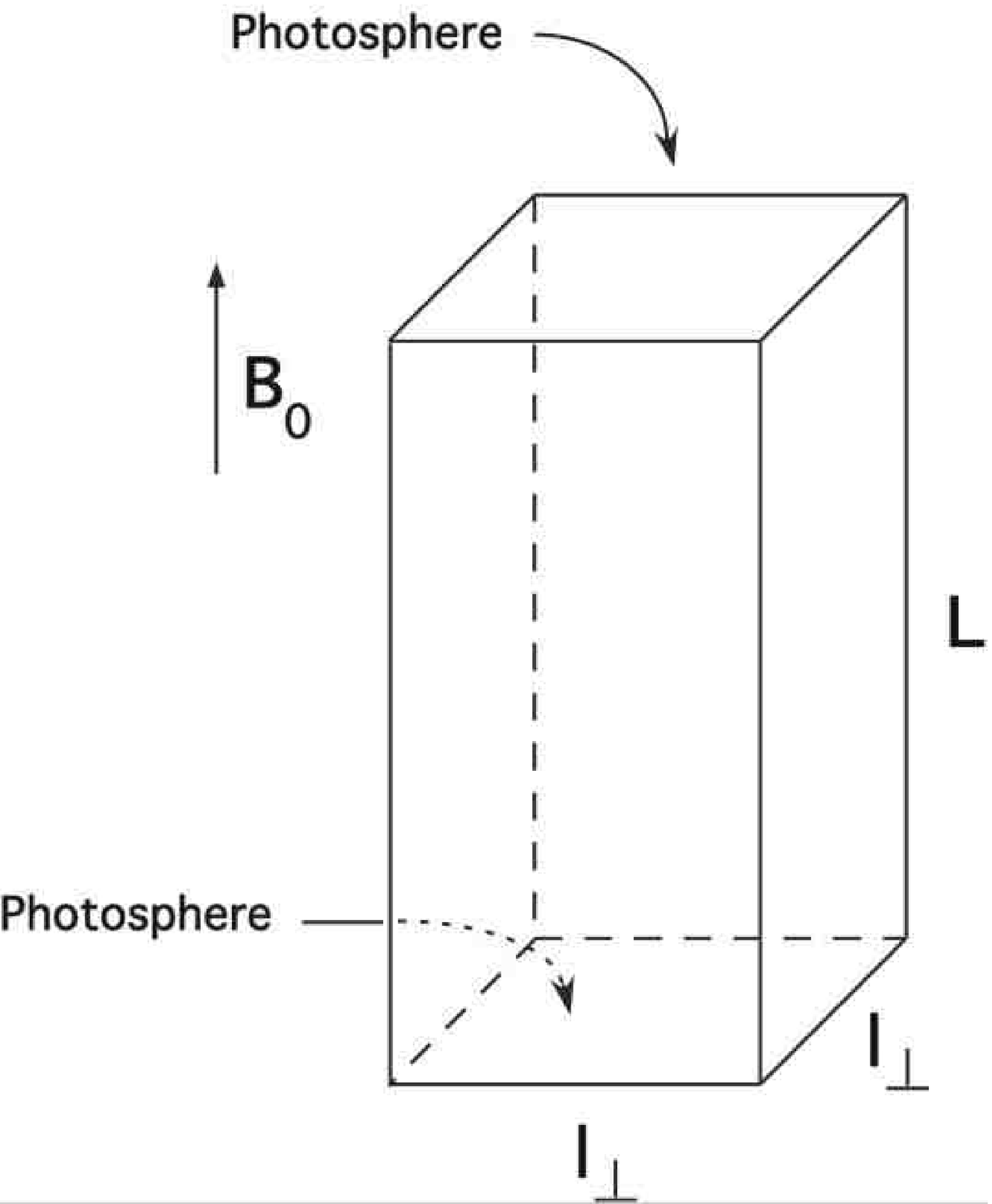}
   \end{minipage}
   \caption{\emph{Left}: Image  of coronal loops over the eastern limb of the Sun 
   in the 171 \AA \ pass band ($\sim 1$ MK) taken on November 6, 1999, at 02:30 UT 
   by the TRACE satellite.
   \emph{Right}: We use a simplified straightened out cartesian geometry to model
   a coronal loop. Top and bottom surfaces represent the two photospheric sections at 
   which the loop is anchored. The plasma is embedded in a uniform and homogeneous
   axial magnetic field $\boldsymbol{B_0}$.
     \label{fig:loopbox}  }
\end{figure}

A coronal loop (Figure~\ref{fig:loopbox}) is a closed magnetic structure threaded by a 
strong axial field, with the footpoints rooted in the photosphere.
This makes it a strongly anisotropic system; the measure of 
this anisotropy is given by the relative magnitude of the Alfv\'enic velocity 
$v_{\mathcal A} \sim 1000\ km\, s^{-1}$ compared to the typical photospheric velocity
$u_{ph} \sim 1\ km\, s^{-1}$. So the photospheric velocity, that is the amplitude of the 
Alfv\'en waves that are launched into the corona, is very
small compared to the axial Alfv\'en wave velocity.

To investigate the properties and dynamics of a so complex system such as a coronal
loop, we make a simplified model which captures the essential features of a real loop.
The main (indeed defining) feature of a loop is its strong axial field which serves as a 
guide for the Alfv\'en waves excited by photospheric motions. It is just the dynamics of
these waves, propagating along the guide field, that we want to study.
We first assume a simplified geometry, neglecting any curvature effect, and model a 
coronal loop as a ``straightened out'' cartesian box (Figure~\ref{fig:loopbox}), i.e.\  as 
a parallelepiped with an orthogonal square cross section of size $\ell_{\perp}$, and an 
axial length $L$ embedded in an axial homogeneous uniform magnetic field 
$\boldsymbol{B_0}$. For a quantitative numerical study, we next adopt the so-called
``reduced MHD'' equations to model the dynamics of the plasma 
(Kadomtsev \& Pogutse~\cite{kp74}, Strauss~\cite{stra76} and 
Montgomery~\cite{mont82,mont87}).
Magnetohydrodynamics (MHD) is used to study long-scale low-frequency phenomena
in plasma physics. When a plasma is embedded in a strong magnetic field,
a further simplified set of equations (``reduced'') are derived from the full set of MHD 
equations, to model the dynamics of the plasma.

In \S~\ref{sec:rmhd} reduced MHD equations will be described in detail.
In \S~\ref{sec:amhdt} we give a brief review of current anisotropic MHD 
turbulence theory,
a phenomenon naturally arising in an anisotropic environment such
as a coronal loop
threaded by a strong axial magnetic field. At last, in \S~\ref{sec:nc} we 
describe the 
numerical parallel code that we have developed to solve the reduced MHD 
equations 
and  that we have used to perform our numerical simulations.

\section{Reduced Magnetohydrodynamics} \label{sec:rmhd}

The equations of reduced MHD have been derived in two different research fields. 
Kadomtsev \& Pogutse~\cite{kp74} and Strauss~\cite{stra76} have derived these equations
in the context of fusion research to model the dynamics of  a plasma embedded in a 
strong magnetic field. They were specifically thinking about tokamaks, but their derivation 
is not strictly limited to the geometry of these devices.  
Montgomery~\cite{mont82,mont87} has derived, in a different way, the same set of equations
to study MHD turbulence in the presence of a strong magnetic field. 

The equations derived by all these authors are exactly the same. This is a clear 
indication that when a plasma is embedded in a strong axial magnetic field the turbulence
which naturally develops is strongly affected by the magnetic field, acquiring its anisotropic 
features (Montgomery~\cite{mont82,mont87}, Shebalin~et~al.~\cite{she83}), 
and that the overall dynamics is well described by the equations of reduced MHD 
(Kadomtsev \& Pogutse~\cite{kp74}, Strauss~\cite{stra76}).  
We now derive these equations following Montgomery~\cite{mont87}.

The equations of incompressible resistive MHD are:{\setlength\arraycolsep{-2pt}
\begin{eqnarray}
& &\rho_0 \left( \frac{\partial \boldsymbol{u}}{\partial t} + 
\boldsymbol{u} \cdot \boldsymbol{\nabla} \boldsymbol{u} \right) =
- \boldsymbol{\nabla} p
+ \frac{ \boldsymbol{j} \times \boldsymbol{B} }{c}
+ \nu \boldsymbol{\nabla}^2  \boldsymbol{u},  \label{eq:mu} \\
& & \frac{\partial \boldsymbol{B}}{\partial t} =
\boldsymbol{\nabla} \times \left( \boldsymbol{u} \times \boldsymbol{B} \right)
+ \eta \frac{c^2}{4\pi}   \boldsymbol{\nabla}^2 \boldsymbol{B}, \label{eq:mb} \\
& & \boldsymbol{\nabla} \cdot \boldsymbol{u} = 0, \label{eq:divu} \\
& & \boldsymbol{\nabla}\cdot \boldsymbol{B} = 0, \\
& & \boldsymbol{j} = \frac{c}{4\pi} \boldsymbol{\nabla}\times \boldsymbol{B}, \label{eq:ec}
\end{eqnarray}
}where $\rho_0$ is a constant and uniform mass density, $\boldsymbol{u}$ is the velocity
field, $\boldsymbol{B}$ the magnetic field, $\boldsymbol{j}$ the electric current,
$p$ the kinetic pressure,  $c$ the speed of 
light. Using a simplified diffusion model, both the magnetic resistivity ($\eta$) and the
shear viscosity ($\nu$) are constant and uniform.

We now derive the set of equations for a plasma embedded in a strong magnetic field
directed along the axial direction $z$. Let's suppose the following ordering for
the magnetic field:
\begin{equation}
\boldsymbol{B} \sim \frac{1}{\epsilon} \, \boldsymbol{B}_0 + \boldsymbol{B}^{(0)}
 + \epsilon\, \boldsymbol{B}^{(1)} + \mathcal{O}(\epsilon^2),  \qquad \epsilon \ll 1,
\end{equation}
where
\be
\boldsymbol{B}_0 = B_0 \, \bsy{e}_z.
\ee
We also suppose that gradient scales along the axial field direction are longer than 
perpendicular scales $\ell_{\perp} / \ell_{\parallel} \sim \epsilon$, an hypothesis which is 
actually one of the main results of modern anisotropic turbulence theory  
(see \S~\ref{par:bik}). We introduce similar expansions for velocity, electric current 
and kinetic pressure:{\setlength\arraycolsep{2pt}
\begin{eqnarray}
\boldsymbol{u} & \sim & \boldsymbol{u}^{(0)} + \epsilon\, \boldsymbol{u}^{(1)} + 
\mathcal{O}(\epsilon^2), \\
\boldsymbol{j} & \sim & \boldsymbol{j}^{(0)} + \epsilon\, \boldsymbol{j}^{(1)} + 
\mathcal{O}(\epsilon^2), \\
p & \sim & p^{(0)} + \epsilon\, p^{(1)} + \mathcal{O}(\epsilon^2).
\end{eqnarray}
}Using as variables $t, x, y$, and $\zeta = \epsilon z$, the gradient operator can be
written as
\begin{equation}
\boldsymbol{\nabla} = \boldsymbol{e}_x \partial_x + \boldsymbol{e}_y \partial_y + 
\epsilon\, \boldsymbol{e}_z \partial_{\zeta} = 
\boldsymbol{\nabla}_{\perp} + \epsilon\, \boldsymbol{e}_z \partial_{\zeta}.
\end{equation}
The magnetic field $\bsy B$ is expressed in terms of the vector potential $\boldsymbol A$ :
\begin{equation} \label{eq:bexp}
\boldsymbol{B} = \frac{1}{\epsilon} \, B_0 \boldsymbol{e}_z + 
\boldsymbol{\nabla} \times \boldsymbol{A},
\end{equation}
with
\begin{equation}
\boldsymbol{A} \sim \boldsymbol{A}^{(0)} + \epsilon\, \boldsymbol{A}^{(1)} + 
\mathcal{O}(\epsilon^2),
\end{equation} 
and we choose the Coulomb gauge $\boldsymbol{\nabla} \cdot \boldsymbol{A} = 0$ for its 
algebraic convenience.

Introducing the expansion~(\ref{eq:bexp}) in the equation~(\ref{eq:mb}) for the 
magnetic field, after a few algebraic manipulations  we obtain:
\begin{equation} \label{eq:chi1}
\frac{ \partial \boldsymbol{A} }{\partial t} =
\boldsymbol{u} \times \left( \frac{1}{\epsilon} B_0 \boldsymbol{e}_z + 
\boldsymbol{\nabla} \times \boldsymbol{A} \right)
- \eta c \boldsymbol{j} + \boldsymbol{\nabla} \chi,
\end{equation}
where $\chi$ is  a scalar potential, resulting from pulling off a curl operator from 
eq.~(\ref{eq:mb}), for which we assume an expansion of the form
\begin{equation} \label{eq:chi}
\chi \sim \frac{1}{\epsilon} \, \chi_0 + \chi^{(0)} + \epsilon\, \chi^{(1)} + \mathcal{O}(\epsilon^2).
\end{equation} 
The momentum equation~(\ref{eq:mu}) becomes:
\begin{equation} \label{eq:mu2}
\rho_0 \left( \frac{\partial \boldsymbol{u}}{\partial t} + 
\boldsymbol{u} \cdot \boldsymbol{\nabla} \boldsymbol{u} \right) =
- \boldsymbol{\nabla} p
+ \frac{1}{\epsilon} \, B_0 \, \frac{ \boldsymbol{j} }{c} \times \boldsymbol{e}_z
+ \frac{ \boldsymbol{j} }{c} \times 
\left( \boldsymbol{\nabla} \times \boldsymbol{A} \right)
+ \nu \boldsymbol{\nabla}^2  \boldsymbol{u}.
\end{equation}
Considering the leading order contributions $\mathcal{O}(1/\epsilon)$ from 
(\ref{eq:chi1}), (\ref{eq:chi}), (\ref{eq:mu2}), and $\mathcal{O}(1)$ from 
(\ref{eq:divu}), (\ref{eq:ec}) we have{\setlength\arraycolsep{-2pt}
\begin{eqnarray}
& &\boldsymbol{u}^{(0)} \times \boldsymbol{e}_z\, B_0 = 
- \boldsymbol{\nabla}_{\perp} \chi_0, \label{eq:o1} \label{eq:mnt1} \\
& &\boldsymbol{j}^{(0)} \times \boldsymbol{e}_z\, B_0 =  0, \label{eq:mnt2} \\
& &\boldsymbol{\nabla}_{\perp} \cdot \boldsymbol{u}^{(0)} = 0, \label{eq:mnt3} \\
& &\boldsymbol{j}^{(0)} = \frac{c}{4\pi} \boldsymbol{\nabla}_{\perp} \times \boldsymbol{B}^{(0)} = 
- \frac{c}{4\pi} \boldsymbol{\nabla}_{\perp}^2  \boldsymbol{A}^{(0)}. \label{eq:mnt4}
\end{eqnarray}
}Setting $\chi_0 = B_0 \, \varphi$, (\ref{eq:o1}) gives
\begin{equation} \label{eq:mnt5}
\boldsymbol{u}^{(0)}_{\perp} =  \boldsymbol{\nabla}_{\perp} \varphi \times \boldsymbol{e}_z.
\end{equation}
Equation~(\ref{eq:mnt2}) implies that the $\bsy{j}^{(0)}$ has only
 the component along the axial direction
\be
\bsy{j}^{(0)} =  j_z \,\boldsymbol{e}_z.
\ee
This with equation~(\ref{eq:mnt4}) implies that $\bsy{A}^{(0)}$ also has vanishing orthogonal
components
\be
\boldsymbol{A}^{(0)} =  A_z\, \bsy{e}_z.
\ee
Consequently from~(\ref{eq:bexp}) for $\bsy{B}^{(0)}$ we have
\be
\boldsymbol{B}^{(0)} =   \left( \bnabla_{\perp} A_z \right) \times \bsy{e}_z,
\ee
so that $\bsy{B}^{(0)}$ has a vanishing axial component $B^{(0)}_{z} = 0$.

Considering the $\mathcal{O}(1)$ terms in~(\ref{eq:chi1}) we have
\be \label{eq:rpot}
\frac{ \partial A_z }{\partial t}\, \bsy{e}_z =
\bsy{u}_{\perp}^{(1)} \times  \bsy{e}_z \, B_0  + 
\bsy{u}_{\perp}^{(0)} \times  \left[ \left( \bnabla_{\perp} A_z \right) \times \bsy{e}_z \right]
- \eta c \, j_z \, \bsy{e}_z + \bsy{e}_z \, B_0 \, \frac{\de \varphi}{\de \zeta} 
+ \bnabla_{\perp} \chi^{(0)}.
\ee
Introducing the vorticity $\boldsymbol{\omega} \equiv \bnabla \times \bsy{u}$, and following the above
expansion conventions, we have
\be
\boldsymbol{\omega}^{(0)} = \bnabla_{\perp} \times \bsy{u}^{(0)}_{\perp} = 
- \bsy{e}_z \, \bnabla_{\perp}^2 \varphi = \omega \, \bsy{e}_z.
\ee
After some algebra the $\mathcal{O}(1)$ terms of equation~(\ref{eq:mu2})
can be written as
\be \label{eq:rv}
\rho_0 \, \left[ \frac{\de \omega}{\de t} 
+ \left( \bsy{u}^{(0)}_{\perp} \cdot \bnabla_{\perp} \right) \omega \right] \bsy{e}_z =
\frac{1}{c} B_0 \frac{\de j_z}{\de \zeta} \bsy{e}_z + 
\frac{1}{c} \left( \bsy{B}^{(0)}_{\perp} \cdot \bnabla_{\perp} \right) j_z \, \bsy{e}_z \\
 + \nu \bnabla^2_{\perp} \omega \, \bsy{e}_z
\ee
Finally, considering the $z$-components of equations~(\ref{eq:rpot}) and (\ref{eq:rv}), we obtain
the reduced MHD equations:{\setlength\arraycolsep{-2pt}
\begin{eqnarray}
& &\frac{\partial A_z}{\partial t} + 
\left( \boldsymbol{u}_{\perp}^{(0)} \cdot \boldsymbol{\nabla}_{\perp} \right) A_z =
B_0 \, \frac{\partial \varphi}{\partial \zeta} 
+ \eta \, \frac{c^2}{4\pi}\, \boldsymbol{\nabla}_{\perp}^2 A_z \label{eq:rmhd1} \\
& &\rho_0 \left[ \frac{\partial \omega}{\partial t} + 
\left( \boldsymbol{u}_{\perp}^{(0)} \cdot \boldsymbol{\nabla}_{\perp} \right) \omega \right] =
\frac{1}{c} \left( \boldsymbol{B}^{(0)}_{\perp} \cdot \boldsymbol{\nabla}_{\perp} \right) j_z 
+ \frac{1}{c}  B_0 \, \frac{\partial j_z}{\partial \zeta} 
+ \nu \, \boldsymbol{\nabla}_{\perp}^2 \omega \label{eq:rmhd2}
\end{eqnarray}}

Finally,  from $\mathcal{O}(\epsilon)$ of equation~(\ref{eq:mnt3}) we have
\be \label{eq:rpf}
\bnabla_{\perp} \cdot \bsy{u}^{(1)}_{\perp} 
+ \frac{\de u^{(0)}_z}{\de \zeta} = 0,
\ee
and considering the orthogonal component of equation~(\ref{eq:rpot})
\be
\bsy{u}^{(1)}_{\perp} \times \bsy{e}_z B_0 = - \bnabla_{\perp} \chi^{(0)}, \qquad
\textrm{i.e.} \qquad
\bsy{u}^{(1)}_{\perp} = \bnabla_{\perp} \left( \frac{\chi^{(0)}}{B_0} \right) \times \bsy{e}_z.
\ee
From this last equation it follows $\bnabla_{\perp} \cdot \bsy{u}^{(1)}_{\perp} = 0$, 
which inserted in (\ref{eq:rpf}) implies that, in the absence of a parallel flow at the
boundaries, we have
\be \label{eq:vacn}
u_z^{(0)} = 0.
\ee

\subsection{Dimensionless Form and Boundary Conditions} \label{sec:dfbc}

The equations of reduced MHD~(\ref{eq:rmhd1})-(\ref{eq:rmhd2}) written for the 
transverse fields $\boldsymbol{u}_{\perp}$
and $\boldsymbol{b}_{\perp}$ in dimensional form are:{\setlength\arraycolsep{-7pt}
\begin{eqnarray}
& &\rho_0 \left( \frac{\partial \boldsymbol u_{\perp}}{\partial t} + \left(  \boldsymbol u_{\perp} \cdot \boldsymbol{\nabla}_{\perp} \right) \boldsymbol{u_{\perp}} \right) = - \boldsymbol{\nabla}_{\perp} 
\left( p + \frac{\boldsymbol b_{\perp}^2}{8\pi} \right) +   
\frac{ \left(  \boldsymbol b_{\perp} \cdot \boldsymbol{\nabla}_{\perp} \right) \boldsymbol{b_{\perp}} }{4\pi} + \frac{B_0}{4\pi}
\frac{\partial \boldsymbol b_{\perp}}{\partial z} + \nu
\boldsymbol{\nabla}^2_{\perp} \boldsymbol u_{\perp} \\
& &\frac{\partial \boldsymbol b_{\perp}}{\partial t} =  \left(  \boldsymbol b_{\perp} \cdot \boldsymbol{\nabla}_{\perp} \right) \boldsymbol{u_{\perp}} -  \left(  \boldsymbol u_{\perp} \cdot \boldsymbol{\nabla}_{\perp} \right) \boldsymbol{b _{\perp}} + B_0 \frac{\partial \boldsymbol u_{\perp}}{\partial z} + \eta \frac{c^2}{4\pi} \boldsymbol{\nabla}^2_{\perp} 
\boldsymbol b_{\perp}\\
& &\boldsymbol{\nabla}_{\perp} \cdot \boldsymbol u_{\perp} = 0\\
& &\boldsymbol{\nabla}_{\perp} \cdot \boldsymbol b_{\perp} = 0
\end{eqnarray}
}where $\rho_0$ is a constant and uniform mass density. To render the previous
equations in dimensionless form, we notice that the magnetic
fields $\boldsymbol{b}_{\perp}$ and $B_0$ can be expressed in velocity dimensions
by dividing by $\sqrt{4\pi \rho_0}$, i.e.\ considering the Alfv\'enic velocities. 
As characteristic quantities we use the perpendicular length of the computational
box $\ell_{\perp}$, the typical photospheric velocity $u_{ph}$, and the related
crossing time $t_{\perp} = \ell_{\perp} / u_{ph}$.
The dimensionless equations are then given by:{\setlength\arraycolsep{-6pt}
\begin{eqnarray}
& &\frac{\partial \boldsymbol u_{\perp}}{\partial t} + \left(  \boldsymbol u_{\perp} \cdot \boldsymbol{\nabla}_{\perp} \right) = - \boldsymbol{\nabla}_{\perp} \left( p + \frac{\boldsymbol b_{\perp}^2}{2} \right) +   \left(  \boldsymbol b_{\perp} \cdot \boldsymbol{\nabla}_{\perp} \right) \boldsymbol{b_{\perp}} + v_{\mathcal{A}}\, \frac{\partial \boldsymbol b_{\perp}}{\partial z} + 
\frac{1}{\mathcal R} \boldsymbol{\nabla}^2_{\perp} \boldsymbol u_{\perp} \label{eq:adim1} \\
& &\frac{\partial \boldsymbol b_{\perp}}{\partial t} =  \left(  \boldsymbol b_{\perp} \cdot \boldsymbol{\nabla}_{\perp} \right) \boldsymbol{u_{\perp}} -  \left(  \boldsymbol u_{\perp} \cdot \boldsymbol{\nabla}_{\perp} \right) \boldsymbol{b _{\perp}} + v_{\mathcal{A}}\, 
 \frac{\partial \boldsymbol u_{\perp}}{\partial z} + \frac{1}{\mathcal R_m} \boldsymbol{\nabla}^2_{\perp} 
\boldsymbol b_{\perp} \label{eq:adim2} \\
& &\boldsymbol{\nabla}_{\perp} \cdot \boldsymbol u_{\perp} = 0 \label{eq:adim3}\\
& &\boldsymbol{\nabla}_{\perp} \cdot \boldsymbol b_{\perp} = 0 \label{eq:adim4}
\end{eqnarray}
}where $v_{\mathcal A}$ is the ratio between the axial Alfv\'enic velocity and
the photospheric velocity, i.e.
\begin{equation}
v_{\mathcal A} = \frac{B_0}{\sqrt{4 \pi \rho_0}} \cdot \frac{1}{u_{ph}},
\end{equation}
and 
\begin{equation} \label{r1}
\frac{1}{\mathcal R} = \frac{\nu}{\rho_0\, l_{\perp} u_{ph}}, \qquad
\frac{1}{\mathcal R_m} = \frac{\eta c^2}{4\pi \rho_0 \, l_{\perp} u_{ph} }
\end{equation}
are the kinetic and magnetic Reynolds numbers.

Introducing the velocity and magnetic potentials $\varphi$ and $\psi$
\begin{equation} \label{eq:intpot}
\boldsymbol u_\perp = \boldsymbol{\nabla} \times \left( \varphi\, \boldsymbol{e}_z \right)  \qquad 
\boldsymbol b_\perp = \boldsymbol{\nabla} \times \left( \psi\, \boldsymbol{e}_z \right) 
\end{equation}
\begin{equation}
\omega = \left( \boldsymbol{\nabla} \times \boldsymbol u_\perp \right)_3 = - \boldsymbol{\nabla}^2_\perp \varphi \ 
\end{equation}
\begin{equation}
j = \left( \boldsymbol{\nabla} \times \boldsymbol b_\perp \right)_3 = - \boldsymbol{\nabla}^2_\perp \psi \ 
\end{equation}
equations~(\ref{eq:adim1})--(\ref{eq:adim2}) in terms of potentials are written as:
\begin{eqnarray}
& &\frac{\partial \psi}{\partial t} = v_{\mathcal A}\, \frac{\partial \varphi}{\partial z} + 
\left[ \varphi, \psi \right] + \frac{1}{\mathcal R_m} \boldsymbol{\nabla}^2_\perp 
\psi,  \label{pot1} \\
& &\frac{\partial \omega}{\partial t} = v_{\mathcal A}\, \frac{\partial j}{\partial z} + 
\left[ j , \psi \right] - \left[ \omega , \varphi \right] 
+  \frac{1}{\mathcal R} \boldsymbol{\nabla}^2_\perp \omega, \label{pot2} 
\end{eqnarray}
where the poisson bracket of two functions $f$ and $g$ is defined as:
\begin{equation}
\left[ f, g \right] = \frac{\partial f}{\partial x}\, \frac{\partial g}{\partial y} -
\frac{\partial g}{\partial x}\, \frac{\partial f}{\partial y}.
\end{equation}
Using equations~(\ref{eq:intpot}) it can be shown that:
\be
\left[ \varphi, \psi \right] = - \left( \bsy{u}_{\perp} \cdot \bnabla_{\perp} \right) \psi, \qquad
\left[ j, \psi \right] =  \left( \bsy{b}_{\perp} \cdot \bnabla_{\perp} \right) j, \qquad
\left[ \omega, \varphi \right] = \left( \bsy{u}_{\perp} \cdot \bnabla_{\perp} \right) \omega.
\ee
Using these relations, equations~(\ref{pot1})-(\ref{pot2}) can be rewritten as:
\begin{eqnarray}
& &\frac{\partial \psi}{\partial t} + \left( \bsy{u}_{\perp} \cdot \bnabla_{\perp} \right) \psi =
v_{\mathcal A}\, \frac{\partial \varphi}{\partial z} 
+ \frac{1}{\mathcal R_m} \boldsymbol{\nabla}^2_\perp 
\psi,  \label{pot3} \\
& &\frac{\partial \omega}{\partial t} + \left( \bsy{u}_{\perp} \cdot \bnabla_{\perp} \right) \omega = 
\left( \bsy{b}_{\perp} \cdot \bnabla_{\perp} \right) j
+ v_{\mathcal A}\, \frac{\partial j}{\partial z} 
+  \frac{1}{\mathcal R} \boldsymbol{\nabla}^2_\perp \omega, \label{pot4}
\end{eqnarray}
which are the dimensionless version of equations~(\ref{eq:rmhd1})-(\ref{eq:rmhd2}).

In turbulence theory the fundamental variables are the Els\"asser variables
\begin{equation} \label{eq:elsv}
\boldsymbol{z}^{\pm} = \boldsymbol{u}_{\perp} \pm \boldsymbol{b}_{\perp},
\end{equation}
in terms of which, and supposing the Reynolds numbers to have the same values 
$\mathcal{R}_m = \mathcal{R}$, the reduced MHD 
equations~(\ref{eq:adim1})-(\ref{eq:adim4}) are:{\setlength\arraycolsep{-2pt}
\begin{eqnarray}
& &\frac{\partial \boldsymbol z^+}{\partial t}  =
- \left(  \boldsymbol z^- \cdot \boldsymbol{\nabla}_{\perp} \right) \boldsymbol{z}^+ 
+ v_\mathcal{A} \frac{\partial \boldsymbol z^+}{\partial z}
+ \frac{1}{\mathcal R}   \boldsymbol{\nabla}^2_{\perp} \boldsymbol z^+ 
 - \boldsymbol{\nabla}_{\perp} P \label{eq:els1} \\
& &\frac{\partial \boldsymbol z^-}{\partial t}  =
- \left(  \boldsymbol z^+ \cdot \boldsymbol{\nabla}_{\perp} \right) \boldsymbol{z}^- 
- v_\mathcal{A} \frac{\partial \boldsymbol z^-}{\partial z}
+ \frac{1}{\mathcal R}   \boldsymbol{\nabla}^2_{\perp} \boldsymbol z^- 
 - \boldsymbol{\nabla}_{\perp} P \label{eq:els2} \\
& &\boldsymbol{\nabla}_{\perp} \cdot \boldsymbol{z}^{\pm} = 0 \label{eq:els3}
\end{eqnarray}
}where $P = p + \boldsymbol{b}_{\perp}^2/2 $ is the total pressure, and is
linked to the nonlinear terms by incompressibility~(\ref{eq:els3}):
\begin{equation}  \label{eq:els4}
\boldsymbol{\nabla}_{\perp}^2 P = 
- \sum_{i,j=1}^2 \Big( \partial_i z_j^- \Big) \Big( \partial_j z_i^+ \Big)
\end{equation}

An analysis of equations~(\ref{eq:els1})-(\ref{eq:els3}) gives us a qualitative preview
of the results which will be obtained both numerically and analytically in the following
chapters.  The linear terms in equations~(\ref{eq:els1})-(\ref{eq:els3})
\begin{equation}
\frac{\partial \boldsymbol z^{\pm}}{\partial t}  =
\pm v_\mathcal{A} \frac{\partial \boldsymbol z^{\pm}}{\partial z}
\end{equation}
show that $\boldsymbol{z}^{\pm}$ fields present an Alfv\'en wave propagation along
the axial field direction. In particular $\boldsymbol{z}^-$ describes waves propagating
in the direction of $\boldsymbol{B}_0$, and  $\boldsymbol{z}^+$ in the opposite 
direction; both at the Alfv\'en wave velocity $v_{\mathcal A}$. This wave propagation
is present also when the nonlinear terms become important, and transport energy
from the photospheric boundaries into the system.

As boundary conditions at the photospheric surfaces ($z=0,L$) we impose a velocity
pattern which mimics  photospheric motions. In terms of the Els\"asser variables 
the velocity is the sum 
\begin{equation}
\boldsymbol{z}^{+} + \boldsymbol{z}^{-} = 2\, \boldsymbol{u}_{\perp}.
\end{equation} 
In terms of characteristics this gives rise to a ``reflection'' of the Alfv\'en waves at the 
boundaries, where we can only impose a condition on the incoming wave (alternately 
$\boldsymbol{z}^{+}$ and $\boldsymbol{z}^{-}$). 
To mimic photospheric motions we impose a velocity pattern on the top and bottom
planes. In terms of the Els\"asser variables to impose a velocity pattern
means using the constraint
\begin{equation}
\boldsymbol{z^{+}} + \boldsymbol{z^{-}} = 2 \, \boldsymbol{u}_{ph} 
\end{equation}
Since  $\bsy{z}^+$ and $\bsy{z}^-$ are, respectively, waves propagating toward the 
inside and the outside of the computational box, this is  a ``reflection" condition on 
these waves, i.e.
\begin{equation} \label{eq:bc0}
\boldsymbol{z^{-}}  = - \boldsymbol{z^{+}} + 2 \, \boldsymbol{u}_{0} 
\quad \textrm{at} \ z=0 
\end{equation}
\begin{equation} \label{eq:bcL}
\boldsymbol{z^{+}} = - \boldsymbol{z^{-}} + 2 \, \boldsymbol{u}_{L}  
\quad \textrm{at} \ z=L 
\end{equation}
At the boundary the value of the incoming wave is equal to the negative value of  the 
outgoing wave plus twice the value of the velocity at the photosphere.

A fundamental feature of the nonlinear terms 
$\left(  \boldsymbol z^{\mp} \cdot \boldsymbol{\nabla}_{\perp} \right) \boldsymbol{z}^{\pm}$
(and also the pressure term~(\ref{eq:els4})) is the absence of self-coupling, i.e.\ the nonlinear
interaction depends by the counter-propagating fields $\boldsymbol z^{\pm}$, and if 
one of the two fields were zero there would be no nonlinear dynamics at all.
This is the basis of the so-called Alfv\'en effect, which is described in \S~\ref{sec:ae}, and is 
the basis of anisotropic turbulence phenomenology.

\subsection{Conservation Laws} \label{sec:cl}

MHD theory provides a number of conservation laws which play an important role in
turbulence theory. We are mainly interested in the Energy, Cross Helicity, and Magnetic
Helicity. Turbulence is usually studied with the hypothesis of periodicity in all three spatial
directions. In our case along the direction of the axial field this condition breaks down.
The flux terms which are usually neglected become important. In this paragraph we
write the conservation laws for the three aforementioned quantities, including the 
flux terms and restricting our attention to the reduced MHD equations.

Multiplying the momentum equation~(\ref{eq:adim1}) by $\bsy u_{\perp}$ and the
magnetic field equation~(\ref{eq:adim2}) by $\bsy b_{\perp}$ 
we obtain:{\setlength\arraycolsep{2pt}
\begin{eqnarray}
& &\frac{\de }{\de t} \left( \frac{1}{2}  \bsy u_{\perp}^2 \right) = 
\bsy u_{\perp} \cdot \bigg[ -  \left(  \bsy u_{\perp} \cdot \bnabla \right) 
\bsy u_{\perp} - \bnabla_{\perp} \left( p + \frac{\bsy b_{\perp}^2}{2} \right) 
+   \left(  \bsy b_{\perp} \cdot \bnabla \right) \bsy{b_{\perp}} + \nonumber\\
& & \qquad \qquad \qquad \qquad \qquad \qquad \qquad \qquad
\qquad \qquad \qquad \quad v_{\mathcal A} \frac{\de \bsy b_{\perp}}{\de z} 
+ \frac{1}{\mathcal R} \bnabla^2_{\perp} \bsy u_{\perp} \bigg] \label{eq:en1} \\
& &\frac{\de }{\de t} \left( \frac{1}{2}  \bsy b_{\perp}^2 \right) = 
\bsy b_{\perp} \cdot \left[   \left(  \bsy b_{\perp} \cdot \bnabla \right) \bsy{u_{\perp}} 
-  \left(  \bsy u_{\perp} \cdot \bnabla \right) \bsy{b _{\perp}} 
+ v_{\mathcal A} \frac{\de \bsy u_{\perp}}{\de z} 
+ \frac{1}{\mathcal R_m} \bnabla^2_{\perp} \bsy b_{\perp} \right] \qquad \label{eq:en2}
\end{eqnarray}
}On the other hand from
\be \label{eq:dec}
\bsy B = v_{\mathcal A}\, \bsy{e}_z + \bsy b_{\perp}, \qquad 
\bsy u = \bsy u_{\perp}, \qquad \bnabla \cdot \bsy B = 
\bnabla \cdot \bsy b_{\perp} = \bnabla \cdot \bsy u_{\perp}  = 0,
\ee
it follows that{\setlength\arraycolsep{-2pt}
\begin{eqnarray}
& &\left( \bnabla \times \bsy B \right) \times \bsy B =
- \frac{1}{2} \bnabla \bsy B^2 + \left( \bsy B \cdot \bnabla \right) \bsy B 
= - \frac{1}{2} \bnabla \bsy b_{\perp}^2
+ v_{\mathcal A} \frac{\de \bsy b_{\perp}}{\de z}
+ \left( \bsy b_{\perp} \cdot \bnabla \right) \bsy b_{\perp} \\
& &\left( \bnabla \times \bsy u_{\perp} \right) \times \bsy u_{\perp} =
- \frac{1}{2} \bnabla \bsy u_{\perp}^2 
+ \left( \bsy u_{\perp} \cdot \bnabla \right) \bsy u_{\perp} \\
& &\bnabla \times \left( \bsy u_{\perp} \times \bsy B \right) =
\left( \bsy B \cdot \bnabla \right) \bsy u_{\perp} -
\left( \bsy u_{\perp} \cdot \bnabla \right) \bsy B =
v_{\mathcal A} \frac{\de \bsy u_{\perp}}{\de z} + \nonumber\\
& & \qquad \qquad \qquad \qquad \qquad \qquad \qquad \qquad \qquad 
 \qquad  + \left( \bsy b_{\perp} \cdot \bnabla \right) \bsy u_{\perp}
- \left( \bsy u_{\perp} \cdot \bnabla \right) \bsy b_{\perp}
\end{eqnarray}
}In this way equations~(\ref{eq:en1})-(\ref{eq:en2}) can be 
rewritten as{\setlength\arraycolsep{-2pt}
\begin{eqnarray}
& &\frac{\de }{\de t} \left( \frac{1}{2}  \bsy u_{\perp}^2 \right) = 
\bsy u_{\perp} \cdot \bigg[ 
-  \left(  \bnabla \times \bsy u_{\perp} \right) \times \bsy u_{\perp} 
- \bnabla_{\perp} \left(  p  +\frac{1}{2} \bsy u_{\perp}^2 \right) 
+  \left(  \bnabla \times \bsy B \right) \times \bsy B + \nonumber\\
& & \qquad \qquad \qquad \qquad \qquad \qquad \qquad \qquad \qquad 
\qquad \qquad \qquad  \qquad \qquad \qquad 
\frac{1}{\mathcal R} \bnabla^2_{\perp} \bsy u_{\perp} \bigg] \label{en3} \\
& &\frac{\de }{\de t} \left( \frac{1}{2}  \bsy b_{\perp}^2 \right) = 
\bsy b_{\perp} \cdot \left[   
\bnabla \times  \left(  \bsy u_{\perp}  \times \bsy B \right) + 
\frac{1}{\mathcal R_m} \bnabla^2_{\perp} \bsy b_{\perp} \right] \label{en4}
\end{eqnarray}
}The first term between square brackets in the right hand side of
equation~(\ref{en3}) is orthogonal to $\bsy u_{\perp}$ and hence its 
scalar product with $\bsy u_{\perp}$ is zero. Furthermore
\be
\bnabla \cdot \left[ \bsy B \times \left( \bsy u_{\perp} \times \bsy B \right) \right]
= \left( \bsy u_{\perp} \times \bsy B \right) 
\cdot \left( \bnabla \times \bsy B \right)
- \bsy B \cdot \left[ \bnabla \times  \left(  \bsy u_{\perp}  \times \bsy B \right) \right]
\ee
and
\be
\left( \bsy u_{\perp} \times \bsy B \right) 
\cdot \left( \bnabla \times \bsy B \right)
= - \bsy u_{\perp} \cdot \left[\left( \bnabla \times \bsy B \right) \times \bsy B \right]
\ee
\be
- \bsy B \cdot \left[ \bnabla \times  \left(  \bsy u_{\perp}  \times \bsy B \right) \right]
= - \bsy b_{\perp} 
\cdot \left[ \bnabla \times  \left(  \bsy u_{\perp}  \times \bsy B \right) \right]
\ee
We can now write for the energy density, summing 
equations~(\ref{en3}) and~(\ref{en4})
\begin{multline} \label{en5}
\frac{\de }{\de t} \left( \frac{1}{2}  \bsy u_{\perp}^2  
+ \frac{1}{2}  \bsy b_{\perp}^2   \right) = 
- \bnabla \cdot \left[ \bsy B \times \left( \bsy u_{\perp} \times \bsy B \right) \right]
-\bsy u_{\perp} \cdot  \bnabla_{\perp} \left(  p  +\frac{1}{2} \bsy u_{\perp}^2 \right) + \\
   \frac{1}{\mathcal R} \bsy u_{\perp} \cdot \bnabla^2_{\perp} \bsy u_{\perp} 
+ \frac{1}{\mathcal R} \bsy b_{\perp} \cdot \bnabla^2_{\perp} \bsy b_{\perp}
\end{multline}
The following relations hold:
\be
\bnabla \cdot \left[ \left( p +\frac{1}{2} \bsy u_{\perp}^2 \right) \bsy u_{\perp} \right] 
= \bsy u_{\perp} \cdot  \bnabla_{\perp} 
\left(  p  +\frac{1}{2} \bsy u_{\perp}^2 \right)
\ee
\be
\bsy u_{\perp} \cdot \bnabla^2_{\perp} \bsy u_{\perp} \sim 
\bsy u_{\perp} \cdot \bnabla^2 \bsy u_{\perp}
= \bnabla \cdot  \left( \bsy u_{\perp} \times \boldsymbol{\omega} \right) - \boldsymbol{\omega}^2
\ee
\be
\bsy b_{\perp} \cdot \bnabla^2_{\perp} \bsy b_{\perp} \sim 
\bsy b_{\perp} \cdot \bnabla^2 \bsy b_{\perp}
= \bnabla \cdot  \left( \bsy b_{\perp} \times \boldsymbol{j} \right) - \boldsymbol{j}^2
\ee
where
\be
\boldsymbol{\omega} = \bnabla \times \bsy u_{\perp} 
\qquad
\boldsymbol{j} = \bnabla \times \bsy b_{\perp} 
\ee
We can now write
\begin{multline} \label{en6}
\frac{\de }{\de t} \left( \frac{1}{2}  \bsy u_{\perp}^2  
+ \frac{1}{2}  \bsy b_{\perp}^2   \right)  
+ \bnabla \cdot \left[ \bsy B \times \left( \bsy u_{\perp} \times \bsy B \right) 
+  \left( p +\frac{1}{2} \bsy u_{\perp}^2 \right) \bsy u_{\perp}
- \frac{1}{\mathcal R} \left( \bsy u_{\perp} \times \boldsymbol{\omega} 
+ \bsy b_{\perp} \times \boldsymbol{j} \right) \right]  \\ 
= - \frac{1}{\mathcal R} \left( \boldsymbol{\omega}^2 + \boldsymbol{j}^2 \right)
\end{multline}
The Poynting Flux $\bsy S$ is given by:
\be \label{eq:pf}
\bsy S = \bsy B \times \left( \bsy u \times \bsy B \right) = {\bsy{B}}^2 \bsy u
               - \left( \bsy B \cdot \bsy u \right) \bsy B 
\ee
So using equation~(\ref{eq:dec}) we  have
\be               
\bsy S = \left( v_{\mathcal A}^2 + \bsy b_{\perp}^2 \right) \bsy u_{\perp}
               - \left( \bsy b_{\perp} \cdot \bsy u_{\perp} \right) \left( v_{\mathcal A}\, \bsy{e}_ z +
               \bsy b_{\perp} \right)
\ee
and the component along the axial direction is
\be
S_z = \bsy S \cdot \bsy{e}_ z = - v_{\mathcal A} \left( \bsy b_{\perp} \cdot \bsy u_{\perp} \right) 
\ee
At last we can write
\be \label{eq:poy}
\frac{\de }{\de t} \left( \frac{1}{2}  \bsy u_{\perp}^2  
+ \frac{1}{2}  \bsy b_{\perp}^2   \right)  
+ \bnabla \cdot \bsy{S}
= - \frac{1}{\mathcal R} \left( \boldsymbol{\omega}^2 + \boldsymbol{j}^2 \right)
\ee
Calling the total Energy $E$, the Ohmic dissipation rate $J$, and the viscous
dissipation rate  $\Omega$ (which is the \emph{enstrophy} divided by the
Reynolds number $\mathcal R$){\setlength\arraycolsep{2pt}
\begin{eqnarray}
E            & \equiv &  \frac{1}{2}  \int_V \! \ud^{3}x 
               \left( \bsy u_{\perp}^2 + \bsy b_{\perp}^2   \right), \label{eq:endef} \\
J             & \equiv &  \frac{1}{\mathcal R}  \int_V \! \ud^{3}x\, \bsy{j}^2,  
               \label{eq:ohmdef} \\
\Omega & \equiv &  \frac{1}{\mathcal R}  \int_V \! \ud^{3}x\, \bsy{\omega}^2, 
              \label{eq:visdef}
\end{eqnarray}
}and $S$ the integral of the Poynting flux given by
\be \label{eq:pfint}
S = - \int_V \! \ud^{3}x\,  \bnabla \cdot \bsy{S} = \oint  \! \ud a\, \bsy{S} \cdot \bsy{n} = 
+ v_{\mathcal A} \int_{z=L} \! \ud a\,  \left( \bsy u_{\perp} \cdot \bsy b_{\perp} \right)
-  v_{\mathcal A} \int_{z=0} \! \ud a\,  \left( \bsy u_{\perp} \cdot \bsy b_{\perp} \right),
\ee
where the signs have been chosen so that $S$ is positive when we have energy
entering the system and negative when it leaves.
We can write the integral of equation~(\ref{eq:poy})
\be
\frac{\partial E}{\partial t} = S - \left( \Omega + J \right)
\ee

Another important conserved quantity is cross helicity $H^C$, defined as:
\be
H^C \equiv \int_V \! \ud^{3}x\ \bsy{u}_{\perp} \cdot \bsy{b}_{\perp}
\ee
Multiplying the momentum equation~(\ref{eq:adim1}) by $\bsy{b}_{\perp}$ and the 
magnetic field equation~(\ref{eq:adim2}) by $\bsy{u}_{\perp}$, after a few algebraic
manipulations we obtain 
\begin{multline}
\frac{\partial}{\partial t} \left( \bsy{u}_{\perp} \cdot \bsy{b}_{\perp} \right) =
- \bnabla \cdot \bigg[ 
\bsy{u}_{\perp} \times \left( \bsy{u}_{\perp} \times \bsy{b}_{\perp} \right) 
+ \bigg( p + \frac{1}{2} \bsy{u}_{\perp}^2 \bigg) \bsy{b}_{\perp}
+ \frac{1}{2} \bigg(  \bsy{u}_{\perp}^2 + \bsy{b}_{\perp}^2 \bigg) 
v_{\mathcal A}\, \bsy{e}_{z} \bigg] \\
+ \frac{1}{\mathcal R} \bigg( \bsy{b}_{\perp} \cdot \bnabla_{\perp}^2 \bsy{u}_{\perp}
+ \bsy{u}_{\perp} \cdot \bnabla_{\perp}^2 \bsy{b}_{\perp} \bigg)
\end{multline}
When taking the integral over the volume the first and second terms in square brackets do not
contribute because their only components lie in the orthogonal plane where we have periodic
boundary conditions. But the third term cannot be neglected and we thus obtain
\begin{multline}
\frac{\partial}{\partial t} \int_V \! \ud^{3}x\  \bsy{u}_{\perp} \cdot \bsy{b}_{\perp} =
\frac{v_{\mathcal A}}{2} \int \! \ud a\ 
\bigg[ \left(   \bsy{u}_{\perp}^2 +  \bsy{b}_{\perp}^2 \right) (x,y,z=L)
          -\left(   \bsy{u}_{\perp}^2 +  \bsy{b}_{\perp}^2 \right) (x,y,z=0) \bigg] \\
+ \frac{1}{\mathcal R}  \int_V \! \ud^{3}x\  \bigg( \bsy{b}_{\perp} \cdot \bnabla_{\perp}^2 \bsy{u}_{\perp}
+ \bsy{u}_{\perp} \cdot \bnabla_{\perp}^2 \bsy{b}_{\perp} \bigg)
\end{multline}

Given a magnetic field $\bsy B$ and its vector potential $\bsy A$ for which
$\bsy B = \bnabla \times \bsy A$, in MHD the magnetic helicity $H^M$ is usually defined as
\be \label{eq:hm}
H^M = \int_V  \! \ud^{3}x\ \bsy{A} \cdot \bsy{B},
\ee
but this definition is, in general, not gauge invariant. In fact taking a gauge transformation
$\bsy A' = \bsy A + \bnabla \chi$ gives
\be \label{eq:hm1}
H^{M'} - H^M = \int_V  \! \ud^{3}x\ \bsy{B} \cdot \bnabla \chi = 
\oint_S \! \ud a\, \chi\, \bsy{B} \cdot \bsy{\hat{n}}
\ee
Now equation~(\ref{eq:hm}) is gauge invariant, but only when the surface integral~(\ref{eq:hm1})
vanishes. This condition is satisfied when the normal component of the magnetic field vanishes
at the boundary surface. For our coronal loop this condition does not apply, and 
definition~(\ref{eq:hm}) cannot be used. 
An alternative expression has been proposed by Finn and Antonsen~\cite{finn85} (see also
Berger and Field~\cite{berg84})
\be \label{eq:hmalt}
H^M_{alt} = \int_V  \! \ud^{3}x\ \left( \bsy{A} + \bsy{A}_0 \right) \cdot \left(  \bsy{B} - \bsy{B}_0 \right)
\ee
where $\bsy{B}_0 = \bnabla \times \bsy{A}_0$ is a reference field to be chosen suitably.
To satisfy gauge invariance $\bsy{B}$ and $\bsy{B}_0$ should have the same normal component
at the surface boundary.

In the reduced MHD case we show that, even if it were possible to give a gauge
invariant definition of the magnetic helicity, it would not have a physical meaning.
In reduced MHD the field is decomposed as $\bsy{B} = \bsy{B}_0 + \bsy{b}_{\perp}$, 
where $\bsy{B}_0 = B_0\, \bsy{e}_z$ is constant and uniform.
We choose the vector potential $\bsy{A}_0 = B_0\, x\, \bsy{e}_y$ and
$\bsy{b}_{\perp} = \bnabla \times \bsy{A}_{\perp} = \bnabla \times ( \psi\, \bsy{e}_z )$.
We have only 4 terms which may contribute to an expression like~(\ref{eq:hmalt}),
but only one of them is not null. In fact
\begin{eqnarray}
& &\bsy{A}_0 \cdot \bsy{B}_0 = B_0\, x\, \bsy{e}_y \cdot B_0\, \bsy{e}_z = 0 \\
& &\bsy{A}_0 \cdot \bsy{b}_{\perp} =  B_0\, x\, \bsy{e}_y \cdot \bsy{b}_{\perp} = 
B_0\, x\, b_y \label{eq:hm2}\\
& &\bsy{A}_{\perp} \cdot \bsy{b}_{\perp} = \psi\, \bsy{e}_z \cdot \bsy{b}_{\perp} = 0 \\
& &\bsy{A}_{\perp} \cdot \bsy{B}_0 = B_0\, \psi \label{eq:hm4} 
\end{eqnarray}
Integrating over the volume equation~(\ref{eq:hm2}) vanishes because of the 
periodic boundary conditions of $b_y$, so only equation~(\ref{eq:hm4}) 
does not vanish. On the other hand, when integrating over the volume, 
this term is the zero frequency component of $\psi$, i.e.\ a constant which can
always be subtracted through a gauge transformation.

In full ideal MHD magnetic helicity is a conserved quantity, but in reduced
MHD it does not seem to have a physical meaning. While 
in 2D MHD magnetic helicity is zero but any moment of $\psi$ is conserved,
in reduced MHD this does not apply. We rewrite for convenience 
equation~(\ref{pot1}) with $[\varphi, \psi] = - \bnabla \cdot ( \psi\, \bsy{u}_{\perp} )$:
\be
\frac{\partial \psi}{\partial t} = 
- \bnabla \cdot \left( \psi\, \bsy{u}_{\perp} \right)+ 
v_{\mathcal A}\, \frac{\partial \varphi}{\partial z} + 
\frac{1}{\mathcal R_m} \boldsymbol{\nabla}^2_\perp \psi.
\ee
When the $z$ derivative term is zero, this equation becomes the 2D MHD
equation, and it easily follows that any moment of $\psi$ is conserved
when $\mathcal{R}_m \rightarrow \infty$. In the reduced MHD case the
divergence gives a vanishing contribute (as it does for 2D MHD):
\be
\int_V  \! \ud^{3}x\ \bnabla \cdot \left( \psi\, \bsy{u}_{\perp} \right) =
\oint_S  \! \ud a\  \psi\, \bsy{u}_{\perp} \cdot \bsy{\hat{n}}
\ee
because the normal component of the velocity is zero at the photospheric
surfaces, and the remaining boundary surfaces are periodic.
The presence of the $z$ derivative term breaks the conservation of 
$\psi^{\alpha}$, except for $\alpha = 1$, but in this case it is again
a function which can be subtracted from $\psi$ through the gauge
transformation
\be
\psi' = \psi - v_{\mathcal A} \int \! \ud t\ \frac{\partial \varphi}{\partial z} 
\ee
so that for $\mathcal{R}_m \rightarrow \infty$
\be
\int_V  \! \ud^{3}x\ \psi = \textrm{const}
\ee

\section{Anisotropic MHD Turbulence} \label{sec:amhdt}

Wherever a fluid is set into motion turbulence tends to develop.
When the fluid is electrically conducting, turbulent motions are accompanied
by magnetic field fluctuations. Although plasmas are abundant in the 
universe (it is said that 99\% of the baryonic material in the universe is in the
plasma state), conducting fluids are rare on earth, where electrical 
conductors are usually solid. Hence it is not surprising that magnetohydrodynamic
turbulence (Biskamp~\cite{bisk03}) has received attention only recently,
after that hydrodynamic turbulence has been studied at length (Frisch~\cite{fri95}).

A milestone in turbulence theory is the work by Kolmogorov~\cite{kol41}
on the scaling properties of hydrodynamic turbulence, where he finds
the well-known $k^{-5/3}$ power spectrum for total energy.
The presence of the magnetic field strongly affects the properties of turbulence.
While it is possible through a Galilean transformation
to  subtract a mean (or local) velocity field, this transformation has no 
effect at all on the magnetic field (a mean global or a local one).

Since the pioneering work of  Iroshnikov~\cite{iro64} and 
Kraichnan~\cite{kra65} (hereafter IK) there has been a lot of debate 
on which are the main properties of MHD turbulence. 
The Alfv\'en effect, which arises from the fact that only oppositely propagating
Alfv\'en waves interact, and the hypothesis of 
homogeneity and isotropy lead to a $k^{-3/2}$ scaling
for the energy spectrum, which differs from the Kolmogorov $k^{-5/3}$ scaling
in the hydrodynamic case (Kolmogorov~\cite{kol41}, Obukhov~\cite{obu41}).

The anisotropy of MHD turbulence is one the properties characterizing the recent debate 
(Shebalin~et~al.~\cite{she83}, Sridhar \& Goldreich~\cite{sg94},
Goldreich \& Sridhar~\cite{gs95}, Montgomery \& Matthaeus~\cite{mm95}, 
Goldreich \& Sridhar~\cite{gs97}, Cho \& Vishniac~\cite{cv00}, 
Maron \& Goldreich~\cite{mg01}, Cho~et~al.~\cite{cv02}). 
There is broad agreement that the anisotropy of MHD strongly affects its properties,
simply due to the presence of a magnetic field, and
that the hypothesis of homogeneity and isotropy must be relaxed.

Shebalin~et~al.~\cite{she83} have shown that  the energy cascade takes place mainly
in the plane orthogonal to the static (DC) magnetic field, while it is weaker in the parallel direction. 
Sridhar \& Goldreich~\cite{sg94} and Goldreich \& Sridhar~\cite{gs95,gs97} have shown that the 
anisotropy gets stronger at large wave-numbers, i.e.\ whilst the cascade takes place. 
These results have been numerically investigated and confirmed by Cho \& Vishniac~\cite{cv00}, 
Cho~et~al.~\cite{cv02}.

\subsection{Turbulent Cascade and Phenomenology of the Inertial Range}

A characteristic property of fully developed turbulence is the presence of a 
broad range of different scales. Relevant physical quantities, such as energy,
are excited within a certain spectral range $k \sim k_{in}$, called the 
\emph{injection scale}. Nonlinear interactions transfer these quantities in $k$-space
towards larger wavenumbers up to the \emph{dissipation scale} $k \sim k_{d}$, where
a dissipative physical process is supposed to act as a sink for this energy flux.

The region in Fourier space between the injection and the dissipation scale
\begin{equation} \label{eq:dir}
k_{in} \ll k \ll k_d
\end{equation}
is called the `inertial range'. In this spectral range 
the turbulence develops solely under the influence of the internal nonlinear dynamics 
without being directly influenced by either the external injection of energy or by dissipative
processes. Spectra in the inertial range exhibit power-laws, and  the inertial range can be
defined as the wavenumber range within which the spectrum
has a power-law behavior. 

What we have just described is called a \emph{direct} cascade, but sometimes
a flux of energy in the opposite direction occurs, i.e.\ from the injection scale $k \sim k_{in}$
towards smaller wavenumbers. This process is called \emph{inverse cascade}, 
and when it occurs a second inertial range, besides~(\ref{eq:dir}), is found
\begin{equation} \label{eq:iir}
L^{-1} \ll k \ll k_{in}
\end{equation}
The lower limit $L^{-1}$ is usually determined by the size of the system $L$.

We now briefly summarize the Kolmogorov phenomenology of the inertial range (K41).
We explicitly consider the case of a direct cascade and of an isotropic system.
The results that we will describe in the next chapter depart substantially from 
the K41 theory, but many concepts, ideas and notations are used also in anisotropic
turbulence theory. Considering \emph{isotropic} turbulence we can define the 
angle-integrated spectra in the following way:
\be
E_k = \int \! \ud\Omega_{\bsy k}  \ E_{\bsy k}, \qquad
E = \int_0^{\infty} \! \ud k  \  E_k
\ee
where thanks to isotropy we consider only the scalar wavenumber 
$k = \sqrt{ k_x^2 + k_y^2 + k_z^2 }$.

The dynamics of turbulence is controlled by the rate $\epsilon_{in}$ at which energy is 
injected into the system at the injection scale $k_{in}$, it is subsequently 
scattered along the inertial range with the transfer rate $\epsilon_{t}$, and finally
swept away from the system at the dissipation scale $k \sim k_{d}$ with the
dissipation rate $\epsilon_{d}$. For stationary turbulence all these spectral
energy fluxes are equal
\be \label{eq:balrate}
\epsilon_{in} = \epsilon_{t} = \epsilon_{d} = \epsilon.
\ee
This equality still holds approximately  when the injection energy rate changes
in time, because the more rapid dynamics at the small scales in the inertial and 
dissipation ranges adjust the spectrum rapidly compared to the slower dynamics
of the large scales.  For convenience we divide the inertial range into a discrete 
number of scales $l_n = k_n^{-1}$,
\be
l_0 > l_1 > \ \cdots \ > l_N, \qquad \textrm{i.e.}  \qquad k_0 < k_1 < \ \cdots \ < k_N,
\ee
and the division is taken on a logarithmic scale $l_n = 2^{-n}\, l_0$, where $l_0$ is of the 
order of the large scale $L$. The time taken for the transfer of energy between two 
neighboring scales $l_n$ and $l_{n+1}$ is given by $\tau_n$, the so-called 
\emph{eddy turnover time} of the eddy $\delta v_{l_n}$, for simplicity $\delta v_n$
\be \label{eq:k1}
\tau_n \sim \frac{l_n}{\delta v_n}
\ee
Since the energy flux $\epsilon$ is constant across the inertial range, we can write
\be \label{eq:ek41}
\epsilon \sim \frac{E_n}{\tau_n} \sim \frac{\delta v_n^3}{l_n}
\ee
From this equality we can find the scaling
\be
\delta v_n \sim \epsilon^{1/3} \, l_n^{1/3}
\ee
To obtain the energy spectrum we identify the eddy energy with the band-integrated
Fourier spectrum
\be
\delta v_n^2 \sim E_n \sim  \int_{k_n}^{ k_{n+1} } \! \ud k  \  E_k 
\sim k_n \, E_{k_n}
\ee
from which we obtain, substituting $k_n \to k$, the $-5/3$ Kolmogorov spectrum
\be \label{eq:k41}
E_k \sim \epsilon^{2/3} \, k^{-5/3}
\ee

We remark again that the hypothesis of isotropy has been essential to obtain 
the K41 spectrum~(\ref{eq:k41}).

\subsection{The Alfv\'en Effect} \label{sec:ae}

In the hydrodynamic case isotropy is normally
justified, but for plasmas a magnetic field is always present. This
introduces an anisotropy of the system which cannot be eliminated. 

Four decades after Iroshnikov~\cite{iro64} and Kraichnan~\cite{kra65} presented
their ideas on MHD turbulence, the debate on which physical mechanisms
drive it is still active. Rewriting the equations of incompressible 
MHD~(\ref{eq:mu})-(\ref{eq:ec}) expressing magnetic field in velocity units, i.e.\ 
$\bsy{B} \rightarrow \bsy{b} = \bsy{B}/\sqrt{4\pi\rho_0}$, and making them non
dimensional choosing a characteristic velocity $u^{\ast}$, a characteristic 
length $l^{\ast}$, and the related crossing time $t^{\ast} = l^{\ast} / u^{\ast}$,
we have in terms of the Els\"asser variables
$\boldsymbol{z}^{\pm} = \boldsymbol{u} \pm \boldsymbol{b}$:
\begin{eqnarray}
& &\frac{\de \bsy{z}^{\pm}}{\de t} = - \left( \bsy{z}^{\mp} \cdot \bnabla \right) \bsy{z}^{\pm}
- \bnabla P + \frac{1}{\mathcal R} \bnabla^2 \bsy{z}^{\pm}  \label{eq:elm1}  \\
& &\bnabla \cdot \bsy{z}^{\pm} = 0  \label{eq:elm2}
\end{eqnarray}
where we assumed the kinetic ($\mathcal R$) and magnetic ($\mathcal R_m$) 
Reynolds numbers, defined as 
\be
\frac{1}{\mathcal R} = \frac{\nu}{\rho_0\, l^{\ast} u^{\ast}}, \qquad
\frac{1}{\mathcal R_m} = \frac{\eta c^2}{4\pi \rho_0 \, l^{\ast} u^{\ast} },
\ee
to be equal ($\mathcal R_m = \mathcal R$). $P$ is the total
pressure $P = p/\rho_0 + \bsy{B}^2/8\pi\rho_0$ in  dimensionless form, and it is tied
to the $\bsy{z}^{\pm}$ fields by incompressibility~(\ref{eq:elm2}):
\be \label{eq:pen}
\bnabla^2 P = - \sum_{i,j=1}^3 \Big( \partial_i z_j^- \Big) \Big( \partial_j z_i^+ \Big)
\ee
In the following analysis we ignore the dissipative terms, because dissipation provided 
by $\mathcal R$, supposed it has a high value, takes place only at small spatial scales. 
A constant and uniform magnetic field in absence of fluid motions:
\be \label{eq:ic}
\bsy{u}_0 = 0, \qquad \bsy{b}_0 = v_{\mathcal A} \, \bsy{e}_z
\ee
is a homogeneous solution
of equations~(\ref{eq:elm1})-(\ref{eq:elm2}). A linear analysis of incompressible MHD
shows that we have only Alfv\'en waves, in particular linearizing 
equations~(\ref{eq:elm1})-(\ref{eq:elm2}) with the equilibrium~(\ref{eq:ic})  yields:
\be \label{eq:elml}
\frac{\de \bsy{z}^{\pm}}{\de t}  \mp v_{\mathcal A} \frac{\de \bsy{z}^{\pm}}{\de z} = 0,
\qquad 
\bnabla \cdot \bsy{z}^{\pm} = 0.
\ee
Equations~(\ref{eq:elml}) show that $\bsy{z}^-$ describes Alfv\'en waves  propagating
toward positive $z$ at the speed $v_{\mathcal A}$, and $\bsy{z}^+$ describes Alfv\'en waves
propagating at the same speed but in the opposite direction. 
A notable property of the Els\"asser fields in equations~(\ref{eq:elm1})-(\ref{eq:elm2})
is the \emph{absence of self-coupling in the nonlinear term}. In fact \emph{there is only 
cross-coupling of  $\bsy{z}^+$ and $\bsy{z}^-$}. This property allows for a 
\emph{nonlinear} generalization of 
the linear Alfv\'en waves described by equations~(\ref{eq:elml}). Singling out
the equilibrium~(\ref{eq:ic}) ($\bsy{z}_0^{\pm} = \pm v_{\mathcal A} \, \bsy{e}_z$
in terms of Els\"asser variables)
\be
\bsy{z}^{\pm} = \bsy{z}_1^{\pm} \pm v_{\mathcal A} \, \bsy{e}_z,
\ee
and requiring the generalized Alfv\'en wave $\bsy{z}_1^{\pm}$ to retain its transversality, 
i.e.\ $\bsy{z}_1^{\pm} \cdot  \bsy{e}_z = 0$, we have for the nonlinear term
\be \label{eq:nlt}
\bsy{z}^{\mp} \cdot \bnabla \bsy{z}^{\pm} = \bsy{z}_1^{\mp} \cdot \bnabla \bsy{z}_1^{\pm}
\mp v_{\mathcal A} \frac{\de  \bsy{z}_1^{\pm}}{\de z}
\ee
This means that if one of the two Els\"asser fields $\bsy{z}_1^{\pm}$ is zero the nonlinear
term in equation~(\ref{eq:nlt}) vanishes and equations~(\ref{eq:elm1})-(\ref{eq:elm2})
assume the linear form~(\ref{eq:elml}) for the remaining Els\"asser field, even if
its amplitude is not small.
If at some time, like $t=0$, $\bsy{z}_1^+ = 0$ and $\bsy{z}_1^- = \bsy{f}(x,y,z)$ from 
equations~(\ref{eq:elml}) we have that the solution is $\bsy{z}_1^+ = 0$, 
$\bsy{z}_1^- = \bsy{f}(x,y,z - v_{\mathcal A}\, t)$. If at time $t=0$ we had 
$\bsy{z}_1^- = 0$ and $\bsy{z}_1^+ = \bsy{f}(x,y,z)$, then the solution would 
be $\bsy{z}_1^- = 0$ and $\bsy{z}_1^+ = \bsy{f}(x,y,z + v_{\mathcal A}\, t)$.
These \emph{nonlinear} solutions are Alfv\'en wave packets of arbitrary form propagating
nondispersively in the direction of the main field $\bsy{b}_0 = v_{\mathcal A}\, \bsy{e}_z$,
and in the opposite direction. The dynamics are very simple as long as there is no spatial
overlap (``collision'') between two oppositely moving packets $\bsy{z}^+$ and $\bsy{z}^-$.
Hence only Alfv\'en waves propagating in opposite directions along the guide
field interact. This is the basis of the \emph{Alfv\'en effect} introduced independently
by Iroshnikov~\cite{iro64} and Kraichnan~\cite{kra65} who noted that the cascade of 
energy in MHD turbulence occurs as a result of collisions between oppositely directed 
Alfv\'en wave packets. This result is quite general for MHD, in fact the guide field need not 
be an external static field, but can also be the field in the large-scale energy-containing 
eddies.

\subsection{The Iroshnikov-Kraichnan Formulation}

To derive the results of IK theory we consider a statistically steady, \emph{isotropic}
excitation of amplitude $\delta z_l^{\pm} \ll v_{\mathcal A}$, at the injection scale $l$ 
of the equilibrium defined by equations~(\ref{eq:ic}). The turbulent cascade produces 
Alfv\'en wave packets at scales $\lambda < l$ traveling in opposite directions along 
the large-scale field. A fundamental hypothesis made by both Iroshnikov~\cite{iro64} 
and Kraichnan~\cite{kra65} is that the energy transfer in Fourier space is \emph{local}
and \emph{isotropic}. At this point we restrict the discussion to weak velocity-magnetic-field 
correlation $\delta z_{\lambda}^+ \sim \delta z_{\lambda}^- \sim \delta z_{\lambda}$, 
which is the condition that applies to our model for a coronal loop and that we will
discuss in more detail in the next chapter. 

We distinguish between two important dynamical time scales, the time for distortion 
of a wave packet $\delta z_{\lambda}^{\pm}$ at scale $\lambda$ by a similar eddy 
$\delta z_{\lambda}^{\mp}$, i.e.\ the eddy turnover time
\be \label{eq:ik1}
\tau_{\lambda} = \frac{\lambda}{\delta z_{\lambda}},
\ee
and the Alfv\'en time $\tau_{\mathcal A} = {\lambda}/v_{\mathcal A}$, which is the 
interaction time of the two oppositely moving wave packets. In general 
$\tau_{\mathcal A} \ll \tau_{\lambda}$, so the interaction time of the two wave 
packets is much shorter than the non-magnetic eddy turnover time 
$\tau_{\lambda}$. The change in amplitude $\Delta \delta z_{\lambda}$ 
is small during a single collision of two wave packets since it is 
proportional to the interaction time:
\be \label{eq:faik}
\frac{\Delta \delta z_{\lambda}}{\delta z_{\lambda}} \sim 
\frac{\tau_{\mathcal A}}{\tau_{\lambda}} \ll 1.
\ee
During successive collisions  these perturbations add with random phases and,
given the diffusive nature of the process, the number of collisions for the small
perturbations to build up to order unity 
(i.e.\ $\Delta \delta z_{\lambda} \sim \delta z_{\lambda}$) is
\be \label{eq:ikn}
N_{\lambda} \sim \left( \frac{\delta z_{\lambda}}{\Delta \delta z_{\lambda}} \right)^2
\sim \left( \frac{\tau_{\lambda}}{\tau_{\mathcal A}} \right)^2 
\sim \left( \frac{v_{\mathcal A}}{\delta z_{\lambda}} \right)^2 \gg 1.
\ee
The energy-transfer time $T_{\lambda}$, which in hydrodynamic turbulence 
is just $\tau_{\lambda}$, is longer
\be
T_{\lambda} \sim N_{\lambda} \tau_{\mathcal A} \sim 
\frac{ \left( \tau_{\lambda} \right)^2 }{\tau_{\mathcal A}}.
\ee
Making the substitution $\tau_{\lambda} \rightarrow T_{\lambda}$ in 
equation~(\ref{eq:ek41}) we obtain for the spectral energy flux
\be
\epsilon \sim \frac{E_{\lambda}}{T_{\lambda}} \sim 
\frac{ \delta z_{\lambda}^4\, \tau_{\mathcal A} }{\lambda^2}.
\ee
From this we get the scaling
\be \label{eq:iks}
\delta z_{\lambda} \sim \left( \epsilon\, v_{\mathcal A}\, \lambda \right)^{1/4},
\ee 
and identifying the eddy energy with the band-integrated Fourier spectrum
$\delta z_{\lambda}^2 \sim k\, E_k$ (where $k \sim \lambda^{-1}$) we obtain the 
Iroshnikov-Kraichnan (IK) spectrum for MHD turbulence
\be \label{eq:ik}
E_k \sim \left( \epsilon v_{\mathcal A} \right)^{1/2} k^{-3/2},
\ee 
which is less steep than the $k^{-5/3}$ Kolmogorov spectrum~(\ref{eq:k41}).
Using the scaling~(\ref{eq:iks}) in (\ref{eq:ikn}) we obtain for the number of collisions
per energy transfer time:
\be
N_{\lambda} \sim \left( \frac{v_{\mathcal A}}{\delta z_l} \right)^2 
\left( \frac{l}{\lambda} \right) ^{1/4}.
\ee
As we proceed along the cascade toward smaller scales $\lambda$ the number of 
collisions required for the fractional perturbations to build up to order unity increases,
verifying the hypothesis~(\ref{eq:faik}) that we made at the beginning.
Furthermore, during each collision the fraction of energy that cascades
gets smaller with decreasing scale; in fact, from equations~(\ref{eq:faik}) 
and~(\ref{eq:ikn}) we have that
\be \label{eq:ik2}
\frac{\Delta \delta z_{\lambda}}{\delta z_{\lambda}} \sim 
\frac{1}{\sqrt{N_{\lambda}}}.
\ee
In this sense we say that the cascade ``weakens'' at large wavenumbers.

\subsection{Beyond IK: Fully Anisotropic MHD Turbulence} \label{par:bik}

\emph{Isotropy} is the underlying hypothesis used in the IK derivation of turbulence scaling 
properties~(\ref{eq:ik1})-(\ref{eq:ik2}).
In particular we have imposed the condition that the wave packets are isotropic  
on the length-scale $\lambda$, also along the direction of the equilibrium magnetic field
$\bsy{b}_0 = v_{\mathcal A}\, \bsy{e}_z$. 

Only later it has been understood that the anisotropy due to the presence of 
the main axial field not only acts through the Alfv\'en effect, but has also a 
deep impact on the nonlinear dynamics, producing  two different behaviors 
along the direction of the main field and in the orthogonal plane 
(Shebalin~et~al.~\cite{she83}, Sridhar \& Goldreich~\cite{sg94}, 
Goldreich \& Sridhar~\cite{gs95}, Montgomery \& Matthaeus~\cite{mm95}, 
Goldreich \& Sridhar~\cite{gs97}, Cho \& Vishniac~\cite{cv00}, 
Maron \& Goldreich~\cite{mg01}, Cho~et~al.~\cite{cv02}).

Shebalin~et~al.~\cite{she83} used the reduced MHD equations~(\ref{pot1})-(\ref{pot2}) 
to numerically investigate the cascade properties of a 2D turbulent system
embedded in a strong field $\bsy{B} = B_0\, \bsy{e}_x$ directed along the $x$-axis,
considering the system invariant along the $z$-direction, so to perform 2D numerical
simulations in the $x$-$y$ plane. As initial conditions they considered wave
packets with an isotropic spectral distribution in the $k_x$-$k_y$ plane, as in
the IK theory, so that the isocontours of the spectral densities were circles at the 
beginning of the simulation. 
They found that the spectrum evolves anisotropically by transferring energy to modes
perpendicular to $\bsy{B}$ far more rapidly than to modes with
$\bsy k$ parallel to  $\bsy{B}$.  In this way the initially circular spectral density 
contours elongated in the perpendicular direction. 
Even if it was not initially valid, the evolution proceeded 
toward the reduced MHD approximation. Even if the simulation was started
with an isotropic initial condition, the temporal evolution was strongly anisotropic.

The isotropic hypothesis used for the IK 
phenomenology~(\ref{eq:ik1})-(\ref{eq:ik2}) is therefore neither consistent nor correct,
as pointed out by Sridhar \& Goldreich~\cite{sg94} and Goldreich \& Sridhar~\cite{gs97}.
In this sense Iroshnikov~\cite{iro64} and Kraichnan~\cite{kra65} have only partially
implemented the consequences of anisotropy in MHD turbulence through the
Alfv\'en effect, but the full consequences of anisotropy have been understood only later,
and its elucidation is not yet complete.

To understand why perpendicular transfer is easier we present a simplified
perturbative argument in order to estimate possible energy transfer between 
Alfv\'en wave modes. Introducing in incompressible MHD 
equations~(\ref{eq:elm1})-(\ref{eq:elm2}) the expansion
\be
\bsy{z}^{\pm} = \bsy{z}_0^{\pm} + \epsilon\, \bsy{z}_1^{\pm} +
\epsilon^2\, \bsy{z}_2^{\pm} + \mathcal{O} \left( \epsilon^3 \right)
\ee
where $\bsy{z}_0^{\pm} = \pm v_{\mathcal A} \, \bsy{e}_z$ is the homogeneous 
equilibrium~(\ref{eq:ic}), and $\bsy{z}_1^{\pm}$ are Alfv\'en waves of the form
\be \label{eq:law}
\bsy{z}_1^{\pm} = \sum_{\bsy k} \bsy{A}_{\bsy k}^{\pm}  
\exp \left[ i \left( \bsy{k} \cdot \bsy{x} \pm \omega t \right) \right],
\ee
where $\bsy{k} \cdot \bsy{A}_{\bsy k}^{\pm} =0$, and $\omega = v_{\mathcal A}\, k_z$.
Noting that from equation~(\ref{eq:pen}) the pressure gradient term has 
only second order terms, i.e.
\be
\bnabla P \sim \epsilon^2 \, \bnabla P_2 + \mathcal{O} \left( \epsilon^3 \right),
\ee
at the second order we have
\be
\frac{\de \bsy{z}_2^{\pm}}{\de t} \pm v_{\mathcal A}\, \frac{\de \bsy{z}_2^{\pm}}{\de z}
+ \bnabla P_2 = - \left( \bsy{z}_1^{\mp} \cdot \bnabla \right) \bsy{z}_1^{\pm}.
\ee
This equation has basically the same structure of the wave equation for 
$\bsy{z}_1^{\pm}$, except for an effective driving term on the right, due to the linear 
Alfv\'en waves~(\ref{eq:law}).

The most efficient mechanism for fast energy transfer between modes is resonant 
interactions occurring among triads of modes with wavenumbers $\bsy{k}_1$, 
$\bsy{k}_2$ and $\bsy{k}_3$ related by the conditions
\be \label{eq:tri}
\bsy{k}_1 +  \bsy{k}_2  =  \bsy{k}_3, \qquad \textrm{and} \qquad
\omega_1 + \omega_2  = \omega_3
\ee
where $\omega_j = v_{\mathcal A}\,  \bsy{k}_{j,z}$.
Shebalin~et~al.~\cite{she83} noted that the only nontrivial solution of~(\ref{eq:tri})
requires that the $z$-component of one member of the triad, e.g.\   
$\bsy{k}_{3,z}$, must be zero. 
This implies that waves with values of $k_z$ not present initially cannot be created during collisions between oppositely propagating wave packets. Hence there is no parallel,
i.e.\ along $z$, cascade of energy. Energy will cascade to large wavenumbers in
the orthogonal plane $\bsy{k}_{\perp}$, making the turbulence cascade anisotropic.

We can now derive the anisotropic version of the IK theory~(\ref{eq:ik1})-(\ref{eq:ik2}), 
taking into account that there is no cascade along the direction of the main magnetic 
field $\bsy{b}_0 = v_{\mathcal A}\, \bsy{e}_z$. We suppose again that the system
is excited at the scale $l$ in a statistically steady and isotropic fashion such that 
$\delta z_l \ll v_{\mathcal A}$. The absence of a  parallel cascade implies that wave 
packets belonging to the inertial range have \emph{parallel} scales $l$ and 
\emph{perpendicular} scales $\lambda < l$. As previously supposed by IK, the 
Alfv\'en effect takes place and only counter-propagating  wave packets interact. 
The wave packets are ``long-lived'' and they need many collisions to loose a 
significant amount of energy. We distinguish again between the eddy turnover time
characterizing the cascade in the orthogonal plane,
\be \label{eq:aik1}
\tau_{\lambda} = \frac{\lambda}{\delta z_{\lambda}},
\ee
where $\lambda \sim k_{\perp}^{-1}$, and the Alfv\'en time $\tau_{\mathcal A}$, which 
is the interaction time of two oppositely 
propagating wave packets $\delta z^+_{\lambda}$ and $\delta z^-_{\lambda}$. 
Now, because of the absence of cascade along $z$, \emph{the Alfv\'en time is 
scale-independent}, i.e.\ it does not depend on the scale $\lambda$:
\be  \label{eq:aikta}
\tau_{\mathcal A} = \frac{l}{v_{\mathcal A}}.
\ee
The interaction time is still small compared to the eddy turnover time 
$\tau_{\mathcal A} \ll \tau_{\lambda}$, so the energy loss of the eddy at the scale
$\lambda$ is small during a single collision
\be \label{eq:el1}
\frac{\Delta \delta z_{\lambda}}{\delta z_{\lambda}} \sim 
\frac{\tau_{\mathcal A}}{\tau_{\lambda}} \ll 1
\ee
The number of collision that a wave packet at the scale $\lambda$ must suffer for 
the fractional perturbation to build up to order unity is now
\be \label{eq:nc1}
N_{\lambda} \sim \left( \frac{\delta z_{\lambda}}{\Delta \delta z_{\lambda}} \right)^2
\sim \left( \frac{\tau_{\lambda}}{ \tau_{\mathcal A}} \right)^2
\sim \left( \frac{v_{\mathcal A}}{\delta z_{\lambda}} \frac{\lambda}{l} \right)^2 \gg 1.
\ee
The energy-transfer time $T_{\lambda}$ is again given by
\be
T_{\lambda} \sim N_{\lambda} \tau_{\mathcal A} \sim 
\frac{\left( \tau_{\lambda} \right)^2}{\tau_{\mathcal A}},
\ee
while for the energy flux we have
\be
\epsilon \sim \frac{E_{\lambda}}{T_{\lambda}} \sim 
\frac{\delta z_{\lambda}^4}{\lambda^2} \frac{l}{v_{\mathcal A}}.
\ee
From the previous equations we obtain the following scaling relation
\be
\delta z_{\lambda} \sim \left( \frac{\epsilon v_{\mathcal A}}{l} \right)^{\frac{1}{4}} \, 
\lambda^{\frac{1}{2}}
\ee
and identifying the eddy energy with the band-integrated Fourier spectrum
$\delta z^2_{\lambda} \sim k_{\perp} E_{k_{\perp}}$ 
(where $k^{-1}_{\perp} \sim \lambda$)
yields the anisotropic version of the IK spectrum for MHD turbulence
\be
E_{k_{\perp}} \sim 
\left( \frac{\epsilon v_{\mathcal A}}{l} \right)^{\frac{1}{2}} \, k_{\perp}^{-2},
\ee
which exhibits the characteristic $-2$ spectral index.
Another difference with the IK formulation is given by the behavior of the number of 
collisions at small scales:
\be \label{eq:aik2}
N_{\lambda} \sim 
\left( \frac{v^3_{\mathcal A}}{\epsilon l^3} \right)^{\frac{1}{2}} \, \lambda
\sim \left( \frac{v_{\mathcal A}}{\delta z_l} \right)^2 \, \frac{\lambda}{l}.
\ee
Contrary to IK, $N_{\lambda}$ decreases with decreasing $\lambda$. 
When the number of collisions decreases at small scales we say that 
the turbulence ``strengthens''. At a small enough scale the
conditions~(\ref{eq:el1})-(\ref{eq:nc1}) will not be satisfied, thus limiting 
the spectral range in which the spectrum $E_{k_{\perp}} \sim  k_{\perp}^{-2}$ 
applies. 

Weak perturbation theory (Zakharov et~al.~\cite{zak92}) deals with the effects of the 
nonlinear terms in 
equations~(\ref{eq:elm1})-(\ref{eq:elm2}) in a systematic, perturbative manner.
When the nonlinear terms are ignored, the Fourier amplitudes and phases of the
waves are constant in time. However, the nonlinearity makes the amplitudes
change slowly over many wave periods. It is this slow change in the amplitudes
that determines energy transfer among the linear modes. A kinetic equation for the 
rate of change of energy in a mode with wave-vector $\bsy k$ describes how other
modes in the system affect the energy in this mode. To lowest order in the nonlinearity, 
the kinetic equation takes into account the interactions among modes taken three at a time, 
as shown in~(\ref{eq:tri}). When equations~(\ref{eq:tri}) is satisfied, one 
says that a 3-wave resonant interaction is allowed. When 3-wave resonant interactions 
are forbidden, because the 3-wave resonant coupling coefficients vanish, the effects of 
4-wave resonant interactions must be considered. 

Goldreich \& Sridhar~\cite{gs97} showed that although the anisotropic IK 
formulation~(\ref{eq:aik1})-(\ref{eq:aik2}) describes a weak turbulence, in the sense that
the fractional change in wave amplitude during each wave period is small, strains in the
fluid are so strong that perturbation theory diverges. It turns out that not only 3-wave 
interactions contribute, but also higher order terms.
At lowest order in perturbation theory, wave packets move along field lines. Thus the
breakdown of perturbation theory can be understood physically by studying the geometry
of the divergence of a bundle of field lines. Assume that the mean field lies along the 
$z$ direction, and consider wave packets with longitudinal scale $l$, and transverse
scale $\lambda$, with $\lambda < l$. For the turbulence to be weak we require the
condition~(\ref{eq:el1}) to be satisfied, i.e.
\be \label{eq:chila}
\chi = \frac{l\, \delta z_{\lambda}}{\lambda \, v_{\mathcal A}} = 
\frac{\tau_{\mathcal A}}{\tau_{\lambda}}   \ll 1
\ee
so that $N_{\lambda} \gg 1$. The rms inclination of the local field is 
$\theta_{\lambda} \sim \delta z_{\lambda}/v_{\mathcal A}$, so that after a single collision
between two wave packets taking place along the longitudinal scale $l$, the wave packet
will suffer an orthogonal displacement $\delta \sim l \, \theta_{\lambda}$. 
Given the diffusive nature of the process after $n$ collisions the wave packet has traveled
a distance $z \sim n\, l$ in the longitudinal direction, suffering an rms orthogonal displacement
\be
\Delta^2 \sim n\, \delta^2 \sim n l^2 \theta^2_{\lambda} \sim |z| l \theta^2_{\lambda}.
\ee
The distance along $z$ over which $\Delta$ increases by a factor of order $\lambda$,
i.e.\ $\Delta^2 \sim \lambda^2$, is
\be
L_{\ast} \sim l\, \left( \frac{\lambda \, v_{\mathcal A}}{l \, \delta z_{\lambda}} \right)^2.
\ee
The perturbative expansion converges if the energy spectrum is cut off at small
wavenumbers,  below
$k_z \, L_{\ast} \sim 1$ (Sridhar \& Goldreich~\cite{sg94}, Goldreich \& Sridhar~\cite{gs97}).
In this case it is shown that 3-wave resonant contributions vanish and 4-wave 
interactions must be considered. For a system with a finite longitudinal 
extension $L$, such as a coronal loop, this means that for a sufficiently 
weak perturbation $\delta z_{\lambda}$, we move from the anisotropic IK 
phenomenology~(\ref{eq:aik1})-(\ref{eq:aik2}) to a new one based on 4-wave 
resonant interactions that we now briefly describe. The elementary interactions 
involve scattering of two waves:
\be
\bsy{k}_1 + \bsy{k}_2 =  \bsy{k}_3 + \bsy{k}_4, \qquad 
\omega_1 + \omega_2 = \omega_3 + \omega_4.
\ee
Using $\omega_j = v_{\mathcal A}\, k_{j,z}$ and the $z$-component of the equation 
for $\bsy k$ conservation, Sridhar \& Goldreich~\cite{sg94} proved that this scattering
process leaves the $k_z$'s components unaltered. This implies that waves with values
of $k_z$ that are not present in the external stirring cannot be created by resonant 
4-wave interactions. The absence of a  parallel cascade implies that wave 
packets belonging to the inertial range have parallel scales $l$, and perpendicular 
scales $\lambda < l$. The phenomenology of this new cascade based on 4-wave
interactions is very similar to the anisotropic IK phenomenology describe by 
equations~(\ref{eq:aik1})-(\ref{eq:aik2}), and can be derived in the same way
with a few modifications. The Alfv\'en effect takes still place, so that only counter-propagating  
wave packets interact, and the wave packets need many 
collisions to loose energy significantly.  The eddy turnover time $\tau_{\lambda}$
and the Alfv\'en time $\tau_{\mathcal A}$ are defined in the same way
(see equations (\ref {eq:aik1}) and  (\ref{eq:aikta})) and
have the same meaning, but now the fractional loss is different
from what computed in~(\ref{eq:el1}), and from perturbation theory 
(Sridhar \& Goldreich~\cite{sg94}, Goldreich \& Sridhar~\cite{gs97}) we have that:
\be \label{eq:sg4w1}
\frac{\Delta \delta z_{\lambda}}{\delta z_{\lambda}} \sim 
\left( \frac{\tau_{\mathcal A}}{\tau_{\lambda}} \right)^2 \ll 1
\ee
The number of collision that a wave packet at the scale $\lambda$ must suffer for 
the fractional perturbation to build up to order unity is now
\be
N_{\lambda} \sim \left( \frac{\delta z_{\lambda}}{\Delta \delta z_{\lambda}} \right)^2
\sim \left( \frac{\tau_{\lambda}}{ \tau_{\mathcal A}} \right)^4
\sim \left( \frac{v_{\mathcal A}}{\delta z_{\lambda}} \frac{\lambda}{l} \right)^4 \gg 1,
\ee
so the energy-transfer time $T_{\lambda}$ is 
\be
T_{\lambda} \sim N_{\lambda} \tau_{\mathcal A} \sim 
\frac{\left( \tau_{\lambda} \right)^4}{\left( \tau_{\mathcal A} \right)^3},
\ee
and for the spectral energy flux we obtain
\be
\epsilon \sim \frac{E_{\lambda}}{T_{\lambda}} \sim 
\frac{\delta z_{\lambda}^6}{\lambda^4} \left( \frac{l}{v_{\mathcal A}} \right)^3
\ee
From the previous equations the scaling relation
\be
\delta z_{\lambda} \sim \epsilon^{\frac{1}{6}} 
\left( \frac{v_{\mathcal A}}{l} \right)^{\frac{1}{2}} \, 
\lambda^{\frac{2}{3}}
\ee
or equivalently
\be
\frac{\delta z_{\lambda}}{\delta z_l} \sim 
\left( \frac{\lambda}{l} \right)^{\frac{2}{3}}
\ee
are obtained. Identifying the eddy energy with the band-integrated Fourier spectrum
$\delta z^2_{\lambda} \sim k_{\perp} E_{k_{\perp}}$ where $k_{\perp} \sim \lambda$
we obtain the spectrum for MHD turbulence bases on 4-wave interactions:
\be
E_{k_{\perp}} \sim \epsilon^{\frac{1}{3}}\, 
\frac{v_{\mathcal A}}{l} \, k_{\perp}^{-\frac{7}{3}}.
\ee
As in the case of the anisotropic IK phenomenology~(\ref{eq:aik1})-(\ref{eq:aik2}) 
the number of collisions
\be \label{eq:sg4w2}
N_{\lambda} \sim \left( \frac{v_{\mathcal A}}{\delta z_l} \right)^4 \, 
\left( \frac{\lambda}{l} \right)^{\frac{4}{3}}
\ee
decreases at small scales and the turbulence becomes stronger.

The previous scalings (\ref{eq:aik1})-(\ref{eq:aik2}) and 
(\ref{eq:sg4w1})-(\ref{eq:sg4w2}) are based respectively on 3-waves and 4-waves
resonant interactions. These scalings can be generalized to the case of single 
collisions of $n$-waves. The derivation is very similar to what we have already
done, and we briefly describe it.
We suppose again that the system is weakly excited ($\delta z_l \ll v_{\mathcal A}$) 
at the scale $l$, and that $n$-waves resonant interactions produce a perpendicular
cascade. The interaction time is small compared to the eddy turnover time 
$\tau_{\mathcal A} \ll \tau_{\lambda}$, so that the energy loss of the eddy at the scale
$\lambda$ is small during a single collision, and is given by 
(see Goldreich \& Sridhar~\cite{gs97})
\be \label{eq:eloss}
\frac{\Delta \delta z_{\lambda}}{\delta z_{\lambda}} \Bigg|_n \sim 
\left( \frac{\tau_{\mathcal A}}{\tau_{\lambda}} \right)^{n-2} \ll 1,
\qquad \textrm{where} \qquad n \ge 3.
\ee
The number of collisions  is now
\be \label{eq:fp}
N_{\lambda} \sim \left( \frac{\delta z_{\lambda}}{\Delta \delta z_{\lambda}} \right)^2
\sim \left( \frac{\tau_{\lambda}}{ \tau_{\mathcal A}} \right)^{2\left(n-2\right)}
\sim \left( \frac{v_{\mathcal A}}{\delta z_{\lambda}} \frac{\lambda}{l} \right)^{2\left(n-2\right)},
\ee
the energy-transfer time $T_{\lambda}$
\be
T_{\lambda} \sim N_{\lambda} \tau_{\mathcal A} \sim 
\frac{\left( \tau_{\lambda} \right)^{2n-4}}{\left( \tau_{\mathcal A} \right)^{2n-5}},
\ee
and the spectral energy flux 
\be
\epsilon \sim \frac{E_{\lambda}}{T_{\lambda}} \sim 
\frac{\delta z_{\lambda}^{2n-2}}{\lambda^{2n-4}} \left( \frac{l}{v_{\mathcal A}} \right)^{2n-5}.
\ee
From the previous equations we obtain the following scaling relation
\be
\delta z_{\lambda} \sim \epsilon^{\frac{1}{2\left(n-1\right)}} 
\left( \frac{v_{\mathcal A}}{l} \right)^{\frac{2n-5}{2n-2}} \, 
\lambda^{\frac{n-2}{n-1}}
\ee
or equivalently
\be
\frac{\delta z_{\lambda}}{\delta z_l} \sim 
\left( \frac{\lambda}{l} \right)^{\frac{n-2}{n-1}}
\ee
and the anisotropic spectrum for MHD turbulence based on single $n$-waves
resonant interactions is
\be
E_{k_{\perp}} \sim \epsilon^{\frac{1}{n-1}}\, 
\left( \frac{v_{\mathcal A}}{l} \right)^{\frac{2n-5}{n-1}} \, k_{\perp}^{-\frac{3n-5}{n-1}}.
\ee
The spectral index spans from $-2$ and $-7/3$ for the cases that we have already 
treated ($n=3$ and $n=4$), and has a lower limit of $-3$ as $n \to \infty$.
A common feature for the number of collisions
\be
N_{\lambda} \sim \left( \frac{v_{\mathcal A}}{\delta z_l} \right)^{2\left(n-2\right)} \, 
\left( \frac{\lambda}{l} \right)^{2\frac{n-2}{n-1}}
\ee
is the ``strengthening'' of the turbulence at small scales, for all $n \ge 3$.

A parameter that characterizes weak turbulence is $\chi$ (see eq.~(\ref{eq:chila})):
\be \label{eq:chila1}
\chi = \frac{l\, \delta z_{\lambda}}{\lambda \, v_{\mathcal A}} = 
\frac{\tau_{\mathcal A}}{\tau_{\lambda}},
\ee
which, for weak turbulence, is small $\chi \ll 1$. 
At a sufficiently small scale along the cascade
the turbulence will get strong enough so that $\chi \sim 1$, and just 
$N_{\lambda} \sim 1$ collision (see eq.~(\ref{eq:fp}))
with another wave packet of comparable size will result in a fractional change in wave
amplitude of order one, i.e.\ $\Delta \delta z_{\lambda} \sim \delta z_{\lambda}$ from
equations~(\ref{eq:eloss})-(\ref{eq:fp}).
The same result is obtained if the perturbation at the
injection scale $l$ is strong enough, i.e.\ $\delta z_l \sim v_{\mathcal A}$ so
from eq.~(\ref{eq:chila1}) we have $\chi \sim 1$.
In this case wave packets lose their identity after they travel
one wavelength along the field lines. Consequently the eddy turnover time and
the Alfv\'enic  time are the same, 
\be \label{eq:cb}
k_{\perp}\, \delta z_{\lambda} \sim k_{\parallel}\, v_{\mathcal A},
\ee
where $\lambda \sim k^{-1}_{\perp}$, which is called a  ``critical balance''
(Goldreich \& Sridhar~\cite{gs95}).
Consider an eddy of dimensions $l$ and $l_{\perp}$ along the directions
parallel and orthogonal to the mean magnetic field. Because of the turbulent
transfer to smaller perpendicular scales, $l_{\perp}$ shrinks and the eddy 
becomes more elongated, leading to sheet-like structures limited only 
by dissipation.  The spectral cascade takes place mainly in the orthogonal plane 
with constant energy flux $\epsilon$  across the inertial range
\be
\epsilon \sim \frac{\delta z^3_{\lambda}}{\lambda} \sim \textrm{const}.
\ee
Combining this relation with equation~(\ref{eq:cb}) we obtain
\be
k_{\parallel} \sim \frac{\epsilon^{1/3}}{v_{\mathcal A}} \, k_{\perp}^{2/3}
\sim k_{\perp}^{2/3} \, \mathscr{L}^{-1/3},
\ee
where we define the scale $\mathscr{L} = v^3_{\mathcal A} / \epsilon$.  
This relation shows also that this anisotropy increases toward smaller scales, 
the ratio
\be
\frac{k_{\perp}}{k_{\parallel}} \sim 
\left( k_{\perp} \, \mathscr{L} \right)^{1/3}  \qquad \Longleftrightarrow \qquad
\frac{l}{\lambda} \sim 
\left( \frac{\mathscr{L}}{\lambda} \right)^{1/3}
\ee
gets bigger for larger $k_{\perp}$ (smaller $\lambda$).

For the spectrum the original K41 phenomenology~(\ref{eq:k1})-(\ref{eq:k41})
is valid, with a slight modification to account that the cascade takes place in the 
orthogonal plane. But \emph{this analogy is only formal}, because now we 
discard isotropy; in particular, the Alfv\'en effect takes place and
the cascade occurs mainly in the orthogonal plane. The eddy turnover time 
$\tau_{\lambda}$ and the energy transfer time now are the same
\be
\tau_{\lambda} \sim \frac{\lambda}{\delta z_{\lambda}},
\ee
and since the energy flux $\epsilon$ is constant across the inertial range, we can write
\be
\epsilon \sim \frac{E_{\lambda}}{\tau_{\lambda}} 
\sim \frac{\delta z_{\lambda}^3}{\lambda},
\ee
yielding the scaling
\be
\delta z_{\lambda} \sim \epsilon^{1/3} \, \lambda^{1/3}.
\ee
So from the band-integrated Fourier spectrum in the orthogonal plane 
(i.e.\ with  $\lambda \sim k^{-1}_{\perp}$)
\be
\delta z_{\lambda}^2 \sim k_{\perp} \, E_{k_{\perp}}
\ee
we recover the $-5/3$ Kolmogorov spectrum
\be
E_{k_{\perp}} \sim \epsilon^{2/3} \, k_{\perp}^{-5/3},
\ee
but now \emph{only} in the orthogonal plane.

\chapter{The Numerical Code} \label{sec:nc}

A numerical code, written in Fortran 90 and parallelized with MPI,  has been developed 
to solve the non-dimensional reduced MHD
equations~(\ref{pot1})-(\ref{pot2}), that we rewrite here for convenience:
\begin{eqnarray}
& &\frac{\partial \psi}{\partial t} = v_{\mathcal A}\, \frac{\partial \varphi}{\partial z} + 
\left[ \varphi, \psi \right] + \frac{1}{\mathcal R_m} \boldsymbol{\nabla}^2_\perp 
\psi,  \label{eq:pot1} \\
& &\frac{\partial \omega}{\partial t} = v_{\mathcal A}\, \frac{\partial j}{\partial z} + 
\left[ j , \psi \right] - \left[ \omega , \varphi \right] 
+  \frac{1}{\mathcal R} \boldsymbol{\nabla}^2_\perp \omega, 
\label{eq:pot2} 
\end{eqnarray}
The computational domain is a parallelepiped ($ \ell \times \ell \times L$) with 
an orthogonal ($x$, $y$) square cross section of size $\ell$, and  an axial ($z$) 
length $L$, with the normalization $\ell = 1$ (because of our choice for the 
length-scale  to render
equations dimensionless shown in \S~\ref{sec:dfbc}), and $L > 1$ (see Figure~\ref{fig:loopbox}). 
In the orthogonal planes ($x$ and $y$ directions) periodic boundary conditions are used
coupled with a Fourier pseudo-spectral numerical method (Canuto et~al.~\cite{can88}). 
In the axial direction $z$, a velocity pattern is imposed  at the top and bottom boundaries
(see equations~(\ref{eq:bc0})-(\ref{eq:bcL})), and a central finite difference scheme 
of the second order is used. Time is discretized with a third-order Runge-Kutta method
coupled with an implicit Crank-Nicholson scheme for the diffusive terms. 

When investigating turbulence, the diffusive terms provide a 
sink for the flux of energy at small scales. One of the problems present when performing
a numerical investigation is the limitation on the number of grid points. While
the Reynolds numbers $\mathcal R$ and $\mathcal{R}_m$ have large values
in most physical problems of interest, the number of points is necessarily limited
when performing a numerical simulation which, in turns, restricts the Reynolds numbers to 
low values. Thus, diffusion is not limited to small scales, but also affects the large
scales. If one of the purposes of the numerical investigation is to study the inertial 
range behavior, even with a resolution of $512 \times 512$ points in the planes, 
the inertial range is disturbed by diffusion. In numerical studies of turbulence it is 
often practical to use higher-order diffusion operators, or \emph{hyperdiffusion}:
\be \label{eq:hyp}
\frac{1}{\mathcal R}\, \bnabla^2_{\perp} \quad \longrightarrow \quad 
\frac{1}{\mathcal{R}_{\alpha}}\,  
\left(-1\right)^{\alpha+1}\, \bnabla^{2 \alpha}_{\perp}, \qquad \alpha > 1,
\ee
where the exponent $\alpha$ is called \emph{dissipativity}.
In this way dissipation is strongly concentrated at the small scales, and the 
dissipative pollution of the inertial range is avoided. We have performed simulations
with both normal diffusion ($\alpha = 1$) and hyperdiffusion with 
dissipativity $\alpha = 4$.

\section{Fourier Transform and Spatial Derivatives} \label{sec:ftsd}
 
Suppose $h(x)$ is a complex periodic function of period $\ell$, defined on a uniformly 
spaced grid of $N$ points:
\be \label{eq:grid}
x_k = \Delta \, k, \qquad k = 0, \dots, N-1, \qquad \textrm{where} \qquad \Delta = \ell / N.
\ee
Indicating with $h_k$ the value of the function at the point $x_k$, i.e.\ $h_k = h(\Delta \, k)$,
the discrete Fourier transform $H_n$ at the wavenumber $k_n$ can be defined as 
(see \emph{Numerical} \emph{Recipes}~\cite{nr92}):
\be \label{eq:ft}
H_n = \sum_{k=0}^{N-1} h_k \, e^{- i\, 2 \pi n k / N}, \qquad n = 0, \dots, N-1, 
\quad \textrm{with} \quad k_n = 2\pi n / \ell,
\ee
to which corresponds the inverse Fourier transform
\be \label{eq:ift}
h_k = \frac{1}{N} \, \sum_{n=0}^{N-1} H_n \, e^{i\, 2 \pi n k / N}, \qquad k = 0, \dots, N-1.
\ee
Since the only differences between (\ref{eq:ft}) and (\ref{eq:ift}) are changing the sign 
in the exponential and dividing the result by N, a routine for calculating the discrete 
Fourier transform can also, with slight modifications, calculate the inverse transform. 

Although equation~(\ref{eq:ft}), the discrete Fourier transform, seems to be
an $\mathcal{O}(N^2)$ process (i.e.\ to compute the $N$ values of the function $H_n$
it would require to compute $N^2$ operations), an efficient algorithm called the 
\emph{fast Fourier transform} or FFT (\emph{Numerical} \emph{Recipes}~\cite{nr92}) 
requires only $\mathcal{O}(N \log_2 N)$ operations. The FFT algorithm became 
generally known in the mid-1960s, from the work of Cooley and Tukey, and
nowadays it is broadly used in scientific computing. FFTs are generally available
as library subroutines and the focus is mainly on achieving the best possible performance
on a computing platform. We have chosen to use the FFTW library
(Frigo \& Johnson~\cite{fftw}, see also \texttt{http://www.fftw.org}), which is Free Software
distributed under the GNU General Public License. 
FFTW is typically faster than other publicly-available FFT implementations and is 
competitive with vendor-tuned libraries that are tuned to work efficiently on
specific CPUs. In this way we have achieved our goal to have a cross-platform 
portable code with a good performance.

From equations~(\ref{eq:grid}) and (\ref{eq:ift}), it easily follows that the spatial
derivative $h'_k$ computed at the point $x_k$ is given by
\be \label{eq:dfft}
h'_k = \frac{\ud h}{\ud x} \left( x_k \right) = 
\frac{1}{N} \, \sum_{n=0}^{N-1} \left( i\, \frac{2 \pi n}{\ell} \right)
H_n \, e^{i\, 2 \pi n k / N}, \qquad k = 0, \dots, N-1.
\ee
Hence, to compute the derivative we first compute the Fourier coefficients
$H_n$, then we multiply them by the factor $i\, 2\pi n/ \ell$, and finally we perform
the inverse Fourier transform (\ref{eq:ift}) with these modified coefficients. 
The precision of this \emph{pseudo-spectral} method is much higher than that
of an ordinary finite difference scheme. Even with relatively few grid points 
the computed value of the derivative is almost exact, with the error mainly due
to the precision of the numerical processor (see Canuto et~al.~\cite{can88}).

Time evolution is performed in Fourier space, rather than in coordinate space, so that
we solve the Fourier transform of the reduced MHD equations~(\ref{pot1})-(\ref{pot2}). 
Extending the notations introduced in (\ref{eq:grid}), (\ref{eq:ft}) and (\ref{eq:ift}) to 2 
dimensions, and noting that in our case $\ell = 1$, we can write for a generic function $f$:
\be \label{eq:ft2}
f \left( x, y, z, t \right) = \frac{1}{N^2}\, \sum_{r,s=0}^{N-1} f_{r,s} \left( z, t \right) 
e^{i \left( k_r x + k_s y \right)}, \qquad \textrm{where} \qquad
k_r = 2 \pi r, \quad k_s = 2 \pi s,
\ee
and $r$, $s$ are intergers ranging from $0$ to $N-1$.
Introducing this expansion for the magnetic and velocity potentials, respectively $\psi$
and $\varphi$, in~(\ref{pot1})-(\ref{pot2}) we obtain the reduced MHD equations
in Fourier space:
\begin{eqnarray}
& &\frac{\partial \psi_{r,s}}{\partial t} = v_{\mathcal A}\, \frac{\partial \varphi_{r,s}}{\partial z} + 
\left[ \varphi, \psi \right]_{r,s} 
- \frac{1}{\mathcal R_m} \, k^2_{r,s} \, \psi_{r,s},  
\label{eq:fou1} \\
& &\frac{\partial \varphi_{r,s}}{\partial t} = v_{\mathcal A}\, \frac{\partial \psi_{r,s}}{\partial z} + 
\frac{ \left\{ \left[ j , \psi \right] - \left[ \omega , \varphi \right] \right\}_{r,s} }{k^2_{r,s}}
- \frac{1}{\mathcal R} \, k^2_{r,s} \, \varphi_{r,s}, 
\label{eq:fou2} 
\end{eqnarray}
where $k^2_{r,s} = (2 \pi)^2 \, (r^2+s^2)$ and $(r,s) \ne (0,0)$.
Note that to compute the Fourier components of the Poisson brackets, e.g.\ 
\be
\left[ \varphi, \psi \right]_{r,s} = \left( \frac{\de \varphi}{\de x}\, \frac{\de \psi}{\de y}
- \frac{\de \psi}{\de x}\, \frac{\de \varphi}{\de y} \right)_{r,s}
\ee
we must first compute the inverse Fourier transform
of $\varphi_{r,s}$ and $\psi_{r,s}$, then calculate the Poisson bracket in real space,
and finally obtain its Fourier transform.

When  hyperdiffusion (\ref{eq:hyp}) is used in Fourier space, the diffusive term is always 
linear, and the only thing that changes is the power of the $k$-factor:
\be \label{eq:hypf}
- \frac{1}{\mathcal R}\, k^2_{r,s}\quad \longrightarrow \quad 
- \frac{1}{\mathcal{R}_{\alpha}}\, k^{2\alpha}_{r,s}, \qquad \alpha > 1.
\ee
Since Crank-Nicholson schemes are well-suited for the time advancement of linear terms,
time evolution is performed  with a third-order Runge-Kutta method
(for the nonlinear terms and the $z$-derivatives)
coupled with an implicit Crank-Nicholson scheme for the diffusive terms.
To prevent numerical instabilities due to aliasing of the solutions, we truncate the Fourier
transforms outside a circle 
of radius $2N/3$ in the $k_x$-$k_y$ plane, where $N$ is the number of modes
and $N \times N$ the resolution in the real domain in the $x$-$y$ plane
(e.g.\ equation~(\ref{eq:ft2})). For time integration we require that the time-step 
satisfies the CFL (Courant-Friedrichs-Levy) condition in the $x$-$y$ plane as well as 
in the $z$ direction. This check is performed by a subroutine which dinamically adapts
the value of the time-step.

\section{Message-Passing Interface and Parallel Computing} \label{sec:mpipc}

To investigate some of the most interesting problems in physics, astrophysics, and engineering,
challenging numerical computations are required. In numerical studies of turbulence
we have already remarked that high grid resolutions are a necessity. For instance, if 
we had not performed some numerical simulations using hyperdiffusion (\ref{eq:hyp})
and a numerical grid with $512 \times 512 \times 200$ points (\emph{these are more than
$52$ millions grid points!} $52.428.800$), most of the conclusions
presented in this thesis would have been unattainable. 

Serial computers are not suitable to perform this kind of simulation, and
it is necessary to use a parallel computing system, i.e.\  a computer with more than one 
processor for parallel processing (commonly referred to as high-performance computers 
or supercomputers). These machines are used to perform numerical simulations of 
phenomena too complex to be reliably investigated by analytical methods, such as 
fully nonlinear problems, and/or very difficult or impossible to reproduce in a laboratory
(unfortunately such as a coronal loop or most of the astrophysical problems of interest).

There are many different kinds of parallel computers. One major way to classify parallel 
computers is based on their memory architectures. \emph{Shared memory} parallel computers 
have multiple processors accessing all available memory as a global address space. They can 
be further divided into two main classes based on memory access times: Uniform Memory 
Access (UMA), in which access times to all parts of memory are equal, or Non-Uniform 
Memory Access (NUMA), in which they are not. \emph{Distributed memory} parallel computers 
also have multiple processors, but each of the processors can only access its own local 
memory. There is no global memory address space; but the processors communicate with 
each other through an intercommunication network, which can have many different topologies 
including star, ring, tree, hypercube, fat hypercube (a hypercube with more than one  
processor at a node), an n-dimensional mesh, etc.. Parallel computing systems 
with hundreds of processors are referred to as \emph{massively parallel}. 

The Message Passing Interface (MPI) is a computer communication protocol
(see \cite{mpi,mpi2,mpi99,red99}, and also \texttt{http://www.mpi-forum.org}). 
The message-passing model posits a set of processes that have only local memory 
but are able to communicate with other processes by \emph{sending and receiving 
messages}. It is a defining feature of the message-passing model that data transfer 
from the local memory of one process to the local memory of another requires 
operations to be performed by both processes. Although the specific 
communication network is not part of the computational model
it is one of the more delicate points in parallel computing, often representing
the bottleneck to the performance of a numerical code. The challenge is always to try
building intercommunication networks (also called switches) that keep up with speeds 
of advanced single processors. Faster computers require faster switches to 
enable most applications to take advantage of them.

MPI is an attempt to collect the best features of many message-passing systems that
have been developed over the years and to improve and standardize them. MPI is a library,
not a language. It specifies the names, calling sequences, and results of subroutines to 
be called from Fortran, C and C++ programs. Programs are compiled with 
ordinary compilers and linked with the MPI library.
It is emerging as a standard for communication among the nodes running a parallel program 
on a \emph{distributed memory system}, although MPI can also be used on shared memory
computers. Its advantage over older message 
passing libraries is that it is both portable (because MPI has been implemented for almost 
every distributed memory architecture) and fast (because each implementation is optimized 
for the hardware on which it runs).

The message-passing model fits well on separate processors connected by a communication
network. Thus, it matches the hardware of most of today's parallel supercomputers. Where 
the machine supplies extra hardware to support a shared-memory model, the message-passing
model can take advantage of this hardware to speed up the rate of data transfer.
Message passing has been found to be a useful and complete model in which to express 
parallel algorithms. It provides control when dealing with data locality and, by controlling 
memory references more explicitly than any of the other models, the message-passing
model makes it easier to locate erroneous memory reads and writes. 

But the most compelling reason why message passing will remain a permanent part of 
the parallel computing environment is performance. As modern CPUs have become faster, 
management of their \emph{caches} (divided in many \emph{levels}) and the memory 
hierarchy in general has become the key to getting good performance. Message 
passing provides a way for the programmer to explicitly associate specific data with 
processes and thus allow the compiler and cache-management hardware to function fully. 
Indeed, one advantage distributed-memory computers have over even the largest 
single-processor machines is that they typically provide more memory and more cache. 
Memory-bound applications can exhibit superlinear speedups when ported to such 
machines and, even on shared-memory computers, the message-passing model 
can improve performance by providing more programmer control of data locality in the 
memory hierarchy. 

For these reasons message-passing has emerged as one of the more 
widely used paradigms for implementing parallel algorithms. Although it has shortcomings, 
message-passing comes closer than any other paradigm to being a standard approach 
for the implementation of parallel applications. Message-passing has only recently, however,
become a standard for portability. Before MPI, there were many competing variations on the 
message-passing theme, and programs could only be ported from one system to another 
with difficulty.

\section{Code Parallelization}

The numerical code has been developed to solve equations~(\ref{eq:fou1})-(\ref{eq:fou2}), which
are the reduced MHD equations in Fourier space. As already noted in \S~\ref{sec:ftsd}, at each
time-step the Fourier transforms of the fields and their inverse transforms must be computed
to calculate the Poisson brackets. As shown in (\ref{eq:ft}) and (\ref{eq:ift}) to compute the value
of a Fourier transform at one point along a given direction, we must know the values of the 
function \emph{in all} grid points along that direction. This makes the Fourier transform an 
\emph{intrisically} non-parallel computation. In fact if the values of the function along that 
direction were assigned to different processors, at each time-step all these processors would
have to communicate these values to one processor which would compute the Fourier 
transform. Communications between processors is not very fast and then we want to
minimize it.

Although there are algorithms which parallelize the fast Fourier transform more or less 
efficiently,  as a first step we have avoided this, taking advantage of the fact that we use
a finite difference scheme of the second order along the $z$ direction. These schemes 
are very well suited for parallelization, because to compute the value of the derivative 
of a function in one point we only need to know the values of the function in a few 
neighboring points. In the case of our central finite difference scheme of the second order 
only the values of the \emph{two neighboring points} are required. 
We have decomposed our computational box, which is a parallelepiped 
(see Figure~\ref{fig:loopbox})  of dimensions $ 1 \times 1 \times L $ with a grid of 
$ n \times n \times n_z$ points, into \emph{slices} along the $z$ direction. So the grid points 
lying on a $x$-$y$ plane at $z=\textrm{const}$ are assigned to the same processor and
no communication is required to perform the numerical computations in the plane,
including the FFT. To compute the $z$-derivatives, where the two neighboring points belong 
to the same slice, no communication is needed. This always happens except for the points 
at the top and bottom boundary $x$-$y$ planes of a single slice. From each slice the values 
of the functions in these two boundary planes must be communicated, at each time step,
respectively and exclusively to the processors to which have been assigned the next and 
previous slices of the computational box.

The next task is to decide how to assign processes to each part of the decomposed domain. 
Handling this assignment of processes to regions is one of the services MPI provides to the 
programmer because the best (or even a good) choice of decomposition depends on the 
details of the underlying hardware. As noted in \S~\ref{sec:mpipc}, the processors of a
supercomputer are linked by
an interconnection network through which communication takes place. These 
networks can have many different topologies. Most of these topologies are complicated,
and especially on massively parallel computers, when the code uses many processors
it is very unlikely that each of these processors can communicate directly with all the others.
Given our choice for the decomposition of the computational box, ideally we would like 
the processors which have been assigned to two neighboring slices of the 
computational box could communicate directly between themselves. 
In general the description of how the processes in a parallel computer are connected to 
one another is often called the \emph{topology} of the computer (or more precisely, of the 
interconnection network). In most parallel programs, each process communicates with 
only a few other processes; the pattern of communication is called an \emph{application 
topology} or \emph{virtual topology}. 

It might seem that simply assigning processes  in increasing rank from the 
bottom is the best approach. On some parallel computers, however, this ordering can  
degrade performance. It is hard for anyone but the vendor to know the best way for 
\emph{application} topologies to be fitted onto the \emph{physical} topology of the 
parallel machine. MPI allows the vendor to help optimize this aspect of the programming
through implementation of the MPI \emph{topology functions}.
MPI allows the user to define a particular application, or virtual, topology. An important virtual topology, the \emph{Cartesian topology}, is a decomposition in the natural 
coordinate (e.g.\ $z$) direction. Although topology functions are sometimes treated as an 
exotic and advanced feature of MPI, they make many types of MPI programs easier 
to write. We implemented a one dimensional Cartesian Topology,
and found it very useful. In this way MPI assigns the slices of the decomposed computational 
box to processes such that the topology of the underlying interconnection network and 
the topology of the communication in the numerical code match in the best possible way.

\section{Performance Evaluation}

Two quantities which are commonly used to quantify the performance of a numerical 
code on a parallel machine are the so-called \emph{speedup} and \emph{efficiency}. 

A \emph{parallel system} is defined as the implementation of a parallel algorithm on a 
specific parallel computer. The \emph{dimension of the problem} $W$ is the number
of operations required by the fastest known serial algorithm, and is equivalent to the 
concept of computational complexity. 

We distinguish between the \emph{serial execution time} $T_s$, the time interval between the
beginning and the end of the execution of the program on one single processor, and
the \emph{parallel execution time} $T_p$, the time interval between the beginning 
of the execution and the instant of time in which the last processor ends the execution.
The serial execution time is basically a function of the computational complexity $W$,
while the parallel execution time depends on $W$, the
number of processors used $N_p$, and the kind of interconnection network present 
in the supercomputer.  The speedup $S$ for a parallel system is defined as the ratio of 
the serial to the parallel execution time:
\be \label{eq:su}
S \left( W, N_p \right) \equiv \frac{T_s \left( W \right)}{T_p \left( W, N_p \right)}
\ee
It could be naively thought that if the execution time on a single processor is $T_s$,
then the execution time on $N_p$ processors should be $T_p = T_s / N_p$, in which 
case for the speedup we would obtain the \emph{linear} behavior
\be \label{eq:lsu}
S \left( W, N_p \right) = N_p.
\ee
Unfortunately this relation is rarely found, and when it happens, the reason is not
because $N_p$ processors perfectly distribute among themselves the work-load.

In fact, as already partially remarked, there may be many different sources of overhead.
The first, obvious, source is that when a numerical code is parallelized, it is usually
necessary to change the structure of the code itself, adding some \emph{extra 
computations} with respect to the serial version of the code. Other important sources
of overhead are the communications and the imbalance of the work-load among the processors.
But, especially for computations involving large data-sets (e.g.\ high-resolution
simulations with a lot of grid points), parallelization has the advantage of splitting the computational 
box among many processes so that each process is assigned a relatively small number of 
grid points. In this way each single processor has to execute fewer computations,
but -- more importantly -- since the data occupy a smaller amount of memory, the 
\emph{processors can manage their caches in a more efficient way}. Sometimes the 
cache-management
improves dramatically, leading to so-called \emph{superlinear} speedups, i.e.\ to 
speedup values which exceed the linear behavior~(\ref{eq:lsu}) $S \left( W, N_p \right) > N_p$.

In general the competition between the improved cache-management and the overhead
due to the communication over the interconnection network determines the performance
of a numerical code on a parallel system.

Efficiency is defined as the ratio of the speedup~(\ref{eq:su}) to the number of processors
$N_p$
\be \label{eq:ef}
E \left( W, N_p \right) \equiv \frac{S \left( W, N_p \right)}{N_p}.
\ee
Its value, typically included between zero and one (except for superlinear algorithms for 
which $E > 1$), estimates how efficiently the processors are used, compared to the time 
wasted in communication and synchronization. Algorithms with linear speedup~(\ref{eq:lsu}) 
and those running on a single processor have $E=1$, while many algorithms that are
difficult to parallelize have $E \sim 1/\log N_p$ that approaches zero as the number of processors 
increases.
\begin{figure}[t]
   \centering
   \begin{minipage}[c]{0.45\linewidth}
       \centering \includegraphics[width=1\textwidth]{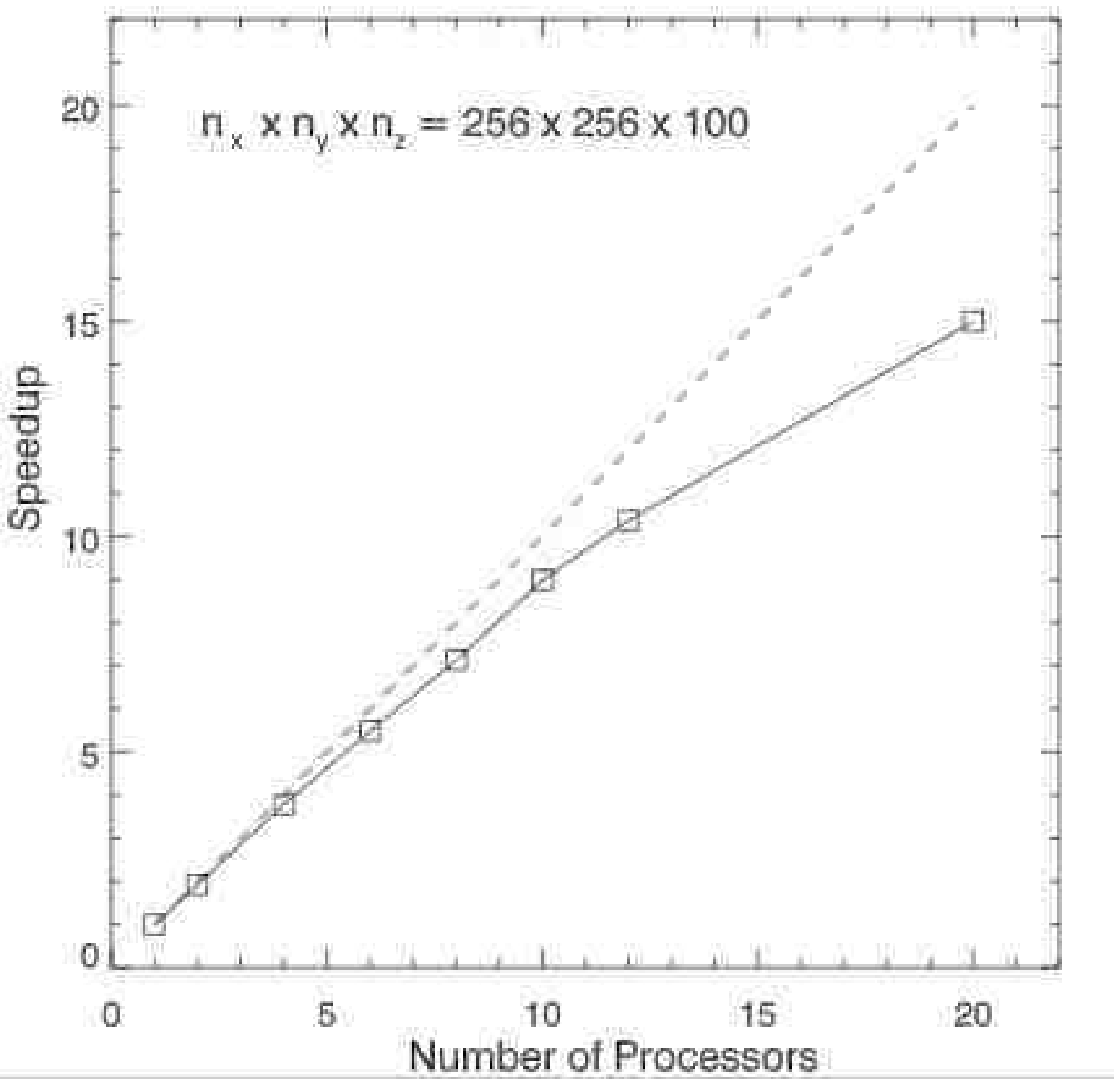}
   \end{minipage}%
   \begin{minipage}[c]{0.45\linewidth}
       \centering \includegraphics[width=1\textwidth]{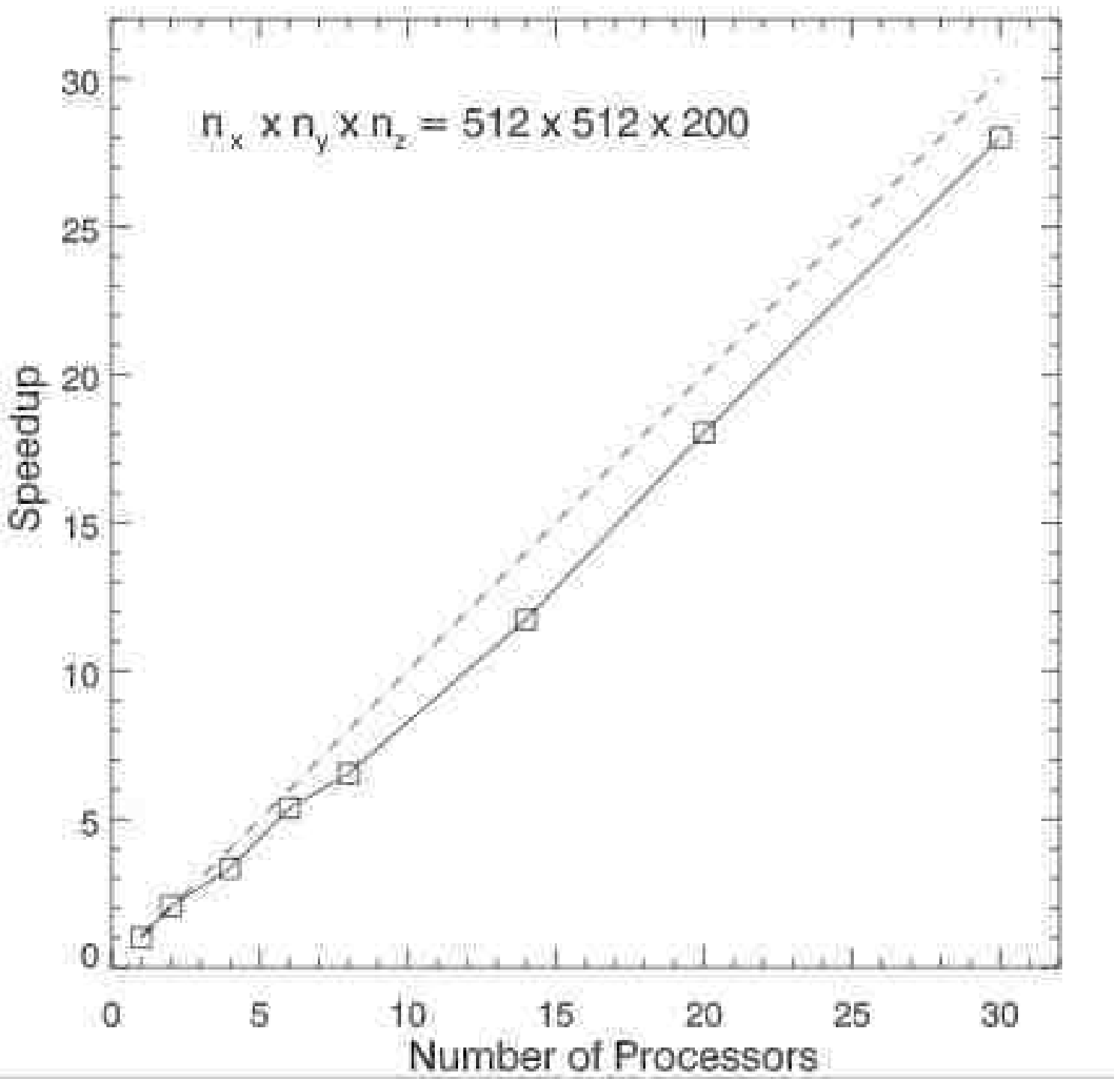}
   \end{minipage}
   \caption{Speedups of our numerical code on the SGI Altix 3000 computer at Naval
                   Research Laboratory (NRL, Washington, DC). The computer has a total of 64 
                   Intel Itanium2 processors at 1.3~GHz of clock speed and 3~MB of cache 
                   memory per processor. The dashed line represents the linear behavior~(\ref{eq:lsu}).
                   On the \emph{left} it is shown the speedup for simulations with a grid of 
                   $n_x \times n_y \times n_z = 256 \times 256 \times 100$ points.
                   On the \emph{right} we show the speedup for simulations with a higher
                   resolution grid with $n_x \times n_y \times n_z = 512 \times 512 \times 200$ 
                   points. The \emph{scalability} of the numerical code increases at higher
                   resolution.
     \label{fig:speedup}  }
\end{figure}

Figure~\ref{fig:speedup} shows two speedup curves for our numerical code relatively to
the SGI Altix 3000 supercomputer at Naval Research Laboratory.
This computer has a total of 64 Intel Itanium2 processors with 1.3~GHz of clock speed 
and 3~MB of cache memory per processor. They are interconnected with a Myrinet network.
In most of the simulations we have performed we have typically used two different numerical
resolutions, a lower resolution grid with $n_x \times n_y \times n_z = 256 \times 256 \times 100$
points and  a higher resolution one with 
$n_x \times n_y \times n_z = 512 \times 512 \times 200$ points. 
The speedup for the lower resolution case is shown on the left in Figure~\ref{fig:speedup}. 
An almost linear behavior is shown up to twelve 
processors, while for an higher number of processors the performance starts to decrease. 
On the other hand using the higher resolution grid, the ratio between computations and
communication grows favorably for computations, and the corresponding speedup
curve, shown on the right in Figure~\ref{fig:speedup}, actually exhibits an approximate 
linear  behavior up to thirty processors. 

The property of a good performance over a wide range of processors is 
called \emph{scalability}. We have achieved our goal of developing a cross-platform portable 
parallel code with a good scalability.

\chapter{Linear Analysis} \label{sec:la}

In this chapter we present the results of our numerical and analytical investigations. 
We make a \emph{Cartesian model} (see \S~\ref{ch:model}) of a coronal loop, i.e.\ the loop 
is ``straightened out'' so that the computational box is a parallelepiped (see 
Figure~\ref{fig:loopbox}) of unit square cross section and axial length $L > 1$ 
(with overall dimensions $1 \times 1 \times L$). The dynamics of the plasma
are studied using the dimensionless equations of \emph{reduced MHD} 
(\ref{pot1})-(\ref{pot2}) that we repeat here for convenience: 
\begin{eqnarray}
& &\frac{\partial \psi}{\partial t} = v_{\mathcal A}\, \frac{\partial \varphi}{\partial z} + 
\left[ \varphi, \psi \right] + \frac{1}{\mathcal R} \boldsymbol{\nabla}^2_\perp 
\psi,  \label{eq:rpot1} \\
& &\frac{\partial \omega}{\partial t} = v_{\mathcal A}\, \frac{\partial j}{\partial z} + 
\left[ j , \psi \right] - \left[ \omega , \varphi \right] 
+  \frac{1}{\mathcal R} \boldsymbol{\nabla}^2_\perp \omega, \label{eq:rpot2} 
\end{eqnarray}
where $v_{\mathcal A}$ is the ratio between the axial Alfv\'enic velocity (associated 
to the strong, uniform, homogeneous field $\bsy{B}_0 = B_0\, \bsy{e}_z$) and 
the typical rms photospheric velocity $u_{ph}$
\be \label{eq:alfs}
v_{\mathcal A} = \frac{B_0}{\sqrt{4\pi \rho_0}} \, \frac{1}{u_{ph}}.
\ee
From observations (see \S~\ref{ch:chp})  we know that the typical photospheric 
magnetic field is of 
the order of $B_0 \sim 10$ \emph{gauss}, the typical numerical electron density
$\sim 10^{10}\, cm^{-3}$ (i.e.\ a mass density 
$\rho_0 \sim 1.7 \cdot 10^{-15}\, g\, cm^{-3}$), and for the photospheric motions
$u_{ph} \sim 1\, km\, s^{-1}$, so the
Alfv\'en velocity is  $\sim 10^3\, km\, s^{-1}$, and the 
dimensionless ratio~(\ref{eq:alfs})  $v_{\mathcal A} \sim 10^3$.

In (\ref{eq:rpot1})-(\ref{eq:rpot2})  we have supposed that the magnetic and kinetic 
Reynolds numbers are equal ($\mathcal{R}_m = \mathcal R$).
The velocity and magnetic \emph{scalar potentials}, respectively $\varphi$ and $\psi$, 
are linked to the orthogonal components of the physical fields by
\be \label{eq:ocpf}
\bsy{u}_{\perp} = \bnabla \times \left( \varphi\, \bsy{e}_z \right), \qquad
\bsy{b}_{\perp} = \bnabla \times \left( \psi\, \bsy{e}_z \right),
\ee
and to the axial component of the electric current ($j$) and of the vorticity ($\omega$) by
\be
\omega = \left( \bnabla \times \bsy{u}_{\perp} \right)_z = - \bnabla^2_{\perp} \varphi, \qquad
            j = \left( \bnabla \times \bsy{b}_{\perp} \right)_z = - \bnabla^2_{\perp} \psi.
\ee
In the orthogonal $x$-$y$ planes periodic boundary conditions
are imposed, while in the $z$ direction at the top ($z=L$) and bottom ($z=0$) 
planes --- that represent the \emph{two photospheric cross-sections} to which a coronal 
loop is anchored --- a velocity pattern is imposed, i.e.\ the velocity potential 
$\varphi$ is specified.

As \textbf{initial conditions} we specify the velocity patterns on the bottom and top  
planes $\varphi(x,y,0)$ and $\varphi(x,y,L)$, and $v_{\mathcal A}$. The potentials 
$\varphi$ and $\psi$ inside the computational domain are set to a small perturbation.
The parallel numerical code (see \S~\ref{sec:nc}) solves the 
Fourier transform of reduced MHD equations (\ref{eq:fou1})-(\ref{eq:fou2}) using
normal diffusion or hyperdiffusion (see equations~(\ref{eq:hyp}) and (\ref{eq:hypf}))
with \emph{dissipativity} $\alpha = 4$.

In \S~\ref{sec:la} we present the linear analysis of our system, which is characterized
by the propagation of Alfv\'en waves, while in \S~\ref{sec:nla} the phenomenology
of the turbulent dynamics which develops in the nonlinear stage is discussed.
Finally in \S~\ref{sec:ns} the numerical simulations are presented.

\section{Boundary Conditions and Linearity}

The velocity forcing imposed at the boundaries produces  Alfv\'en waves that
propagate into the computational box. The scalar fields $\varphi$ and $\psi$ 
associated with these waves are small and since the Poisson brackets are 
quadratic in these terms --- so their contribution is negligible respect to the linear 
terms --- the initial stage of the dynamics is \emph{linear}. Linearizing the 
equations~(\ref{eq:rpot1})-(\ref{eq:rpot2}), and neglecting for 
the moment the dissipative terms, yields
\be \label{eq:rlin} 
\frac{\partial \psi}{\partial t} = v_{\mathcal A}\, \frac{\partial \varphi}{\partial z},  
\qquad
\frac{\partial \varphi}{\partial t} = v_{\mathcal A}\, \frac{\partial \psi}{\partial z}.
\ee
Introducing $z^{\pm}$, the analogs of the Els\"asser variables~(\ref{eq:elsv})
for the scalar potentials:
\be
z^{\pm} = \varphi \pm \psi
\ee
equations~(\ref{eq:rlin}) can be written as
\be \label{eq:zlin}
\frac{\partial z^{\pm}}{\partial t} = \pm v_{\mathcal A}\, \frac{\partial z^{\pm}}{\partial z}  
\ee
These wave equations show that $z^+$ describes a wave propagating towards
negative $z$, while $z^-$ waves propagating towards positive $z$. 
In terms of these Els\"asser variables, the boundary conditions need be expressed 
only for the incoming
wave, i.e.\ $z^-$ at the bottom ($z=0$) and $z^+$ at the top ($z=L$). The imposition of
the velocity potential patterns $\varphi^0$ and $\varphi^L$ respectively  in $z=0$ and 
$z=L$  is then achieved by the \emph{reflection condition}:
\begin{eqnarray}
& & z^-  |_{z=0} = \left( - z^+  + 2 \varphi \right) |_{z=0} = 
- z^+ |_{z=0} + 2 \varphi^0(x,y,t),  \label{eq:vpb1}   \\
& & z^+ |_{z=L} = \left( - z^-   + 2 \varphi \right) |_{z=L} = 
- z^- |_{z=L} +2 \varphi^L(x,y,t), \label{eq:vpb2}
\end{eqnarray}
Introducing the column-vector $\bsy U$ and the square matrix $\bsy A$
\be
\bsy{U} = 
\begin{pmatrix}
z^+  \\
z^-
\end{pmatrix},
\qquad \qquad
\bsy{A} = v_{\mathcal A}
\begin{pmatrix}
1 &  0 \\
0 & -1 
\end{pmatrix}
\ee
equations~(\ref{eq:zlin}) can be written as
\be \label{eq:U}
\frac{\partial \bsy U}{\partial t} = \bsy{A} \frac{\partial \bsy{U} }{\partial z}  
\ee
indicating with $\mathcal L$ the operator:
\be
\mathcal{L} = \frac{\partial }{\partial t} - \bsy{A} \frac{\partial  }{\partial z}  
\ee
linear equations~(\ref{eq:U}), or equivalently~(\ref{eq:zlin}) can be written as
$\mathcal{L} \left( \bsy U \right) = 0$.
Because of linearity, if $\bsy U_1$ and $\bsy U_2$ are solutions, then
$\alpha \bsy U_1 + \beta \bsy U_2$ is also a solution.
If $\bsy U_1$, $\bsy U_2$ have respectively the 
stream-functions $\varphi^0_1$, $\varphi^L_1$ and $\varphi^0_2$, $\varphi^L_2$ as
boundary velocity potentials~(\ref{eq:vpb1})-(\ref{eq:vpb2}),
then a simple calculations shows that the boundary conditions satisfied
by the new solution 
\be \label{eq:bsum1}
\bsy U = 
\begin{pmatrix}
z^+  \\
z^-
\end{pmatrix}
= \alpha \bsy U_1 + \beta \bsy U_2 = 
\begin{pmatrix}
\alpha z^+_1 + \beta z^+_2  \\
\alpha z^-_1 + \beta z^-_2
\end{pmatrix}
\ee
are, indicating with $\varphi^0$, $\varphi^L$ the boundary stream-functions of
the solution $\bsy U$:{\setlength\arraycolsep{-2pt}
\begin{eqnarray}
& & \varphi^0 = \frac{1}{2}\, \left( z^+ + z^- \right) |_{z=0} = 
\frac{1}{2}\, \left[ \alpha \left( z^+_1 + z^-_1 \right) + 
\beta \left( z^+_2 + z^-_2 \right) \right]_{z=0} =
\alpha \varphi^0_1 + \beta \varphi^0_2  \label{eq:bd1} \\
& & \varphi^L = \frac{1}{2}\,\left( z^+ + z^- \right) |_{z=L} =  
\frac{1}{2}\, \left[ \alpha \left( z^+_1 + z^-_1 \right) + 
\beta \left( z^+_2 + z^-_2 \right) \right]_{z=L} =
\alpha \varphi^L_1 + \beta \varphi^L_2   \label{eq:bd2}
\end{eqnarray}
}which is simply the linear combination of the boundary conditions for the solutions
$\bsy{U}_1$ and $\bsy{U}_2$. We will use the property~(\ref{eq:bd1})-(\ref{eq:bd2}) many 
times in the 
following paragraphs. In particular, if $\bsy U_1$ is the \emph{bottom plane forced} solution 
with $\varphi^0_1 = \varphi^0$ , $\varphi^L_1 = 0$, and $\bsy U_2$ is the \emph{top plane
forced} solution with $\varphi^0_2 = 0$ , $\varphi^L_2 = \varphi^L$ then the function 
$\bsy U = \bsy U_1 + \bsy U_2$ is the solution that satisfies the boundary conditions 
\be
\left( z^+ + z^- \right) |_{z=0} =  2 \varphi^0, \qquad
\left( z^+ + z^- \right) |_{z=L} =  2 \varphi^L.  \label{eq:bsum2}
\ee
This result allows us to investigate the two \emph{one-sided} solutions separately,
the \emph{two-sided} solution results simply from their sum.

\section{One-sided Problem: Time Independent Forcing} \label{sec:osp}

Considering at first only \emph{time-independent} velocity patterns, calling 
$\tau_{\mathcal A} = L / v_{\mathcal A}$
the crossing time of an Alfv\'en wave for the axial length $L$, 
the solution to the linear equations~(\ref{eq:zlin}) with the top boundary forcing
\be \label{eq:bf1}
\left( z^+ + z^- \right) |_{z=0} =  0, \qquad
\left( z^+ + z^- \right) |_{z=L} = 2 \varphi^L \left( x, y \right),
\ee
is given by
\begin{eqnarray}
& & \textrm{for} \quad 2k\, \tau_{\mathcal A} < t < \left( 2k + 1 \right)\, \tau_{\mathcal A} 
\qquad k=0,1,2,\dots \label{eq:timel} \\
& & z^+ \left( x,y,z,t \right) = 2k\, \varphi^L + 2\varphi^L\, \Theta_{\leftarrow} \label{eq:timel2} \\
& & z^- \left( x,y,z,t \right)  = - 2k\, \varphi^L \label{eq:timel3}
\end{eqnarray}
while
\begin{eqnarray}
& & \textrm{for} \quad \left( 2k + 1 \right)\, \tau_{\mathcal A} < t < \left( 2k + 2 \right)\, \tau_{\mathcal A} 
\qquad k=0,1,2,\dots \label{eq:timer} \\
& & z^+ \left( x,y,z,t \right) =   2k\, \varphi^L + 2\varphi^L  \label{eq:timer2} \\
& & z^- \left( x,y,z,t \right)  = - 2k\, \varphi^L - 2\varphi^L\, \Theta_{\rightarrow}  \label{eq:timer3}
\end{eqnarray}
$\Theta_{\leftarrow}$ and $\Theta_{\rightarrow}$ are step functions ``propagating'' towards
negative and positive $z$, starting from one side and reaching the other one distant
$L$ in the time $\tau_{\mathcal A}$. For example, indicating with $\Theta$ the step 
function
\be
\Theta \left( z -z_0 \right) = \left\{
\begin{array}{ll}
1 & z > z_0 \\
0 & z < z_0
\end{array}
\right.
\ee
referring to the time interval~(\ref{eq:timel})
\be
\Theta_{\leftarrow} \left( z, z_0 \right) = \Theta \left( z - z_0 \right) \qquad \textrm{where} \qquad
z_0 \left( t \right) = L - v_{\mathcal A} \left( t - 2k\tau_{\mathcal A} \right)
\ee
so that the point $z_0$ moves from $z_0 = L$ at time $t = 2k\, \tau_{\mathcal A}$ to
$z_0 = 0$ at time $t = \left( 2k + 1 \right)\, \tau_{\mathcal A}$, at the speed $v_{\mathcal A}$.
In the same fashion, referring to the time interval~(\ref{eq:timer}) 
\be
\Theta_{\rightarrow} \left( z, z_0 \right) = \Theta \left( z_0 - z \right) \qquad \textrm{where} \qquad
z_0 \left( t \right) = v_{\mathcal A} \left[ t - \left( 2k + 1 \right) \tau_{\mathcal A} \right]
\ee
so that the point $z_0$ moves from $z_0 = 0$ at time $t = ( 2k + 1 ) \tau_{\mathcal A}$ to
$z_0 = L$ at time $t= \left( 2k + 2 \right)\, \tau_{\mathcal A}$, at the speed $v_{\mathcal A}$.

To understand why the solution to the problem~(\ref{eq:zlin}) with the one-sided boundary
condition~(\ref{eq:bf1}) has the structure shown in (\ref{eq:timel})-(\ref{eq:timer3}), consider
the first interval of time in equation~(\ref{eq:timel}) $0  < t <  \tau_{\mathcal A}$.
At time $t=0$ the boundary condition  $z^+ = 2\varphi^L$ in $z=L$ 
causes a $z^+$ wave to propagate towards negative $z$, while the boundary condition 
$z^- = 0$ in $z=0$ does not excite any $z^-$ wave to propagate towards positive $z$.
Hence the solution during this first time interval is:
\be
z^+ \left( x,y,z,t \right) = 2\varphi^L\, \Theta_{\leftarrow}\ , \qquad
z^- \left( x,y,z,t \right)  = 0
\ee
which are exactly equations~(\ref{eq:timel2})-(\ref{eq:timel3}) for $k=0$.
When at time $t = \tau_{\mathcal A}$ the $z^+$ wave reaches the $z=0$ boundary
conditions~(\ref{eq:bf1}) read:
\be
z^- |_{z=0} = - z^+  |_{z=0} =  -2\varphi^L, \qquad
z^+ |_{z=L} =  - z^- |_{z=L}  + 2 \varphi^L =  +2 \varphi^L
\ee
which cause the $z^+$ wave to continue propagating towards negative $z$, and 
a new $z^-$ wave to originate from $z=0$ (at time $t=\tau_{\mathcal A}$) and 
propagate towards positive $z$ (reaching $z=L$ at time $t = 2\tau_{\mathcal A}$).
This is described by equations~(\ref{eq:timer2})-(\ref{eq:timer3}) for $k=0$, i.e.\
\be
z^+ \left( x,y,z,t \right) =   2\varphi^L \ ,  \qquad
z^- \left( x,y,z,t \right)  = - 2\varphi^L\, \Theta_{\rightarrow}
\ee
The iteration of these arguments finally leads to equations~(\ref{eq:timel})-(\ref{eq:timer3}).

Solution~(\ref{eq:timel})-(\ref{eq:timer3}) shows that ``reflection'' boundary 
conditions~(\ref{eq:bf1}) result in a wave propagation in our computational
box that causes the amplitude of the scalar fields $z^{\pm}$ to grow of the value
$2\varphi^L$ at each interval of time corresponding to an Alfv\'enic crossing
time $\tau_{\mathcal A}$. In terms of the velocity and magnetic potentials
equations~(\ref{eq:timel})-(\ref{eq:timer3}) yield:
\begin{eqnarray}
& & \textrm{for} \quad 2k\, \tau_{\mathcal A} < t < \left( 2k + 1 \right)\, \tau_{\mathcal A} 
\qquad k=0,1,2,\dots \label{eq:pl1} \\
& & \varphi \left( x,y,z,t \right) = \varphi^L\, \Theta_{\leftarrow} \label{eq:pl2} \\
& & \psi \left( x,y,z,t \right)  = 2k\, \varphi^L + \varphi^L\, \Theta_{\leftarrow} \label{eq:pl3}
\end{eqnarray}
and
\begin{eqnarray}
& & \textrm{for} \quad \left( 2k + 1 \right)\, \tau_{\mathcal A} < t < \left( 2k + 2 \right)\, \tau_{\mathcal A} 
\qquad k=0,1,2,\dots \label{eq:pr1} \\
& & \varphi \left( x,y,z,t \right) =   \varphi^L - \varphi^L\, \Theta_{\rightarrow}  \label{eq:pr2} \\
& & \psi \left( x,y,z,t \right)  = 2k\, \varphi^L + \varphi^L + \varphi^L\, \Theta_{\rightarrow}  \label{eq:pr3}
\end{eqnarray}
From these equations we note that while the velocity potential $\varphi$ is a step
function that bounces back and forth with the amplitude $\varphi^L$, the magnetic
potential is characterized by a propagating wave and grows roughly linearly
in time gaining a factor $\varphi^L$ at each time-interval $\tau_{\mathcal A}$.
From equations~(\ref{eq:ocpf}) we obtain for the velocity and magnetic fields:
\begin{eqnarray}
& & \textrm{for} \quad 2k\, \tau_{\mathcal A} < t < \left( 2k + 1 \right)\, \tau_{\mathcal A} 
\qquad k=0,1,2,\dots \label{eq:plf1} \\
& & \bsy{u}_{\perp} \left( x,y,z,t \right) = \bsy{u}^L \left( x,y \right)\, \Theta_{\leftarrow} 
\label{eq:plf2} \\
& & \bsy{b}_{\perp} \left( x,y,z,t \right)  = 2k\, \bsy{u}^L \left( x,y \right)\, +
 \bsy{u}^L \left( x,y \right)\, \Theta_{\leftarrow} \label{eq:plf3}
\end{eqnarray}
and
\begin{eqnarray}
& & \textrm{for} \quad \left( 2k + 1 \right)\, \tau_{\mathcal A} < t < \left( 2k + 2 \right)\, \tau_{\mathcal A} 
\qquad k=0,1,2,\dots \label{eq:prf1} \\
& & \bsy{u}_{\perp} \left( x,y,z,t \right) =   \bsy{u}^L \left( x,y \right)\, - 
\bsy{u}^L \left( x,y \right)\, \Theta_{\rightarrow}  \label{eq:prf2} \\
& & \bsy{b}_{\perp}  \left( x,y,z,t \right)  = \left( 2k + 1\right) \bsy{u}^L \left( x,y \right)\, + 
\bsy{u}^L \left( x,y \right)\, \Theta_{\rightarrow}  
\label{eq:prf3}
\end{eqnarray}
where $\bsy{u}^L \left( x,y \right)$ is the velocity forcing imposed at the top boundary
plane $z=L$. Equations~(\ref{eq:plf1})-(\ref{eq:prf3}) show that this velocity forcing 
results in a mapping of the
boundary velocity inside the computational box for both the magnetic and velocity fields.
The difference is that while the velocity is bounded to its boundary value $\bsy{u}^L$,
the magnetic field does not satisfy this condition and grows linearly in time, its amplitude
gaining a factor $\bsy{u}^L$ at each Alfv\'enic crossing time $\tau_{\mathcal A}$.

\section{Physical Interpretation} \label{sec:pi}

The \emph{physical interpretation} of equations~(\ref{eq:plf1})-(\ref{eq:prf3}) is simple.
The photospheric motions imposed at the boundary plane cause the footpoints of the
magnetic field-lines (which are almost straight because of the strength of the dominant
axial magnetic field) to move. As a result Alfv\'en waves, of amplitude equal to the
velocity forcing $\bsy{u}^L$, are launched into the computational box and reflected at 
the boundaries. Because of the boundary conditions the magnetic field grows roughly
linearly in time, while the velocity field is bounded to its boundary value.
\emph{Like in the Parker picture (see \S~\ref{sec:lfm}), the shuffling of the field-lines 
footpoints causes the field-lines to braid. But differently from that picture
we make explicit from equations~(\ref{eq:plf1})-(\ref{eq:prf3}) 
that the braiding of the magnetic field-lines takes place through Alfv\'en wave 
propagation.} This is a phenomenon that
naturally happens for a system embedded in a strong axial field, in both the linear
and nonlinear stages. On the other hand this simple observation leads 
to one of the main original results of this thesis: since the field-line braiding is due
to propagating Alfv\'en waves then \emph{the nonlinear
stage dynamics (which will be analized in \S~\ref{sec:nla}) is described by 
anisotropic turbulence (see \S~\ref{par:bik})
due to the ``collisions'' of counter-propagating Alfv\'en waves packets.}

\section{Two-sided Problem: Time Independent Forcing} \label{sec:tsp}

In \S~\ref{sec:osp} we have obtained the solution for the linearized 
equations~(\ref{eq:zlin}) with a velocity forcing at the top boundary
\be
\left( \bsy{z}^+ + \bsy{z}^- \right) |_{z=0} =  0, \qquad
\left( \bsy{z}^+ + \bsy{z}^- \right) |_{z=L} = 2 \bsy{u}^L \left( x, y \right),
\ee
Using the same techniques, the solution with a velocity forcing at the
bottom boundary
\be
\left( \bsy{z}^+ + \bsy{z}^- \right) |_{z=0} =  2 \bsy{u}^0, \qquad
\left( \bsy{z}^+ + \bsy{z}^- \right) |_{z=L} =  0,
\ee
is the following:
\begin{eqnarray}
& & \textrm{for} \quad 2k\, \tau_{\mathcal A} < t < \left( 2k + 1 \right)\, \tau_{\mathcal A} 
\qquad k=0,1,2,\dots \label{eq:pbf1} \\
& & \bsy{u}_{\perp} \left( x,y,z,t \right) = \bsy{u}^0 \left( x,y \right)\, \Theta_{\rightarrow} 
\label{eq:pbf2} \\
& & \bsy{b}_{\perp} \left( x,y,z,t \right)  = - 2k\, \bsy{u}^0 \left( x,y \right)\, 
- \bsy{u}^0 \left( x,y \right)\, \Theta_{\rightarrow} \label{eq:pbf3}
\end{eqnarray}
and
\begin{eqnarray}
& & \textrm{for} \quad \left( 2k + 1 \right)\, \tau_{\mathcal A} < t < \left( 2k + 2 \right)\, \tau_{\mathcal A} 
\qquad k=0,1,2,\dots \label{eq:pbf4} \\
& & \bsy{u}_{\perp} \left( x,y,z,t \right) =   \bsy{u}^0 \left( x,y \right)\, - 
\bsy{u}^0 \left( x,y \right)\, \Theta_{\leftarrow}  \label{eq:pbf5} \\
& & \bsy{b}_{\perp}  \left( x,y,z,t \right)  = - \left( 2k + 1\right) \bsy{u}^0 \left( x,y \right)\,  
- \bsy{u}^0 \left( x,y \right)\, \Theta_{\leftarrow}  
\label{eq:pbf6}
\end{eqnarray}
And finally, the general two-sided solution with velocity patterns $\bsy{u}^0$
at the bottom boundary plane $z=0$ and $\bsy{u}^L$ at the top boundary plane
$z=L$, is given by
\begin{eqnarray}
& & \textrm{for} \quad 2k\, \tau_{\mathcal A} < t < \left( 2k + 1 \right)\, \tau_{\mathcal A} 
\qquad k=0,1,2,\dots \label{eq:ts1} \\
& & \bsy{u}_{\perp} \left( x,y,z,t \right) = 
\bsy{u}^0 \left( x,y \right)\, \Theta_{\rightarrow} 
+ \bsy{u}^L \left( x,y \right)\, \Theta_{\leftarrow} 
\label{eq:ts2} \\
& & \bsy{b}_{\perp} \left( x,y,z,t \right)  = 
 2k\, \left[ \bsy{u}^L \left( x,y \right)\, - \bsy{u}^0 \left( x,y \right)\, \right]  
- \bsy{u}^0 \left( x,y \right)\, \Theta_{\rightarrow} 
+ \bsy{u}^L \left( x,y \right)\, \Theta_{\leftarrow} 
\label{eq:ts3}
\end{eqnarray}
and{\setlength\arraycolsep{-2pt}
\begin{eqnarray}
& & \textrm{for} \quad \left( 2k + 1 \right)\, \tau_{\mathcal A} < t < \left( 2k + 2 \right)\, \tau_{\mathcal A} 
\qquad k=0,1,2,\dots \label{eq:ts4} \\
& & \bsy{u}_{\perp} \left( x,y,z,t \right) =   
\bsy{u}^0 \left( x,y \right)\, + \bsy{u}^L \left( x,y \right)\, 
- \bsy{u}^0 \left( x,y \right)\, \Theta_{\leftarrow}  
- \bsy{u}^L \left( x,y \right)\, \Theta_{\rightarrow}  
\label{eq:ts5} \\
& & \bsy{b}_{\perp}  \left( x,y,z,t \right)  = 
\left( 2k + 1\right) \left[ \bsy{u}^L \left( x,y \right) - \bsy{u}^0 \left( x,y \right) \right]  
- \bsy{u}^0 \left( x,y \right)\, \Theta_{\leftarrow}  
+ \bsy{u}^L \left( x,y \right)\, \Theta_{\rightarrow}  
\label{eq:ts6}
\end{eqnarray}}

\section{One-sided Problem: Time Dependent Forcing} \label{sec:osptd}

Solutions (\ref{eq:timel})-(\ref{eq:timer3})  can be generalized to the case of a 
time-dependent forcing:{\setlength\arraycolsep{2pt}
\begin{eqnarray}
& & \textrm{for} \quad 2k\, \tau_{\mathcal A} < t < \left( 2k + 1 \right)\, \tau_{\mathcal A} 
\qquad k=0,1,2,\dots \label{eq:gl1} \\
& & z^+ \left( x,y,z,t \right) = + \sum_{n=0}^{k-1} 
2\, \varphi^L \left( t -2n\tau_{\mathcal A} - \frac{L-z}{v_{\mathcal A}} \right) \nonumber\\
& & \qquad \qquad \qquad \qquad \qquad \qquad \qquad \qquad \qquad
+ 2\, \varphi^L \left( t -2k\tau_{\mathcal A} - \frac{L-z}{v_{\mathcal A}} \right) 
\Theta_{\leftarrow} \label{eq:gl2} \\
& & z^- \left( x,y,z,t \right)  = - \sum_{n=1}^{k} 
2\, \varphi^L \left( t - \left( 2n - 1 \right) \tau_{\mathcal A} - \frac{z}{v_{\mathcal A}} \right)
 \label{eq:gl3}
\end{eqnarray}
}and{\setlength\arraycolsep{2pt}
\begin{eqnarray}
& & \textrm{for} \quad \left( 2k + 1 \right) \tau_{\mathcal A} < t < \left( 2k + 2 \right)\, \tau_{\mathcal A} 
\qquad k=0,1,2,\dots \label{eq:gl4} \\
& & z^+ \left( x,y,z,t \right) = + \sum_{n=0}^{k} 
2\, \varphi^L \left( t -2n\tau_{\mathcal A} - \frac{L-z}{v_{\mathcal A}} \right)
\label{eq:gl5} \\
& & z^- \left( x,y,z,t \right)  = - \sum_{n=1}^{k} 
2\, \varphi^L \left( t - \left( 2n - 1 \right) \tau_{\mathcal A} - \frac{z}{v_{\mathcal A}} \right) \nonumber\\
& & \qquad \qquad \qquad \qquad \qquad \qquad \qquad \qquad \qquad
- 2\, \varphi^L \left( t - \left( 2k + 1 \right) \tau_{\mathcal A} - \frac{z}{v_{\mathcal A}} \right) 
\Theta_{\rightarrow}
\label{eq:gl6}
\end{eqnarray}
}These equations describe the propagation and reflection of the Alfv\'en waves originating
from the boundary $z=L$. It can be easily checked that when the stream-function $\varphi^L$ 
is time-independent equations~(\ref{eq:gl1})-(\ref{eq:gl6}) coincide with 
(\ref{eq:timel})-(\ref{eq:timer3}).

Most of the energy of photospheric motions is at low frequency ($\sim 3.3\, mHz$) and 
at spatial scales of $\sim 1000\, km$. But a small fraction of their energy is 
also present at smaller spatial scales and higher frequencies.
It is therefore interesting to perform an analysis of a Fourier component
\be \label{eq:famp}
\varphi^L \left( x, y, t \right) = f \left( x, y \right) \cos \left( \omega\, t \right).
\ee
Substituting this expression in the previous equations~(\ref{eq:gl1})-(\ref{eq:gl6})
we note that, in general, the waves which are continuously
injected and reflected in the computational box will be out of phase, so their sum
will be of the order of $f$. The only exception to this behaviour is when they are in phase.
This happens when the terms that are integer multiples of $\omega\, \cdot 2\, \tau_{\mathcal A}$, 
arising in the
cosine terms from the substitution~(\ref{eq:famp}) in (\ref{eq:gl1})-(\ref{eq:gl6}) and which are
the only ones to differentiate among them the terms in the summations, are equal to 
an integer multiple of $2\pi$, so that they can be removed. 
The term $\omega \cdot 2n\tau_{\mathcal A}$ with $n$ integer to be a integer multiple of 
$2\pi$ requires the ``resonant'' condition
\be
\omega_m\, \tau_{\mathcal A} = m \pi, \qquad \textrm{with} \qquad m \in \mathbb{N}.
\ee
to be satisfied. Using $\nu_{\mathcal A} = 1/ \tau_{\mathcal A}$, the corresponding 
frequencies 
$\nu = \omega / 2\pi$ are given by
\be
\nu_m = \frac{m}{2}\, \nu_{\mathcal A} = 0, \ \frac{1}{2}\, \nu_{\mathcal A}, \ \nu_{\mathcal A}, \ \dots,
\qquad \textrm{with} \qquad m \in \mathbb{N}.
\ee
At the frequency $\omega_m\,  = m \pi / \tau_{\mathcal A}$ equations~(\ref{eq:gl1})-(\ref{eq:gl6}) 
yield:{\setlength\arraycolsep{2pt}
\begin{eqnarray}
& & \textrm{for} \quad 2k\, \tau_{\mathcal A} < t < \left( 2k + 1 \right)\, \tau_{\mathcal A} 
\qquad k=0,1,2,\dots \label{eq:ris1} \\
& & z^+ \left( x,y,z,t \right) =  +2\, \left( -1 \right)^m\, \left( k\, + \Theta_{\leftarrow} \right) f \left( x, y \right) 
\cos \left( \omega_m\,  t + \omega_m\,  \frac{z}{v_{\mathcal A}} \right) 
\label{eq:ris2} \\
& & z^- \left( x,y,z,t \right)  = -2\, \left( -1 \right)^m\, k\, f \left( x, y \right) 
\cos \left( \omega_m\,  t - \omega_m\,  \frac{z}{v_{\mathcal A}} \right) 
\label{eq:ris3}
\end{eqnarray}
}while{\setlength\arraycolsep{2pt}
\begin{eqnarray}
& & \textrm{for} \quad \left( 2k + 1 \right) \tau_{\mathcal A} < t < \left( 2k + 2 \right)\, \tau_{\mathcal A} 
\qquad k=0,1,2,\dots \label{eq:ris4} \\
& & z^+ \left( x,y,z,t \right)  = +2\, \left( -1 \right)^m\, \left( k + 1\right) f \left( x, y \right) 
\cos \left( \omega_m\,  t + \omega_m\,  \frac{z}{v_{\mathcal A}} \right) \label{eq:ris5} \\
& & z^- \left( x,y,z,t \right) = - 2\, \left( -1 \right)^m\, \left( k\, + \Theta_{\rightarrow} \right) f \left( x, y \right) 
\cos \left( \omega_m\,  t - \omega_m\,  \frac{z}{v_{\mathcal A}} \right) \label{eq:ris6}
\end{eqnarray}
}In terms of potentials, remembering that $\cos \left( \alpha \pm \beta \right) = \cos \alpha \cos \beta
\mp \sin \alpha \sin \beta$ we have{\setlength\arraycolsep{2pt}
\begin{eqnarray}
& & \textrm{for} \quad 2k\, \tau_{\mathcal A} < t < \left( 2k + 1 \right)\, \tau_{\mathcal A} 
\qquad k=0,1,2,\dots \label{eq:risp1} \\
& & \varphi \left( x,y,z,t \right) =  -2\, \left( -1 \right)^m\, k\, f \left( x, y \right) 
\sin \left( \omega_m\,  t \right) \sin \left( \omega_m\,  \frac{z}{v_{\mathcal A}} \right) \nonumber\\
& & \qquad \qquad \qquad \qquad \qquad \qquad \qquad 
+ \left( -1 \right)^m\, f \left( x, y \right) 
\Theta_{\leftarrow} \, \cos \left( \omega_m\,  t + \omega_m\,  \frac{z}{v_{\mathcal A}} \right)
\label{eq:risp2} \\
& & \psi \left( x,y,z,t \right)  = + 2\, \left( -1 \right)^m\, k\, f \left( x, y \right) 
\cos \left( \omega_m\,  t \right) \cos \left( \omega_m\,  \frac{z}{v_{\mathcal A}} \right) \nonumber\\
& & \qquad \qquad \qquad \qquad \qquad \qquad \qquad 
+ \left( -1 \right)^m\, f \left( x, y \right) 
\Theta_{\leftarrow} \, \cos \left( \omega_m\,  t + \omega_m\,  \frac{z}{v_{\mathcal A}} \right)
\label{eq:risp3}
\end{eqnarray}
}and{\setlength\arraycolsep{2pt}
\begin{eqnarray}
& & \textrm{for} \quad \left( 2k + 1 \right) \tau_{\mathcal A} < t < \left( 2k + 2 \right)\, \tau_{\mathcal A} 
\qquad k=0,1,2,\dots \label{eq:risp4} \\
& & \varphi \left( x,y,z,t \right) =  -2\, \left( -1 \right)^m\, k\, f \left( x, y \right) 
\sin \left( \omega_m\,  t \right) \sin \left( \omega_m\,  \frac{z}{v_{\mathcal A}} \right) \nonumber\\
& & \qquad \qquad \quad
+ \left( -1 \right)^m\, f \left( x, y \right) \left[ \cos  \left( \omega_m\,  t + \omega_m\,  \frac{z}{v_{\mathcal A}} \right)
- \Theta_{\rightarrow} \, \cos \left( \omega_m\,  t - \omega_m\,  \frac{z}{v_{\mathcal A}} \right) \right]
\label{eq:risp5} \\
& & \psi \left( x,y,z,t \right) =  +2\, \left( -1 \right)^m\, k\, f \left( x, y \right) 
\cos \left( \omega_m\,  t \right) \cos \left( \omega_m\,  \frac{z}{v_{\mathcal A}} \right) \nonumber\\
& & \qquad \qquad \quad
+ \left( -1 \right)^m\, f \left( x, y \right) \left[ \cos  \left( \omega_m\,  t + \omega_m\,  \frac{z}{v_{\mathcal A}} \right)
+ \Theta_{\rightarrow} \, \cos \left( \omega_m\,  t - \omega_m\,  \frac{z}{v_{\mathcal A}} \right) \right]
\label{eq:risp6}
\end{eqnarray}
}The main result of these calculations is that, when present,
these so-called \emph{resonances} grow \emph{linearly} in time with
their amplitude proportional to time and to the forcing function~(\ref{eq:famp})
$f \left( x, y \right)$, i.e.\ to the amplitude of the boundary motions at the corresponding
resonant frequency. This implies that \emph{as long as the energy fraction
of the high frequency components is small, their contribution to the linear
dynamics, and hence also to the nonlinear dynamics, is and remain small.}

\section{Effects of Diffusion}

In all the previous calculations we have neglected the effect of diffusion. 
In general diffusion is not very important when the Reynolds numbers
are very high but numerically we use lower values so that it can affect 
the solutions that we have found (even at large scales). In particular diffusion
is more important when the linear stage lasts
for long times. The mathematical algebra is simplified if we consider
the physical fields averaged in time, on scales bigger than the Alfv\'enic crossing time 
$\tau_{\mathcal A}$. For example from equations~(\ref{eq:plf1})-(\ref{eq:prf3}) we have
\begin{equation} \label{eq:flin}
\boldsymbol{u}_{\perp}  \sim  \boldsymbol{u}^{L} 
 \left( x, y \right) \ \frac{z}{L} , 
\qquad
\boldsymbol{b}_{\perp}   \sim  \boldsymbol{u}^{L}
 \left( x, y \right)  \  \frac{t}{\tau_{\mathcal A}}
\end{equation}
While diffusion will only slightly change the shape in $z$ of the velocity 
field in~(\ref{eq:flin}), it has a stronger effect on the magnetic field which
would otherwise grow linearly in time. Here we describe briefly such effect.
Linearizing equation~(\ref{eq:rpot1}) retaining the diffusive term, in terms
of the magnetic and velocity fields we have
\begin{equation} \label{eq:lind2}
\frac{\partial \boldsymbol{b}_{\perp} }{\partial t} 
=  v_\mathcal{A} \frac{\partial  \boldsymbol{u}_{\perp}}{\partial z}
+ \frac{1}{\mathcal R} \boldsymbol{ \nabla}^2_{\perp}  \boldsymbol{b}_{\perp}
\end{equation}
We use the general boundary conditions $\bsy{u}_{\perp} = \bsy{u}^0$, 
$\bsy{u}_{\perp} =  \bsy{u}^L$ 
respectively in $z=0$ and $z=L$, but now we take into account that our
forcing velocity has only components at the injection scale. In dimensionless
form, the cross-section or our numerical box is a unit square of lenght $\ell = 1$,
and the injection scale is $\ell_{in} \sim 1/4$. For example,
one of the forcing patterns we use is $\bsy{u}^L = \bsy{e}_y\, sin(2\pi r\, x + 1)$ 
with $r=4$, corresponding to the wavelength $\lambda = \ell_{in} = 1/4$. For this 
pattern the relation 
$\boldsymbol{\nabla}^2_{\perp}  \boldsymbol{u}^L = - \left( 2 \pi r \right)^2 
\boldsymbol{u}^L$ is exact, but it is still approximately valid
when we consider velocity patterns which result from the linear combination
of components with $\lambda \sim 1/4$, i.e.\ $r \sim 4$.
For the integration of equation~(\ref{eq:lind2}) we use the \emph{ansatz}
$\boldsymbol{\nabla}^2_{\perp}  \boldsymbol{b}_{\perp} = 
- \left( 2 \pi r \right)^2 \boldsymbol{b}_{\perp} $. This is justified 
because the magnetic field is always the result of the mapping of the
boundary velocity forcing, as found in the calculations without diffusion.
Then integrating~(\ref{eq:lind2}) over $z$, dividing by the length $L$, 
using the boundary conditions $\bsy{u}_{\perp} ( z=0) = \bsy{u}^0$ and 
$\bsy{u}_{\perp} ( z=L) = \bsy{u}^L$, we obtain for the field 
$\boldsymbol{b}_{\perp}$ averaged in $z$:
\begin{equation} \label{eq:eqdiff}
\frac{\partial \boldsymbol{b}_{\perp} }{\partial t} 
=  \frac{v_\mathcal{A}}{L} 
\left[ \bsy{u}^L \left( x, y \right) - \bsy{u}^0 \left( x, y \right) \right]
- \frac{\left( 2 \pi r \right)^2}{\mathcal R}  \boldsymbol{b}_{\perp},
\end{equation}
and indicating with $\bsy{u}_{ph} = \bsy{u}^L - \bsy{u}^0$, and with 
$\tau_{\mathcal R} = \mathcal{R}/ \left( 2 \pi  r \right)^2$ the diffusive
time-scale and with $\tau_{\mathcal A} = L / v_{\mathcal A}$ the
Alfv\'enic crossing time,  the solution is given by:
\begin{equation} \label{eq:bdiff}
\boldsymbol{b}_{\perp} \left( x, y, t \right) =
\boldsymbol{u}_{ph} \left( x, y \right)
\frac{\tau_{\mathcal R}}{\tau_{\mathcal A}}
\left[ 1 - \exp  \left( - \frac{t}{\tau_{\mathcal R}} \right) \right]
\end{equation}
\begin{equation}
\left| j \left( x, y, t \right) \right| =
\left| \bsy{u}_{ph} \left( x, y \right) \right|
\left( 2 \pi r \right) 
\frac{\tau_{\mathcal R}}{\tau_{\mathcal A}}
\left[ 1 - \exp  \left( - \frac{t}{\tau_{\mathcal R}} \right) \right]
\end{equation}
So that the magnetic energy $E_M$ and the ohmic dissipation $J$ rate are given by
\begin{equation} \label{eq:emdiff}
E_M =  \frac{1}{2} \int_V \mathrm{d}^3 \boldsymbol{x} \ \bsy{b}_{\perp}^2 = 
\frac{1}{2} \ell^2\, L\, u^2_{ph}\,
\left( \frac{\tau_{\mathcal R}}{\tau_{\mathcal A}} \right)^2
\left[ 1 - \exp  \left( - \frac{t}{\tau_{\mathcal R}} \right) \right]^2
\end{equation}
\begin{equation} \label{eq:hrdiff}
J =  \int_V \mathrm{d}^3 \boldsymbol{x} \ \frac{j^2}{\mathcal R} = 
\ell^2\, v_{\mathcal A}\, u^2_{ph}\,
\frac{\tau_{\mathcal R}}{\tau_{\mathcal A}}
\left[ 1 - \exp  \left( - \frac{t}{\tau_{\mathcal R}} \right) \right]^2
\end{equation}
where $u_{ph}$ is the rms of $\bsy{u}_{ph}$, and with the rms of the
boundary velocities $\bsy{u}^0$ and $\bsy{u}^L$ fixed to $1/2$ we have 
$u_{ph} \sim 1$.
Equation~(\ref{eq:hrdiff}) implies that the heating rate for unitary volume grows 
quadratically in time for 
a time small compared to the resistive time $\tau_{\mathcal R}$, 
while on this diffusive time scale the heating rate 
saturates to a value that is proportional to the value of the Reynolds 
number and the square of the axial Alfv\'enic velocity.

\section{Energy Balance}

As shown in \S~\ref{sec:cl} the energy equation to which our system obeys is
\be \label{eq:eneq}
\frac{\partial E}{\partial t} = S - \left( J +  \Omega \right)
\ee
where{\setlength\arraycolsep{-4pt}
\begin{eqnarray}
& & E = \frac{1}{2}  \int_V \! \ud^{3}x 
                                   \left( \bsy u_{\perp}^2 + \bsy b_{\perp}^2   \right), \qquad
J = \frac{1}{R}  \int_V \! \ud^{3}x\, \bsy{j}^2, \qquad 
\Omega = \frac{1}{R}  \int_V \! \ud^{3}x\, \bsy{\omega}^2,  \\
& & S = \oint  \! \ud a\, \bsy{S} \cdot \bsy{n} = 
+ v_{\mathcal A} \int_{z=L} \! \ud a\,   \bsy{u}^L \cdot \bsy b_{\perp}
-  v_{\mathcal A} \int_{z=0} \! \ud a\,   \bsy{u}^0 \cdot \bsy b_{\perp},
\label{eq:pfb}
\end{eqnarray}
}$E$ is the total energy, $J$ and $\Omega$ are respectively the ohmic and viscous
dissipation rates. $S$ is the surface integral of the Poynting flux $\bsy S$~(\ref{eq:pf})
\be \label{eq:pf1}
\bsy S = \bsy B \times \left( \bsy u \times \bsy B \right) = {\bsy{B}}^2 \bsy u
               - \left( \bsy B \cdot \bsy u \right) \bsy B 
\ee
In reduced MHD (see \S~\ref{sec:rmhd}) the fields are decomposed in orthogonal
and axial components $\bsy{B} = v_{\mathcal A}\, \bsy{e}_z + \bsy b_{\perp}$, 
$\bsy u = \bsy u_{\perp}$. The only dynamical fields are the orthogonal components of the
magnetic and velocity fields $\bsy{b}_{\perp}$ and $\bsy{u}_{\perp}$, and in particular the 
velocity is supposed
to have  a vanishing axial component (at least in the first order contribution of the expansion
described in \S~\ref{sec:rmhd}, in particular see~(\ref{eq:vacn})).
So the only boundaries of the computational box that contribute to the surface integral $S$
of the Poynting flux are
the top ($z=L$, with velocity $\bsy{u}^L$) and bottom ($z=0$, with velocity $\bsy{u}^0$) 
planes, while the contributions from the other planes 
cancel  each other because of periodicity. This implies that the only term on the right hand 
side of equation~(\ref{eq:pf1}) that gives a contribution is the last one, because the other one
has vanishing component along $z$. The axial component of $\bsy S$
is then given by
\be \label{eq:pfz}
S_z = \bsy S \cdot \bsy{e}_ z = - v_{\mathcal A} \left( \bsy b_{\perp} \cdot \bsy u_{\perp} \right) 
\ee
and the total flux of energy from the boundaries $S$ is given by~(\ref{eq:pfb}), where
the signs have been chosen so that $S$ is positive when we have energy entering
the system and negative when it leaves.

One of the results of the anisotropic turbulence theory and an hypothesis of the 
reduced MHD expansion is that the spatial structures are elongated in the axial direction,
because the cascade is strongly reduced in that direction. As shown in the 
linear analysis (see \S~\ref{sec:osp}) \emph{the boundary conditions limit the 
velocity to its boundary value, while the magnetic field grows to  
a higher value (see equations~(\ref{eq:plf1})-(\ref{eq:prf3})), so that in both the
linear and nonlinear stages the magnetic field is the dominant field and 
its variation along the $z$ direction is weak}.

We can so approximate the magnetic field $\bsy{b}_{\perp}$ as uniform
along the axial direction in the flux integral~(\ref{eq:pf1}):
\be
S \sim v_{\mathcal A} \int \! \ud a\,  
\left( \bsy u^{L} - \bsy u^{0} \right) \cdot \bsy b_{\perp}
\ee
and indicating with $\bsy{u}_{ph} = \bsy{u}^{L} - \bsy{u}^{0}$, with $\bsy{u}^L \ne
\bsy{u}^0$ but 
$\int \! \ud a\, \left| \bsy{u}^L \right|^2 =  \int \! \ud a\, \left| \bsy{u}^0 \right|^2 =  1/2$, 
we have
\be \label{eq:pfc}
S \sim v_{\mathcal A} \int \! \ud a\, \bsy{u}_{ph} \cdot \bsy b_{\perp}
\ee
2D spatial periodicity in the orthogonal
planes allows us to expand the velocity and magnetic fields in Fourier series, e.g.\
\be \label{eq:2dfe}
\bsy{u} \left( x, y \right) = \sum_{r,s} \bsy{u}_{r,s}\, e^{i \bsy{k}_{r,s} \cdot \bsy{x}},
\qquad \textrm{where} \qquad \bsy{k}_{r,s} = \frac{2\pi}{\ell} \left( r, s, 0 \right)
\ee
Then using this expansion in~(\ref{eq:pfc}) we have
\begin{equation} \label{eq:inr}
S \sim 
v_{\mathcal A} \sum_{r,s} \left( \bsy{u}_{ph} \right)_{r,s}  \cdot 
\int_0^{\ell} \! \! \! \int_0^{\ell} \mathrm{d}x  \mathrm{d}y \
\boldsymbol{b}_{\perp} e^{i \bsy{k}_{r,s} \cdot \bsy{x}} =
\ell^2\, v_{\mathcal A} \sum_{r,s}  \left( \bsy{u}_{ph} \right)_{r,s} \cdot \bsy{b}_{-r,-s}
\end{equation}
This equation shows that energy flux entering the system is proportional to the
sum of the scalar product of the Fourier amplitudes of the velocity and magnetic
fields at the same absolute wavenumber $k_{r,s}$.
From observations we know
that most of the energy of photospheric motions is at the characteristic scale
of a granule while shorter wavelengths have a reduced power level. Supposing
that a turbulent cascade takes place, this implies that the Fourier components
of the magnetic field $\bsy{b}_{r,s}$ are smaller at higher  wavenumbers.
In this way \emph{both the velocity and magnetic field amplitudes entering in the 
sum~(\ref{eq:inr}) decrease at high wavenumbers, so that the contributions
to the flux $S$ from scales smaller than the injection one are noticeably 
smaller and we neglect them}.
\begin{figure}[t]
\begin{center}
\includegraphics[width=0.6\textwidth]{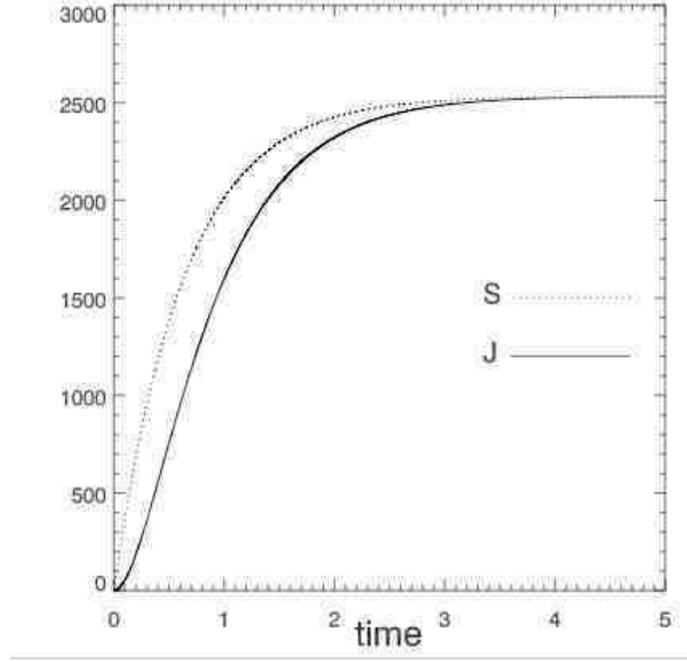}
\caption{Poynting flux integral~S, representing the energy rate entering the box, and
the dissipation rate~J. The parameters used are $\ell = 1$, $L=10$, $u_{ph} = 1$, 
$v_{\mathcal A} = 200$ and $\mathcal{R} = 400$ to which correspond
an Alfv\'enic crossing time $\tau_{\mathcal A} = L / v_{\mathcal A} = 5\cdot 10^{-2}$
and a diffusive time $\tau_{\mathcal R} = 400/(4\pi \cdot 4)^2 \sim 6.3\cdot 10^{-1}$, 
and a saturation level $J_{sat} = S_{sat} \sim 2.5\cdot 10^2$.
 \label{fig:selfdiff}}
\end{center}
\end{figure}

We now use the linear solution~(\ref{eq:ts1})-(\ref{eq:ts6}) for the two-sided problem with 
velocity patterns $\bsy{u}^0$ and $\bsy{u}^L$, respectively in the boundary planes
$z=0$ and $z=L$. Averaging over time-scales bigger than the Alfv\'enic crossing time,
the magnetic field for this solution can be approximated as
\be
\bsy{b}_{\perp}  \left( x,y,z,t \right)  \sim \frac{t}{\tau_{\mathcal A}}\,
\left[ \bsy{u}^L \left( x,y \right) - \bsy{u}^0 \left( x,y \right) \right]  
\sim \frac{t}{\tau_{\mathcal A}}\, \bsy{u}_{ph}
\ee
and from~(\ref{eq:pfb}) we can write for the injection energy rate
\be
S \sim 
v_{\mathcal A} \int \! \ud a\,  \left( \bsy{u}^L - \bsy{u}^0 \right) \cdot \bsy b_{\perp}
\sim  
v_{\mathcal A} \int \! \ud a\,  \left| \bsy{u}^L - \bsy{u}^0 \right|^2 
\frac{t}{\tau_{\mathcal A}} \sim  
\ell^2\, v_{\mathcal A}\, u_{ph}^2\, \frac{t}{\tau_{\mathcal A}}
\ee
so that during the linear stage the injection energy rate is always positive.

Taking into account diffusion, from (\ref{eq:eqdiff})-(\ref{eq:bdiff}) we have that
\be
\bsy{b}_{\perp} \sim \bsy{u}_{ph}\, \frac{\tau_{\mathcal R}}{\tau_{\mathcal A}}
\left[ 1 - \exp \left( - \frac{t}{\tau_{\mathcal R}} \right) \right]
\ee
so that from~(\ref{eq:pfb})
\be
S \sim 
\ell^2\, v_{\mathcal A}\, u_{ph}^2\, \frac{\tau_{\mathcal R}}{\tau_{\mathcal A}}\,
\left[ 1 - \exp \left( - \frac{t}{\tau_{\mathcal R}} \right) \right]
\ee
For times smaller than the diffusive time $t \ll \tau_{\mathcal R}$ the injection
of energy rate $S$ is alway bigger than the dissipation rate $J$~(\ref{eq:hrdiff})
\be
S \sim \ell^2\, v_{\mathcal A}\, u_{ph}^2\, \frac{t}{\tau_{\mathcal A}},
\qquad
J \sim \ell^2\, v_{\mathcal A}\, u_{ph}^2\, \frac{1}{\tau_{\mathcal A}}\,
\frac{t^2}{\tau_{\mathcal R}}
\ee
until for $t \gg \tau_{\mathcal R}$ they both saturate to the value
\be
J_{sat} = S_{sat} =
\ell^2\, v_{\mathcal A}\, u_{ph}^2\, \frac{\tau_{\mathcal R}}{\tau_{\mathcal A}}
\ee
as shown in Figure~\ref{fig:selfdiff}.

\chapter{Numerical Simulations} \label{sec:ns}

Since Parker~\cite{park72,park91} proposed his nano-flare scenario for coronal heating
a number of numerical experiments have been carried out to investigate
this idea, with particular emphasis on its possible relationship 
with the power-law distribution of observed emission events at optical, 
ultraviolet and x-ray wavelengths of the quiet solar corona.
Mikic~et~al.~\cite{mik89}, Longcope \& Sudan~\cite{long94} and 
Hendrix \& Van~Hoven~\cite{hen96} carried out simulations using a 3D 
``straightened out'' loop
and imposing photospheric shearing given by alternate direction flow patterns. 
They showed that a complex coronal magnetic field results from the 
photospheric field line tangling and although the field does not --- strictly 
speaking --- evolve through a sequence of static force-free equilibrium states, 
magnetic energy nonetheless tends to dominate the kinetic energy. In this
system, which may be thought of as evolving in a special regime of MHD 
turbulence, the field is restructured into current sheets elongated along 
the axial direction.

The numerical investigation of the Parker scenario has been challenging
since the beginning. The reasons are the large aspect ratio of a coronal loop,
and because the two different time-scales --- along the axial direction
(the Alfv\'enic crossing time $\tau_{\mathcal A} \sim L/v_{\mathcal A}$) and 
the perpendicular direction (the non-linear time $T_{NL} \sim \ell / u_{ph}$), 
with $\tau_{\mathcal A} \ll T_{NL}$ --- together require a large number of grid points. 
High-resolution simulations are also a necessity when investigating
turbulence since this  takes place at high Reynolds numbers. 
Turbulent systems are in fact characterized by a \emph{transition}
to the turbulent state which is not detected at small Reynolds numbers. 
A limitation of the first simulations carried out during the late 1980s and
the early 1990s was their low resolution. While the results of these simulations
mostly agreed among themselves, the interpretations that were given did not, 
and there was not a clear evidence in favor of one of them.

To investigate the turbulent nature of this model with higher resolution, and to
 carry out long-time simulations necessary to define the 
statistics of heating events, Einaudi~et~al.~\cite{ein96} first carried out 2D numerical 
simulations of incompressible magnetohydrodynamics (MHD) turbulence using 
a random large scale magnetic forcing function to mimic the forcing exerted in 
three dimensions by the photosphere. The simulations displayed flux 
tube behaviour similar to the first 3D simulations, confirmed 
also in 2D simulations by Dmitruk~et~al.~\cite{dim98}. These 2D simulations  were 
extended to longer times by  Georgoulis~et~al.~\cite{georg98}, and showed for the 
first time how the magnetically dominated turbulence in the 2D system displays 
bursts in the dissipation (\emph{intermittency})
 which follow a power law behaviour in total energy, peak 
dissipation, and duration with indices similar to those determined
observationally in X-rays. 

Recently Gudiksen \& Nordlund~\cite{gud02,gud05} have performed numerical
simulations of the Parker scenario, modeling a large part of the solar corona
with a more realistic geometry. While this approach has advantages when 
investigating the coronal loop dynamics within its neighboring coronal
region, modeling a larger part of the solar corona drastically reduces the number
of points occupied by the coronal loops. Thus, these  simulations have not been 
able to shed further light on the \emph{physical mechanism} responsible
for the coronal heating.

In order to further investigate the underlying physical processes at work, 
we return to the simpler cartesian 3D model, performing numerical simulations
with the highest resolution and longest duration to date.
In section \S~\ref{sec:vvp} we describe the numerical simulations performed with
a vortex-like velocity pattern and in \S~\ref{sec:svp} those performed with
a sheared velocity pattern. In \S~\ref{sec:nla} we analyze in the framework
of anisotropic turbulence theory the results of the simulations.
\begin{table}[t]
\begin{center}
\begin{tabular}{c c r r c r c r}
\hline\hline
Run No. &    v, s &      $v_{\mathcal A}$           &$n_x \times n_y \times n_z$ 
& n, h              & $\mathcal{R}$,  $\mathcal{R}_4$    & $\mathcal{R}_{\parallel}$           
& $t_{max}/\tau_{\mathcal A}$ \\
\hline
1   & v   &    200   &   $512 \times 512 \times 200$   &   n    &     800              & $\infty$  &    508 \\
2   & v   &    200   &   $256 \times 256 \times 100$   &   n    &     400              & $\infty$  &  1061 \\
3   & v   &    200   &   $128 \times 128 \times   50$   &   n    &     200                   & 10     &   2172 \\
4   & v   &      50   &   $512 \times 512 \times 200$   &   h    & $3\cdot10^{20}$ & 10      &    196 \\
5   & v   &    200   &   $512 \times 512 \times 200$   &   h    &    $10^{19}$         & 10     &    453 \\
6   & v   &    400   &   $512 \times 512 \times 200$   &   h    &    $10^{20}$         & 10     &      77 \\
7   & v   &   1000  &   $512 \times 512 \times 200$   &   h    &    $10^{19}$         & 10     &    502 \\
8   & s   &    200   &   $512 \times 512 \times 200$   &   n    &     800              & $\infty$  &    551 \\
9   & s   &    200   &   $256 \times 256 \times 100$   &   n    &     400              & $\infty$  &  4123 \\
10 & s   &    200   &   $128 \times 128 \times   50$   &   n    &     200              & $\infty$  &  2356 \\
\hline
\hline
\end{tabular}
        \caption{Summary of the simulation runs. v or s indicate the kind of boundary velocity forcing
        pattern: vortex or shear. $v_{\mathcal A}$ is the axial Alfv\'enic velocity value and 
        $n_x \times n_y \times n_z$ is number of points for the numerical grid.
        n, h indicate normal or hyperdiffusion. $\mathcal{R}$ or  $\mathcal{R}_4$ indicates respectively
        the Reynolds number value and the hyperdiffusion coefficient, while $\mathcal{R}_{\parallel}$
        is the Reynolds number along the axial direction where always normal diffusion is used.
        The duration of the simulation $t_{max}/\tau_{\mathcal A}$ is given in Alfv\'enic crossing 
        time unit $\tau_{\mathcal A} = L / v_{\mathcal A}$.
        \label{tab:r}}
\end{center}
\end{table}

%\section{Vortex-like Velocity Pattern: Time Independent Forcing} \label{sec:vvp}

\section{Vortex-like Velocity Pattern} \label{sec:vvp}

In this paragraph we show the results of 3D numerical simulations modeling
a coronal layer driven by a forcing velocity pattern constant in time. On the 
bottom and top planes (respectively $z=0$ and $z=L$) we impose two 
independent velocity forcings, resulting from the linear combination
of large-scale eddies.  The velocity potential at each boundary is given by
\be \label{eq:frand}
\varphi (x,y) = \frac{1}{\sqrt{ \sum_{mn} \alpha_{mn}^2}} \ \sum_{k,l}
            \frac{\alpha_{kl}}{2\pi \sqrt{k^2+l^2}} \ 
            \sin \left[ 2 \pi \left( kx+ly \right) + 2\pi \, \xi_{kl} \right]
\ee
where $ 0 \le x, y \le 1 $ and $ 0 \le z \le L$, with $L=10$ (corresponding
 aspect ratio equal to $10$).
The wave number values $(k,l)$ and $(m,n)$ used are all those in the range
$3 \le \left( k^2 + l^2 \right)^{1/2} \le 4$, so that the injection wavenumber is
$k_{in} \sim 4$ and the corresponding injection scale $\ell_{in} \sim 1/4$.
$\alpha_{kl}$ and $\xi_{kl}$ are two sets of random numbers whose values
range between 0 and 1
\be \label{eq:rampl}
0 \le \alpha_{kl}, \xi_{kl} < 1,
\ee
and are independently chosen for the two boundary surfaces. 
\begin{figure}[t]
      \centering
      %%---- start ----
      \subfloat[]{
               \label{fig:atsfv:a}             %% label for subfigure
               \includegraphics[width=0.45\linewidth]{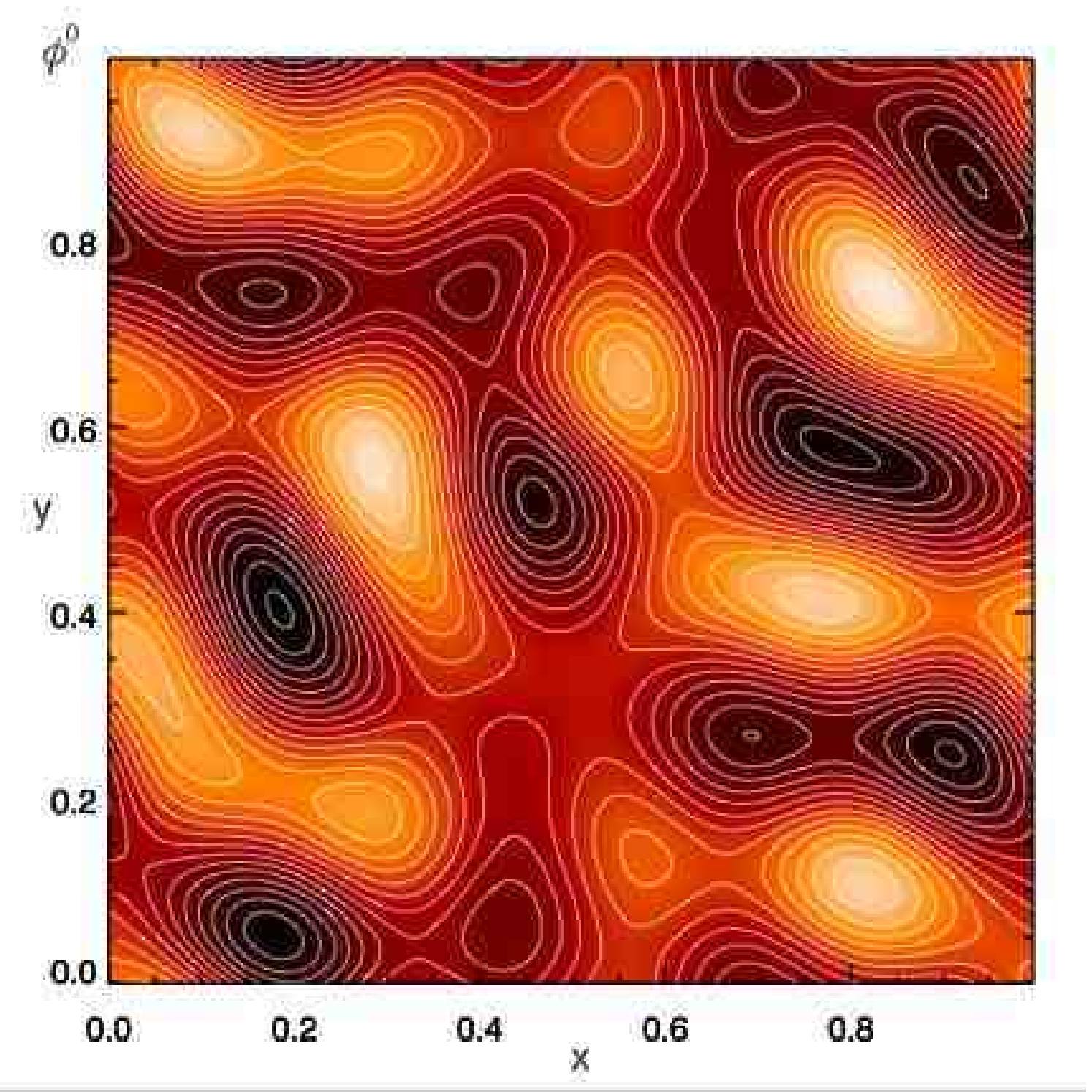}}
      \hspace{0.01\linewidth}
     %%---- start ----
      \subfloat[]{
               \label{fig:atsfv:b}             %% label for subfigure
               \includegraphics[width=0.45\linewidth]{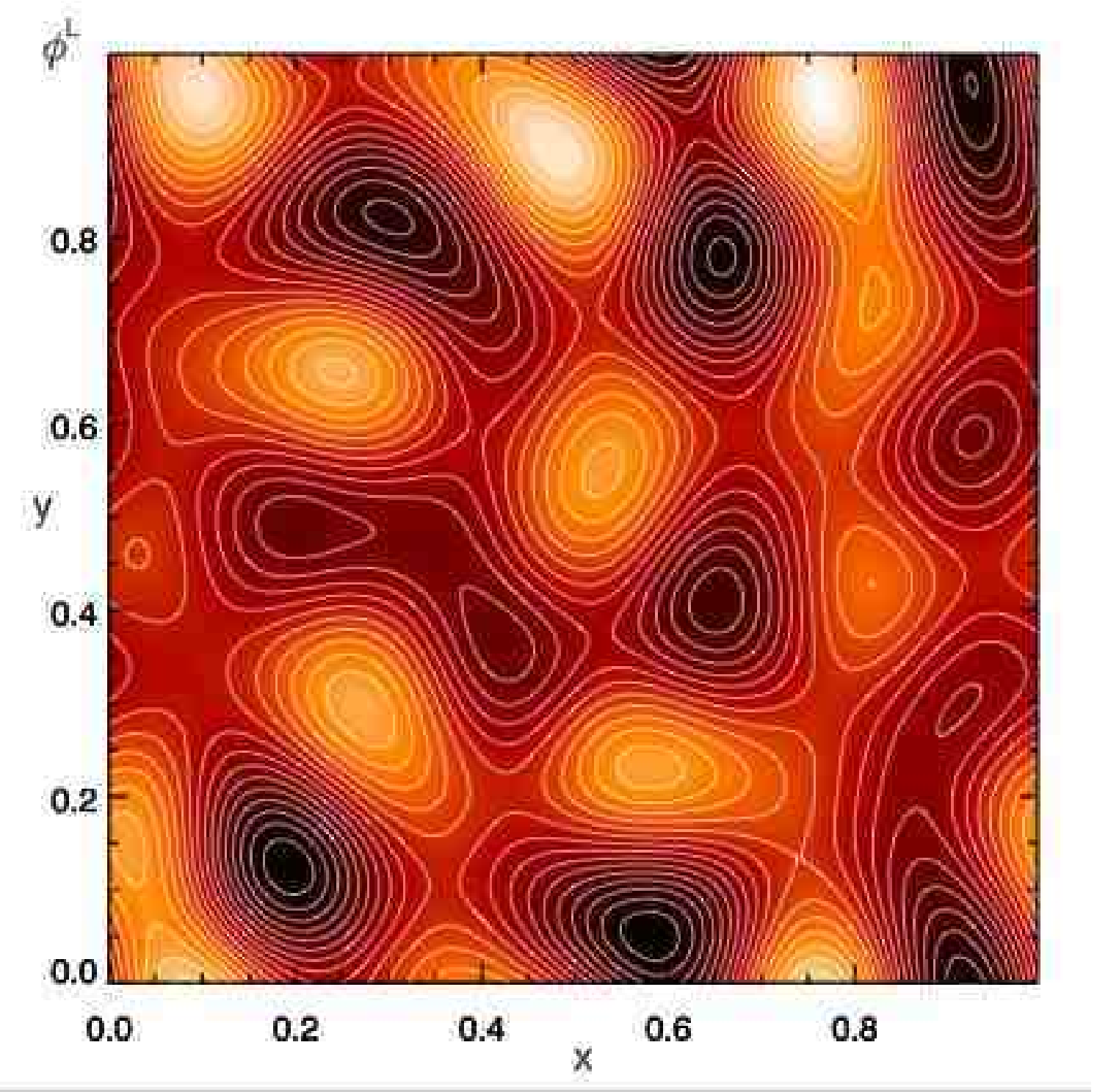}}
   \caption{Boundary conditions for our numerical simulations are
   specified imposing the velocity potentials $\varphi^0 (x,y)$ in the bottom
   plane $z=0$ and $\varphi^L (x,y)$ in the top plane $z=L$. 
   These result from the linear combination of large-scale eddies 
   (wavelength $\sim 1/4$) with random amplitudes~(\ref{eq:frand}),
   normalized so that the velocity rms is $1/\sqrt{2}$.
   Figure shows the contours of the velocity potential in $z=0$ (a) and
   $z=L$ (b) for the numerical simulation with $v_{\mathcal A} = 200$
   and $\mathcal R = 800$. In lighter vortices the velocity field is 
   directed anti-clockwise while in darker vortices it is directed
   clockwise.
     \label{fig:atsfv}}             %% label for entire figure
\end{figure}
The velocity field $\bsy{u}_{\perp}$ follows from equation~(\ref{eq:ocpf}), and the
normalization in equation~(\ref{eq:frand}) has been chosen so that the rms value
of the corresponding velocity is $1/\sqrt{2}$, i.e.\
\be \label{eq:rmscon}
\int_0^{1} \! \! \! \int_0^{1} \mathrm{d}x\,  \mathrm{d}y \
\left( u_x^2 + u_y^2 \right)  = \frac{1}{2}
\ee
The initial condition is given by the strong uniform axial field specified by the 
value of the corresponding Alfv\'enic speed $v_{\mathcal A}$~(\ref{eq:alfs}).
We have performed simulations with different values of $v_{\mathcal A}$.

\subsection{Simulations with $v_{\mathcal A} = 200$}

We start showing the results of a numerical simulation 
performed with $v_{\mathcal A} = 200$, a numerical grid with 
$n_x \times n_y \times n_z = 512 \times 512 \times 200$ points and a Reynolds 
number $\mathcal R = 800$ and $\mathcal R_{\parallel} = 10$. 
In Figure~\ref{fig:atsfv} we show the two boundary velocity 
patterns (equations~(\ref{eq:frand})) for this simulation which result from 
a specific random choice of the amplitudes~(\ref{eq:rampl}).
The total duration is roughly $500$ axial Alfv\'enic crossing times 
($\tau_A = L / v_{\mathcal A}$). To remark how challenging these kind
of simulations are this simulation has used $52.428.800$ grid points
and $814.215$ time-steps which, using a $3^{rd}$ order runge-kutta,
correspond to $2.442.645$ substeps in which the derivatives have been
computed in all the grid points.
\begin{figure}[t]
      \centering
      %%---- start ----
      \subfloat[]{
               \label{fig:atsendiss:a}             %% label for subfigure
               \includegraphics[width=0.45\linewidth]{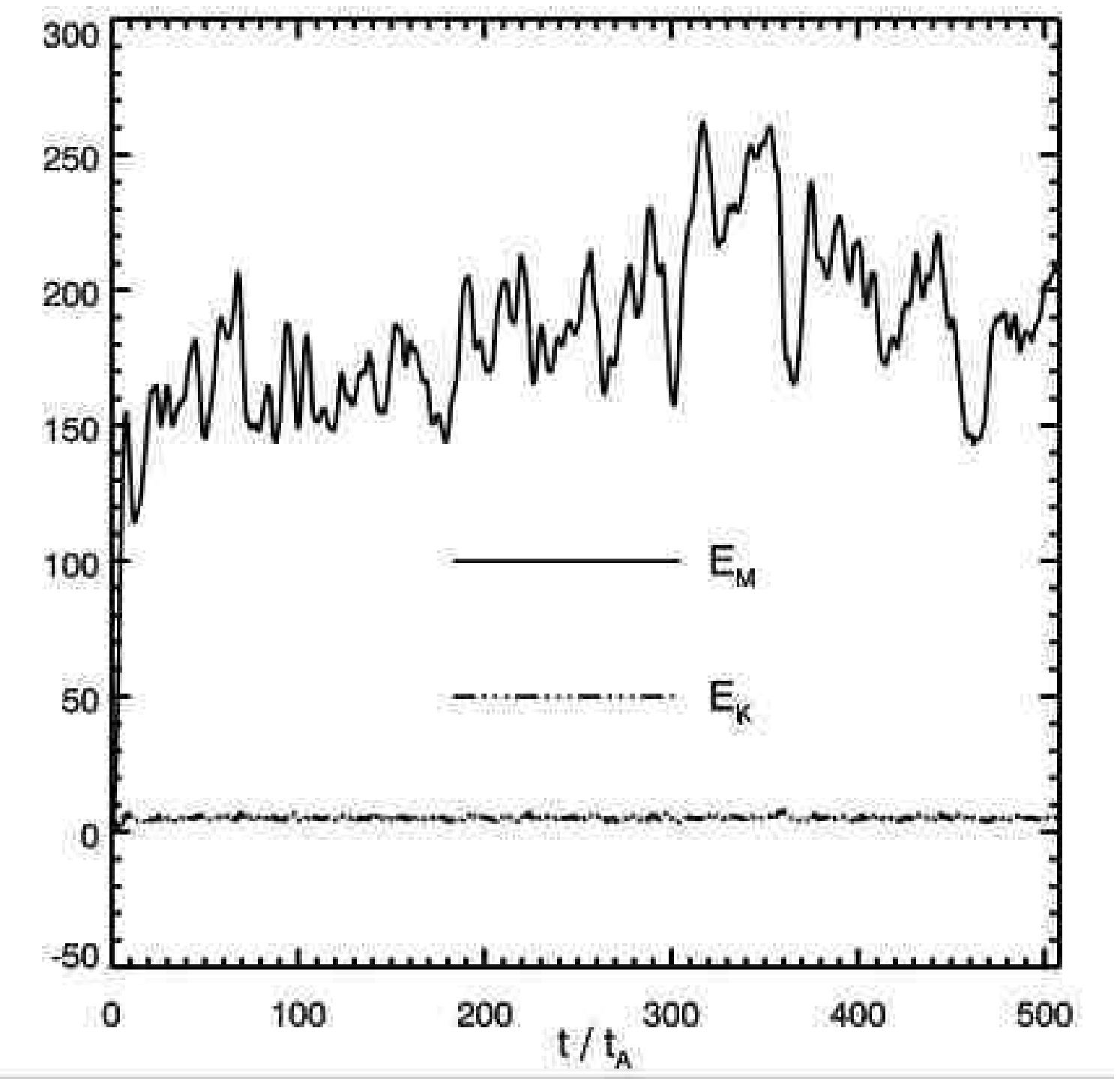}}
      \hspace{0.01\linewidth}
     %%---- start ----
      \subfloat[]{
               \label{fig:atsendiss:b}             %% label for subfigure
               \includegraphics[width=0.45\linewidth]{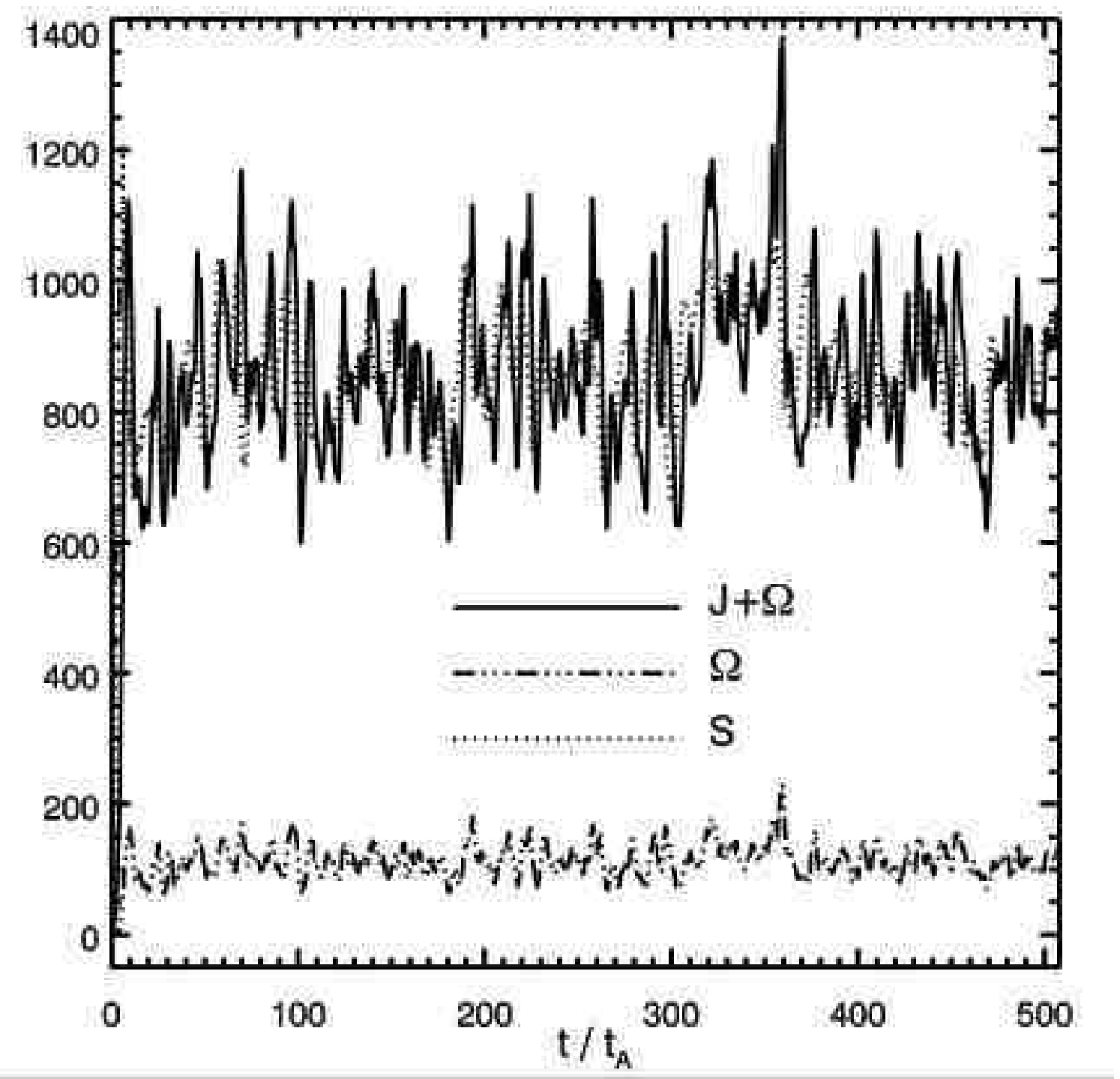}}
   \caption{High-resolution simulation with $v_{\mathcal A} = 200$, 512x512x200 
                grid points and $\mathcal{R}=800$, $\mathcal{R}_{\parallel}=10$.
                (a)~Magnetic ($E_M$) and kinetic ($E_K$) energies as a function 
                of time ($\tau_{\mathcal A}=L/v_{\mathcal A}$ is the axial Alfv\'enic 
                crossing time).  
                Kinetic energy is always a small fraction of total energy.
                (b)~Poynting flux $S$ dynamically balances the 
                Ohmic (J) and viscous ($\Omega$) dissipation.
                 Similar to kinetic energy dynamics, also
                enstrophy ($\Omega$) is always a small fraction of total dissipation,
                confirming that the system is magnetically dominated.
     \label{fig:atsendiss}}             %% label for entire figure
\end{figure}
Fig.~\ref{fig:atsendiss:a} shows the total magnetic ($E_M$) and kinetic ($E_K$) energies 
in the loop as a function of time, and Fig.~\ref{fig:atsendiss:b} shows both ohmic ($J$) and 
viscous ($\Omega$) dissipations and the Poynting flux $S$ (the energy that for unit time is 
injected into the system). 
All these quantities at first grow following roughly the linear behaviour described
by the equations~(\ref{eq:ts1})-(\ref{eq:ts6}), until time $t \sim 6 \tau_{\mathcal A}$ when
nonlinearity sets in. The system results
magnetically dominated both for  energy (magnetic energy $E_M$ is $\sim 35$ times
bigger than kinetic energy $E_K$) and dissipation (ohmic dissipation rate $J$ is
$\sim 6.5$ times viscous dissipation $\Omega$). 

In the fully nonlinear regime a statistically steady state is reached,
in which the Poynting flux $S$, i.e. the energy which is continuously injected for unit 
time into the system at the boundaries by the field-line tangling due to the photospheric 
forcing, balances the energy which is dissipated for unity  of time. On the other hand, as 
previously stated, the
Poynting flux, depends not only on the forcing velocity imposed at the boundaries,
but also on the value of the magnetic field generated inside the system.
This is the reason for which this system must be considered a self-organized 
system. In fact the balance between Poynting flux and dissipation is reached 
by letting the magnetic field grow at the opportune value.

In the absence of dynamical evolution, magnetic energy and ohmic dissipation
would follow the curves~(\ref{eq:emdiff})-(\ref{eq:hrdiff}) saturating on a 
time-scale given by $\tau_{\mathcal R} = \mathcal{R} / (2 \pi n)^2 \sim 25 \tau_{\mathcal A}$
respectively to the values $E_M \sim 3200$ and $J \sim 5000$, well
beyond the levels reached in the simulation.
\begin{figure}[!t]
      \centering
      %%---- start ----
      \subfloat[]{
               \label{fig:atsendisslong:a}             %% label for subfigure
               \includegraphics[width=0.45\linewidth]{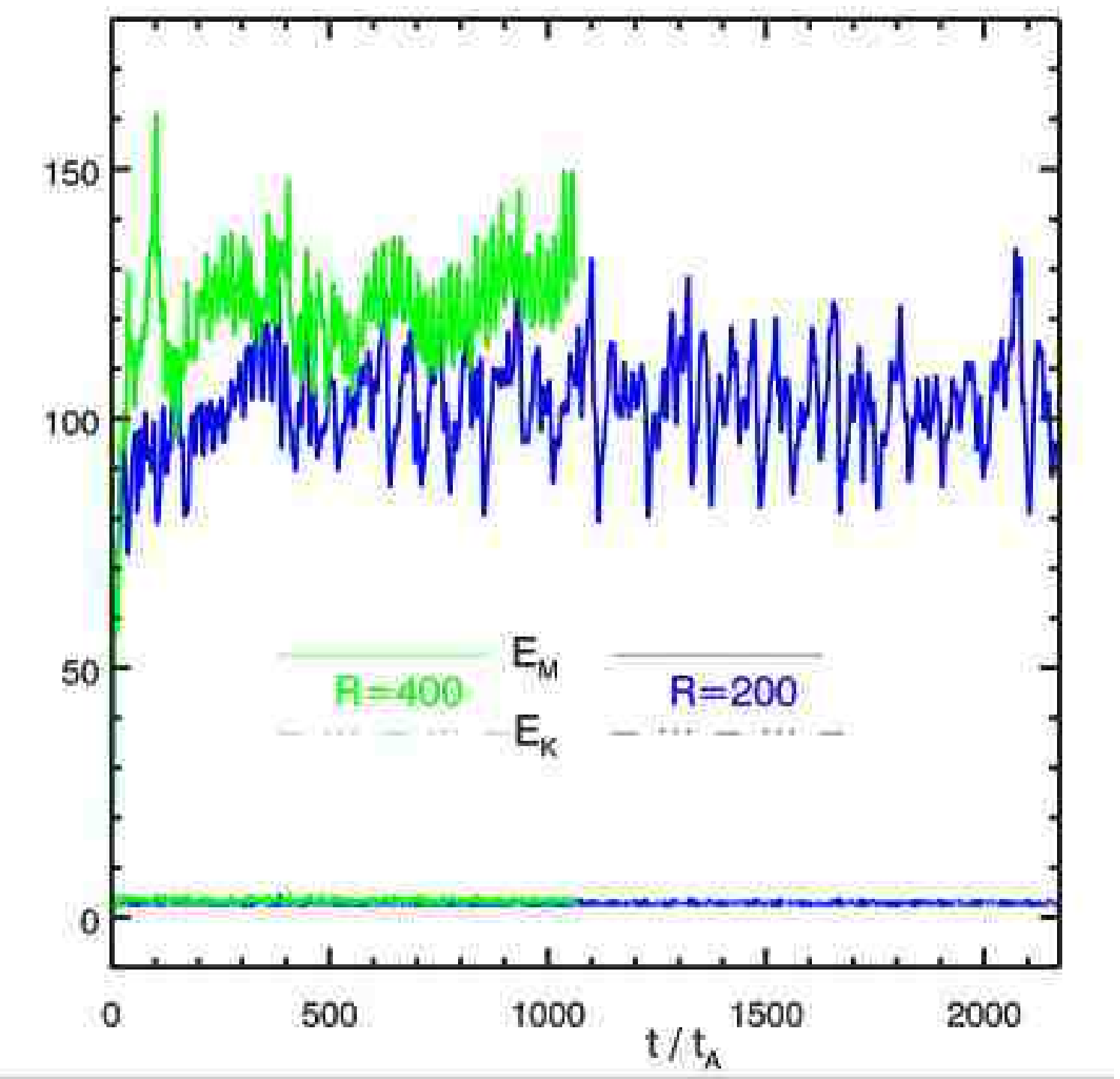}}
      \hspace{0.01\linewidth}
     %%---- start ----
      \subfloat[]{
               \label{fig:atsendisslong:b}             %% label for subfigure
               \includegraphics[width=0.45\linewidth]{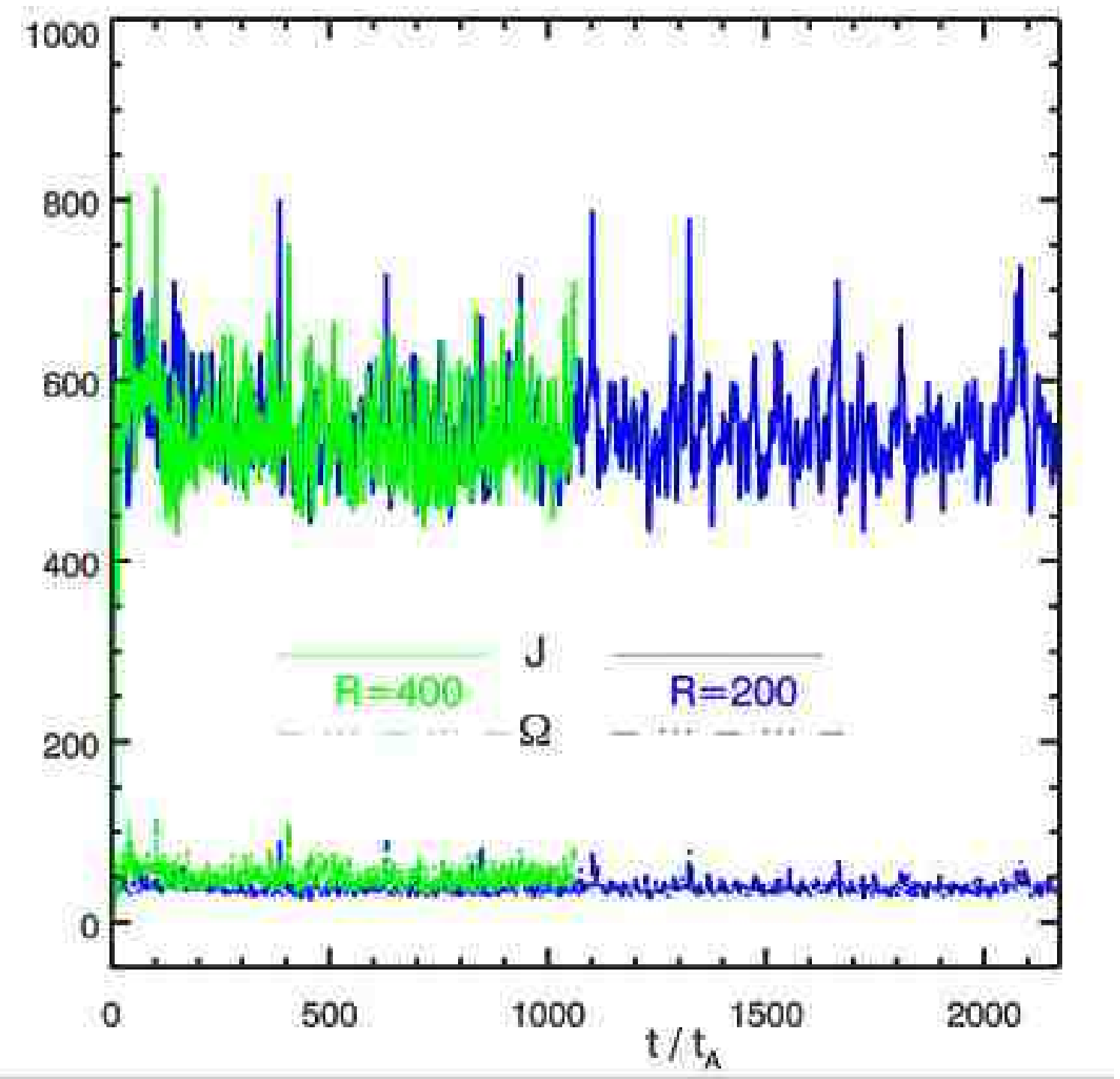}}
   \caption{Two simulations with Alfv\'enic velocity $v_{\mathcal A} = 200$ 
   but performed with lower resolutions to reach longer durations.
   In blue lines are shown physical quantities for a simulation with
   $\mathcal R = 200$ and $128 \times 128 \times 50$ grid points,
   while in green those for $\mathcal R = 400$ and $256 \times 256 \times 100$.
   Both simulations do not have parallel diffusion $\mathcal{R}_{\parallel} = \infty$.
                (a)~Magnetic ($E_M$) and kinetic ($E_K$) energies as a function 
                of time.  Kinetic energy is always a small fraction of total energy.
                (b)~Ohmic (J) and viscous ($\Omega$) dissipation rates.
                 Similar to kinetic energy dynamics,  ($\Omega$) is always a small 
                 fraction of total dissipation. Not shown in Figure the Poynting flux $S$
                 always balances the dissipation rate.
                        \label{fig:atsendisslong}}             %% label for entire figure
\end{figure}

In order to verify the temporal stability of the steady state found in the fully 
nonlinear regime we have performed other numerical simulations, always with
the value $v_{\mathcal A}=200$ for the axial Alfv\'enic speed, but  with
lower resolutions and Reynolds numbers to reach longer durations, 
namely $n_x \times n_y \times n_z = 256 \times 256 \times 100$
with $\mathcal R =400$ and 
$n_x \times n_y \times n_z = 128 \times 128 \times 50$ with $\mathcal R = 200$.
In this earlier simulations no diffusion along the axial direction was used,
i.e.\ $\mathcal{R}_{\parallel} = \infty$.
Time durations are respectively $\Delta t \sim 1100 \, \tau_{\mathcal A}$ and 
 $\Delta t \sim 2200 \, \tau_{\mathcal A}$. 
In Fig.~\ref{fig:atsendisslong} we show
the magnetic and kinetic energies, and the ohmic and viscous dissipation 
rates as a function of time for these simulations. 
The results of the previous simulation,
which was carried out with a higher resolution, but for a shorter duration,
are fully confirmed, and in particular in the nonlinear regime a 
steady state is reached. In particular the integral quantities, i.e.\
total energies and dissipation rates, are intermittent in time.

Fig.~\ref{fig:atsendisslong:b} shows another very
interesting result. Ohmic and viscous dissipation rates for the
simulations with $\mathcal R = 200$ and $400$ roughly overlap.
Having a dissipation rate independent of the
Reynolds number is a property of turbulent systems where 
the injection, transfer and dissipation rates are equal 
(see eq.~(\ref{eq:balrate})) and independent of the 
Reynolds numbers \emph{beyond a threshold}. The value of the threshold
is determined by the diffusion time at the large scale, which must be 
larger than the nonlinear time-scale. The diffusion time at the scale
$\lambda$ is $\tau_D \sim \mathcal{R}\, \lambda^2$,
so that at the injection scale $\ell_{in} \sim 1/4$ for our lowest resolution
simulation with $\mathcal{R} = 200$ the diffusion time is  
$\tau_D \sim 250 \tau_{\mathcal A}$ which we can suppose to be beyond
the nonlinear timescale. On the other hand both this time-scales depend
quadratically on $\lambda$, so that with our lowest resolution we might be just 
beyond the threshold. 

We use \emph{hyperdiffusion} (see (\ref{eq:hyp})) to eliminate the diffusive effects 
at the large scales with a dissipativity exponent $\alpha = 4$. 
In this case the diffusive time at the scale $\lambda$ is given by 
$\tau_{\alpha} \sim \mathcal{R}_{\alpha}\, \lambda^{2\alpha}$. Numerically we require
the diffusion time at the resolution scale $\lambda_{min} = 1/N$,
where $N$ is the number of grid points, to be the same for both normal
and hyperdiffusion, i.e.\
\be
\frac{\mathcal R}{N^2} \sim \frac{\mathcal{R}_{\alpha}}{N^{2\alpha}} \qquad
\longrightarrow \qquad 
\mathcal{R}_{\alpha} \sim \mathcal{R}\, N^{2 \left( \alpha - 1 \right)}
\ee 
So for a grid with 512 points --- which has required a Reynolds number 
$\mathcal R = 800$ with normal diffusion  --- and  a dissipativity $\alpha = 4$, 
the required hyperdiffusion gives $R_4 \sim 10^{19}$. 
From equation~(\ref{en5}) we find that for a generic dissipativity $\alpha$,
including also a diffusive term along the axial direction $z$ with Reynolds
number $\mathcal R_{\parallel}$, dissipation rate is given by
\be
\begin{split}
\mathcal{D} = \mathcal{D}^{\alpha}_{\perp} + \mathcal{D}_{\parallel} = 
& - \frac{1}{\mathcal{R}_{\alpha}}\,  \int_V \! \ud^{3}x\, 
 \left(  \bsy u_{\perp} \cdot \bnabla^{2 \alpha}_{\perp} \bsy u_{\perp} 
+ \bsy b_{\perp} \cdot \bnabla^{2 \alpha}_{\perp} \bsy b_{\perp} \right) \\
& - \frac{1}{\mathcal{R}_{\parallel}}\,  \int_V \! \ud^{3}x\, 
\left(   \bsy u_{\perp} \cdot \de_z \bsy u_{\perp} 
+ \bsy b_{\perp} \cdot \de_z \bsy b_{\perp} \right)
\end{split}
\ee
In general because of the anisotropy of the system 
$\mathcal{D}_{\parallel} \ll  \mathcal{D}^{\alpha}_{\perp}$, for
example in the high-resolution simulation shown in Figure~\ref{fig:atsendiss}
we have $\mathcal{D}_{\parallel} \sim 10^{-2}  \mathcal{D}^{\alpha}_{\perp}$
even though the values of the Reynolds numbers used are 
$\mathcal{R}_{\parallel} = 10 \ll \mathcal{R} = 800$.
For normal diffusion $\alpha = 1$ the dissipation rate (see 
equations~(\ref{en5})-(\ref{en6})) is approximated by 
$\mathcal{D} \sim \mathcal{D}^{\alpha}_{\perp} \sim  J + \Omega$,
with the ohmic and viscous dissipation rates as defined in 
(\ref{eq:ohmdef})-(\ref{eq:visdef}).

\begin{figure}[t]
      \centering
      %%---- start ----
      \subfloat[]{
               \label{fig:atsmtdiss:a}             %% label for subfigure
               \includegraphics[width=0.45\linewidth]{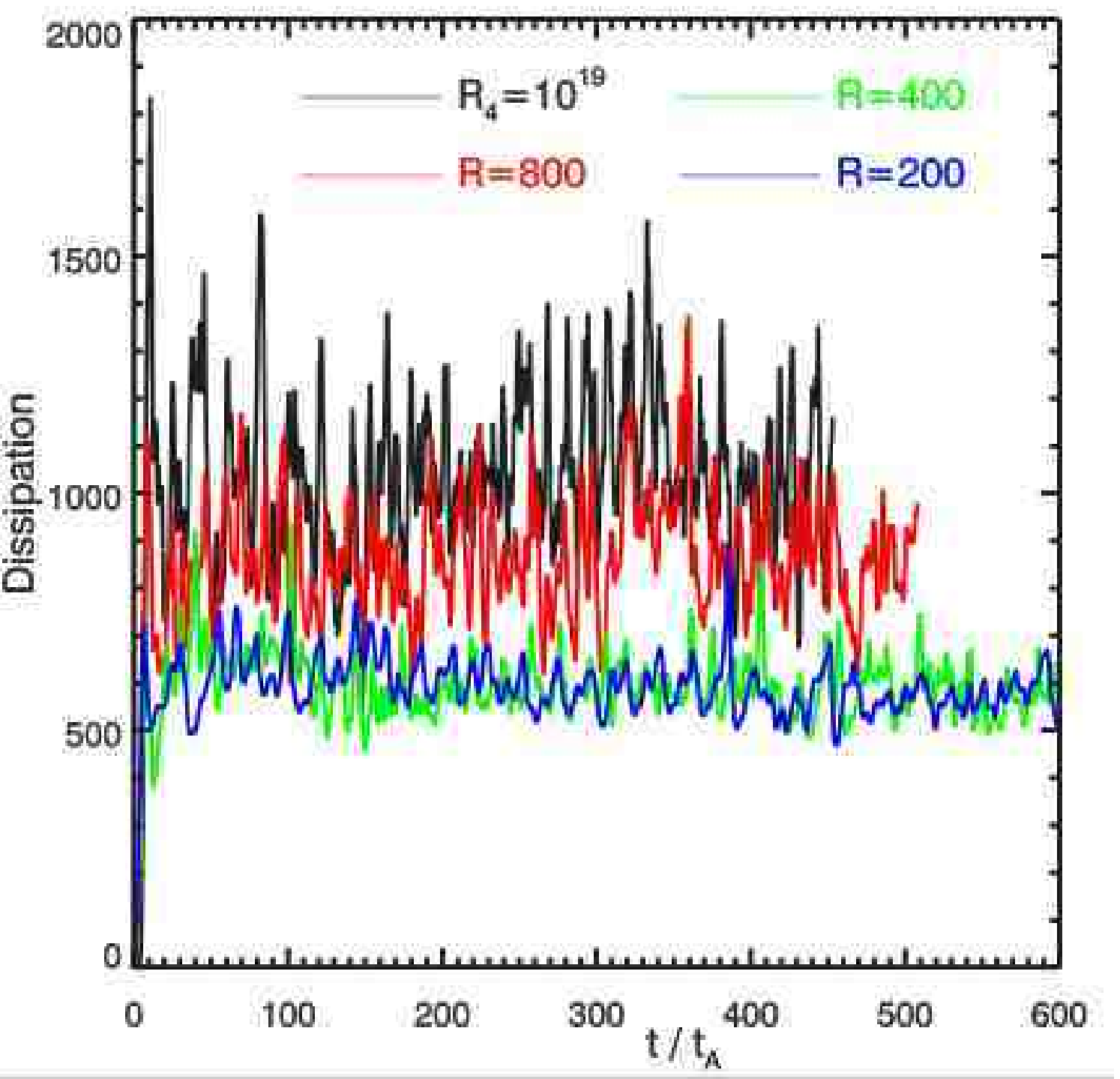}}
      \hspace{0.01\linewidth}
     %%---- start ----
      \subfloat[]{
               \label{fig:atsmtdiss:b}             %% label for subfigure
               \includegraphics[width=0.45\linewidth]{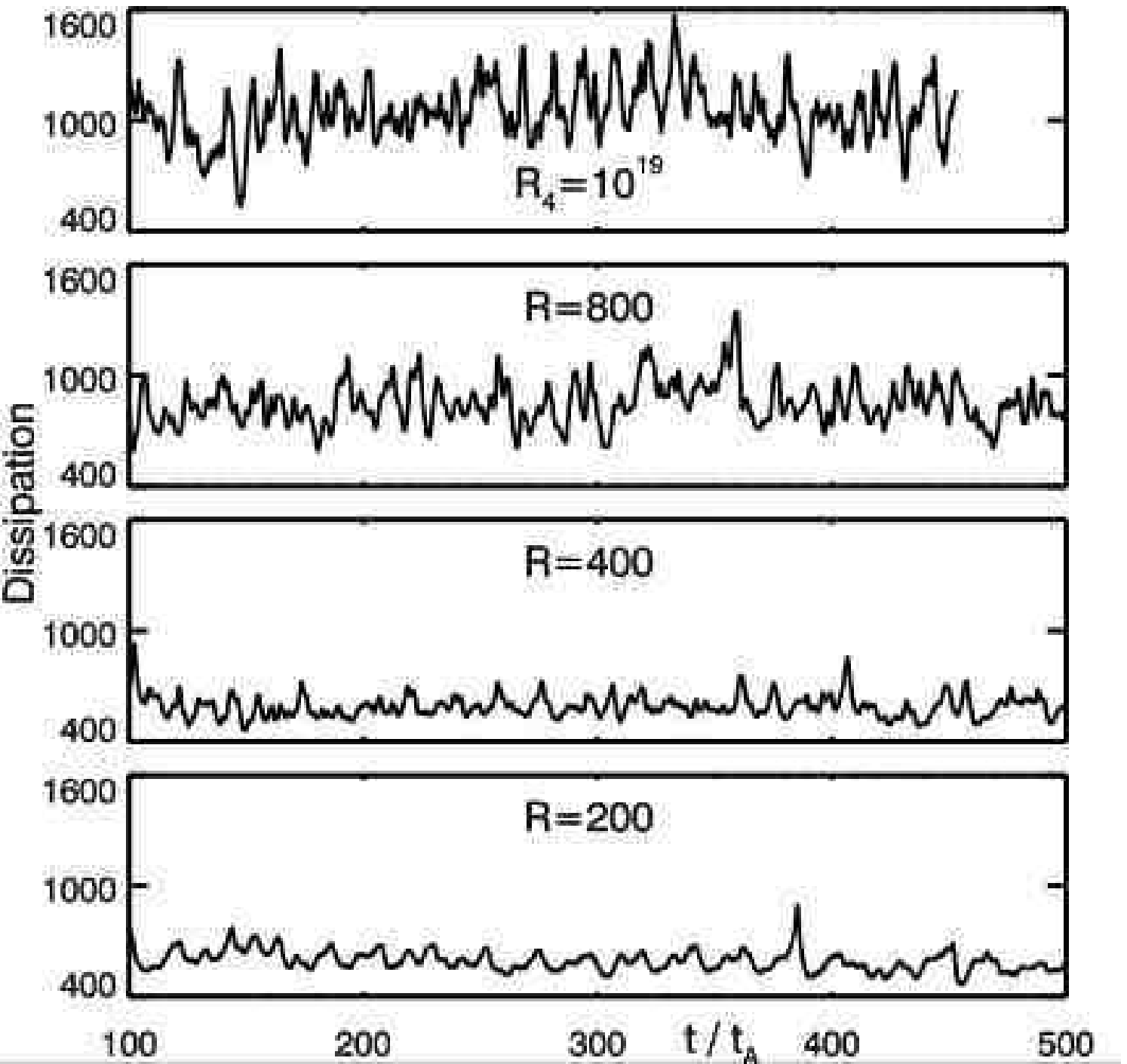}}
   \caption{(a) Dissipation rates as a function of time for four numerical 
   simulations with different Reynolds numbers and hyperdiffusion with 
   dissipativity $\alpha = 4$.  Their close values over a very large range
   of diffusion times at the large scales show that dissipation is independent
   of the Reynolds numbers, a characteristic of turbulent systems.
   (b) Close-up of dissipation rates. With increasing Reynolds number
   shorter temporal variations develop.
        \label{fig:atsmtdiss}}             %% label for entire figure
\end{figure}

We have then performed a numerical simulation using hyperdiffusion, with
dissipativity $\alpha = 4$, $n_x \times n_y \times n_z = 512 \times 512 \times 200$ 
grid points. After a few tests we have used $R_4 \sim 10^{19}$ and a Reynolds
number for the parallel diffusion $\mathcal{R}_{\parallel}=10$. This
simulation confirms the results with ordinary diffusion.
In Fig.~\ref{fig:atsmtdiss} we plot the dissipation rates for the 4 simulations
with 3 different Reynolds numbers and hyperdiffusion.
The simulations with $\mathcal R =200$ and $400$ do not use parallel
diffusion ($\mathcal{R}_{\parallel} = \infty$) while those with $\mathcal{R} =800$
and hyperdiffusion have $\mathcal{R}_{\parallel} = 10$. 
The dissipation rates roughly overlap, in particular those performed with
the same numerical methods, i.e.\ those which  implement or not
parallel diffusion. We also note that these 
four simulations do not differ solely for the value of the Reynolds numbers,
the number of grid points, and the implementation of parallel diffusion:
in each simulation the coefficients that define the spatial shape of the  
forcing boundary velocities (see eqs.~(\ref{eq:frand})-(\ref{eq:rampl})) are 
chosen randomly, and are always different.
Thus, the four simulations shown in 
Fig.~\ref{fig:atsmtdiss:a} refer to four different velocity patterns, although
all of them have an rms value of $1/\sqrt{2}$ (see (\ref{eq:rmscon})).
The small differences are probably due to this difference more than
to the different Reynolds numbers. To further investigate the independence
of the dissipation rate on Reynolds numbers it will be necessary to
perform these numerical simulations using the same spatial pattern
for the boundary velocities and parallel diffusion for all of them.
The diffusion time at the large scales has a tremendous 
increase between $\mathcal R = 200$ and $\mathcal{R}_4 = 10^{19}$,
going from $\tau_D \sim 250\, \tau_{\mathcal A}$ up to 
$\tau_{D,4} \sim 3\cdot 10^{15}\, \tau_{\mathcal A}$, so that the results
shown in Figure~\ref{fig:atsmtdiss} demonstrate that there is at most a very
small variation in the dissipation rates once the diffusion time at 
the large scales is beyond a threshold. 
\begin{figure}[t]
\begin{center}
\includegraphics[width=0.5\textwidth]{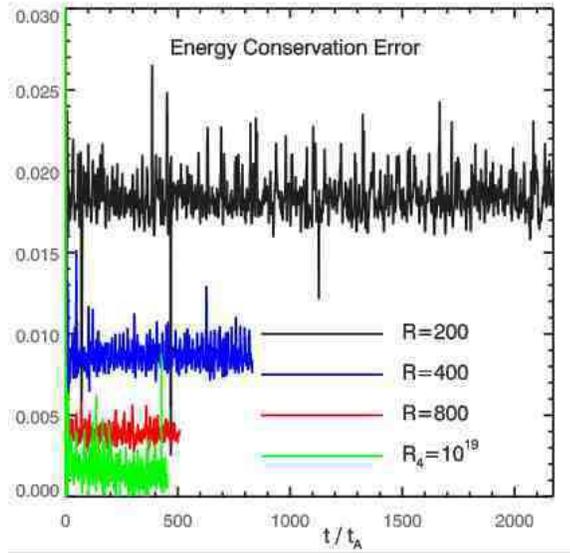}
\caption{Numerical energy conservation error $|\Delta/S|$ (see eq.(\ref{eq:delta}))
as a function of time.
 \label{fig:atsenerr}}
\end{center}
\end{figure}

Figure~\ref{fig:atsmtdiss:b} shows a close-up of
the dissipation rates as a function of time at different Reynolds numbers and
with hyperdiffusion.
At increasing higher Reynolds numbers the function clearly exhibits more
fine structures at  smaller time-scales. \emph{This is a typical feature of a 
transition to turbulence.}

In order to check the validity of these results it is important that the energy
equation~(\ref{eq:eneq}) is numerically well satisfied. Although this equation
is not directly integrated by the numerical code, it follows from the
equations that \emph{are} integrated, so that there could be a numerical discrepancy. 
Indicating with $\Delta$ the quantity
\be \label{eq:delta}
\Delta = \frac{\partial E}{\partial t} - \left[ S - \mathcal{D} \right],
\ee
which analytically should be equal to $0$, we define as the \emph{numerical 
energy conservation error} the absolute value of the normalized quantity 
$\Delta / S$. In Figure~\ref{fig:atsenerr} we plot $|\Delta / S|$ as a function of time
for the numerical simulations just described.  The value of $|\Delta / S|$ is at most 
2.5\%, with an average of 1.8\% for $\mathcal R = 200$ decreasing to 
0.15\% for $\mathcal{R}_4 = 10^{19}$, while for $\mathcal R = 400$ and $800$ 
the average errors are respectively  0.9\% and 0.4\%. 
Hence the energy conservation equation~(\ref{eq:eneq}) is numerically 
very well satisfied.

\begin{figure}[p]
\begin{center}
\includegraphics[width=0.8\textwidth]{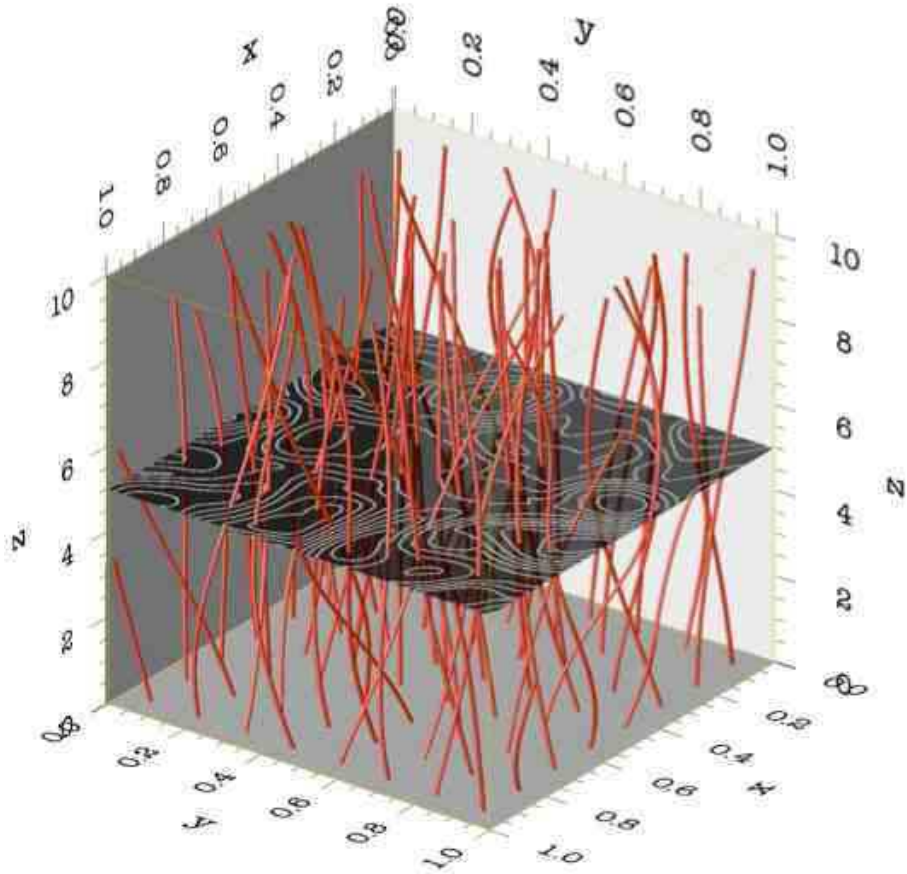}
\caption{Field lines of the total magnetic field (orthogonal plus axial)
                at time $t = 18.5\, \tau_{\mathcal A}$ for the high-resolution
                numerical simulation with $v_{\mathcal{A}}=200$, 
                $512 \times 512 \times 200$ grid points, $\mathcal R =800$
                and $\mathcal{R}_{\parallel}=10$. 
               \emph{Mid-plane:} field lines of the orthogonal magnetic field. 
               The orthogonal magnetic field magnitude fluctuates around a value 
               which is roughly the $3 \%$ of the axial component, well within
               the reduced MHD ordering. This is reflected in the slight bending
               of the magnetic field lines. For an improved visualization the box size
               has been rescaled. But the axial length of the box is ten times longer 
               than the orthogonal one. The resize of the box artificially enhances the field 
               line bending. The orthogonal magnetic fields is structured in magnetic 
               islands, and is mostly homogeneous in the axial direction ($z$). 
               \label{fig:atsbfside}}
\end{center}
\end{figure}

\begin{figure}[p]
\begin{center}
\includegraphics[width=0.8\textwidth]{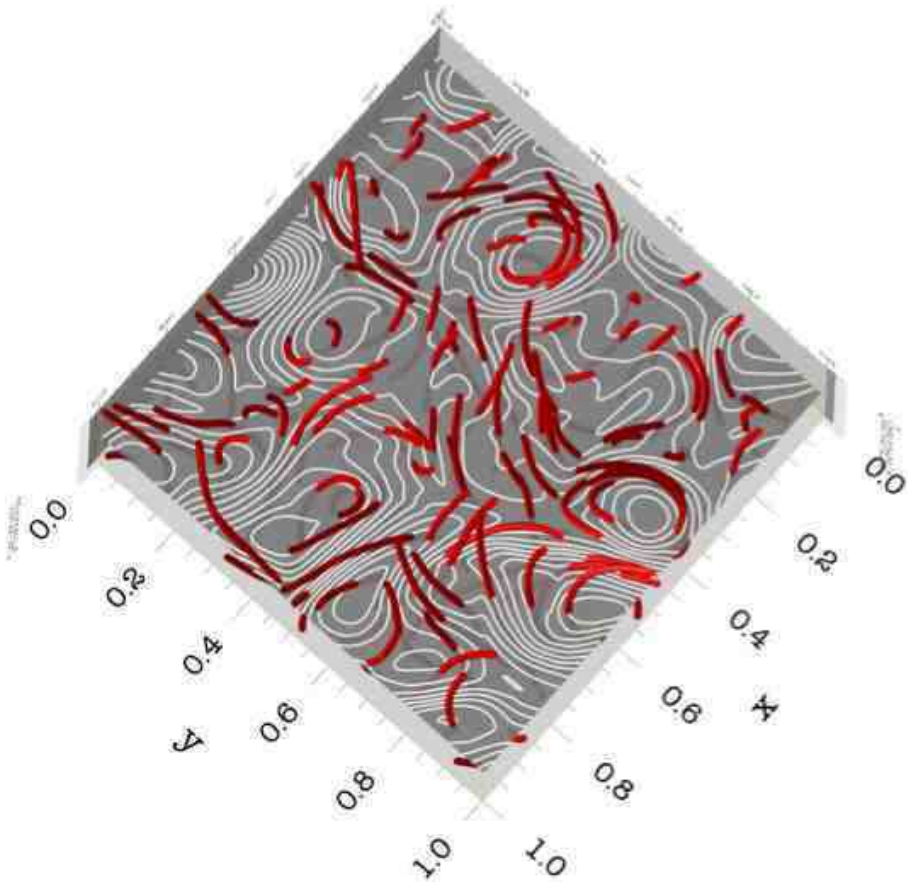}
\caption{A top view of the magnetic field lines shown in
                Figure~\ref{fig:atsbfside}.
                \emph{Mid-plane:} field lines of the orthogonal magnetic field. 
                The orthogonal magnetic field is structured in magnetic islands, and is 
                mostly homogeneous in the axial direction ($z$). Because of the 
                quasi-homogeneity of the orthogonal field, the field lines of the
                total magnetic field are more bent when they are at the outskirts of
                magnetic islands, while at different heights they always have the 
                orthogonal component of the magnetic field roughly in the same
                direction.
                \label{fig:atsbftop}}
\end{center}
\end{figure}

\begin{figure}[p]
\begin{center}
\includegraphics[width=0.8\textwidth]{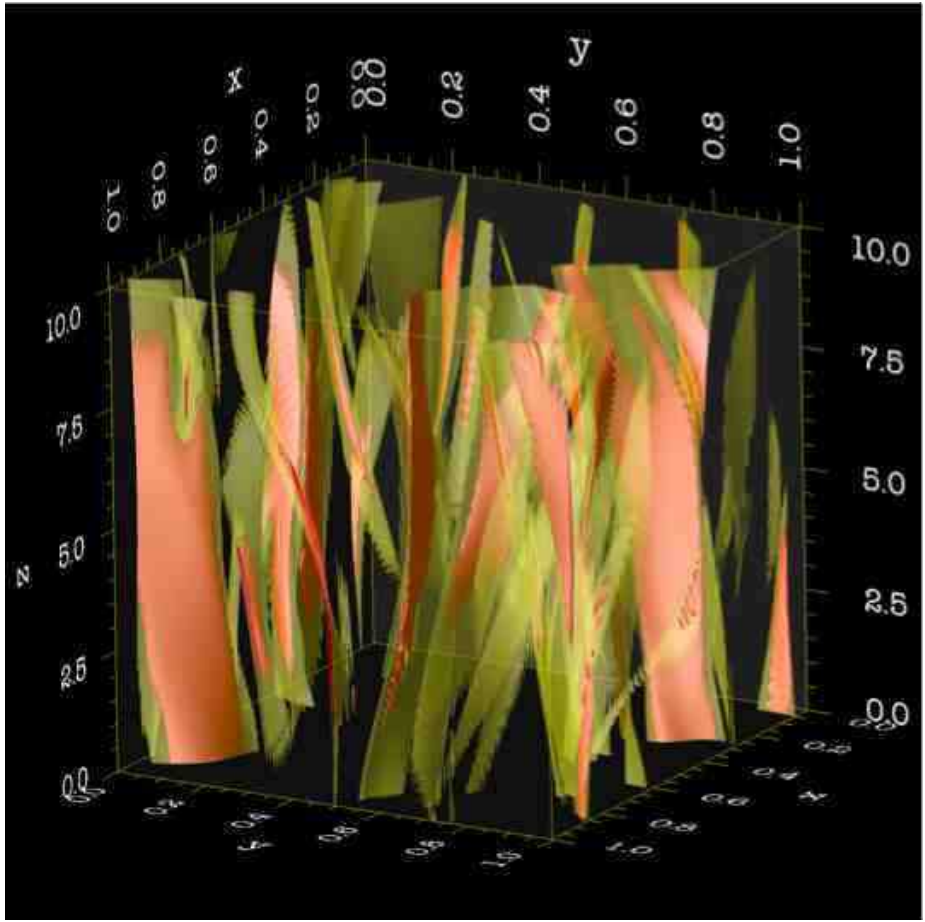}
\caption{Isosurfaces of the squared current $j^2$: \emph{side view}.
               Two isosurfaces of the squared current at time $t \sim 18.5 \, \tau_A$ 
               for the numerical simulation with $v_{\mathcal A} = 200$,
                512x512x200 grid points and Reynolds
               numbers $\mathcal R =800$, $\mathcal{R}_{\parallel} =10$ is represented.  
               The isosurface at the value $j^2 = 2.8 \cdot 10^5$ is represented in partially 
               transparent yellow,  while  red displays
               the isosurface with  $j^2 = 8 \cdot 10^5$, well below the value of the 
               maximum of the squared current that at this time is $j^2 = 8.4 \cdot 10^6$.
               N.B.: The red isosurface is always nested inside the yellow one, and
               appears pink in the figure.
               \label{fig:atsisoside}}
\end{center}
\end{figure}

\begin{figure}[p]
\begin{center}
\includegraphics[width=0.8\textwidth]{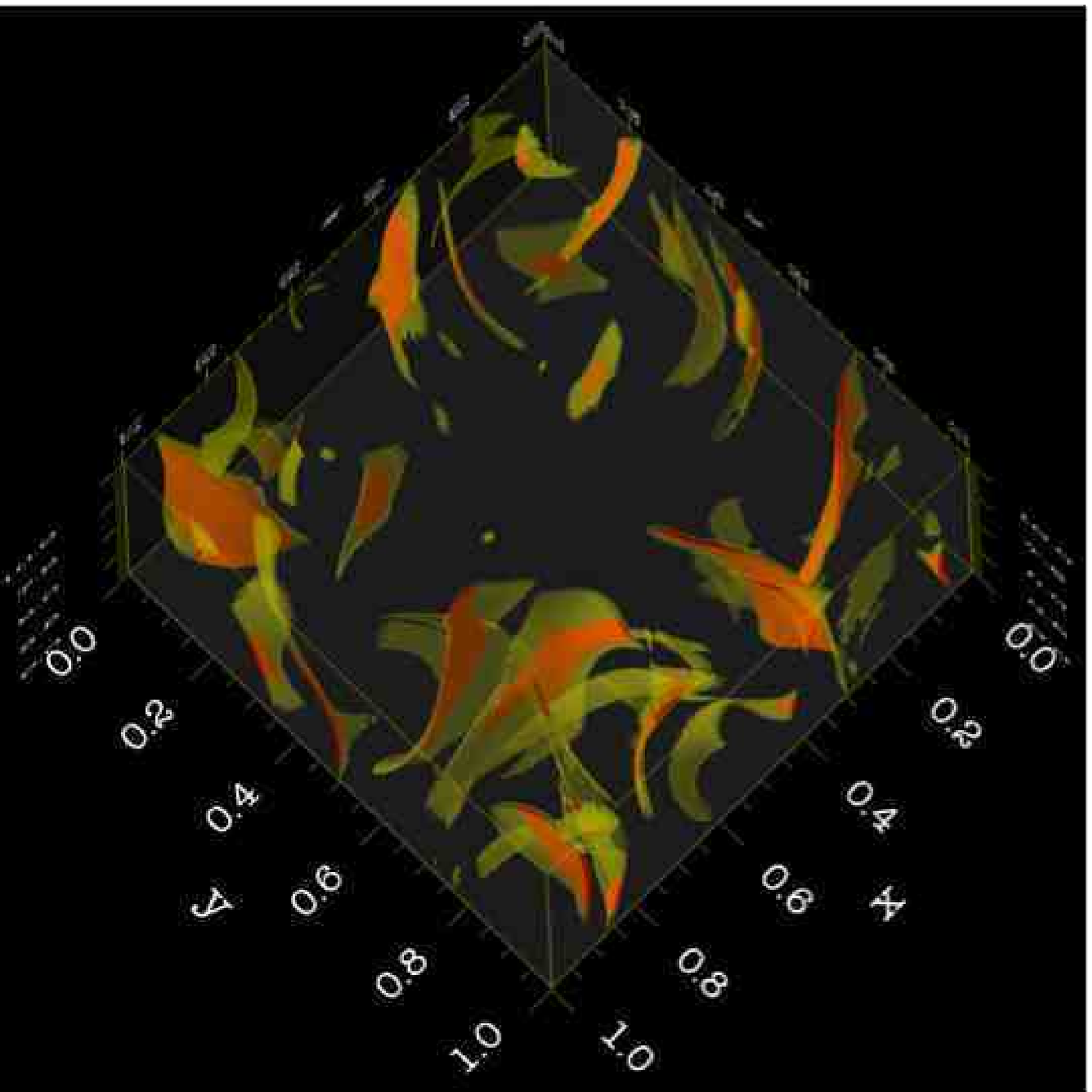}
\caption{\emph{Top view} of the isosurfaces of the squared current
                shown in Figure~\ref{fig:atsisoside} using the same color display.
                The isosurfaces are extended along the axial direction, and the
                corresponding filling factor is small.
               \label{fig:atsisotop}}
\end{center}
\end{figure}

We now briefly examine the 3D structure of the physical fields. One of the 
hypothesis of reduced MHD is that during time evolution the orthogonal 
magnetic field is small compared to the axial component.
The orthogonal field component
always fluctuates around a value which is roughly the $3 \%$ of the the axial 
field (see Figure~\ref{fig:atsendiss:a}) thus verifying this hypothesis.
In Fig.~\ref{fig:atsbfside}-\ref{fig:atsbftop} we show the magnetic field lines, 
a view from the side and from the top of the computational box respectively.
For an improved visualization the box has been rescaled, but it should
be noted that the computational box is ten times longer in the axial
direction ($z$) than in the orthogonal plane.
Consequently it can be verified that the magnetic field lines are only slightly
bent. In both figures we show the field lines of the orthogonal magnetic 
field in the mid-plane. The structure of the orthogonal magnetic field is
almost invariant in the axial direction and it is structured in magnetic
islands. In particular, no boundary layer is present. The lines of the
total magnetic field which bend most are those which happen to be
at the outskirts of  the magnetic islands, where the orthogonal magnetic 
field is enhanced.
Current sheets develop, extended along the axial direction, as shown in
Fig.~\ref{fig:atsisoside}-\ref{fig:atsisotop},  and in these thin sheets
the energy flux is finally dissipated.

\subsection{Spectral Properties}

In this paragraph we discuss  the dynamics of our system using 
its spectral quantities in more detail. We show  how the current sheets shown in 
Figures~\ref{fig:atsisoside}-\ref{fig:atsisotop} are formed and study
the long-time dynamics behaviour. 

\begin{figure}[t]
      \centering
      %%---- start ----
      \subfloat[]{
               \label{fig:atshc:a}             %% label for subfigure
               \includegraphics[width=0.45\linewidth]{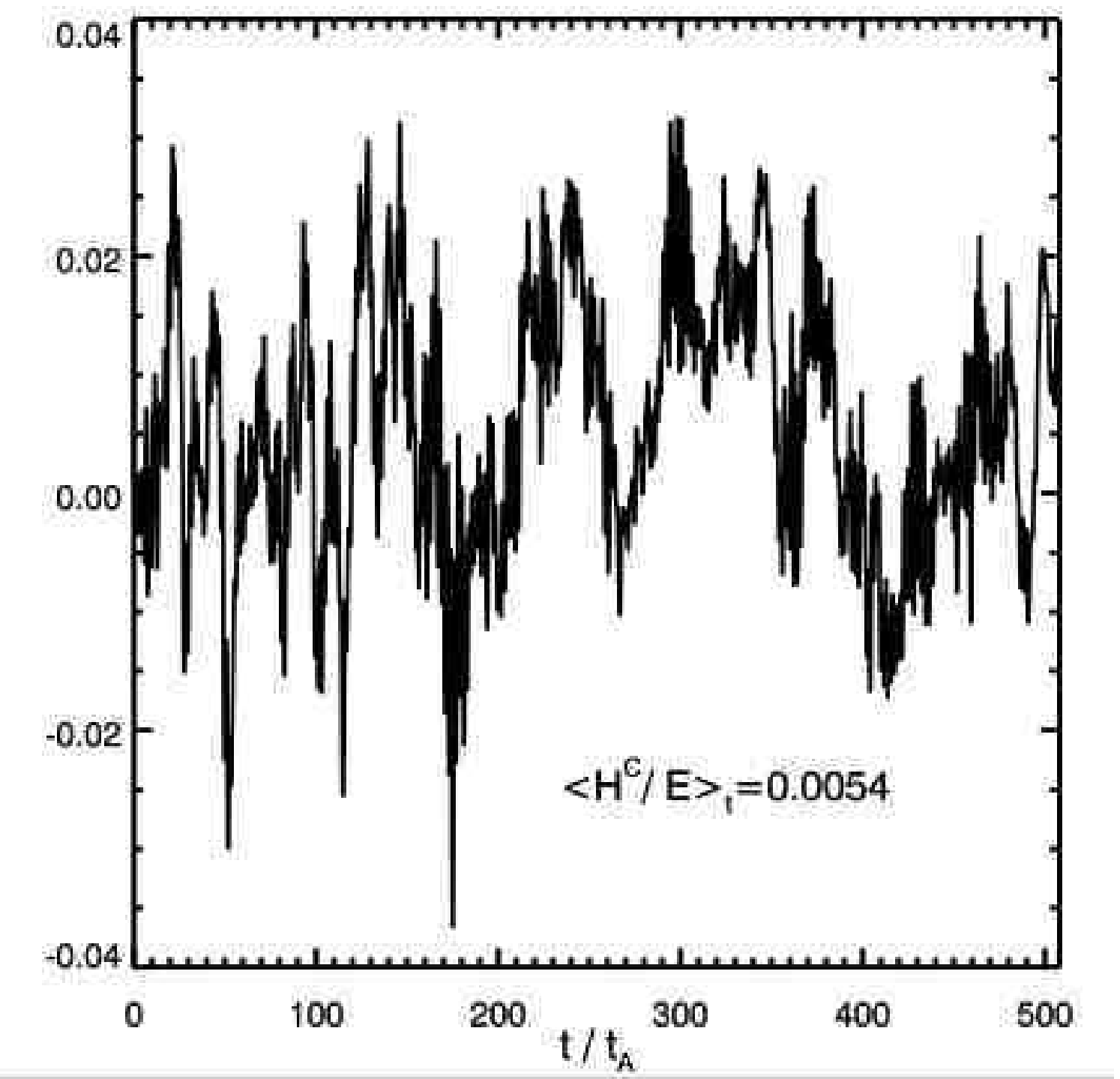}}
      \hspace{0.01\linewidth}
     %%---- start ----
      \subfloat[]{
               \label{fig:atshc:b}             %% label for subfigure
               \includegraphics[width=0.45\linewidth]{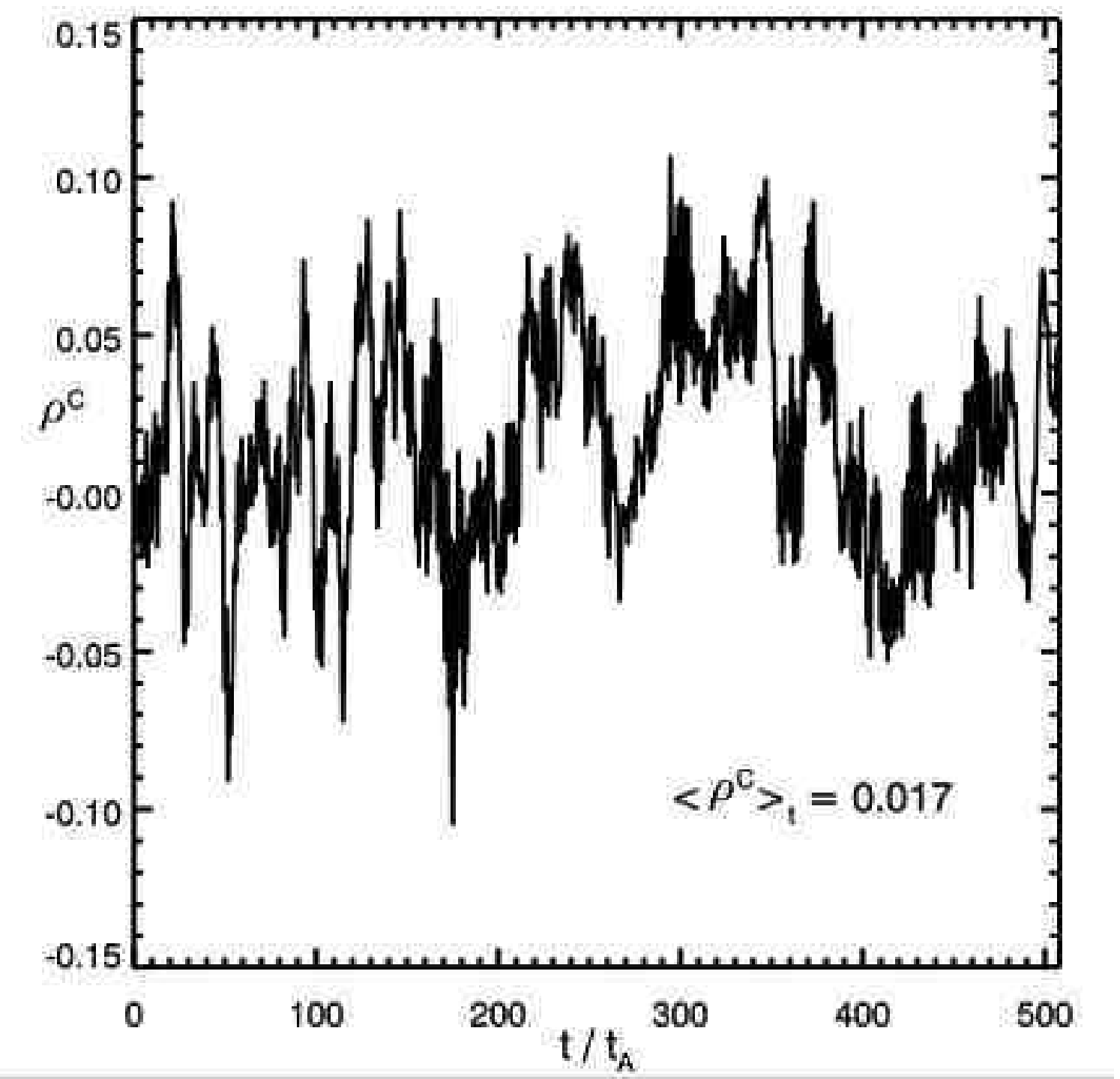}}
   \caption{(a) Ratio between cross helicity $H^C$ and total energy $E$ as a 
   function of time. (b) Velocity-magnetic field correlation $\rho^C$ as a function of time,
   showing that the system is weakly correlated. Both quantities are calculated 
   for run~1.
        \label{fig:atshc}}             %% label for entire figure
\end{figure}

In turbulence the fundamental physical fields are the Els\"asser variables
$\bsy{z}^{\pm} = \bsy{u}_{\perp} \pm \bsy{b}_{\perp}$. The analysis 
of the \emph{weak and strong anisotropic turbulence} presented in
\S~\ref{par:bik} has been carried out  assuming 
\emph{weak velocity-magnetic-field  correlation at all scales}
$\delta z_{\lambda}^+ \sim \delta z_{\lambda}^- \sim \delta z_{\lambda}$.
This property basically follows from the symmetry in $\bsy{z}^{\pm}$
of both the reduced MHD equations~(\ref{eq:els1})-(\ref{eq:els4}) and the boundary 
conditions used~(\ref{eq:bc0})-(\ref{eq:bcL}), as shown in \S~\ref{sec:sos}, and
will be used in all \S~\ref{sec:nla} to perform our \emph{nonlinear analysis}.
The Els\"asser variables are associated with the corresponding energies
\be
E^{\pm} =  \frac{1}{2}\, \int_V \! \ud^{3}x\ \left( \bsy{z}^{\pm} \right)^2 
\ee
related to the kinetic and magnetic energies $E_K$, $E_M$ and
to the cross helicity $H^C$
\be
H^C =  \int_V \! \ud^{3}x\ \bsy{u}_{\perp} \cdot \bsy{b}_{\perp}, \qquad
\ee
by
\be
E^{\pm} = E_K + E_M \pm\,  H^C
\ee
so that the low-correlation of the fields requires $H^C \ll E_K + E_M$. 
Figure~\ref{fig:atshc:a} shows this quantity as a function of time for 
the simulation with $v_{\mathcal A}=200$, $\mathcal R =800$, 
$\mathcal{R}_{\parallel} =10$. The cross helicity fluctuates around zero,
with a variation of at most 4\% with respect to total energy and
an average of 0.54\%. Alternatively for a magnetically or kinetically
dominated systems, in which one of the two energies exceeds
the other, the correlation function $\rho^C$ is defined as
\be
\rho^C = \frac{H^C}{\left( 4 E_K E_M \right)^{1/2}} =
\frac{  \int \! \ud^{3}x\ \bsy{u}_{\perp} \cdot \bsy{b}_{\perp} }
       { \left(  \int \! \ud^{3}x\ \bsy{u}^2_{\perp} \cdot
           \int \! \ud^{3}x\   \bsy{b}^2_{\perp} \right)^{1/2}  }
\ee 
This quantity is shown as a function of time in Figure~\ref{fig:atshc:b}.
It fluctuates around zero with an average of $0.017$ and a maximum
of $\sim 0.1$. Hence the velocity-magnetic-field  correlation is very small
when the whole computational box is considered. We present a more
detailed correlation analysis later in this paragraph.
In Figures~\ref{fig:atssp} we show the spectra for run~1. Unless otherwise
specified the spectra are one-dimensional, averaged in the axial direction.
We define the Fourier transform of a function $f(x,y)$ as
\begin{equation}
f \left( \boldsymbol{x} \right) = \sum_{\boldsymbol{k}} 
\hat{f} \left( \boldsymbol{k} \right) e^{i \boldsymbol{k} \cdot \boldsymbol{x} }, \qquad
\hat{f} \left( \boldsymbol{k} \right) = \frac{1}{\ell^2} 
\iint_{0\ \ \ }^{\ell\ \, } \displaylimits \! \mathrm{d}x \, \mathrm{d}y \
f \left( \boldsymbol{x} \right) e^{- i \boldsymbol{k} \cdot \boldsymbol{x} },
\end{equation}
where
\begin{equation*}
\boldsymbol{k}_{ij} = \frac{2\pi}{\ell} \left( i, j \right), \qquad 
i,j= \pm 1, \pm 2, \ldots 
\end{equation*}
The discrete form of Parseval's theorem gives:
\begin{equation}
\iint_{0\ \ \ }^{\ell\ \, } \displaylimits \! \mathrm{d}x \, \mathrm{d}y \,
\left| f \left( \boldsymbol{x} \right) \right|^2 =
\ell^2 \sum_{\boldsymbol{k}} \left| \hat{f} \left( \boldsymbol{k} \right) \right|^2
\end{equation}
For example considering the $\bsy{z}^{\pm}$ energy spectra
\begin{equation}
E^{\pm} = 
\frac{1}{2}\, \int_0^L \! \ud z\,\iint_{0\ \ \ }^{\ell\ \, } \displaylimits \! \ud x\, \ud y\, 
\left( \bsy{z}^{\pm} \right)^2 = 
\frac{1}{2}\, \int_0^L \! \ud z\, \ell^2 \sum_{\bsy{k}}
\left| \bsy{\hat{z}}^{\pm} \right|^2 \left( \bsy{k}, z \right) 
\ee
Even if $\boldsymbol k$ is discrete in our case, it is useful to use a 
continuum formalism through the substitution
\begin{equation}
\left( \frac{2\pi}{\ell} \right)^2  \sum_{\boldsymbol k} \to 
\int \mathrm{d}k_x \mathrm{d}k_y 
\end{equation}
In this way we can write the energies as:
\begin{equation} \label{eq:fe1}
E^{\pm} =
\frac{1}{2} \int_0^L \! \mathrm{d}z \ 
\left( \frac{\ell^2}{2 \pi} \right)^2 \iint \! \mathrm{d}k_x \mathrm{d}k_y \ 
\left| \bsy{\hat{z}}^{\pm} \right|^2 \left( \bsy{k},  z \right) =
\frac{1}{2} \left( \frac{\ell^2}{2 \pi} \right)^2 \int_0^L \mathrm{d}z \ 
2 \pi \int \! \mathrm{d}k \ k \ 
\left| \bsy{\hat{z}}^{\pm} \right|^2 \left( k, z \right)
\end{equation}
and finally integrating along $z$ we have
$E^{\pm} = \int \! \mathrm{d}k \ E^{\pm}_k$, and in the discrete case
\be \label{eq:fe2}
E^{\pm} = \sum_n E^{\pm}_n, \qquad n=1,2,\dots
\ee
In Figure~\ref{fig:atssp:a} we show the spectrum $E^-_n$ for run~1. 
\begin{figure}[t]
      \centering
      %%---- start ----
      \subfloat[]{
               \label{fig:atssp:a}             %% label for subfigure
               \includegraphics[width=0.45\linewidth]{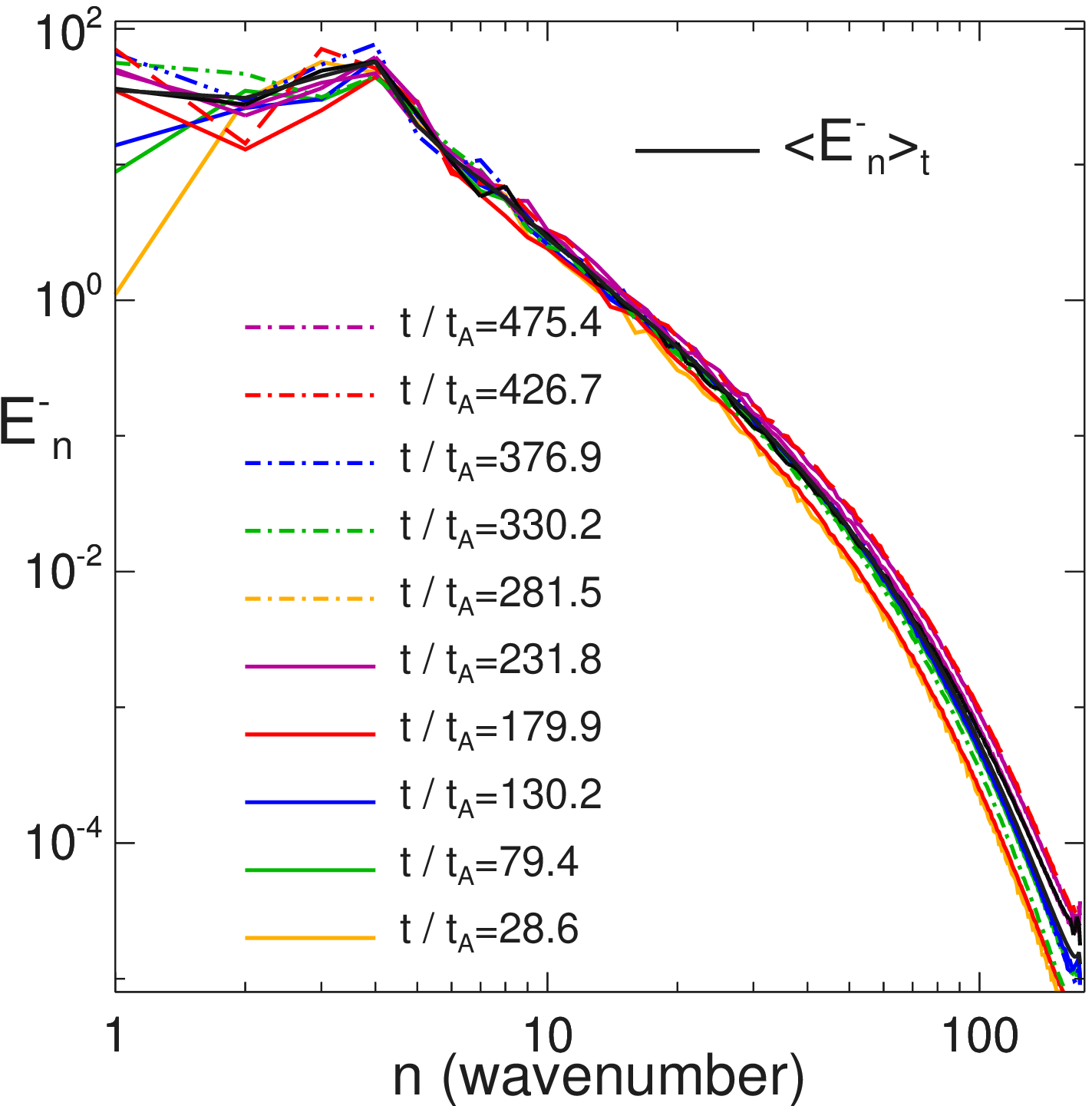}}
      \hspace{0.01\linewidth}
     %%---- start ----
      \subfloat[]{
               \label{fig:atssp:b}             %% label for subfigure
               \includegraphics[width=0.45\linewidth]{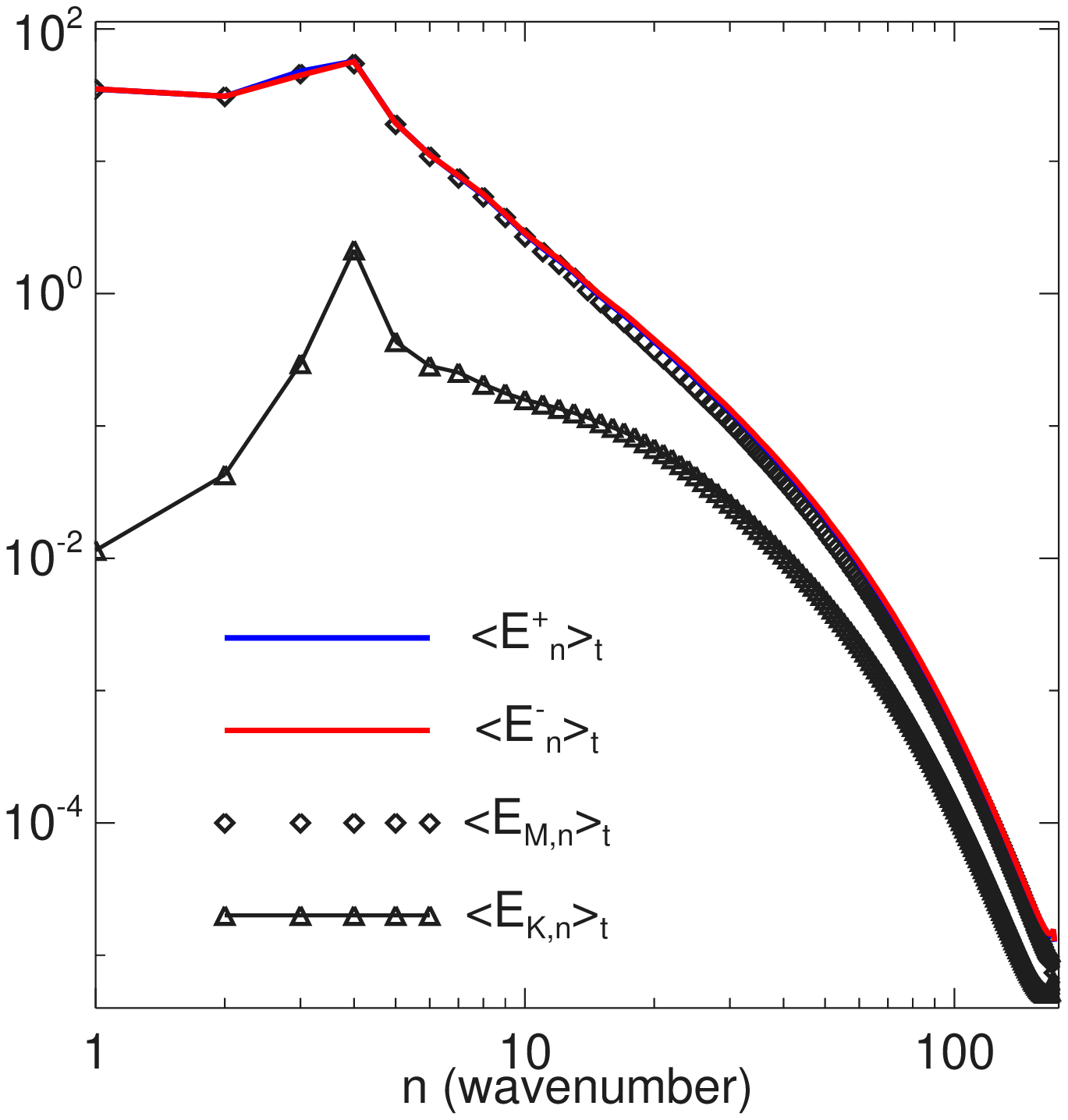}}
   \caption{(a) $E^{-}$ energy spectra at different times are shown in color.
   Their time average, performed over a duration of $\sim 500$ Alfv\'enic crossing
   times, is shown in black. (b) Time averaged spectra for magnetic ($E_M$), 
   kinetic ($E_K$) and $E^{\pm}$ energies.
        \label{fig:atssp}}             %% label for entire figure
\end{figure}
In different colors the spectrum at different times is displayed, while the black line 
shows the time average of the spectrum performed on a total duration of 
$\sim 500$ Alfv\'enic crossing times. A well-developed spectrum is formed
with a peak at $n=4$, which is the injection scale. \emph{This is a clear proof
that this system is turbulent and that an energy cascade exists.} 
Figure~\ref{fig:atssp:b} shows the time averages spectra 
of the kinetic, magnetic and  $E^{\pm}$ energies.
As expected $E^{\pm}$ almost completely overlap, while they are
barely distinguishable from the magnetic energy spectrum. The kinetic
energy spectrum has noticeably  smaller values, and in particular the 
mode $n=4$ at the injection scale is fixed by the boundary forcing velocity
as we show later in this paragraph.  

Figure~\ref{fig:atsspf} shows a close-up of the time-averaged $E^-$ spectrum.
Even if run~1, for which this spectrum has been calculated, used
a grid with $512 \times 512 \times 200$ points and a Reynolds number
$\mathcal R =800$ an inertial range is barely formed and diffusion
affects the spectrum at least up to $n \sim 10$. On the other hand 
energy is injected  at the scale  $n\sim 4$  at small wavenumbers, 
so in this kind of simulation it is very difficult 
to obtain a well-developed inertial range because very few points 
are left. Two lines corresponding to energy spectra $k_{\perp}^{-5/3}$
and $k_{\perp}^{-2}$ are drawn for reference.
\begin{figure}[t]
\begin{center}
\includegraphics[width=0.5\textwidth]{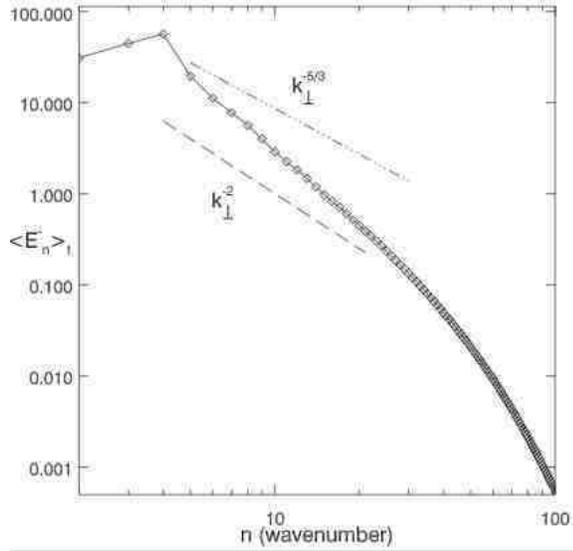}
          \caption{Close-up of time-averaged $E^-$ energy spectrum. 
          An inertial range is barely formed, even using  
          $512 \times 512 \times 200$ grid points corresponding to 
            a Reynolds number $\mathcal R =800$.
               \label{fig:atsspf}}
\end{center}
\end{figure}
In order to study the properties of the inertial range we have used
hyperdiffusion, that we will describe in a moment. But now
we want to analyze how the current sheets elongated along
the axial direction shown in Figures~\ref{fig:atsisoside}-\ref{fig:atsisotop} 
are generated.

\begin{figure}[p]
      \centering
      %%---- start ----
      \subfloat[]{
               \label{fig:atsspnk:a}             %% label for subfigure
               \includegraphics[width=0.45\linewidth]{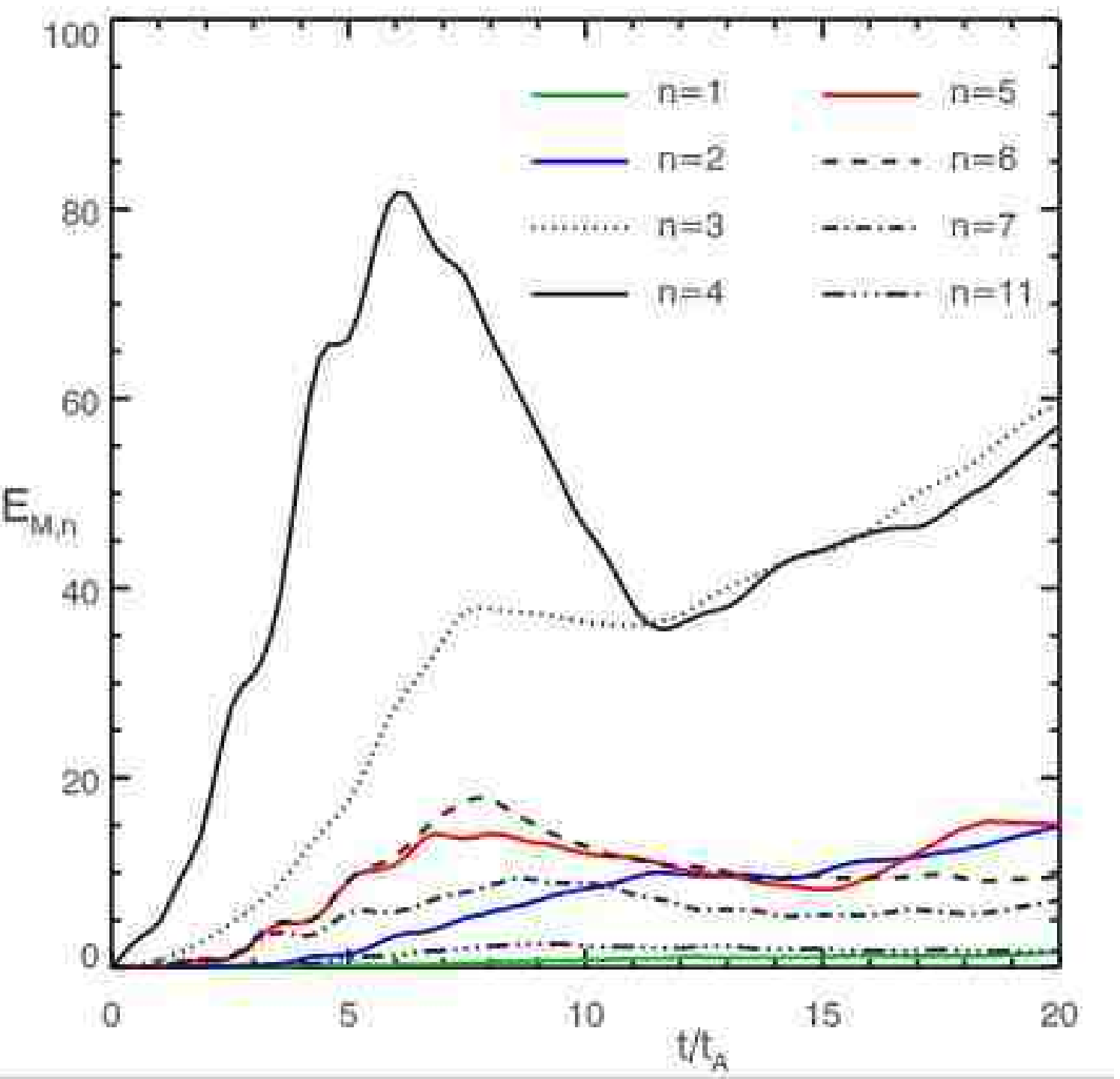}}
      \hspace{0.01\linewidth}
     %%---- start ----
      \subfloat[]{
               \label{fig:atsspnk:b}             %% label for subfigure
               \includegraphics[width=0.45\linewidth]{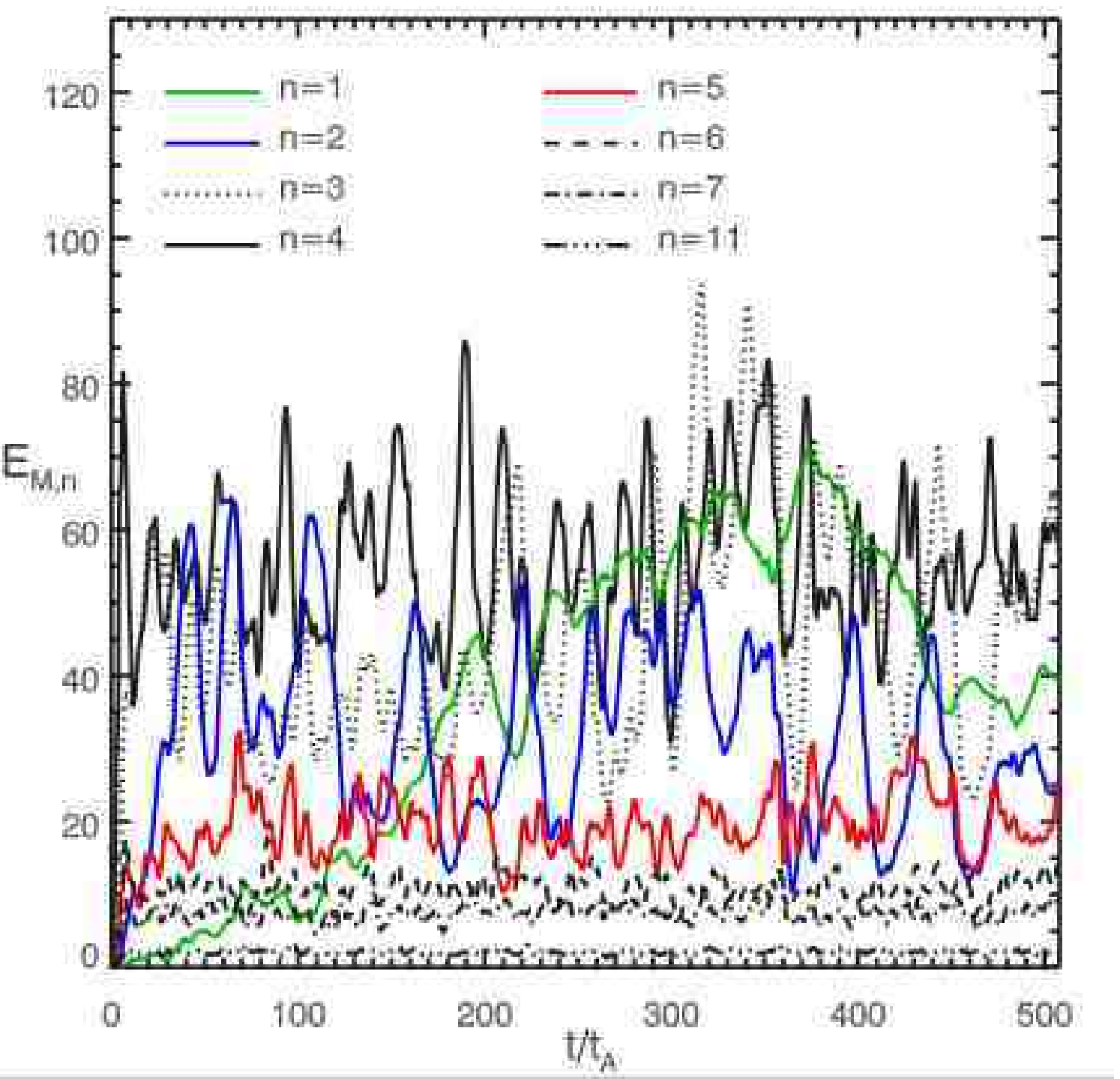}}\\[20pt]
      %%---- start ----
      \subfloat[]{
               \label{fig:atsspnk:c}             %% label for subfigure
               \includegraphics[width=0.45\linewidth]{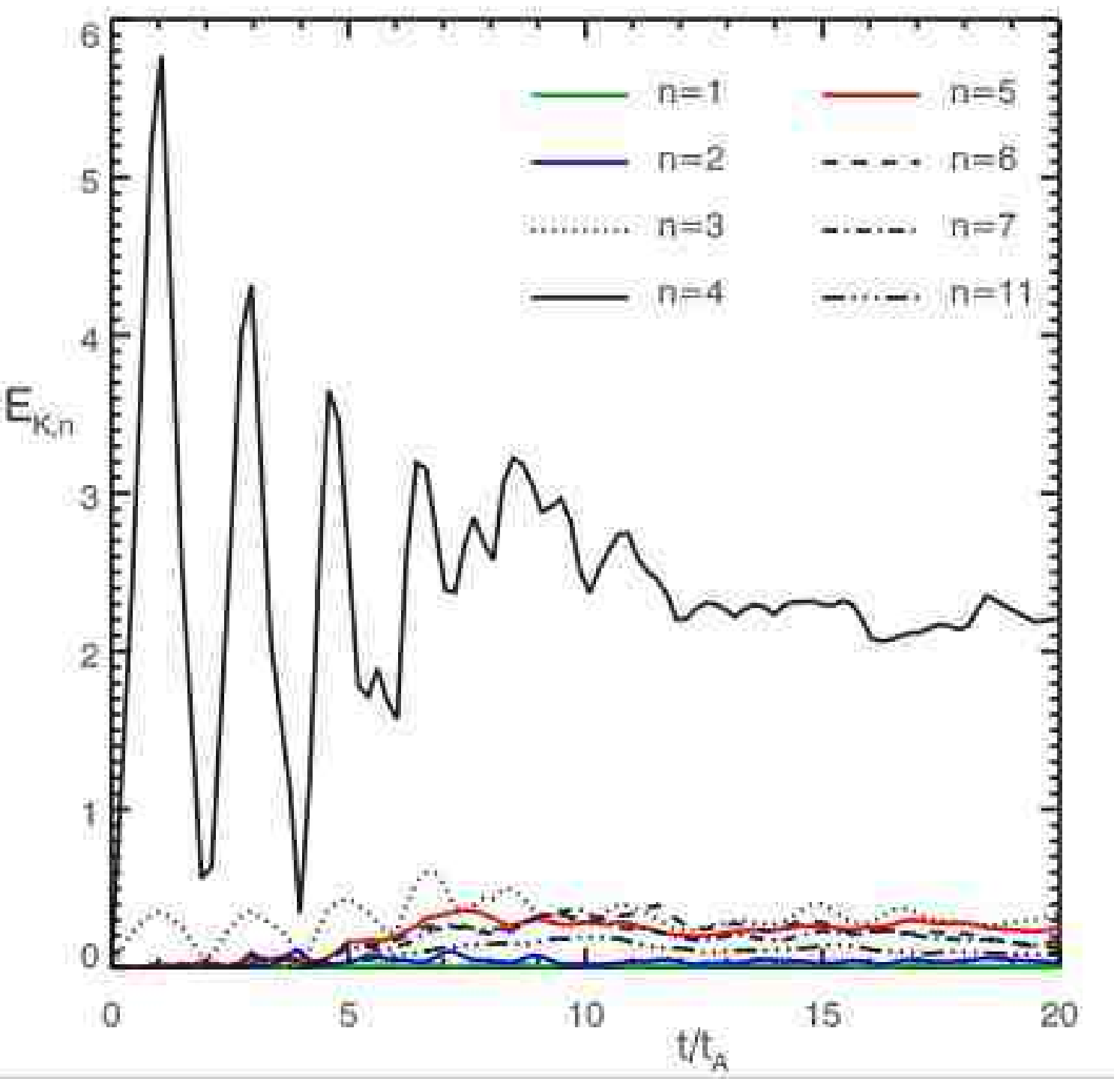}}
      \hspace{0.01\linewidth}
     %%---- start ----
      \subfloat[]{
               \label{fig:atsspnk:d}             %% label for subfigure
               \includegraphics[width=0.45\linewidth]{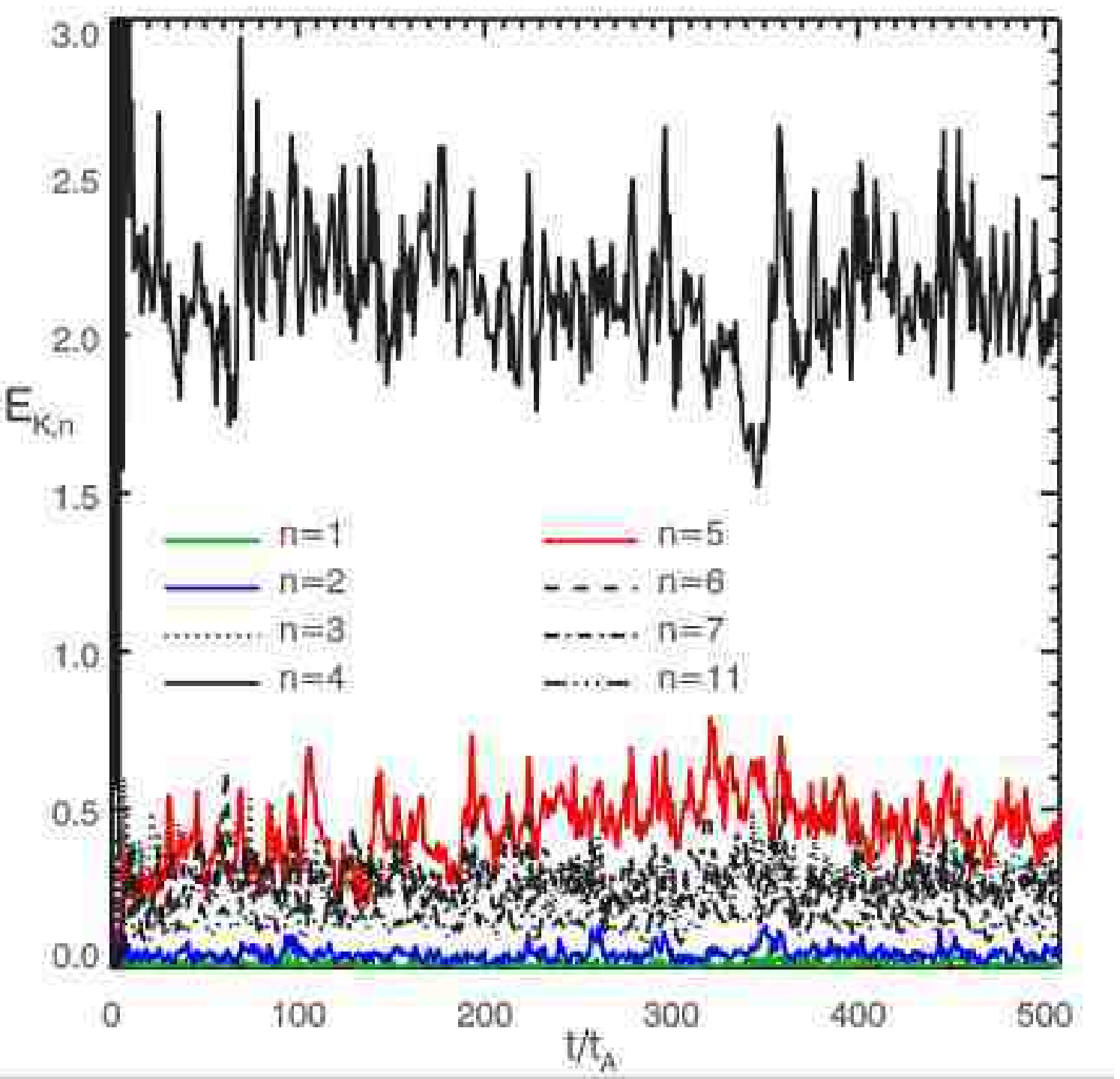}}
   \caption{Magnetic (a)-(b) and kinetic (c)-(d) energy modes as a function
   of time. (a)-(c) are a close-up, showing the first $20$ Alfv\'enic crossing times
   $\tau_{\mathcal A}$, of the respective figures (b)-(d) which show the dynamic
   of this modes for the whole duration of run~1 $\sim 500\, \tau_{\mathcal A}$.
        \label{fig:atsspnk}}            %% label for entire figure
\end{figure}
\begin{figure}[p]
      \centering
      %%---- start ----
      \subfloat[]{
               \label{fig:atsjfl:a}             %% label for subfigure
               \includegraphics[width=0.46\linewidth]{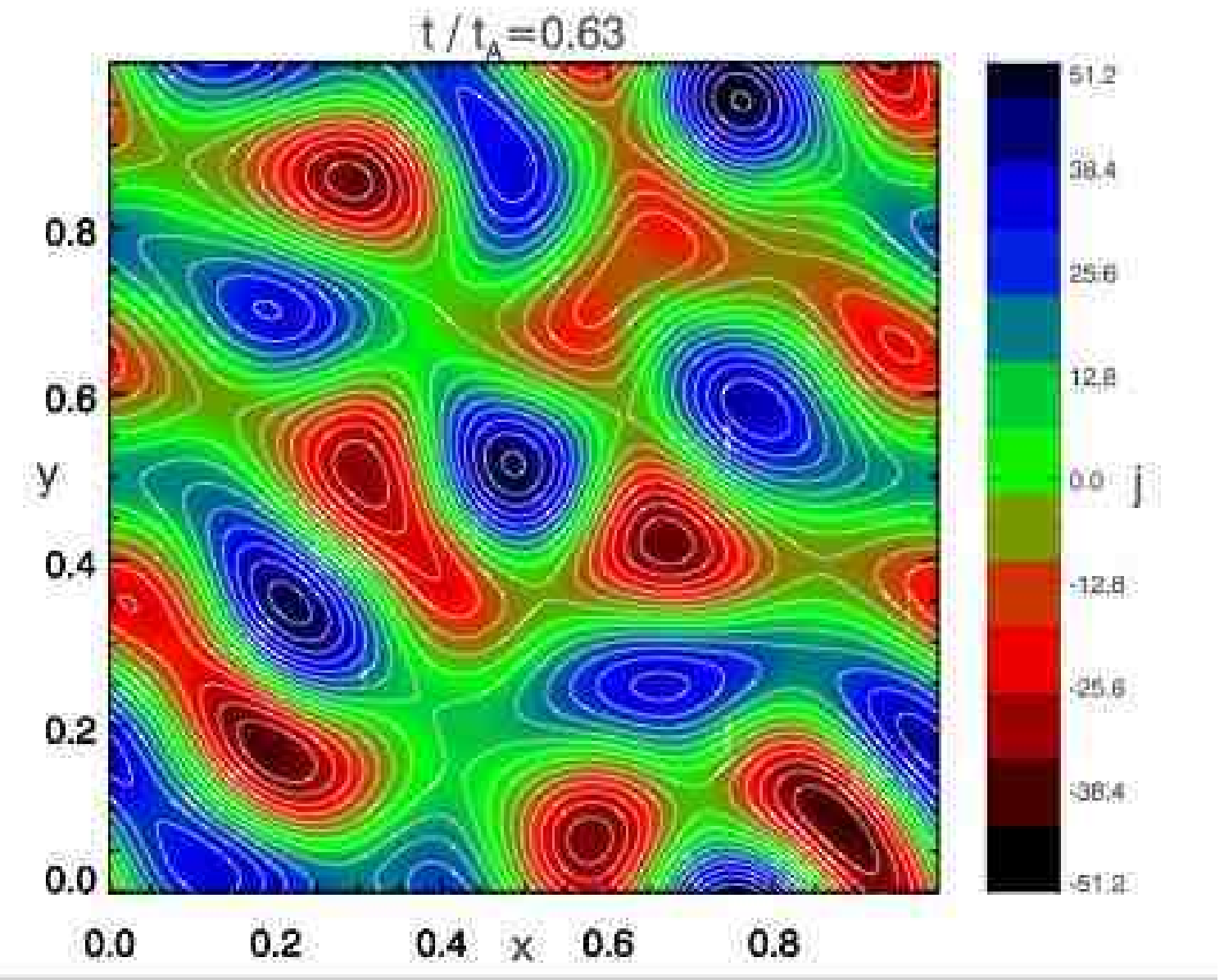}
               }
      \hspace{0.01\linewidth}
      %%---- start ----
      \subfloat[]{
               \label{fig:atsjfl:b}             %% label for subfigure
               \includegraphics[width=0.46\linewidth]{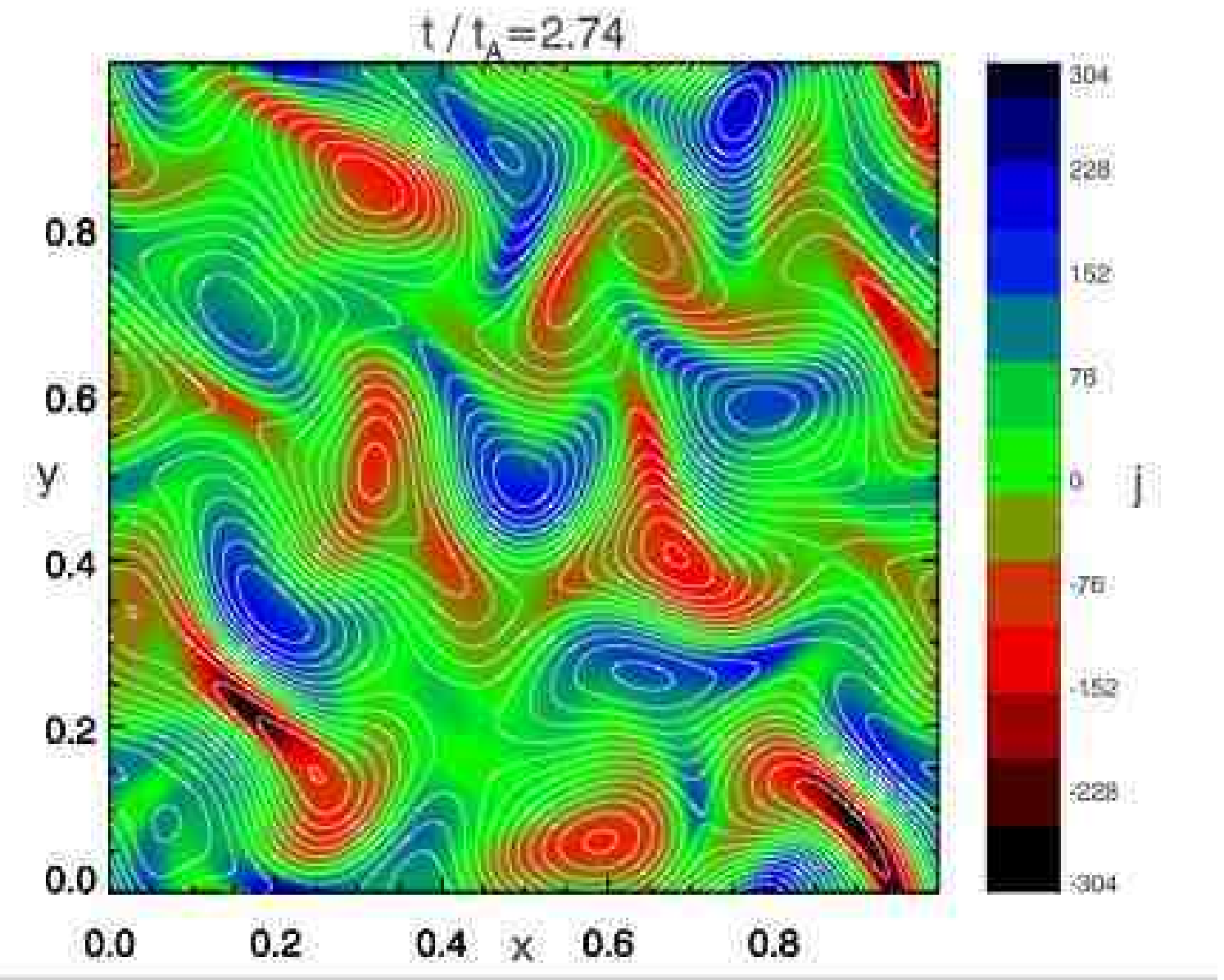}
               }\\[20pt]
      %%---- start ----
      \subfloat[]{
               \label{fig:atsjfl:c}             %% label for subfigure
               \includegraphics[width=0.46\linewidth]{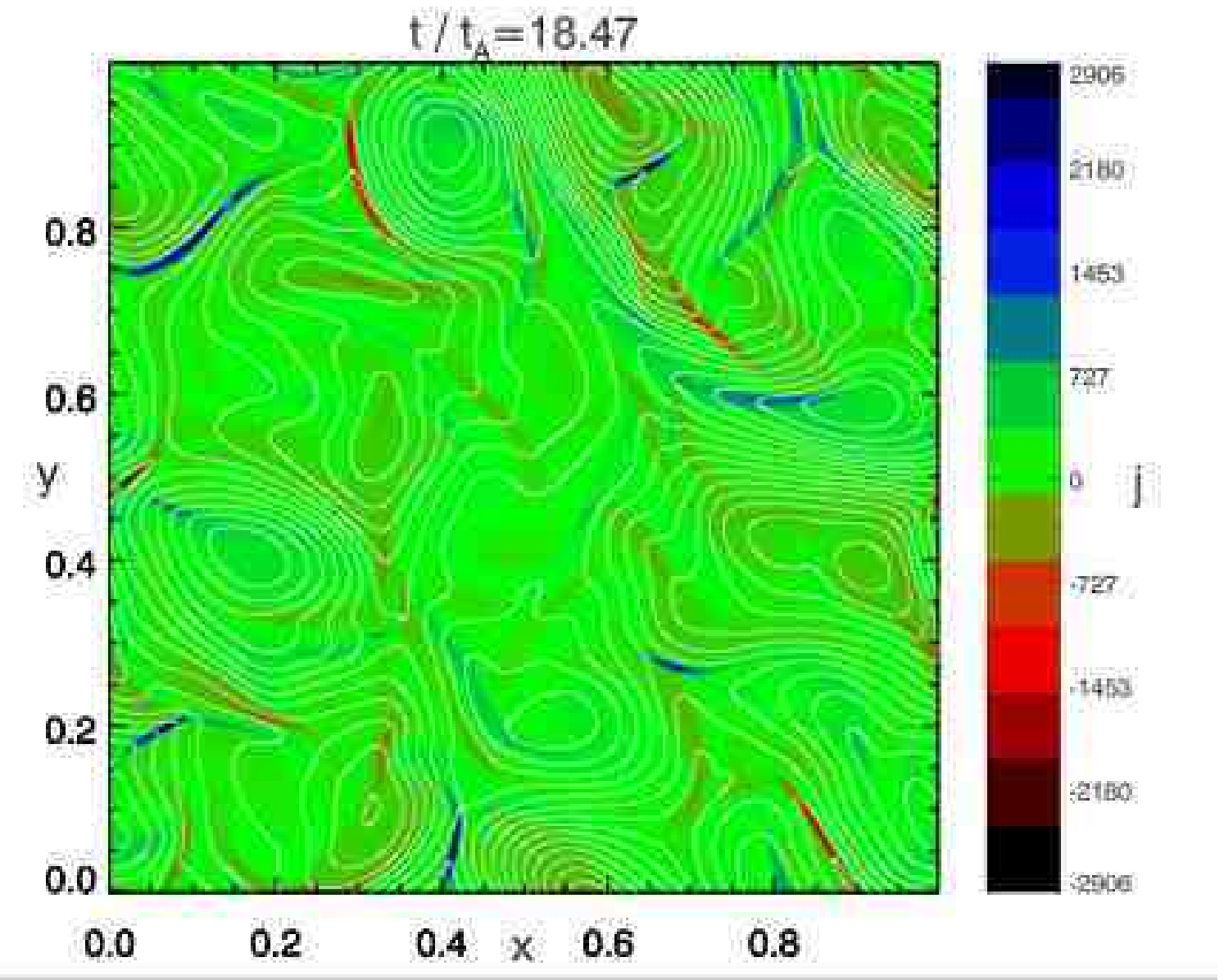}
               }
      \hspace{0.01\linewidth}
      %%---- start ----
      \subfloat[]{
               \label{fig:atsjfl:d}             %% label for subfigure
               \includegraphics[width=0.46\linewidth]{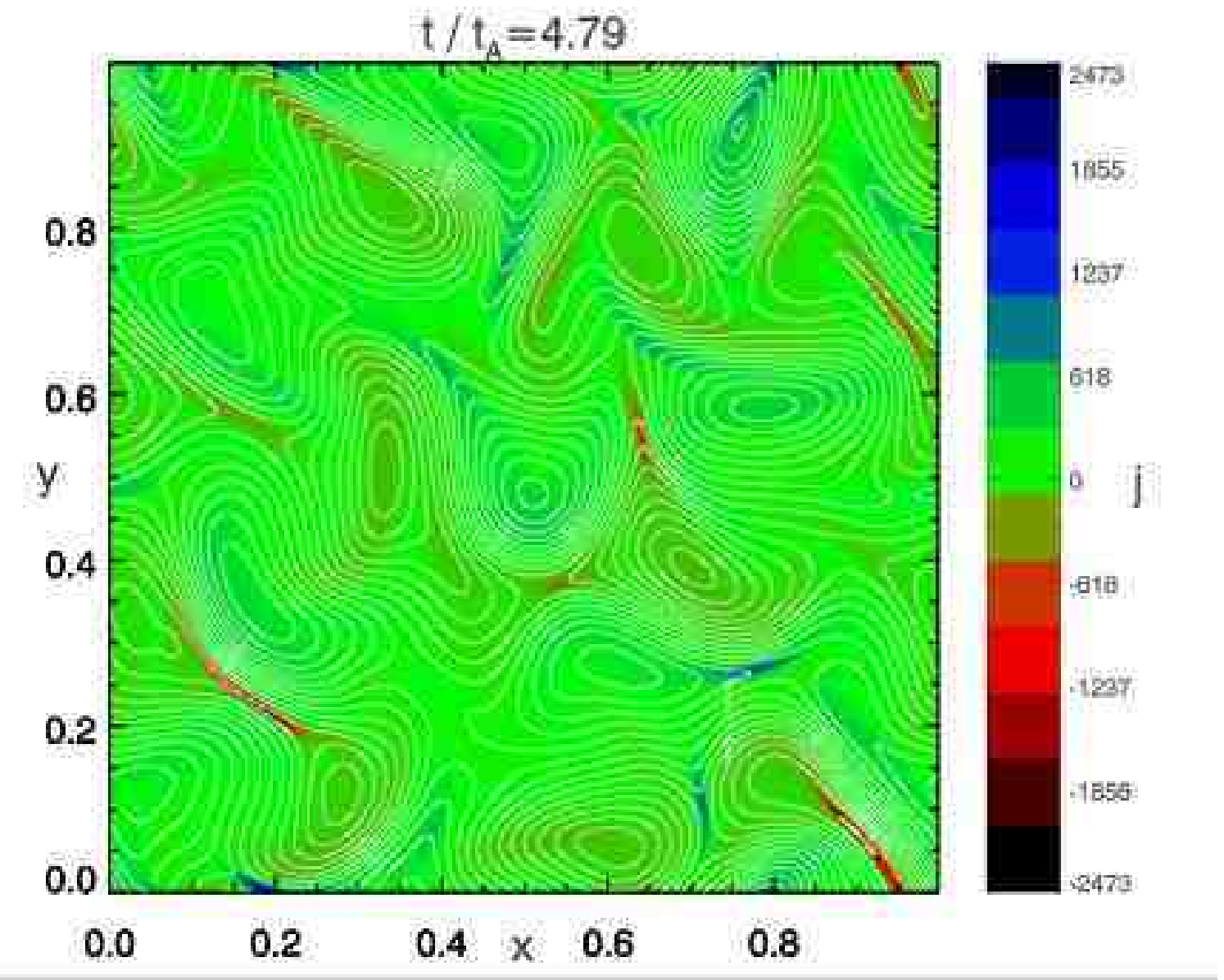}
               }
      \caption{Axial component of the current $j$ (in color) and field-lines of the
      orthogonal magnetic field in the mid-plane ($z=0$), at selected times
      covering the time interval shown in Figures~\ref{fig:atsspnk:a} and \ref{fig:atsspnk:c}. 
      \label{fig:atsjfl}}                           %% label for entire figure
\end{figure}
\begin{figure}[p]
      \centering
      %%---- start ----
      \subfloat[]{
               \label{fig:atswfl:a}             %% label for subfigure
               \includegraphics[width=0.46\linewidth]{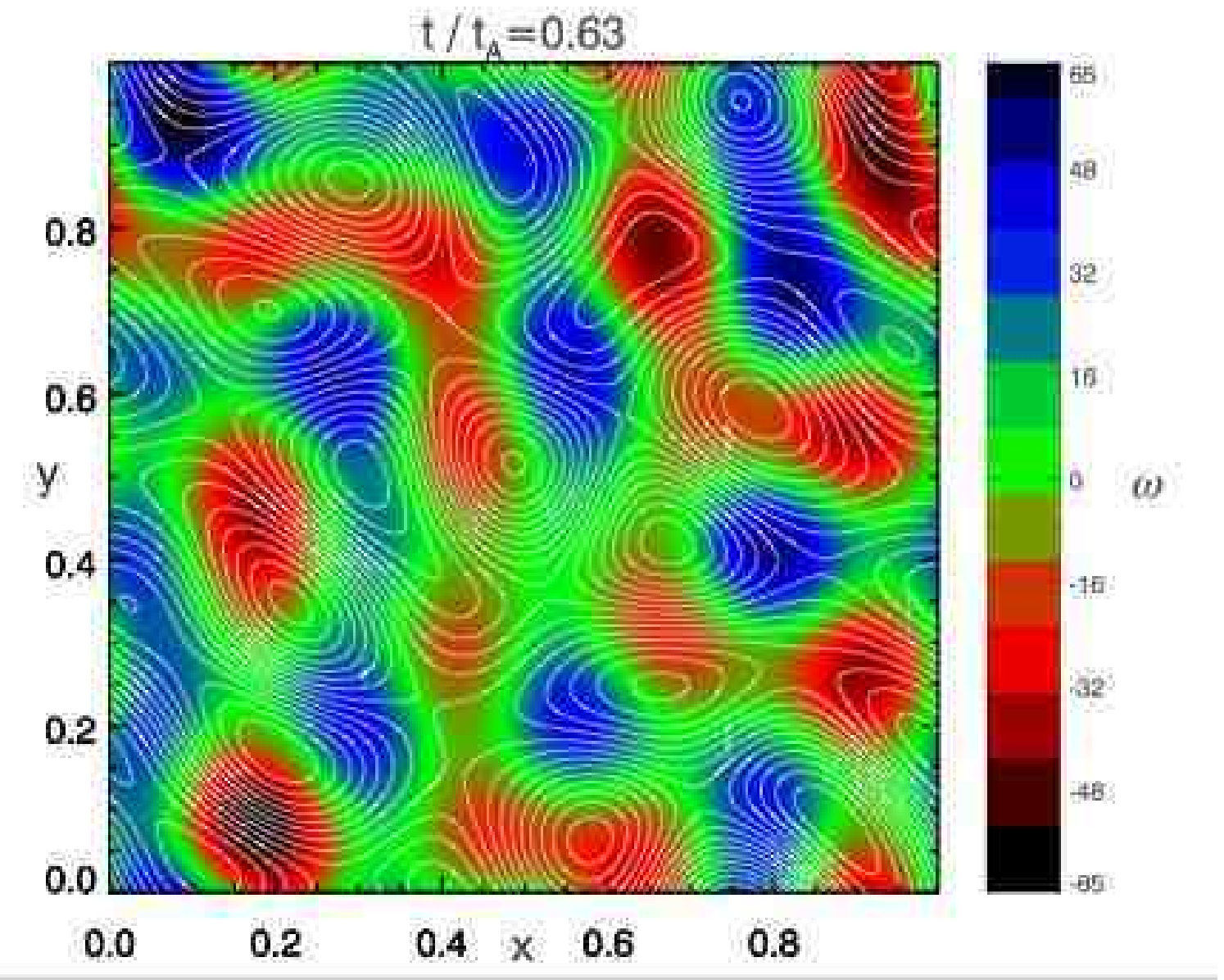}
               }
      \hspace{0.01\linewidth}
      %%---- start ----
      \subfloat[]{
               \label{fig:atswfl:b}             %% label for subfigure
               \includegraphics[width=0.46\linewidth]{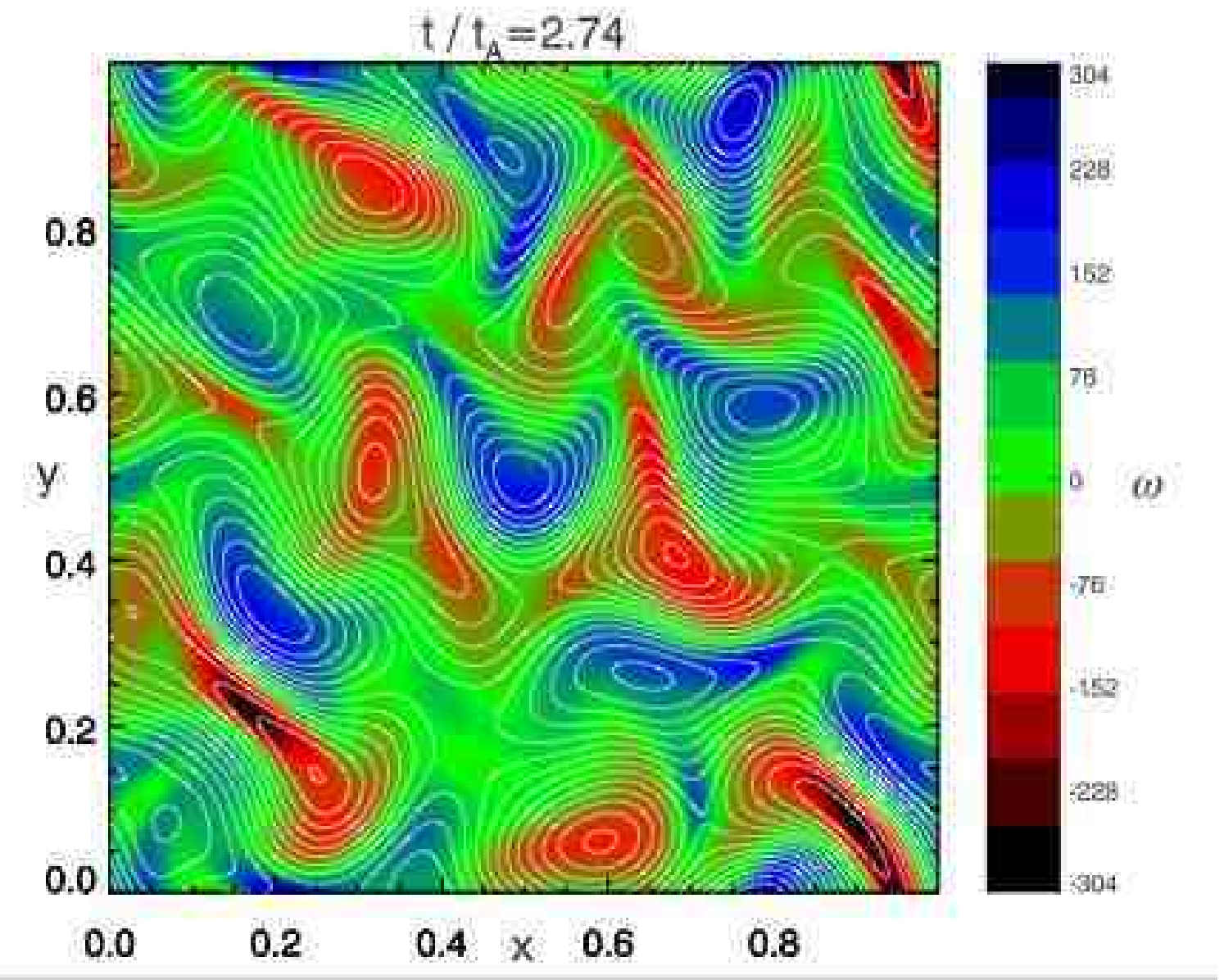}
               }\\[20pt]
      %%---- start ----
      \subfloat[]{
               \label{fig:atswfl:c}             %% label for subfigure
               \includegraphics[width=0.46\linewidth]{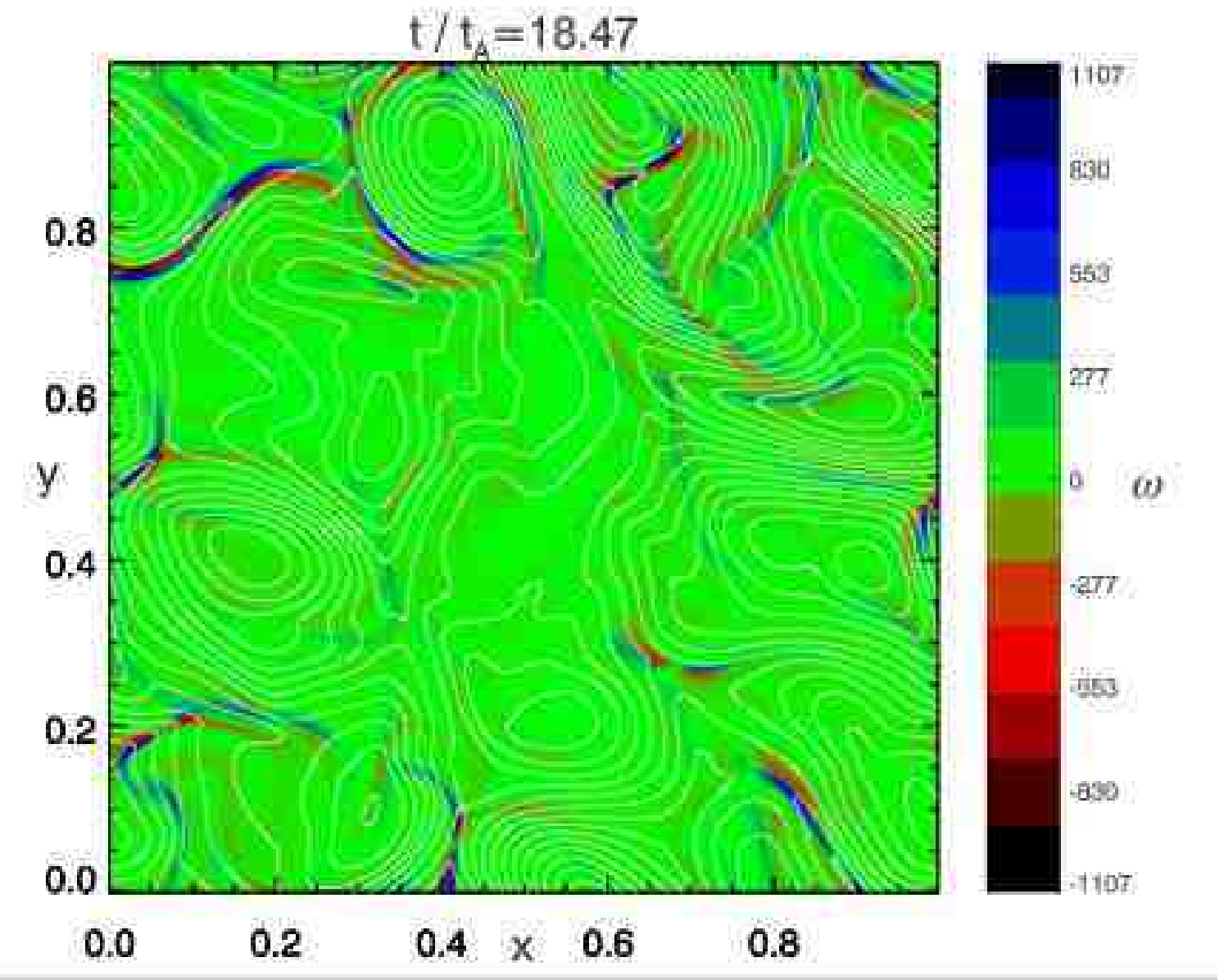}
               }
      \hspace{0.01\linewidth}
      %%---- start ----
      \subfloat[]{
               \label{fig:atswfl:d}             %% label for subfigure
               \includegraphics[width=0.46\linewidth]{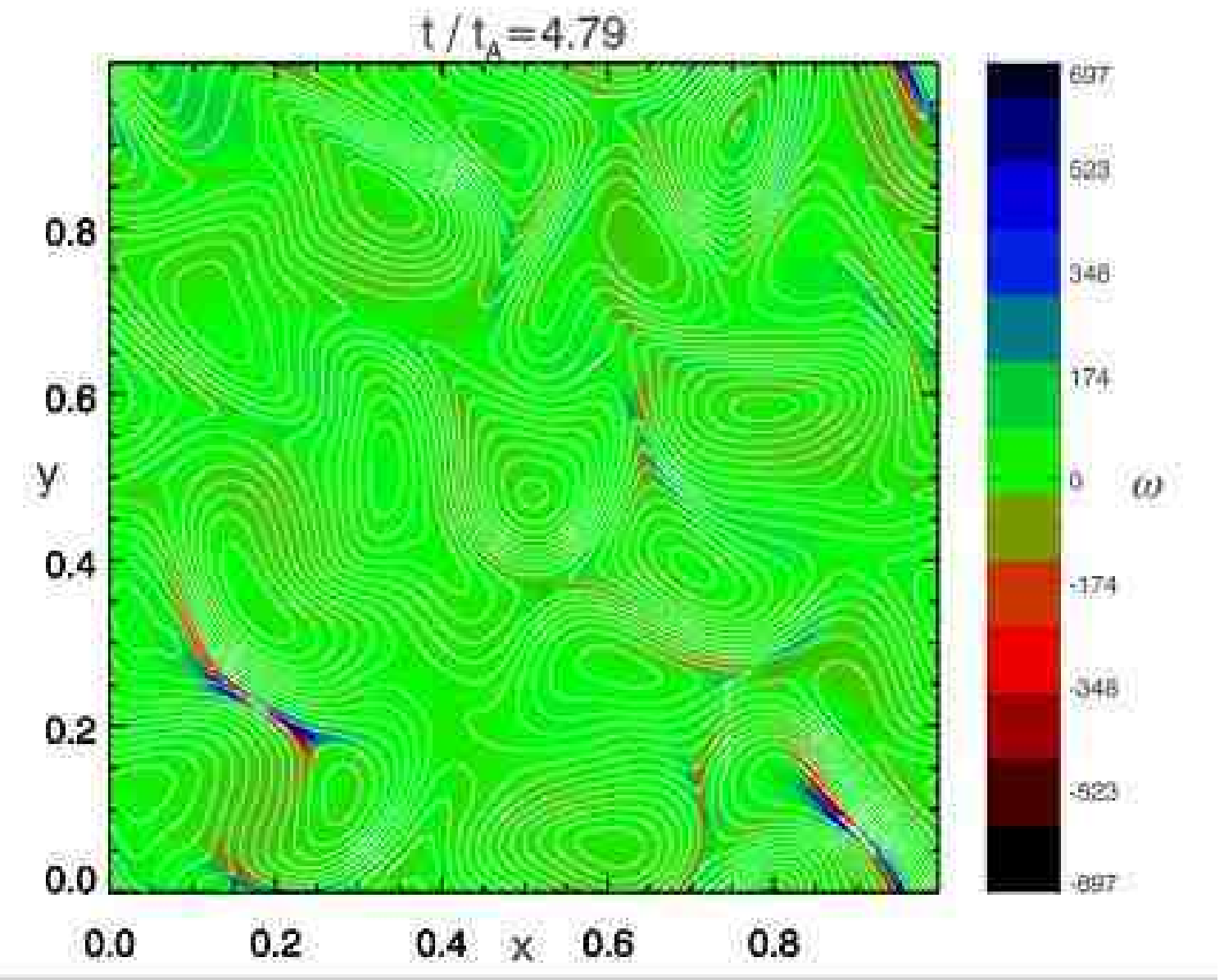}
               }
      \caption{Axial component of the vorticity $\omega$ (in color) and field-lines of the
      orthogonal magnetic field in the mid-plane ($z=0$), at selected times
      covering the time interval shown in Figures~\ref{fig:atsspnk:a} and \ref{fig:atsspnk:c}.
      \label{fig:atswfl}}                           %% label for entire figure
\end{figure}

In Figures~\ref{fig:atsspnk:a} and \ref{fig:atsspnk:c} we show the first $7$
modes and mode $n=11$ for the magnetic and kinetic energy spectra,
for the first $20$ Alfv\'enic crossing times $\tau_{\mathcal A}$.
In Figures~\ref{fig:atsjfl} and \ref{fig:atswfl} we show at selected times,
covering the same time-interval, the axial component
of the current $j$ and vorticity $\omega$ in the midplane $z=5$. 
The  modes at the injection
scale $n=4$ follow the linear dynamics described by 
equations~(\ref{eq:ts1})-(\ref{eq:ts6}) until time $\sim 6 \tau_{\mathcal A}$.
In particular the magnetic field and the current at time 
$t \sim 0.63 \tau_{\mathcal A}$ --- i.e.\, just after time 
$t = 0.5\, \tau_{\mathcal A}$ when the two counterpropagating
front-waves $\bsy{z}^{\pm}$ reach the midplane ---
 are mapping the photospheric
velocities $\bsy{u}^L - \bsy{u}^0$ as predicted by our linear analys.
Figures~\ref{fig:atsfv:a}  and \ref{fig:atsfv:b} show the two photospheric
velocity patterns separately, while for comparison with  Figure~\ref{fig:atsjfl:a}
we show the contour of the linear combination $\bsy{u}^L - \bsy{u}^0$
in Figure~\ref{fig:atsavz:a}.
\begin{figure}
      \centering
      %%---- start ----
      \subfloat[]{
               \label{fig:atsavz:a}             %% label for subfigure
               \includegraphics[width=0.45\linewidth]{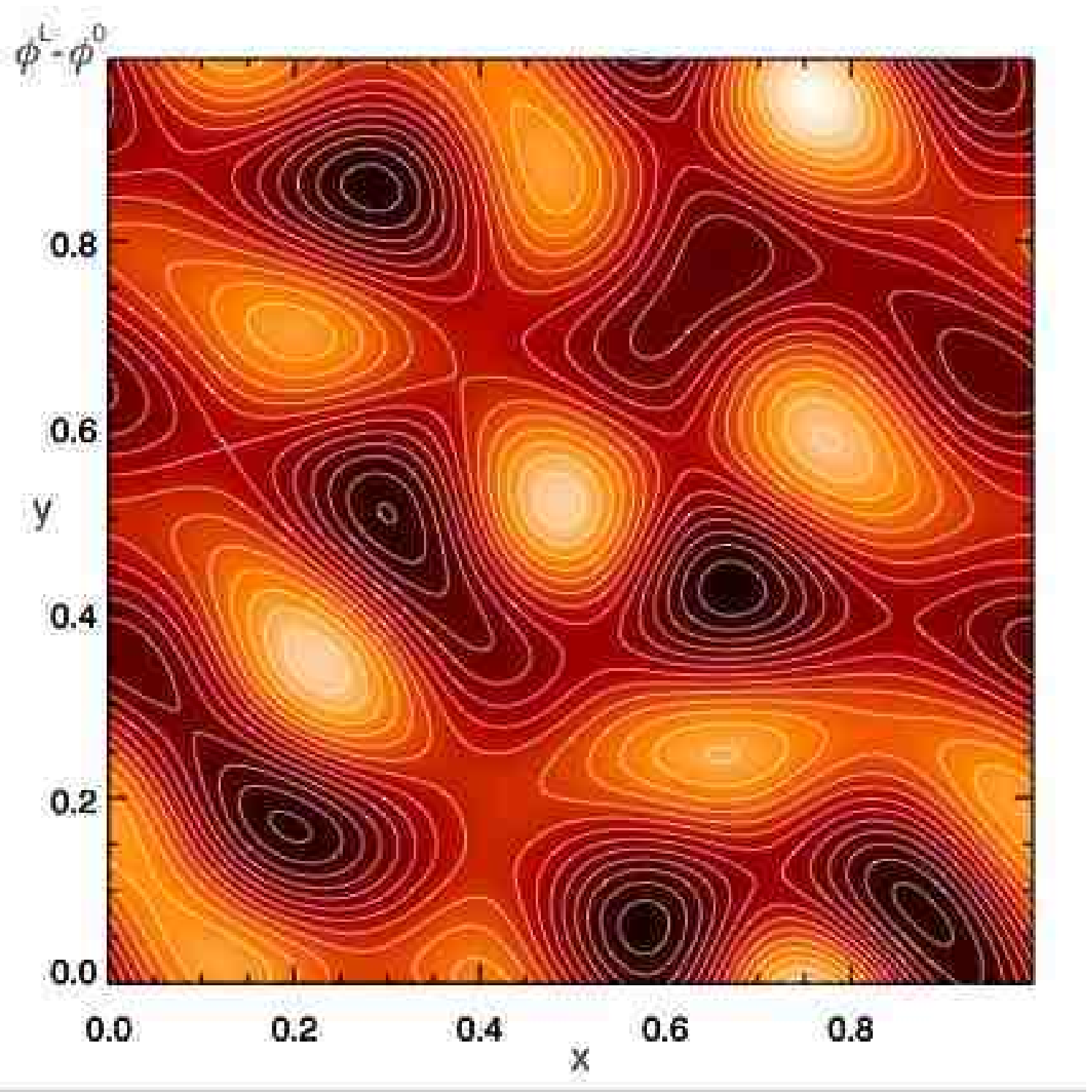}}
      \hspace{0.01\linewidth}
     %%---- start ----
      \subfloat[]{
               \label{fig:atsavz:b}             %% label for subfigure
               \includegraphics[width=0.45\linewidth]{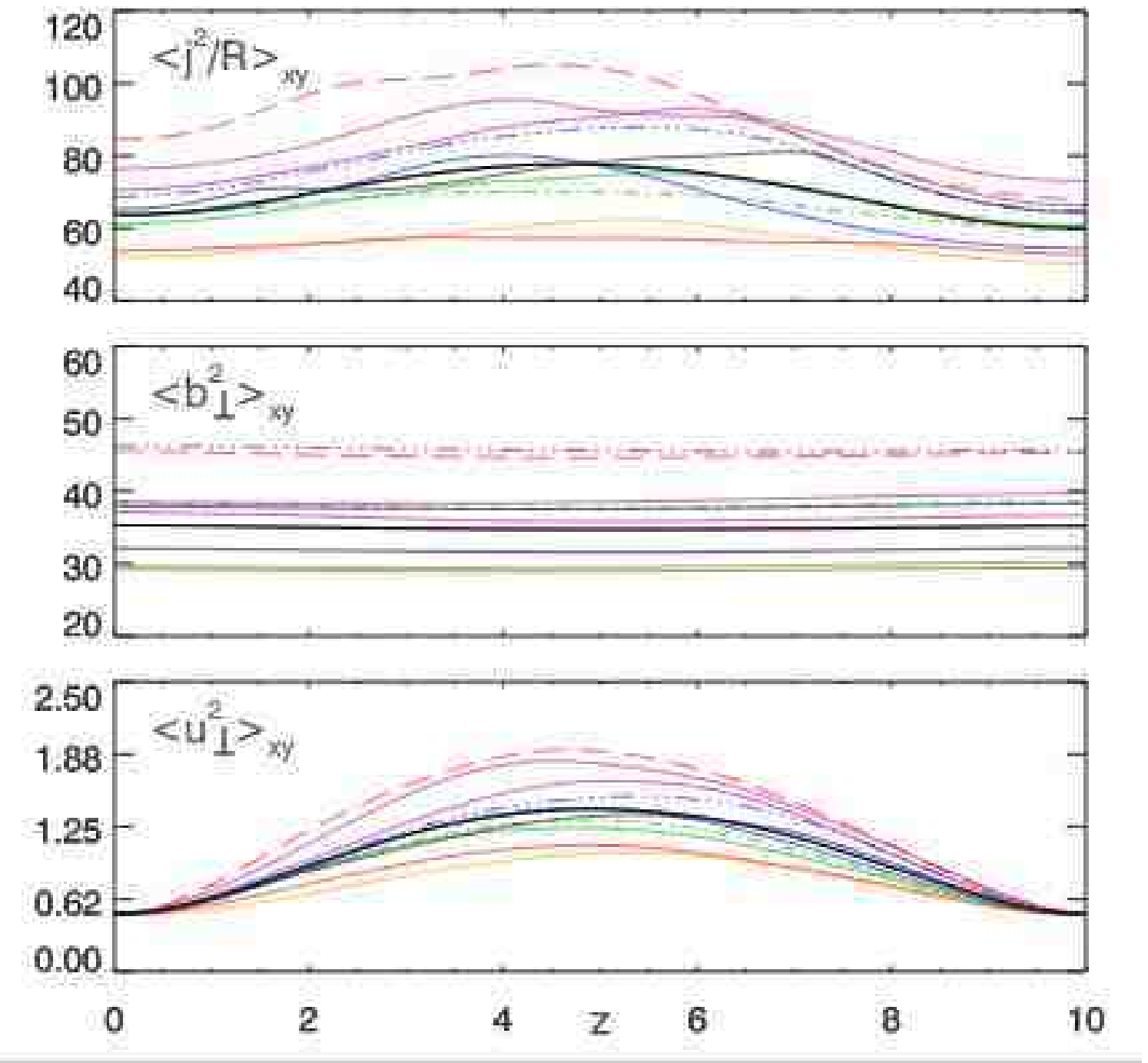}}
   \caption{(a) Contour lines of the linear combination of the boundary
   velocity fields $\bsy{u}^L - \bsy{u}^0$. (b) Current density $j$ squared
   averaged in the orthogonal $x-y$ planes as a function of $z$ at selected
   times. Same averages for the orthogonal magnetic and velocity fields. 
        \label{fig:atsavz}}             %% label for entire figure
\end{figure}

In particular the magnetic energy mode $n=4$ grows roughly quadratically
in time, while the $n=4$ kinetic energy mode exhibits a sawtooth structure
with a period $\tau_{\mathcal A}$. The dynamics are  described by a
cascade towards smaller scales. No instability of any kind is detected.
In physical space this cascade 
corresponds to the formation of small scales which are organized in
vortex-current sheets. This picture remains unaltered throughout  the rest of
the nonlinear stage. Energy is injected at the large scales ($n \sim 4$) by photospheric 
motions and an energy cascade develops, which transports this energy from the
large to the small scales where it is finally dissipated. 
Figures~\ref{fig:atsjfl:c} and \ref{fig:atswfl:c} show that during the nonlinear stage
($t \sim 18.5\, \tau_{\mathcal A}$) the magnetic field exhibits a complex topology
and that distorted current sheets are almost always present together with
quadrupolar vorticity structures. These current sheets are subsequently
subject to tearing, which leads to suppose that the dissipation mechanism
is \emph{nonlinear magnetic reconnection in a turbulent environment.}

An important question is what is the dissipation rate and how much it is
influenced by the magnetic reconnection rate.
Longcope \& Sudan~\cite{long94} suppose that the reconnection proceeds
at a rate independent from the turbulent state of the system, which they 
completely neglect.

A similar behavior, i.e.\ microcurrent sheets formed by a turbulent cascade
and then subject to tearing, has been detected in decaying 2D turbulence 
e.g.\ by Politano et~al.~\cite{poli89}. Biskamp \& Welter~\cite{bisk89} and
Biskamp \& Schwarz~\cite{bisk01} have also studied its influence on
dissipation rates. In particular Biskamp \& Schwarz~\cite{bisk01} have
performed the highest resolution 2D MHD simulations 
finding that the tearing mode occasionally occurs but its influence
is rather weak. This is because this reconnection is taking place
in a turbulent environment, so that the current sheets dynamics
is influenced by the larger-scale dynamics.

Our numerical simulations leads us to suppose that
whether or not the reconnection rate is  influenced by the turbulent dynamics
the tearings happen continuously, so that the energy flux
trasported by the cascade is dissipated in many ``microbursts'',
so that the turbulent energy flux is dissipated in a statistically steady
fashion.   

In particular, the fact that the spectrum which is developed
(see Figure~\ref{fig:atssp}) does not show any anomalous structure
at smaller scales means that the rate at which energy is dissipated 
by magnetic reconnection is that at which energy flows along
the inertial range. This scenario is reinforced by the fact that
when we use hyperdiffusion the spectra that are found are
consistent with those expected theoretically for weak anisotropic
turbulence, as shown in the following paragraphs.

The only difference detected at longer times is the growth of 
the  $n=1$ and $n=2$ modes for the magnetic energy 
(see Figures~\ref{fig:atsspnk:b}), which show a tendency towards an
inverse cascade, even though a complete inverse cascade never
fully develops. This is a difference with the corresponding 2D
simulations (Einaudi \& Velli~\cite{ein:1999}), where at long times a
complete inverse cascade leads to the coalescence of the magnetic
islands. The kinetic energy mode at the injection scale $n=4$
fluctuates around an equilibrium value which is the average 
of its value during the linear regime, and indicates that the
dynamics for the modes at $n=4$ do not depart dramatically
from the linear behavior (\ref{eq:ts1})-(\ref{eq:ts6}).

The inhibition of an inverse cascade is probably due to the stiffness
of the magnetic field lines, which are only slightly bent along the
axial direction. In fact  the magnetic field line tension associated with
these field lines inhibits the coalescence of the magnetic islands
that form in the orthogonal planes. In Figure~\ref{fig:spn} we plot
the first 4 modes for the magnetic energy with two different
values of the axial Alfv\'enic velocity, respectively $v_{\mathcal A} = 50$
and $v_{\mathcal A} = 1000$. At $v_{\mathcal A} = 50$ the field line
tension is smaller than for $v_{\mathcal A} = 1000$, and in fact
in the first case an inverse cascade is clearly present (Figure~\ref{fig:spn:a})
while in the second it is not (Figure~\ref{fig:spn:b}).
\begin{figure}[t]
      \centering
      %%---- start ----
      \subfloat[$v_{\mathcal A} = 50$]{
               \label{fig:spn:a}             %% label for subfigure
               \includegraphics[width=0.45\linewidth]{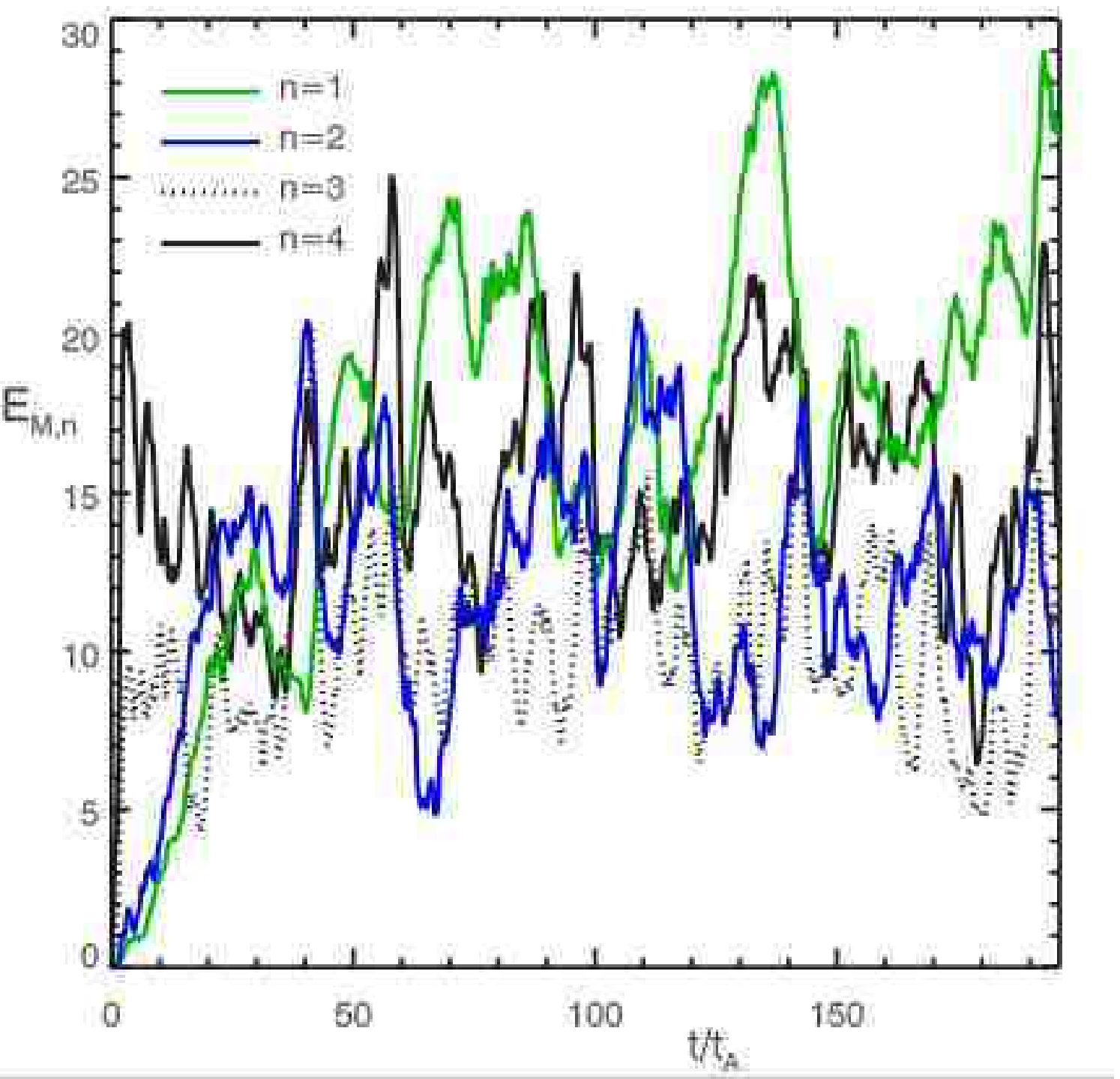}}
      \hspace{0.01\linewidth}
     %%---- start ----
      \subfloat[$v_{\mathcal A} = 1000$]{
               \label{fig:spn:b}             %% label for subfigure
               \includegraphics[width=0.45\linewidth]{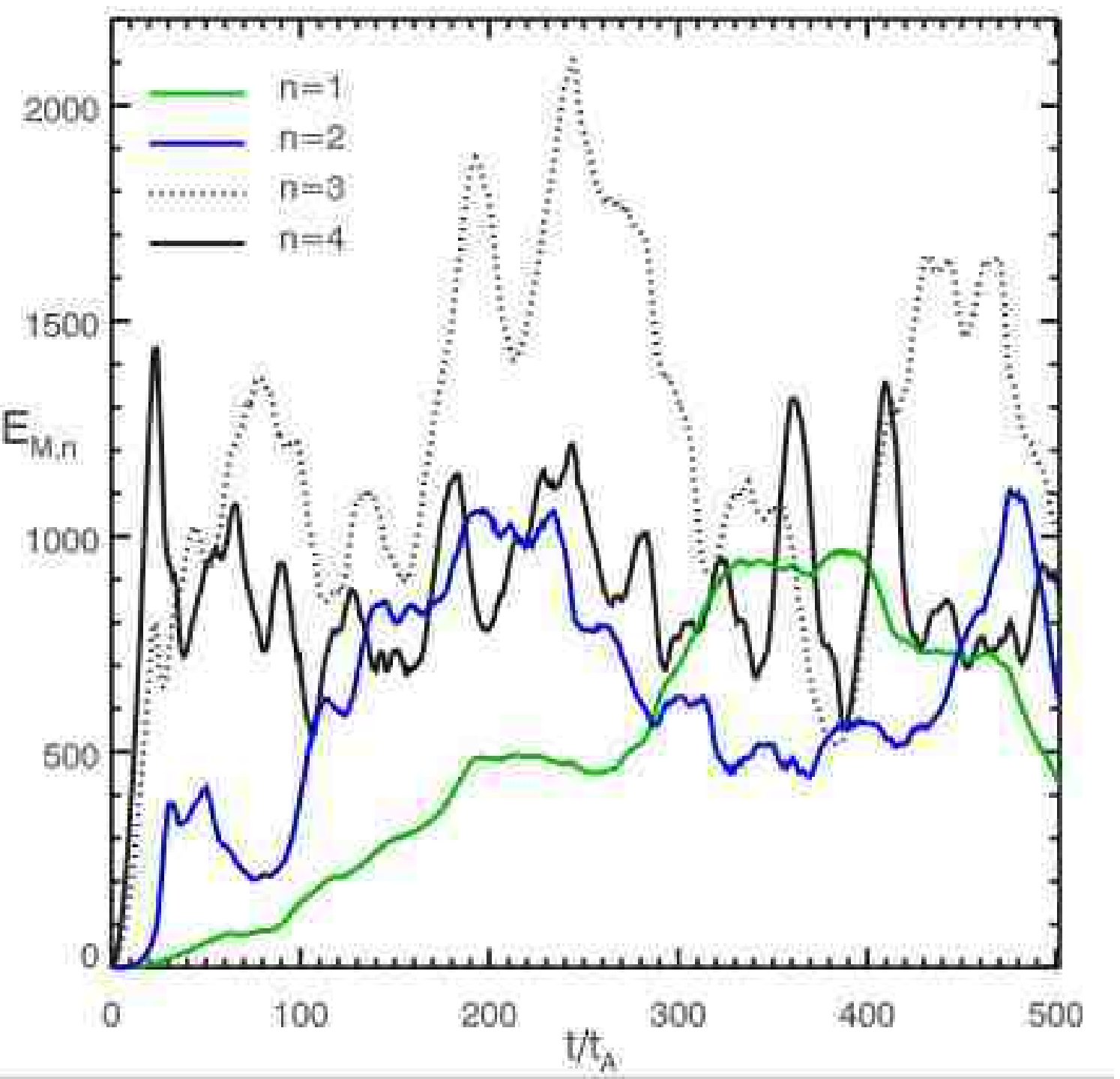}}
   \caption{Magnetic energy modes $n=1, 2, 3, 4$ 
   ($n=4$ is the injection wavenumber) as a function of time
   for run~4 (a) and run~7 (b), respectively with $v_{\mathcal A} = 50$
   and $v_{\mathcal A} = 1000$, showing that a bigger
   axial magnetic field inhibits through magnetic field line tension
   an inverse cascade.
        \label{fig:spn}}             %% label for entire figure
\end{figure}

We show in Figure~\ref{fig:atsavz:b}  some physical quantities  averaged
in the $x-y$ planes as a function of the axial direction $z$.
Both the magnetic and velocity fields are elongated along
the axial direction,  as is the current. 

\begin{figure}[t]
      \centering
      %%---- start ----
      \subfloat[]{
               \label{fig:atsxy:a}             %% label for subfigure
               \includegraphics[width=0.45\linewidth]{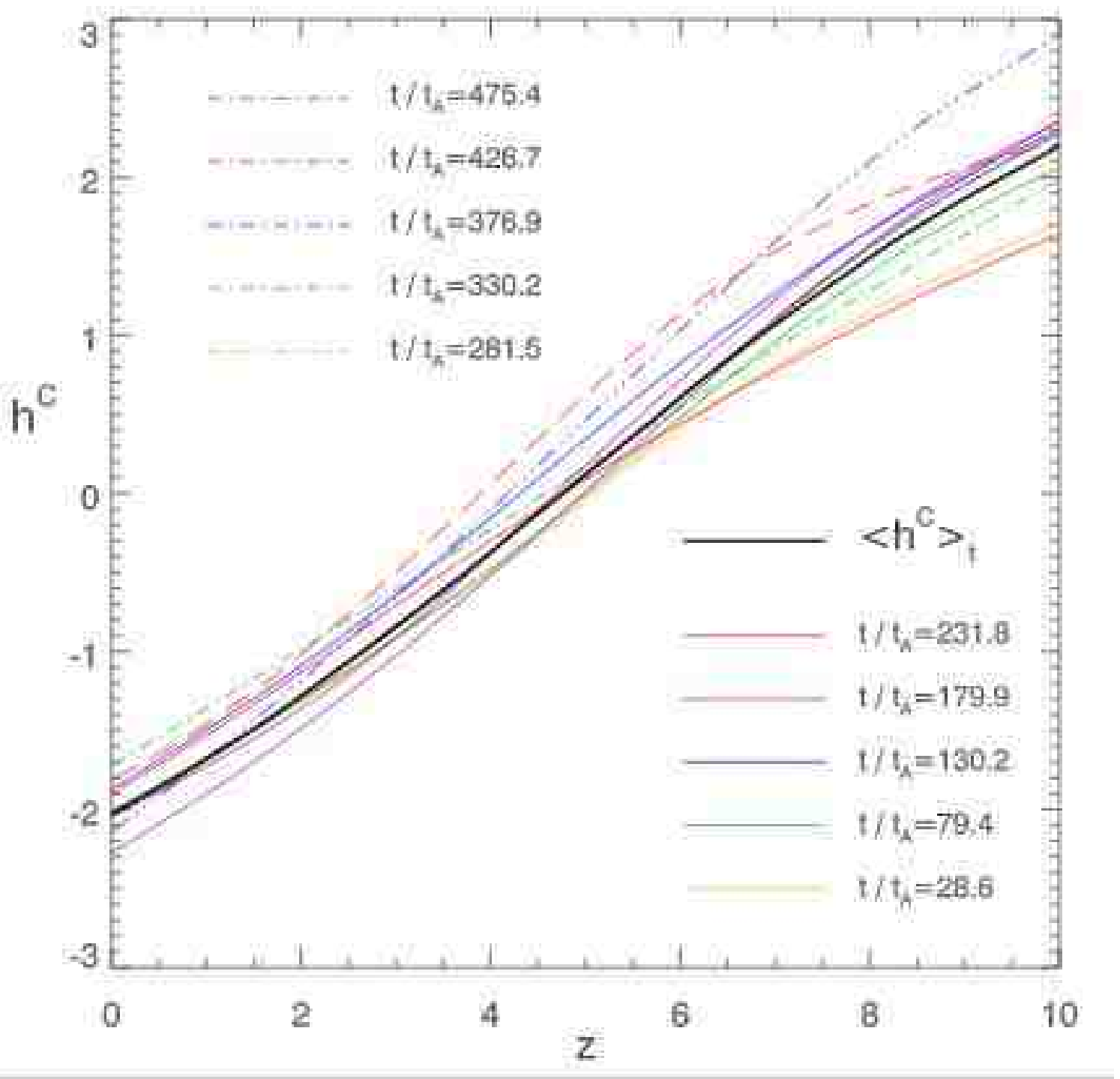}}
      \hspace{0.01\linewidth}
     %%---- start ----
      \subfloat[]{
               \label{fig:atsxy:b}             %% label for subfigure
               \includegraphics[width=0.45\linewidth]{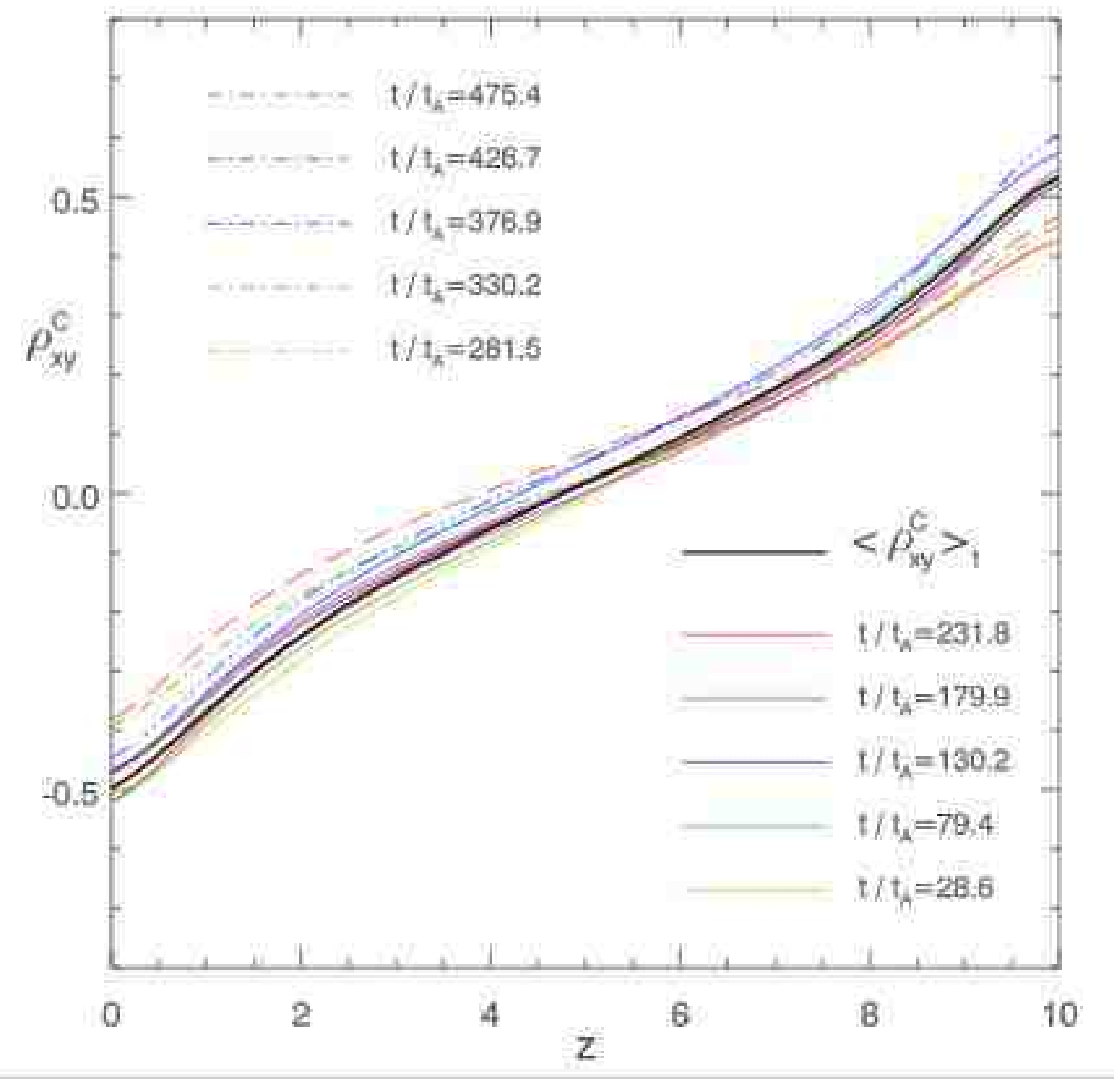}}
   \caption{(a) Cross helicity density $h^C$ as a function of $z$ at 
   selected times.
   (b) Velocity-magnetic field correlation in the orthogonal planes
   as a function of $z$ at selected times.
        \label{fig:atsxy}}             %% label for entire figure
\end{figure}
Finally in Figure~\ref{fig:atsxy:a} we show the magnetic helicity density
integrated in the $x-y$ planes as a function of $z$ at selected times
\be
h^C \left( z, t \right) = \iint_{0\ \ \ }^{\ell\ \, } \displaylimits \! \ud x \, \ud y\, 
\bsy{u}_{\perp} \cdot \bsy{b}_{\perp}
\ee
The figure shows an almost linear behavior of $h^C$ as a function of $z$,
confirming the homogeneity of this system in the axial direction.
$h^C$  is the axial component of the Poynting vector $\bsy S$ divided 
by the Alfv\'enic velocity $v_{\mathcal A}$ (see eq.~(\ref{eq:pfz})), i.e.\
\be
h^C \left( z, t \right) = - \frac{1}{v_{\mathcal A}}\, 
\iint_{0\ \ \ }^{\ell\ \, } \displaylimits \! \ud x \, \ud y\ \bsy{S} \cdot \bsy{e}_z
\ee
so that energy is flowing from the boundaries into the computational box
at a uniform rate. The energy equation~(\ref{eq:eneq}) still holds for
any slice of the computational box included between two planes $z=const$
so  for any two slices of equal volume the rate of injection of energy and
the dissipation rate are the same.

The velocity-magnetic field correlations in the $x-y$ planes $\rho^C_{xy}$,
which is closely related to the cross helicity density $h^C$, is defined as
\be
\rho^C_{xy} \left( z, t \right) =
\frac{ \iint_0^{\ell} \! \ud x \, \ud y\, \bsy{u}_{\perp} \cdot \bsy{b}_{\perp} }
        {\left( \iint_0^{\ell} \! \ud x \, \ud y\, \bsy{u}^2_{\perp} \cdot 
          \iint_0^{\ell} \! \ud x \, \ud y\, \bsy{b}^2_{\perp} \right)^{1/2}}
\ee
In Figure~\ref{fig:atsxy:b} we show this quantity as a function of $z$ at selected
times. As expected the velocity and magnetic field are more correlated
at the boundaries, where the velocity is imposed as a boundary condition,
than in the interior of the computational box.

\subsection{Weak Turbulence Spectra}

In order to resolve the inertial range we perfomed a series of 
high-resolution simulations using hyperdiffusion. The numerical grid was
$512 \times 512 \times 200$, and the axial Alfv\'enic velocities are
$v_{\mathcal A} = 50,\ 200,\ 400,\ 1000$. In Figure~\ref{fig:hypsp} we
show the time-averaged $E^-_n$ energy spectrum for  these simulations.
The time averages have been performed  over $\sim 200,\
450,\ 77$ and $500\, \tau_{\mathcal A}$ respectively.
\begin{figure}[p]
      \centering
      %%---- start ----
      \subfloat[]{
               \label{fig:hypsp:a}             %% label for subfigure
               \includegraphics[width=0.46\linewidth]{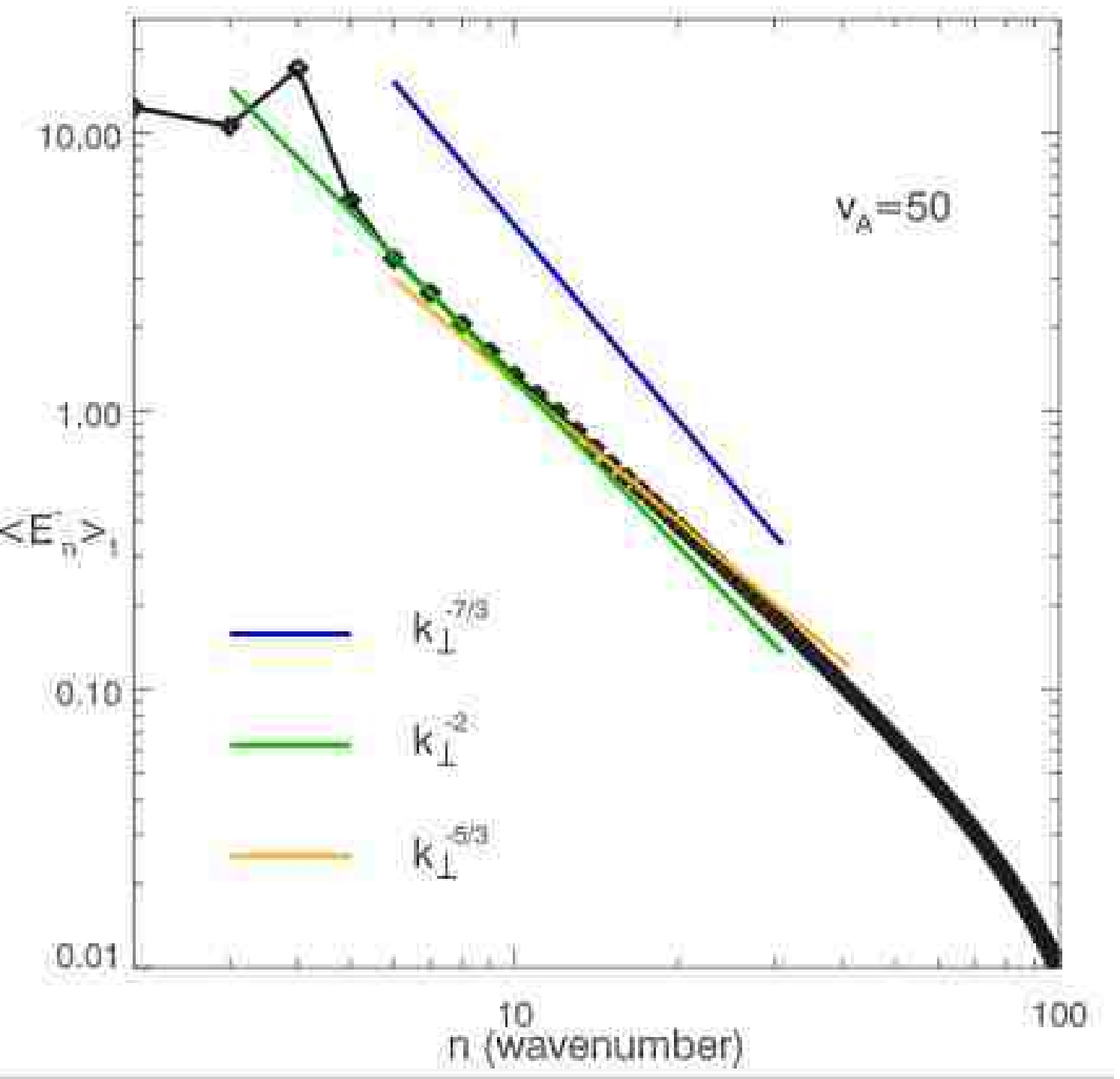}
               }
      \hspace{0.01\linewidth}
      %%---- start ----
      \subfloat[]{
               \label{fig:hypsp:b}             %% label for subfigure
               \includegraphics[width=0.46\linewidth]{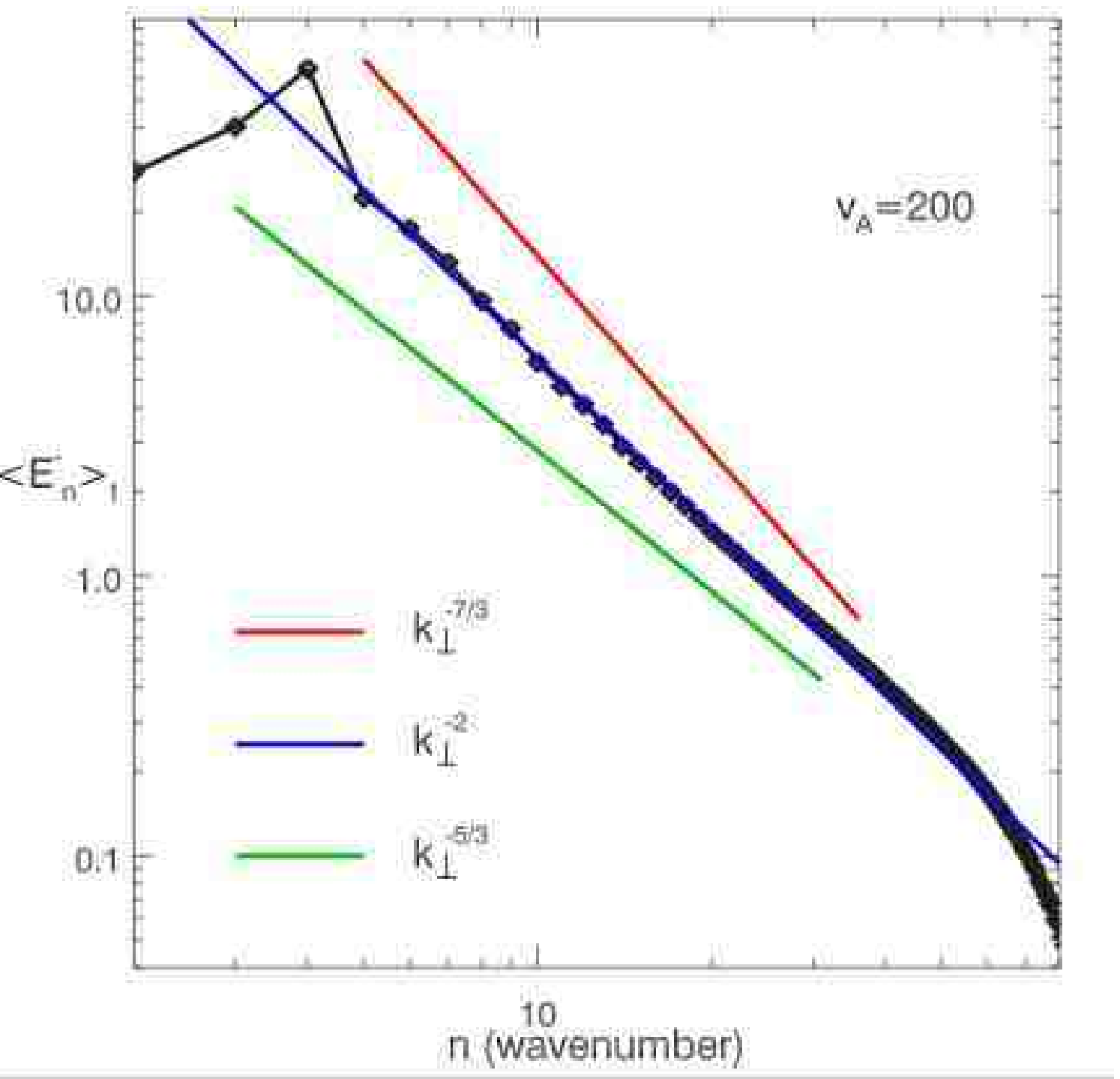}
               }\\[20pt]
      %%---- start ----
      \subfloat[]{
               \label{fig:hypsp:c}             %% label for subfigure
               \includegraphics[width=0.46\linewidth]{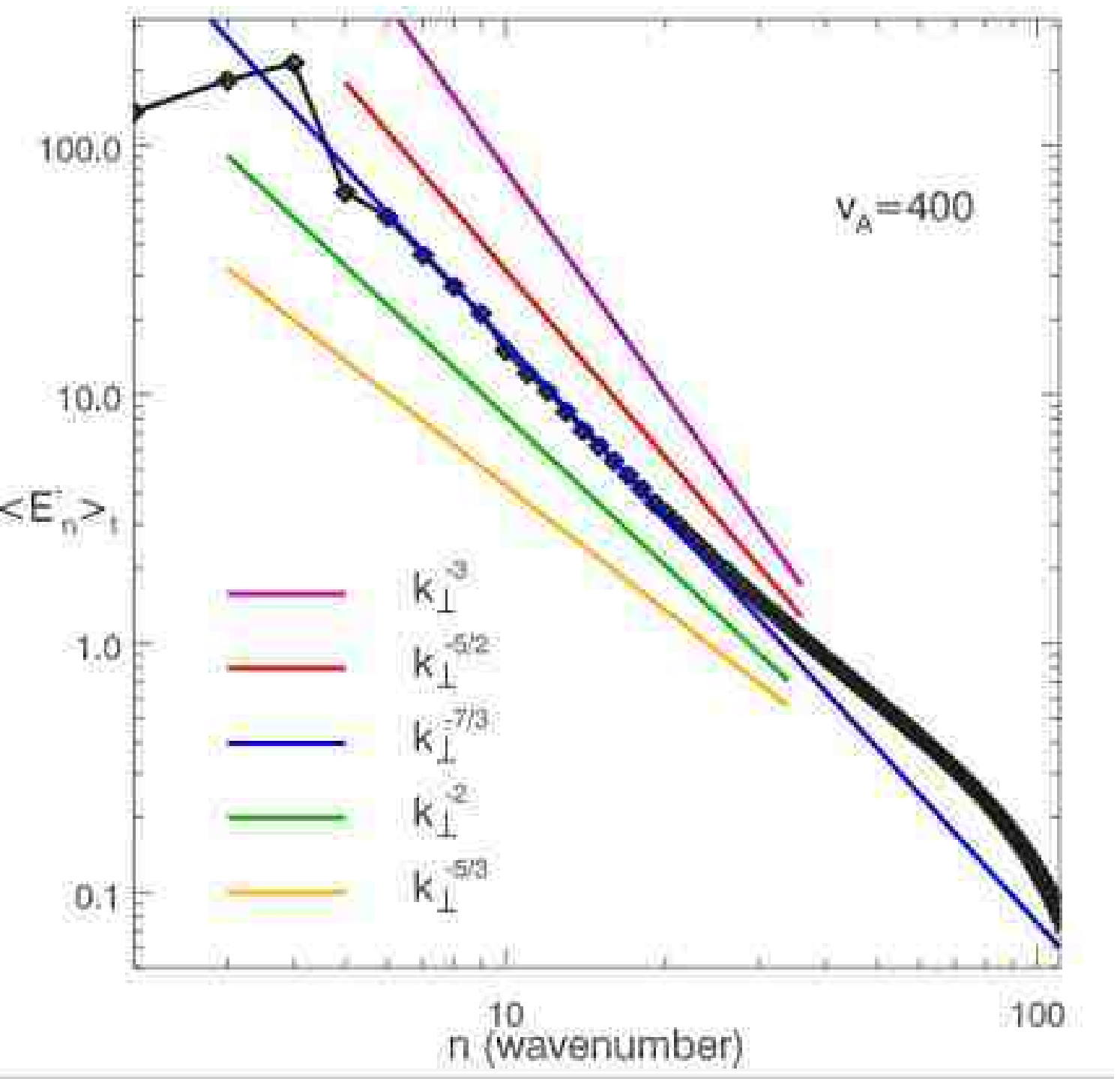}
               }
      \hspace{0.01\linewidth}
      %%---- start ----
      \subfloat[]{
               \label{fig:hypsp:d}             %% label for subfigure
               \includegraphics[width=0.46\linewidth]{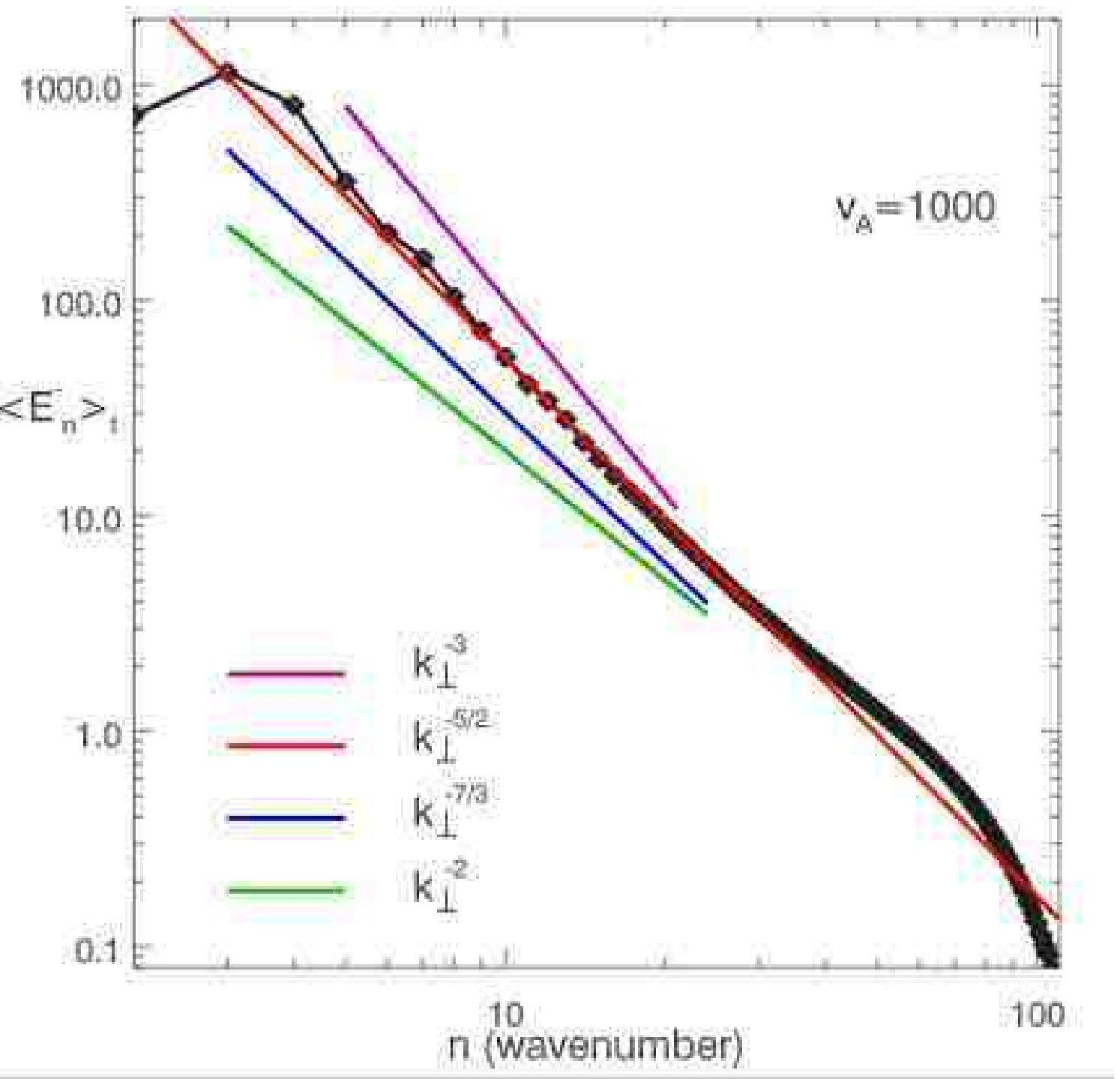}
               }
      \caption{$E^-_n$ energy spectra for 4 simulations with 
      $v_{\mathcal A} = 50,\ 200,\ 400,\ 1000$. Hyperdiffusion with dissipativity 
      $\alpha = 4$ has been used for all the simulations with respectively
      $\mathcal{R}_4 = 3 \cdot 10^{20}$, $1 \cdot 10^{20}$, $1 \cdot 10^{19}$,
      $1 \cdot 10^{19}$. All the simulations use a numerical grid with
      $512 \times 512 \times 200$ grid points.
      \label{fig:hypsp}}                           %% label for entire figure
\end{figure}
Comparing Figure~\ref{fig:hypsp:b} with Figure~\ref{fig:atssp:a} we notice 
that a well-developed inertial range is now present, in contrast to the 
analogous case with normal diffusion. 
In the framework of the weak anisotropic turbulence
theory that we have briefly surveyed in \S~\ref{par:bik}
the spectral slopes are expected to increase
from $-5/3$ for strong turbulence, i.e.\ for turbulent regimes in which the 
perturbation and the mean field are similar,
to steeper exponents as long as we proceed towards weaker regimes of 
turbulence, i.e.\ turbulent regimes in which the ratio of the perturbation
over the mean field decreases. For weak turbulence the
exponents are $\alpha_n = -(3n-5)/(n-1)$, i.e.\ $-2,\ -7/3,\ -5/2, \ldots$.
To distinguish between spectral slopes is usually a difficult task because
the differences among them are relatively small. In the case of weak
turbulence at increasing $n$ the difference between neighboring slopes
$\alpha_n$ becomes very small.

Consider Figure~\ref{fig:hypsp:b}, which is the run with the same value
of the Alfv\'enic velocity that we have extensively considered in the previous
paragraphs $v_{\mathcal A} = 200$. This is the easiest  case:  
while the $k^{-2}$ line fits well the inertial range, the two neighboring 
slopes $k^{-5/3}$ and $k^{-7/3}$ clearly do not.

A very well known problem with hyperdiffusion is the so-called 
\emph{bottleneck effect} (Falkovich~\cite{fal94}, 
Lithwick \& Goldreich~\cite{lg03}): the spectrum on scales slightly larger 
than the dissipation scales becomes flatter (in bilogarithmic scale
a small bump appears). The energy, in effect, is backed up and the origin
of this reflection is the rapid increase of diffusion at the small scales.
Hyperdiffusion virtually removes diffusion from the large scales,
but while normal diffusion slowly decreases the diffusion time towards 
the small scales, hyperdiffusion has a very steep decrease which
increases with increasing dissipativity $\alpha$. We have used a mild
exponent $\alpha = 4$ but the bottleneck effect is present visible,
especially in the simulation with $v_{\mathcal A} = 400$ 
(Figure~\ref{fig:hypsp:c}) and $1000$ (Figure~\ref{fig:hypsp:d}).
In Figure~\ref{fig:hypsp:c} the spectrum is steeper than for
$v_{\mathcal A} =200$ and the slope that fits better the inertial range is
$k^{-7/3}$ while  $-5/3$ and $-3$ certainly do not.
Consider now Figure~\ref{fig:hypsp:d} where the spectrum for a 
simulation with the higher value $v_{\mathcal A} =1000$ is shown.
Between $-5/2$ and $-3$, the first slope fits  better in the inertial
range but $-7/3$ and $-2$ clearly do not fit the inertial range.
The case with the lowest value of the Alfv\'enic
velocity $v_{\mathcal A} = 50$ shown in Figure~\ref{fig:hypsp:a}.
The spectrum shown is an average over $196$ Alfv\'enic crossing times.
The $-7/3$ slope fails in the inertial range, but  $-2$ and
$-5/3$ seem to fit the first part and the  last part of the 
inertial range respectively. Actually we could consider if in this case it does develop 
an inertial range. The slope change which  develops around $n \sim 10$
is in a very advanced position towards the small wavenumbers respect to 
the others simulations to be the bottleneck effect. But we have no objective 
method to say if this is really a bottleneck effect. There is another possibility.
As shown in \S~\ref{par:bik} all the weak turbulence scalings share
the property that at smaller scales the turbulence gets stronger, and 
Goldreich \& Sridhar~\cite{gs97} have proposed the possibility
of multiple inertial ranges, where starting, for example, from a $-2$
slope the system transitions at smaller scales to a stronger $-5/3$ 
regime. If it is already difficult to distinguish numerically one inertial range,
it is certainly far more difficult to distinguish among two.
Nevertheless in Figure~\ref{fig:hypsp:a} either we have a bottleneck effect
or it is this transition, and there is no objective method to decide.
It would be certainly interesting to perform a numerical simulation
with the same parameters but doubling the resolution.
\begin{figure}
      \centering
      %%---- start ----
      \subfloat[]{
               \label{fig:hypdz:a}             %% label for subfigure
               \includegraphics[width=0.45\linewidth]{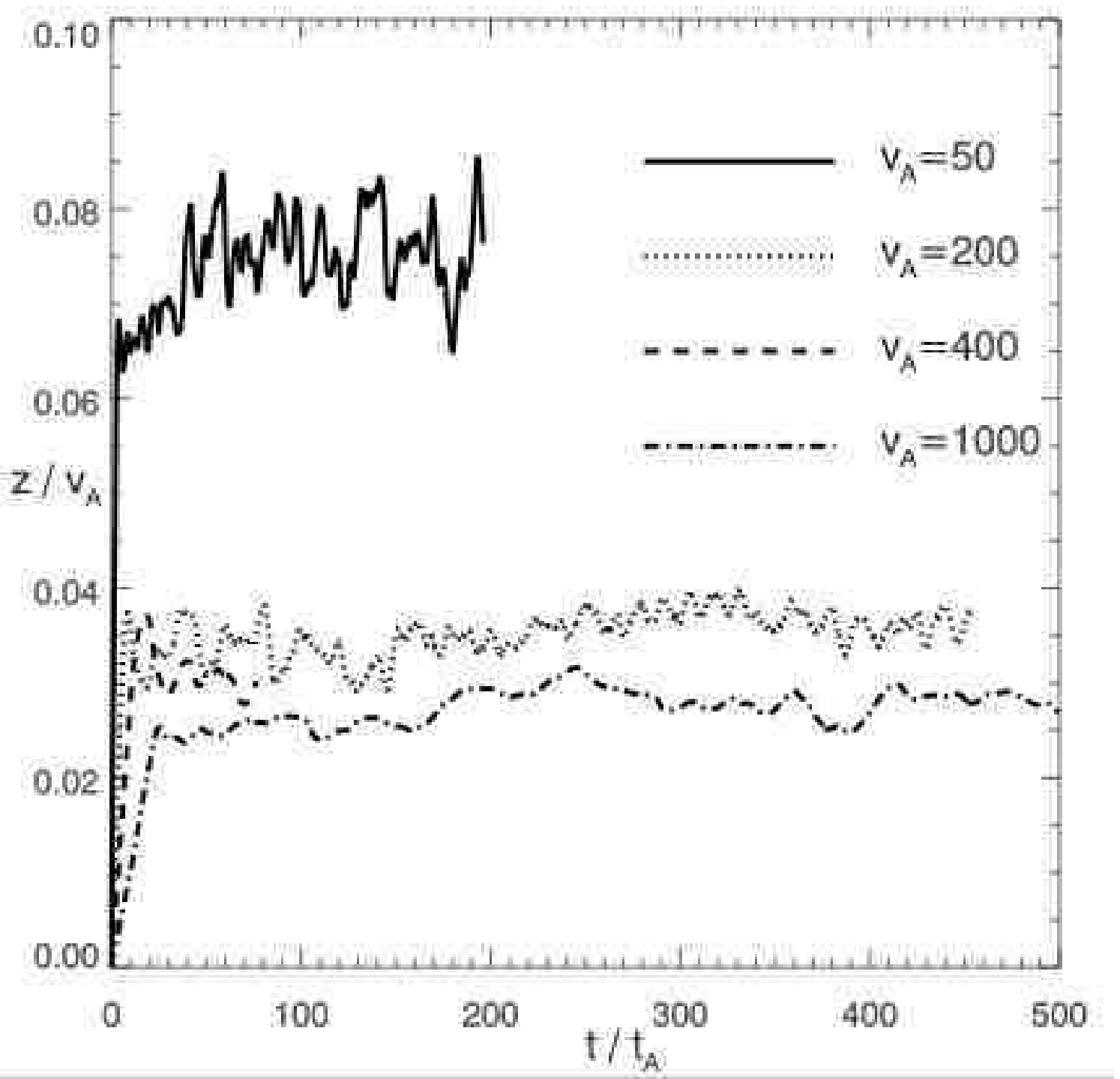}}
      \hspace{0.01\linewidth}
     %%---- start ----
      \subfloat[]{
               \label{fig:hypdz:b}             %% label for subfigure
               \includegraphics[width=0.45\linewidth]{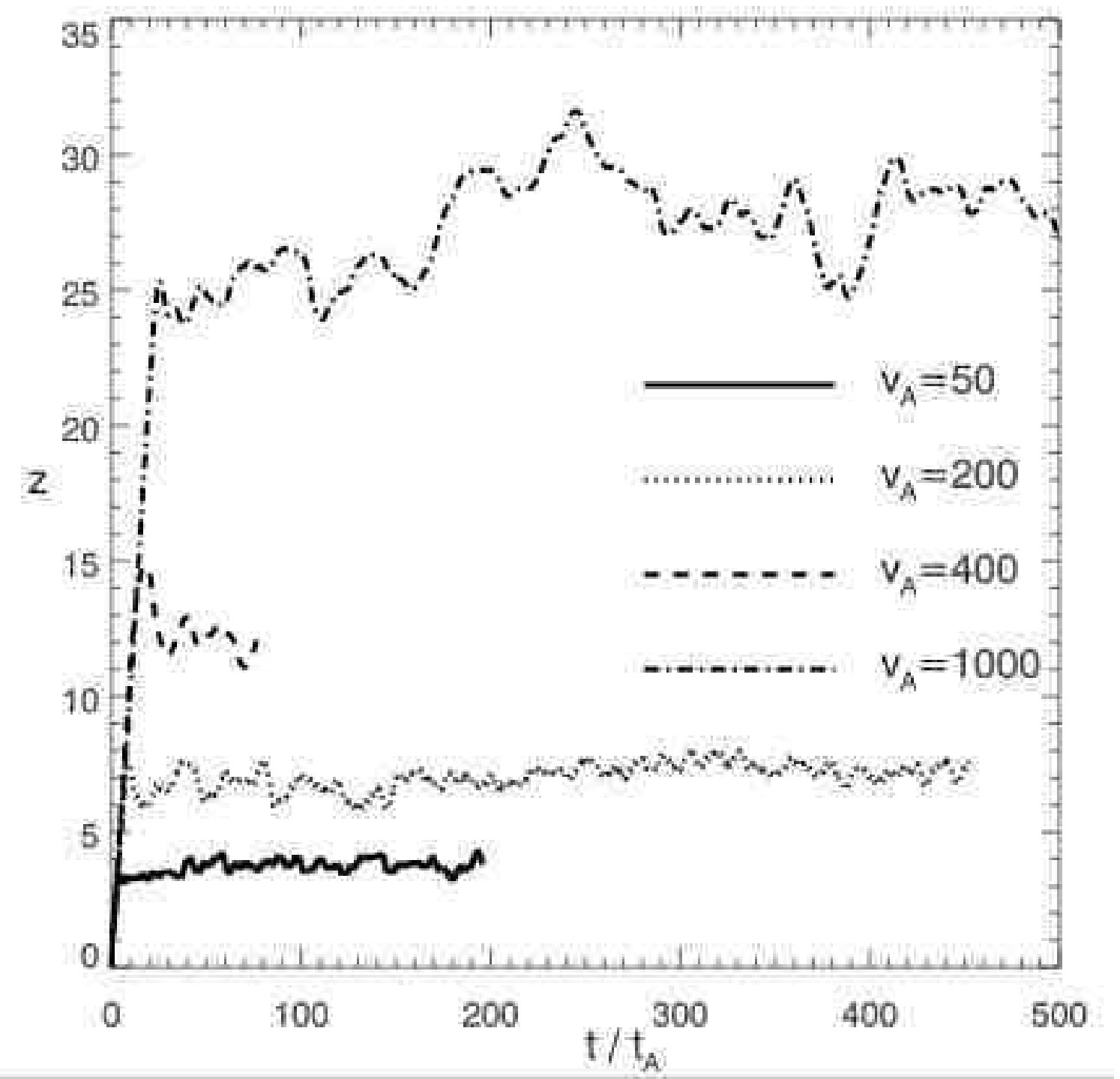}}
   \caption{The rms of the $\bsy{z}^{\pm}$ fields, which is roughly the
   same for the two fields and that we generically indicate with $z$, 
   is shown as a function of time for the four simulations with
   $v_{\mathcal A} = 50,\ 200,\ 400$ and $1000$. In (a) the relative
   value $z/v_{\mathcal A}$ is shown as a function of time, while in (b)
   it is shown $z$.
        \label{fig:hypdz}}             %% label for entire figure
\end{figure}

As already shown, for this system $H^C \sim 0$ so that $E^+ \sim E^-$.
Hence the rms of the fields $\bsy{z}^{\pm}$ are also roughly the same,
and  we now consider the integral quantity 
\be
z = \sqrt{  \frac{2 \left( E_K + E_M \right)}{V} } 
\sim \sqrt{ \frac{1}{V} \int_V \! \ud^3 x \left| \bsy{z}^{\pm} \right|^2 }
\ee
Figure~\ref{fig:hypdz:a} shows the relative ratio $z/v_{\mathcal A}$ as a function
of time while Figure~\ref{fig:hypdz:b} show $z$ for the four simulations considered
in  Figure~\ref{fig:hypsp}. While the value of $z$ increases at higher values
of $v_{\mathcal A}$, the relative ratio $z/v_{\mathcal A}$ decreases.
Hence the results shown in Figures~\ref{fig:hypsp}-\ref{fig:hypdz}
show that the \emph{our system is described by a weak anisotropic
turbulence regime which gets weaker for higher values of the
axial Alfv\'enic velocity $v_{\mathcal A}$}.

The physical mechanism by which the system dynamically choses a value 
$z$ for different values of $v_{\mathcal A}$ is self-organization. Both this
mechanism and its consequences for coronal heating scalings are examined
in \S~\ref{sec:nla}. We conclude this paragraph showing the dissipation
rate as a function of time for the four simulations in Figure~\ref{fig:hypmdiss}: 
the dissipation rate is higher for higher values of $v_{\mathcal A}$.
\begin{figure}
\begin{center}
\includegraphics[width=0.5\textwidth]{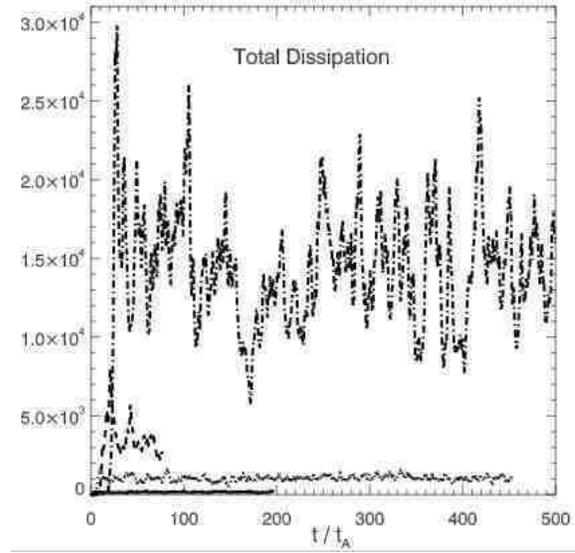}
          \caption{Dissipation rates for the four simulation with
          $v_{\mathcal A} = 50,\ 200,\ 400,\ 1000$.
               \label{fig:hypmdiss}}
\end{center}
\end{figure}

\subsection{Strong Turbulence and Dissipation}

In previous paragraphs we have shown the spectra of many physical 
quantities always considering the orthogonal $k_{\perp}$ variable,
and summing over the axial $k_z$ direction.
\begin{figure}[t]
      \centering
      %%---- start ----
      \subfloat[]{
               \label{fig:spz:a}             %% label for subfigure
               \includegraphics[width=0.45\linewidth]{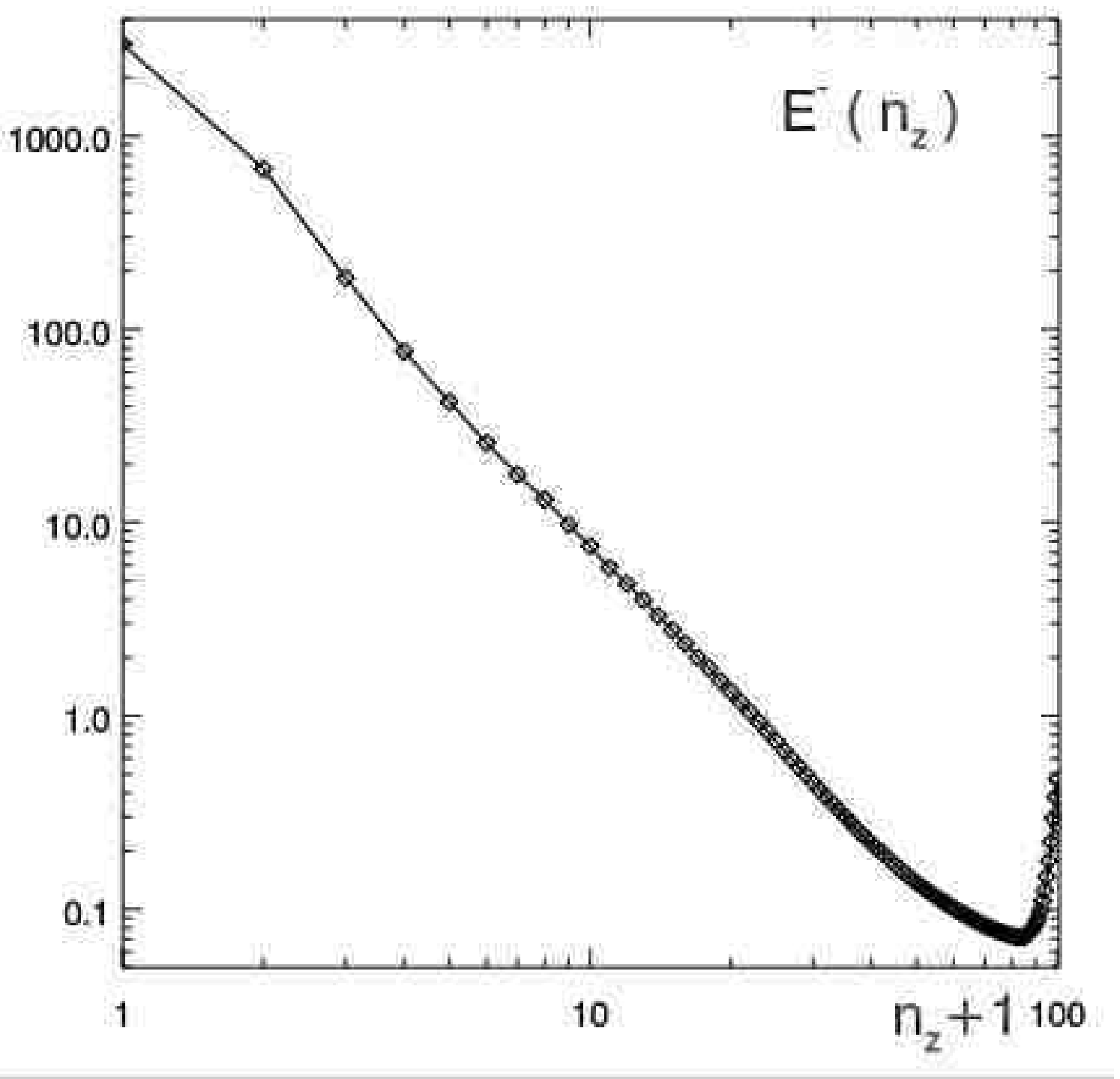}}
      \hspace{0.01\linewidth}
     %%---- start ----
      \subfloat[]{
               \label{fig:spz:b}             %% label for subfigure
               \includegraphics[width=0.45\linewidth]{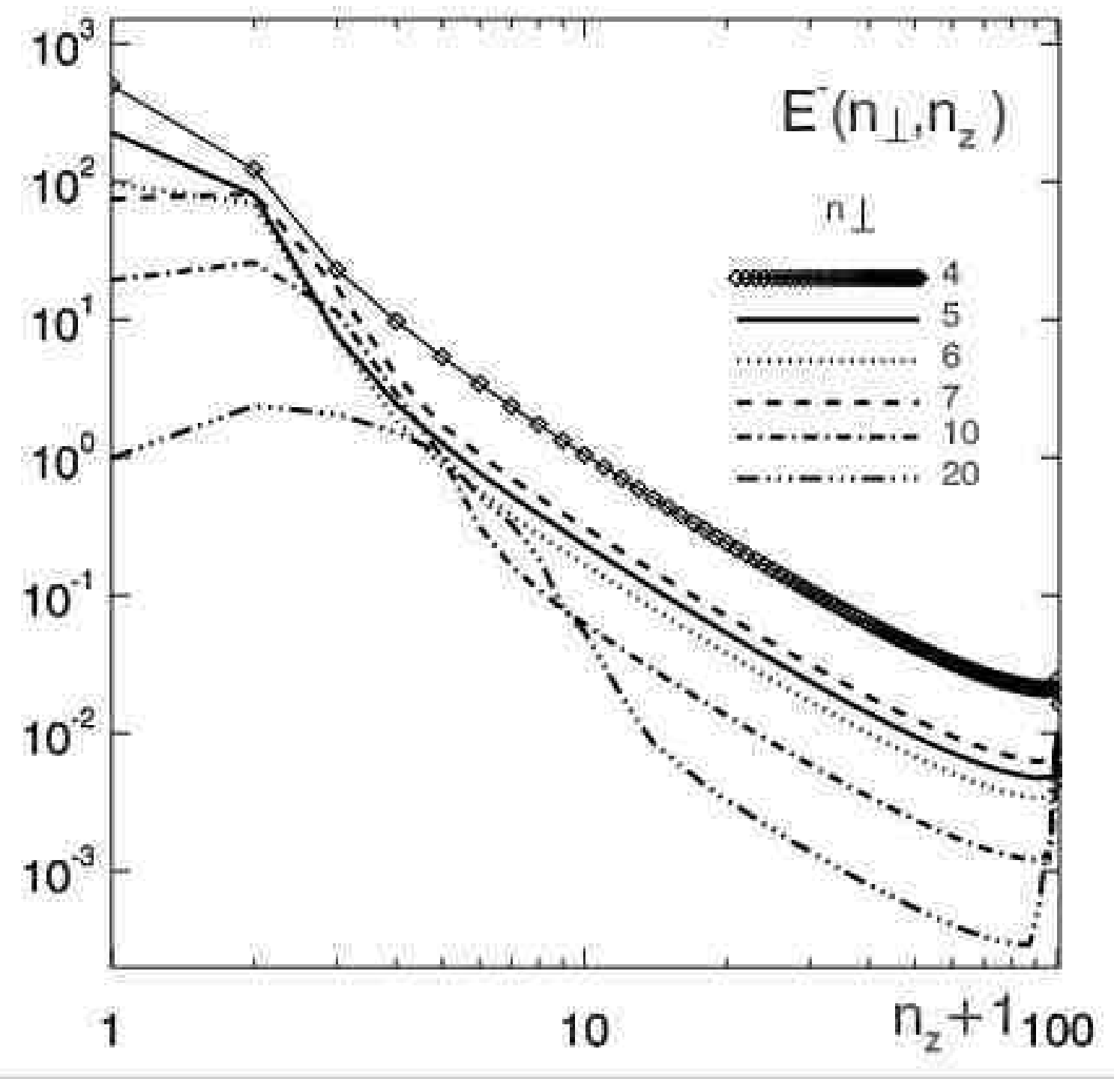}}
   \caption{(a) $E^{-} (n_z)$ spectrum as a function of $n_z+1$. The small
   scale increase is due to aliasing. 
   (b)  $E^{-} ( n_{\perp}, n_z )$ as a function of $n_z+1$ at selected values of
   $n_{\perp}$. Spectra are calculated for run~7 
   ($R_4 = 10^{19}, v_{\mathcal A} = 1000$) at time 
   $t \sim 500\, \tau_{\mathcal A}$.
        \label{fig:spz}}             %% label for entire figure
\end{figure}
In our simulations we impose non-periodic boundary conditions,
along the axial direction $z$, but nevertheless the system results
almost periodic, i.e.\ there are not strong variations along the
axial direction. It is then possible to perform a fourier transform
of the physical quantities also along the axial direction.
As the functions are not strictly periodic we have to expect
some ``aliasing'' at the small scales.

Consider the Fourier expansion~(\ref{eq:fe1}) for the 
$E^{\pm}$ energies. If instead of summing along $z$ we perform
the fourier transform we obtain the 2D analogue of the 
1D spectra~(\ref{eq:fe2}), and we can now write
\be \label{eq:fe3}
E^{\pm} = \sum_{n_{\perp}, n_z} E^{\pm} \left( n_{\perp}, n_z \right), \qquad
n_{\perp} = 1,2,\dots \quad n_z = 0,1,2,\dots
\ee
where $n_{\perp}$ is the orthogonal wavenumber, and $n_z$ is the 
axial wavenumber for which also the $n_z=0$ component does not
vanish. From the 2D spectra $E^{\pm} \left( n_{\perp}, n_z \right)$
we easily obtain the 1D orthogonal and axial spectra
$E^{\pm} ( n_{\perp} )$ and  $E^{\pm} ( n_z )$ summing over
the remaining variable, i.e.\
\be
E^{\pm} \left( n_{\perp} \right) = \sum_{n_z} E^{\pm} \left( n_{\perp}, n_z \right),
\qquad
E^{\pm} \left( n_z \right) = \sum_{n_{\perp}} E^{\pm} \left( n_{\perp}, n_z \right),
\ee

In Figure~\ref{fig:spz} the spectra $E^{-} (n_z)$ and 
$E^{-} ( n_{\perp}, n_z )$ at selected values of $n_{\perp}$ are shown.
Figure~\ref{fig:spz:a} shows that already  the first few modes have a sharp
decrease in their values. Figure~\ref{fig:spz:b} shows that at least up
to $n_{\perp} \sim 20$ the presence of axial ($z$) modes is 
negligible. 
\begin{figure}[t] \label{fig:spzl}
\begin{center}
\includegraphics[width=0.7\textwidth]{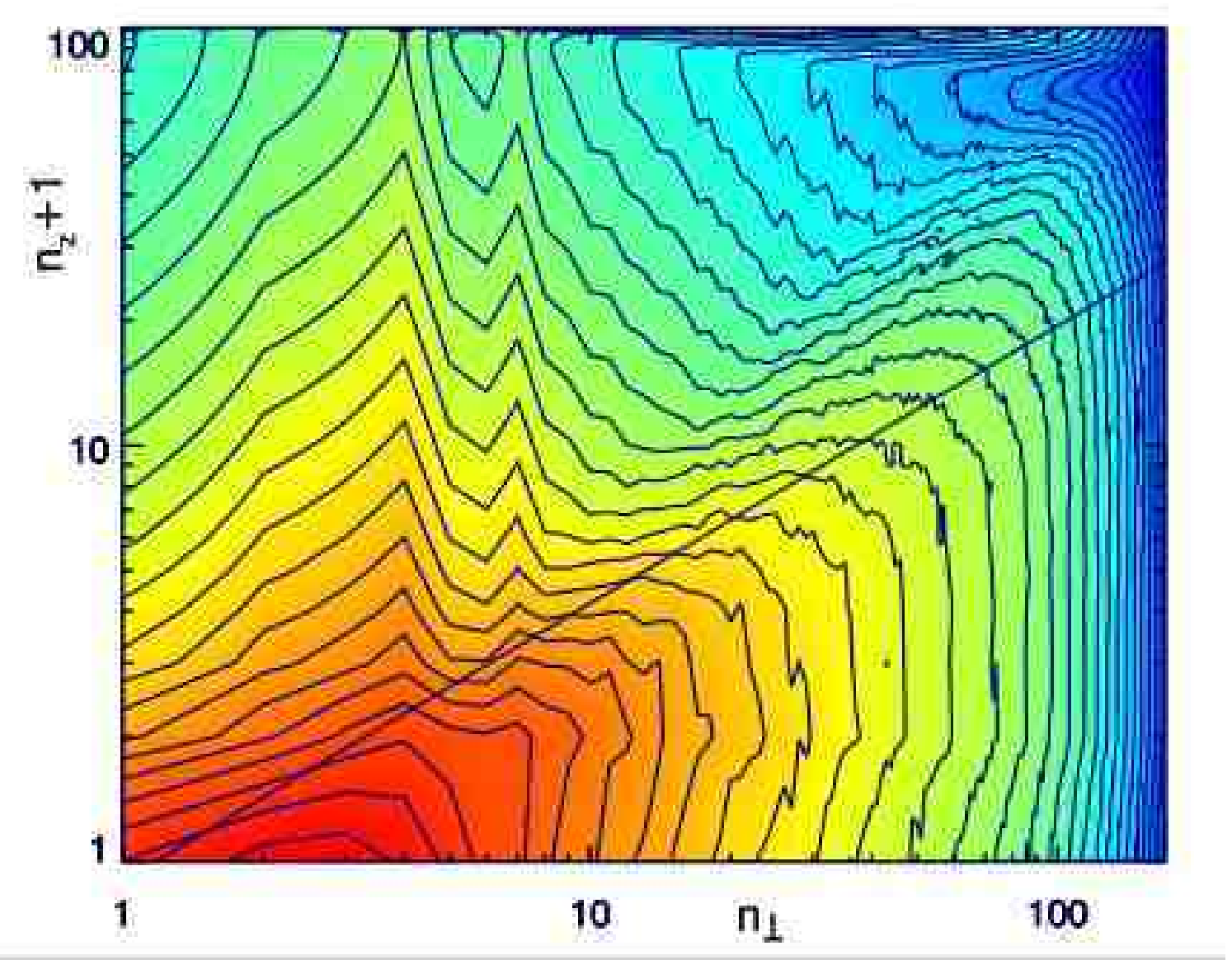}
          \caption{Logarithmic contour of the $E^- ( n_{\perp}, n_z )$ spectrum at time
          $t \sim 500\, \tau_{\mathcal A}$ for run7. At sufficiently high
          orthogonal wavenumbers $n_{\perp} > 20$ the formation of small scales 
          along the axial direction becomes significant. The over-plotted line
          represents the curve $n_z \propto n_{\perp}^{2/3}$.
               \label{fig:spzp}}
\end{center}
\end{figure}

In Figure~\ref{fig:spzl} are shown the contours of the natural logarithm of
the 2D $E^{-} ( n_{\perp}, n_z )$ spectra. It is clearly shown that
at small scales strong turbulence sets in, with the formation of
small scales along the axial direction.

\section{Sheared Velocity Pattern} \label{sec:svp}

In this section we consider the 3D evolution of a coronal layer driven by a 
time independent forcing pattern.  A vanishing photospheric flow is 
imposed on the bottom plane ($z=0$)
\begin{equation}  \label{eq:bd0}
\boldsymbol u_{\perp} (x,y,z=0,t) = 0 ,
\end{equation}
while on the top plane ($z=L$) the velocity pattern is a stationary photospheric 
shear flow aligned along the $y$ direction
\begin{equation} \label{eq:bdL2}
\boldsymbol u_{\perp} (x,y,z=L,t) = \sin ( 2 \pi n x + 1)  \ \bsy{e}_{y}
\end{equation}
where $n=4$, $ 0 \le x, y \le 1 $ and $ 0 \le z \le L$, with an aspect ratio of $10$, i.e. $L=10$. 
The initial condition is given by the strong uniform axial field with very 
small amplitude noise in the perpendicular velocity and magnetic fields. This 
case is interesting because the growing magnetic field induced in the 
corona  (see equations~(\ref{eq:bdiff})-(\ref{eq:hrdiff}))
by such motions is  an equilibrium field at every instant. 
In other words, any dynamical evolution beyond the slow 
increase of magnetic energy with time must be due, at first, to instabilities arising in the 
system. A similar configuration was used by \cite{hey92}, who developed a 
semi-analytical quasi self-consistent ``turbulent'' model of coronal heating. 
The most interesting result from these simulations is that 
after the first energetic burst (due to a tearing instability), which releases the energy
accumulated during the linear stage, the dynamics are turbulent
once the nonlinear stage sets in and are similar for all to those which develop
with a vortex-like forcing velocity pattern.

If there were no perturbation at all, at longer times diffusive terms would not be
negligible even at large scales and the solution would be
modified over diffusive time-scales.
While diffusion will only slightly change the shape in $z$ of the velocity 
field, it has a stronger effect on the magnetic field which
would otherwise grow linearly in time. We describe briefly such effect.
From which the following expressions for the total magnetic energy and 
ohmic dissipation are obtained:
\begin{equation} \label{eq:em}
E_M = \frac{1}{2} \, 
\int_V \mathrm{d}^3 \boldsymbol{x} \ \boldsymbol{b}_{\perp}^2 = 
\frac{l^2 L}{4}
\left( \frac{ v_{\mathcal A} \mathcal{R} }{L  ( 2 \pi n )^2 } \right)^2
\left[ 1 - \exp  \left( - \frac{\left( 2 \pi n \right)^2}{\mathcal R} t \right) \right]^2,
\end{equation}
\begin{equation} \label{eq:ohm}
J = \frac{1}{\mathcal R} \int_V \mathrm{d}^3 \boldsymbol{x} \ j^2 = 
\frac{l^2 L}{2}
\left(\frac{v_{\mathcal A}}{L} \right)^2 \frac{\mathcal R}{\left( 2 \pi n \right)^2 }
\left[ 1 - \exp  \left( - \frac{\left( 2 \pi n \right)^2}{\mathcal R} t \right) \right]^2.
\end{equation}
At first, for a time short compared to the resistive time 
$\tau_{\mathcal R} = \mathcal{R}/ \left( 2 \pi  n \right)^2 $, 
both quantities  grow quadratically in time while they saturate on the diffusive 
time-scale.  If non-linearity did
not set in, the heating rate would saturate at a value that is proportional to the 
Reynolds number and the square of the axial Alfv\'enic velocity.
\begin{figure}
   \centering
      %%---- start ----
      \subfloat[]{
               \label{fig:ensjs:a}             %% label for subfigure
               \includegraphics[width=0.45\linewidth]{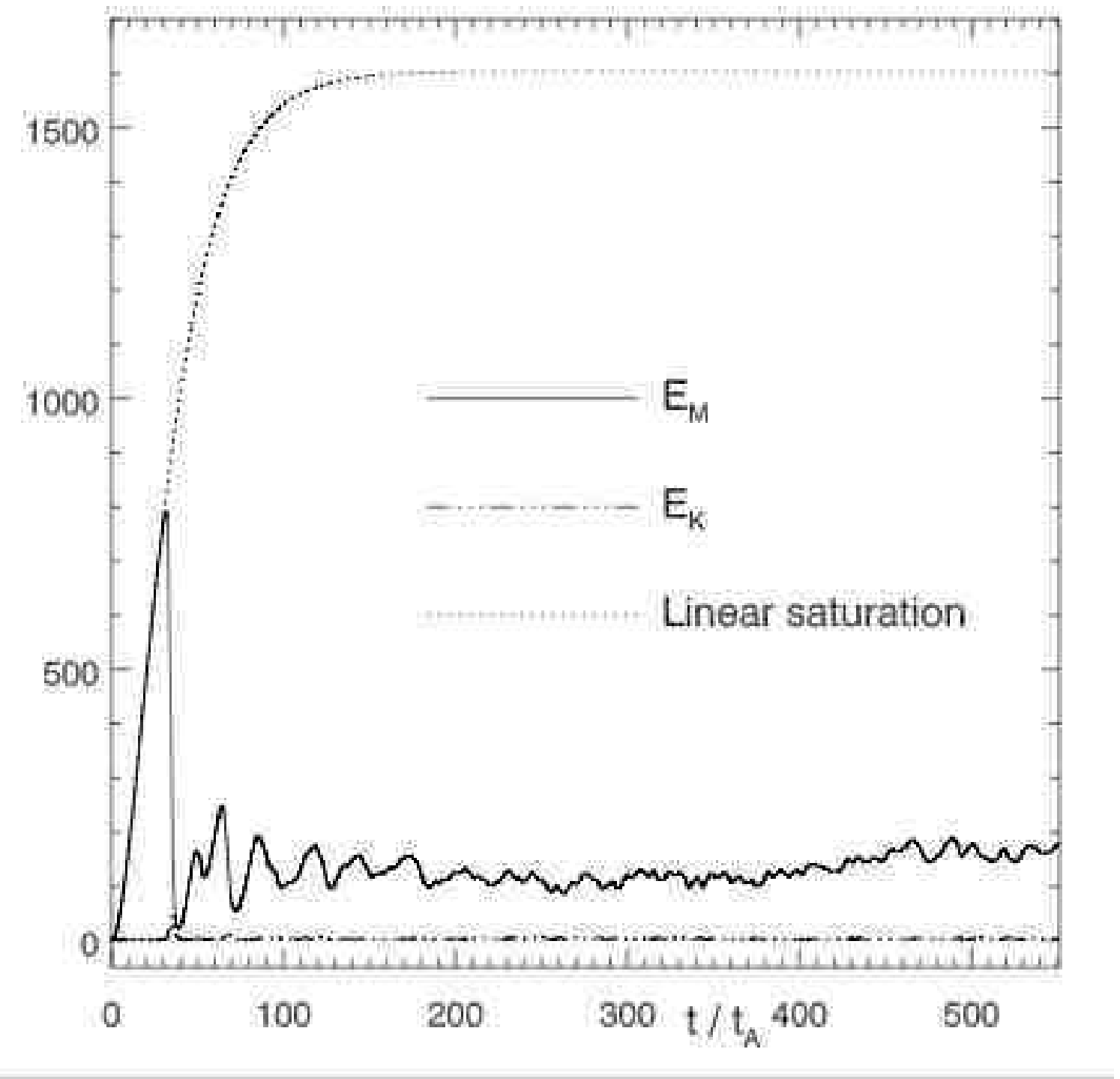}}
      \hspace{0.01\linewidth}
     %%---- start ----
      \subfloat[]{
               \label{fig:ensjs:b}             %% label for subfigure
               \includegraphics[width=0.45\linewidth]{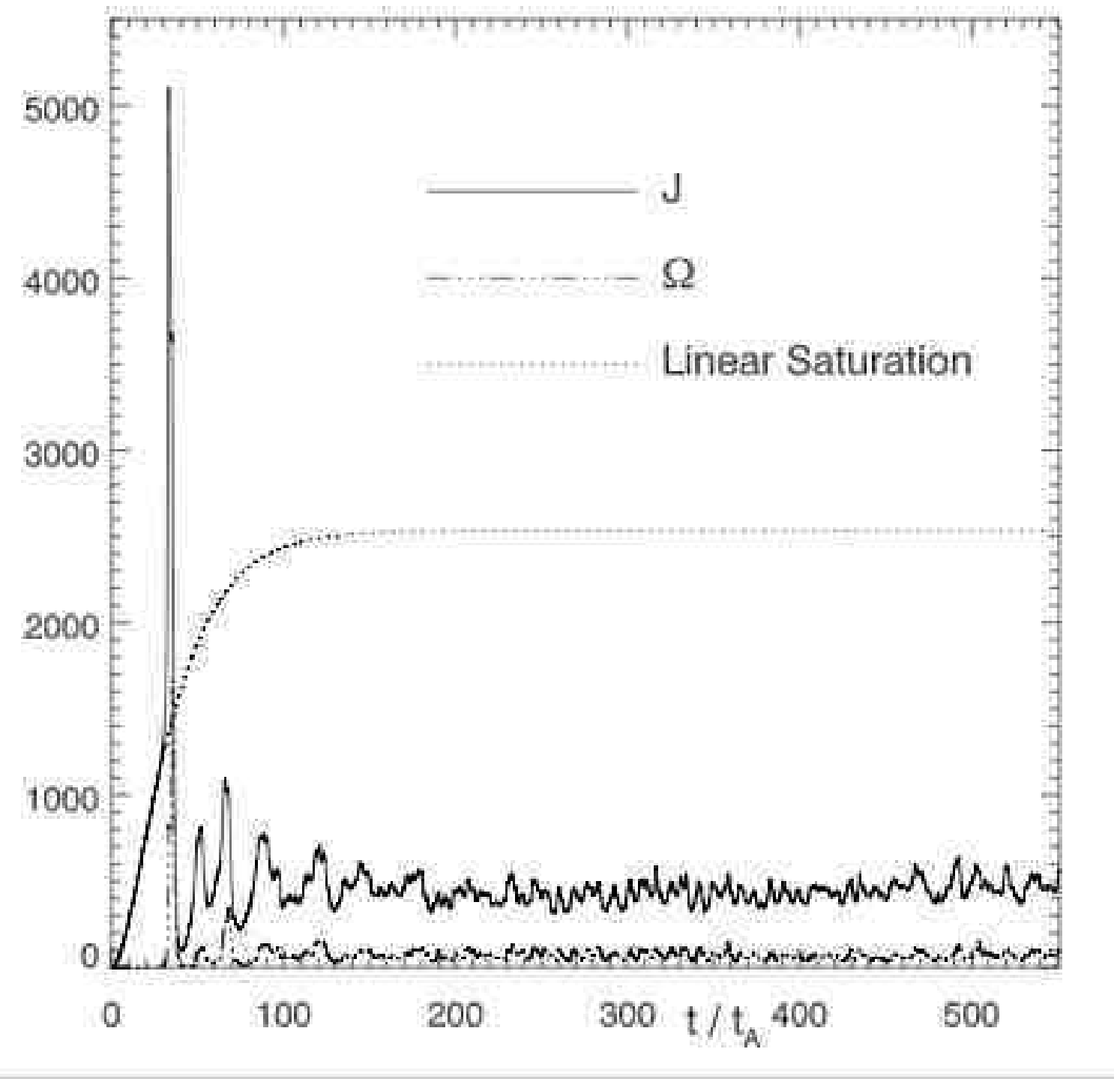}}
   \caption{High-resolution simulation with 512x512x200 
                grid points and $\mathcal{R}=800$. \emph{Left}: 
                Magnetic ($E_M$) and kinetic ($E_K$) energies as a function 
                of time. In dotted line curve~(\ref{eq:em}), which represents the
                linear saturation of the magnetic energy if there was solely the linear
                dynamics. At time $t \sim 30 \, \tau_A$
                a reconnection event abruptly decreases the value of magnetic
                energy. Kinetic energy is always a small fraction of total energy.
                Until time $t \sim 30 \, \tau_A$ magnetic energy follows the linear
                saturation curve~(\ref{eq:em}).
                \emph{Right}: Ohmic (J) and viscous ($\Omega$) dissipation as 
                a function of time.  Curve~(\ref{eq:ohm}) represents the
                linear saturation of the ohmic dissipation and  is drawn in dotted line.
                 At time $t \sim 30 \, \tau_A$
                the reconnection event abruptly releases the energy previously stored
                in the orthogonal magnetic field. Similar to kinetic energy dynamics,        
                also
                enstrophy ($\Omega$) is always a small fraction of total dissipation,
                confirming that the system is magnetically dominated.
                Until time $t \sim 30 \, \tau_A$ ohmic dissipation follows the linear
                saturation curve~(\ref{eq:ohm}).     
     \label{fig:ensjs}}
\end{figure}

The results of a numerical simulation 
performed with $n_x \times n_y \times n_z = 512 \times 512 \times 200$ grid 
points, a Reynolds number $\mathcal R = 800$ and carried on for roughly
$550$ axial Alfv\'enic crossing times ($\tau_A = L / v_{\mathcal A}$)
are shown in Figs.~\ref{fig:ensjs} and \ref{fig:poy}.
Fig.~\ref{fig:ensjs:a} shows the total magnetic and kinetic energies in the loop as a 
function of time, Fig.~\ref{fig:ensjs:b} shows ohmic and viscous dissipation, and
Fig.~\ref{fig:poy} shows the total dissipation and the Poynting flux.
Magnetic energy grows quadratically at first, reaching a maximum  at 
$t \sim 30 \, \tau_A$
\begin{figure}
\begin{center}
\includegraphics[width=0.5\textwidth]{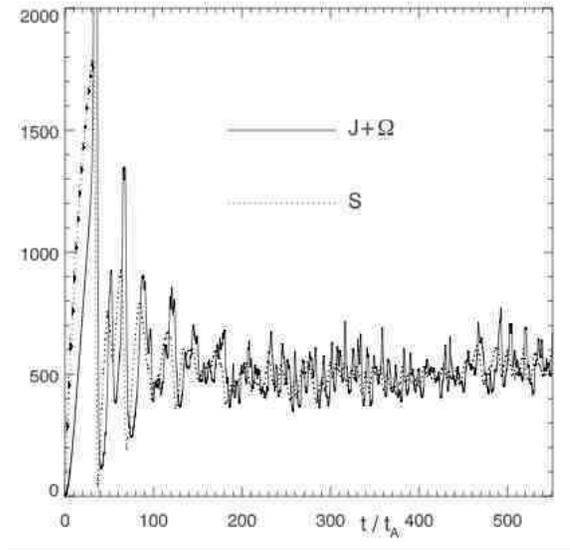}
\caption{Total dissipation $J+\Omega$ (ohmic plus viscous dissipations) 
               and incoming Poynting flux $S$ as a function of time for a high-resolution 
               simulation with a grid of 512x512x200 points and $R=800$. Sign convention 
               is chosen so that flux is positive when energy goes into the system and 
               negative when it goes out. After the reconnection event 
               at $t \sim 30 \, \tau_A$ a statistically steady state is reached where
               energy which is continuously injected for unity time into the system at
               the boundaries through the field-line tangling, due to the photospheric 
               forcing, balances the energy which is dissipated. \label{fig:poy}}
\end{center}
\end{figure}
while the kinetic energy is much smaller and remains limited at longer times. 
In the absence of dynamical evolution, magnetic energy and ohmic dissipation
would follow the curves~(\ref{eq:em})-(\ref{eq:ohm}) (as they actually do until
$t \sim 30 \, \tau_{\mathcal A}$ when nonlinearity sets in), saturating on the 
diffusive time-scale  $\tau_{\mathcal R}$, 
and shown by the dashed lines in Fig.~\ref{fig:ensjs}.
What occurs at $t \sim 30 \, \tau_A$ is that reconnection begins in 
the sheared induced field. Given the periodicity of the system, the actual 
instability is a multiple tearing mode, which grows faster than the classical tearing 
mode. Until $t \sim 30 \, \tau_A$ energy is stored in the large-scale magnetic
field, while the kinetic energy is only a small fraction of the total energy. 
This energy is abruptly released in the reconnection event that gives rise to 
the first big burst in the heating rate that consequently lowers the
magnetic energy.
After this event a statistically steady state is reached. In this state the Poynting
Flux balances the rate of energy dissipation, as shown in Fig.~\ref{fig:poy}. 
On the other hand, as previously
stated, the rate of energy injection into the system, i.e. the
Poynting flux, depends on both the forcing velocity imposed at the boundaries
and  the magnetic field generated inside the system.
This is why this system is self-organizing. The balance between 
Poynting flux input and dissipation rate is reached 
by letting the magnetic field grow at the ``opportune'' value, as it is shown
in \S~\ref{sec:nla}.

Now we examine briefly the 3D structure of the physical fields. One  
hypothesis of reduced MHD is that during time evolution the orthogonal 
magnetic field is always small compared to the axial component.
After the first big burst, where the ratio of the rms of the orthogonal 
field over the axial field reaches the $6 \%$, this 
component always fluctuates around a value which is roughly the $3 \%$.
Thus  the reduced MHD requirement is satisfied.
In Fig.~\ref{fig:flside}-\ref{fig:fltop} we show the magnetic field lines, 
\begin{figure}
\begin{center}
\includegraphics[width=0.7\textwidth]{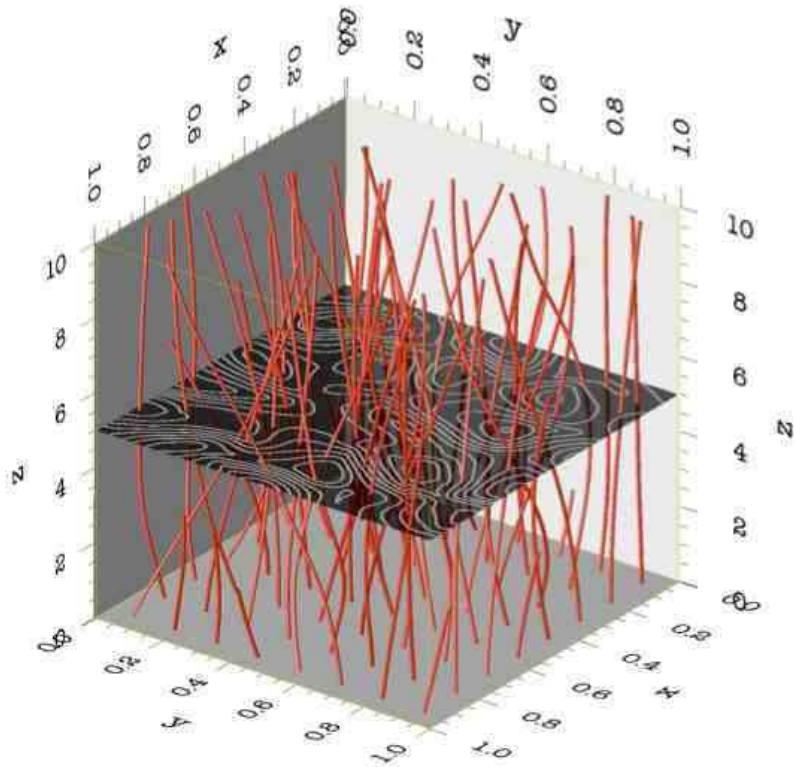}
\caption{Field lines of the total magnetic field, orthogonal plus axial. 
               \emph{Mid-plane:} field lines of the orthogonal magnetic field. 
               The orthogonal magnetic field magnitude fluctuates around a value 
               which is roughly the $3 \%$ of the axial component, well within
               the reduced MHD ordering. This is reflected in the slight bending
               of the magnetic field lines. For an improved visualization the box size
               has been rescaled. But the axial length of the box is ten times longer 
               than the orthogonal one. The resize of the box artificially enhances the field 
               line bending. The orthogonal magnetic fields is structured in magnetic 
               islands, and is mostly homogeneous in the axial direction ($z$). 
               \label{fig:flside}}
\end{center}
\end{figure}
\begin{figure}
\begin{center}
\includegraphics[width=0.7\textwidth]{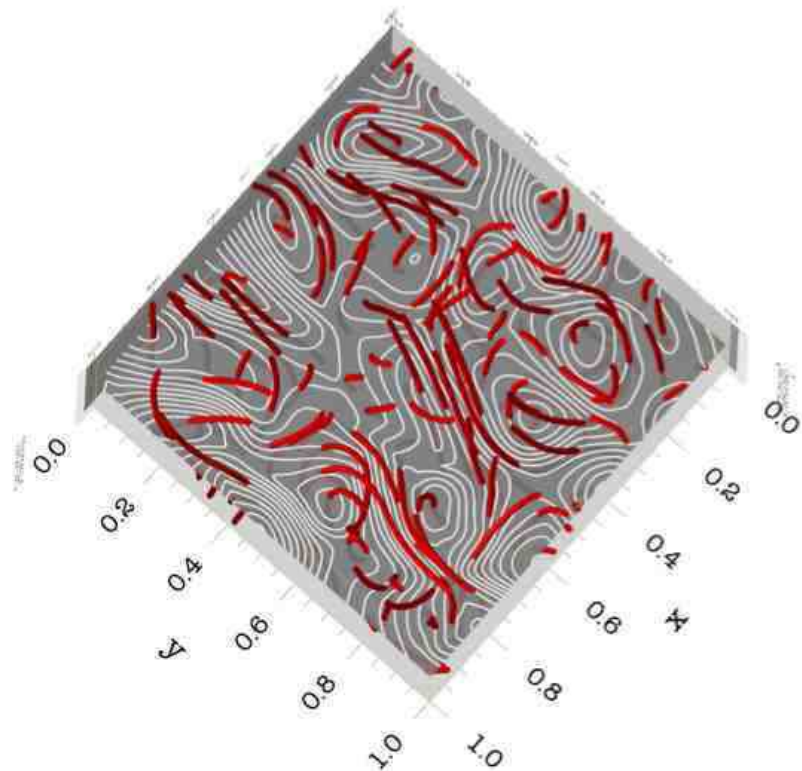}
\caption{A view from the top of the field lines of the total magnetic field (orthogonal
                plus axial). \emph{Mid-plane:} field lines of the orthogonal magnetic field. 
                The orthogonal magnetic field is structured in magnetic islands, and is 
                mostly homogeneous in the axial direction ($z$). Because of the 
                quasi homogeneity of the orthogonal field, the field lines of the
                total magnetic field are more bent when they are at the outskirts of
                the magnetic islands, while at different heights they always have the 
                orthogonal component of the magnetic field roughly in the same
                direction.
                \label{fig:fltop}}
\end{center}
\end{figure}
respectively a view from the side and from the top of the computational box.
For an improved visualization the box has been rescaled. But it should
be noted that the axial length is ten times the orthogonal length of the box.
The magnetic field lines are only slightly
bent. In both figures we show the field lines of the orthogonal magnetic 
field in the mid-plane. The structure of the orthogonal magnetic field is
almost invariant in the axial direction and it is structured in magnetic
islands. In particular no boundary layer is present. The field lines of the
total magnetic field which bend more are those which happen to be
inside the magnetic islands where the orthogonal magnetic field is enhanced.
The magnetic island structure for the magnetic field gives rise to
current sheets elongated along the axial direction, as shown in
Fig.~\ref{fig:isoside}-\ref{fig:isotop}, where the energy flux is finally
dissipated.
\begin{figure}
\begin{center}
\includegraphics[width=0.7\textwidth]{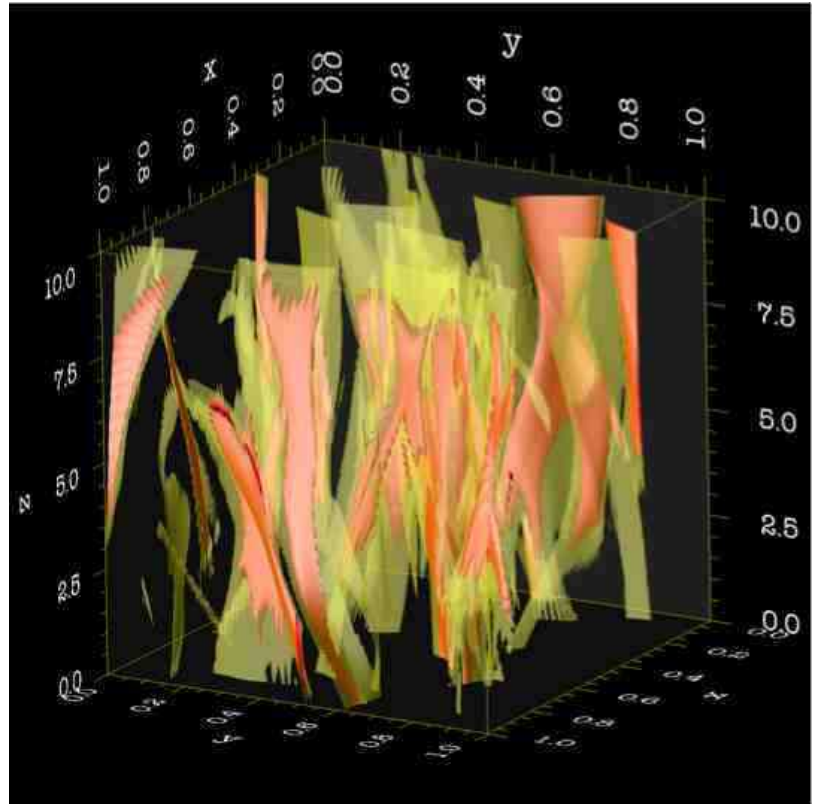}
\caption{Isosurfaces of the squared current: \emph{side view}.
               Two isosurfaces of the squared current at time $t \sim 550 \, \tau_A$ 
               for a numerical simulation with 512x512x200 grid points and a Reynolds
               number $\mathcal R =800$ are shown. In partially transparent yellow is
               represented the isosurface at the value $j^2 = 2.8 \cdot 10^5$ while in red is
               the isosurface with  $j^2 = 8 \cdot 10^5$, well below the value of the 
               maximum of the squared current that at this time is $j^2 = 3.6 \cdot 10^7$.
               The red isosurface is always nested inside the yellow one, and in fact it
               appears as pink.
               \label{fig:isoside}}
\end{center}
\end{figure}
\begin{figure}
\begin{center}
\includegraphics[width=0.7\textwidth]{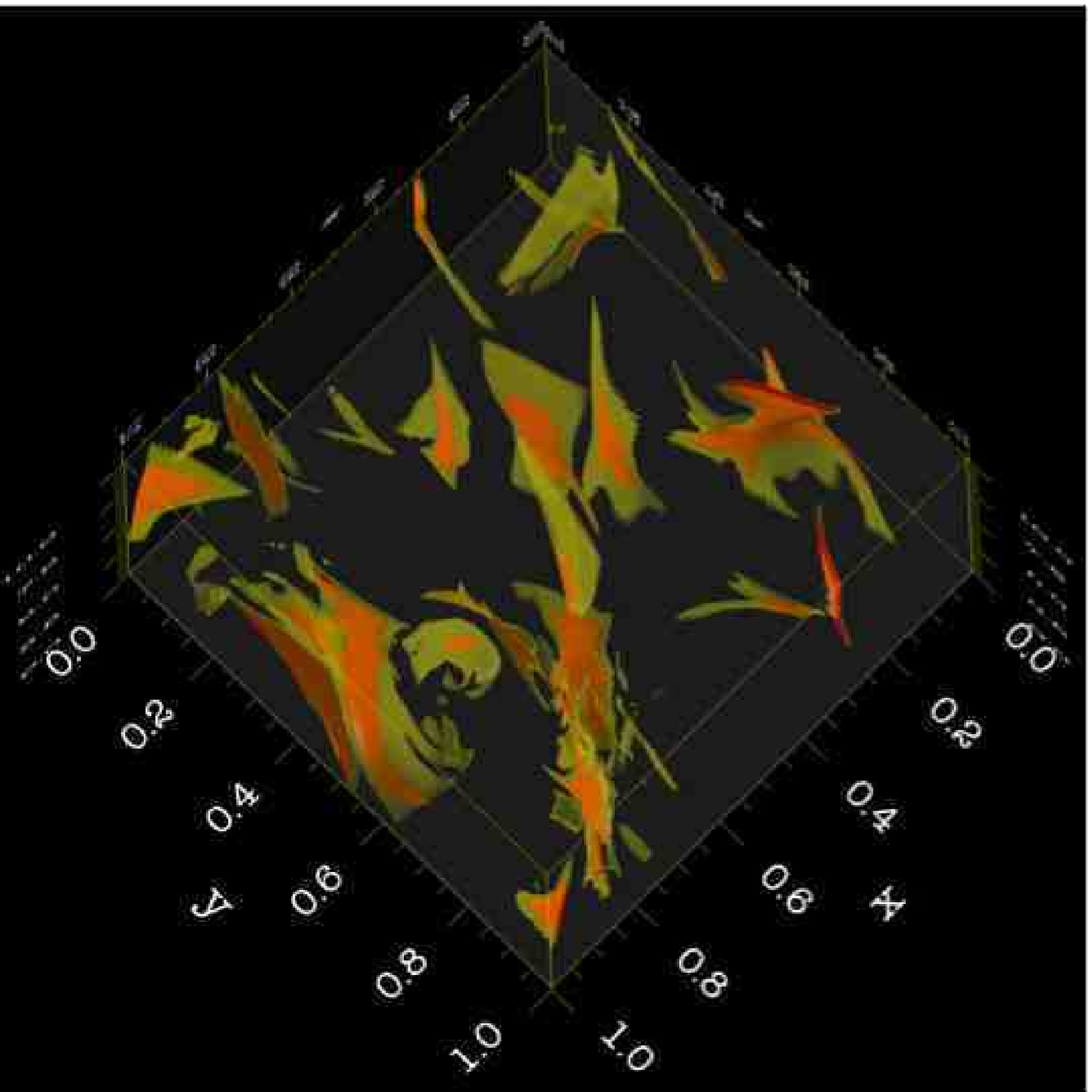}
\caption{Isosurfaces of the squared current: \emph{top view}.
               The same two isosurfaces of the previous figure are shown.         
               In partially transparent yellow is
               represented the isosurface at the value $j^2 = 2.8 \cdot 10^5$ while in red is
               the isosurface with  $j^2 = 8 \cdot 10^5$, well below the value of the 
               maximum of the squared current that at this time is $j^2 = 3.6 \cdot 10^7$.
               The isosurfaces are elongated along the axial direction, and the
               corresponding filling factor is small.
               \label{fig:isotop}}
\end{center}
\end{figure}

We have also performed other numerical simulations with different Reynolds
numbers and spatial resolutions. In Figs.~\ref{fig:enjl} we show the 
\begin{figure}[t]
   \centering
   \begin{minipage}[c]{0.45\linewidth}
       \centering \includegraphics[width=1\textwidth]{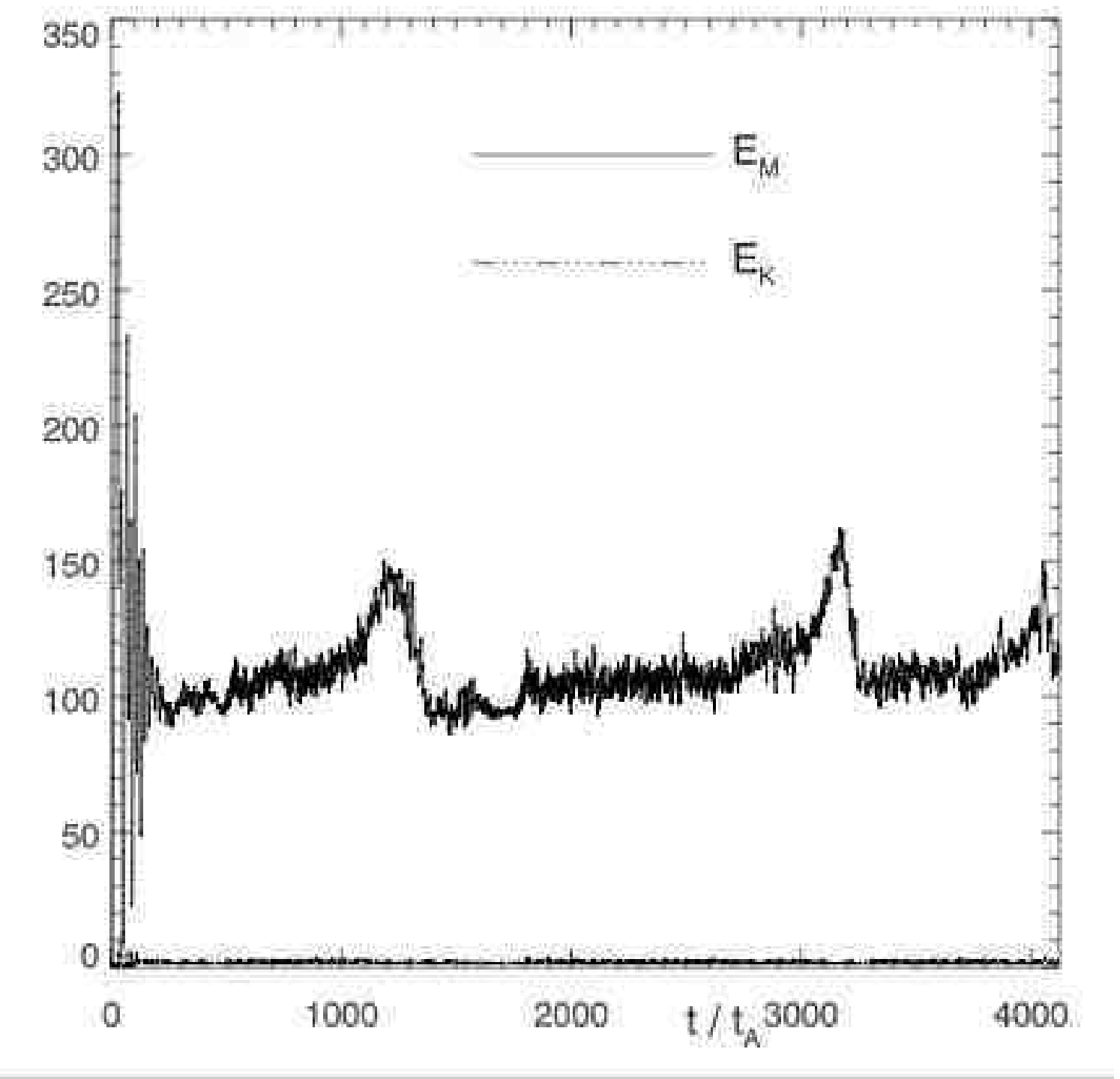}
   \end{minipage}%
   \begin{minipage}[c]{0.45\linewidth}
       \centering \includegraphics[width=1\textwidth]{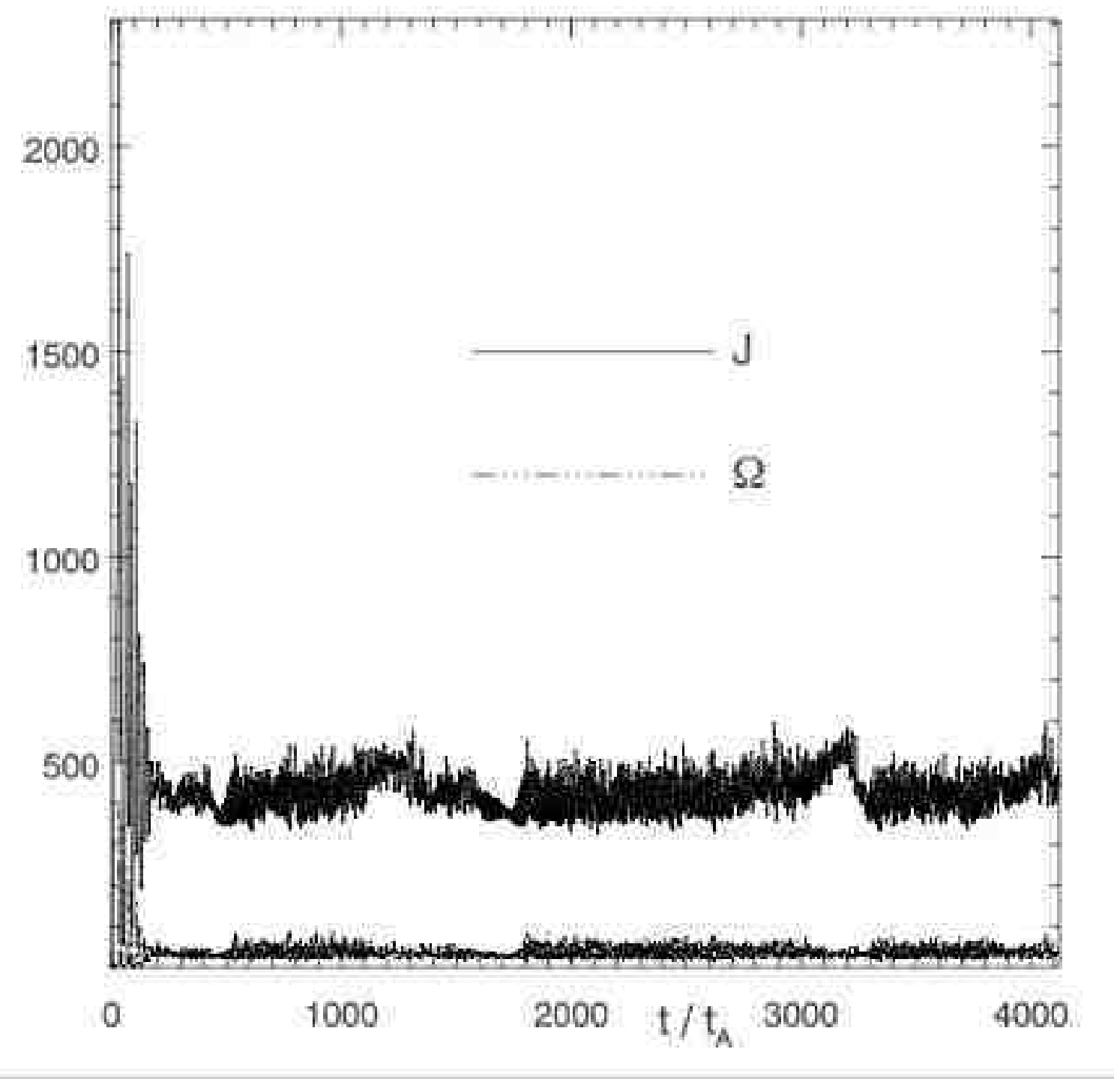}
   \end{minipage}
   \caption{\emph{Left}: Magnetic ($E_M$) and kinetic ($E_K$) energies as a function of time 
                for a long-time simulation with $R=400$ and 256x256x100 grid points.
                After the initial reconnection event a statistically steady state is reached.
                \emph{Right}: Ohmic (J) and viscous ($\Omega$) dissipation as a function of time 
                for a long-time simulation with $R=400$ and 256x256x100 grid points. 
                 After the initial reconnection event a statistically steady state is reached.
     \label{fig:enjl}}
\end{figure}
results of a very long simulation with 
$n_x \times n_y \times n_z = 256 \times 256 \times 100$ grid points, 
a Reynolds number $\mathcal R = 400$ and carried on for more than
$4000\, \tau_{\mathcal A}$. The results of the previous simulation,
which was carried out with a higher resolution, but for a shorter duration,
are fully confirmed. In particular the longer interval allows us to confirm that 
after the first reconnection event  a steady state is finally established.

We have another very interesting result.
In Fig.~\ref{fig:jm} we plot ohmic dissipation for 3 numerical simulations 
\begin{figure}
\begin{center}
\includegraphics[width=0.8\textwidth]{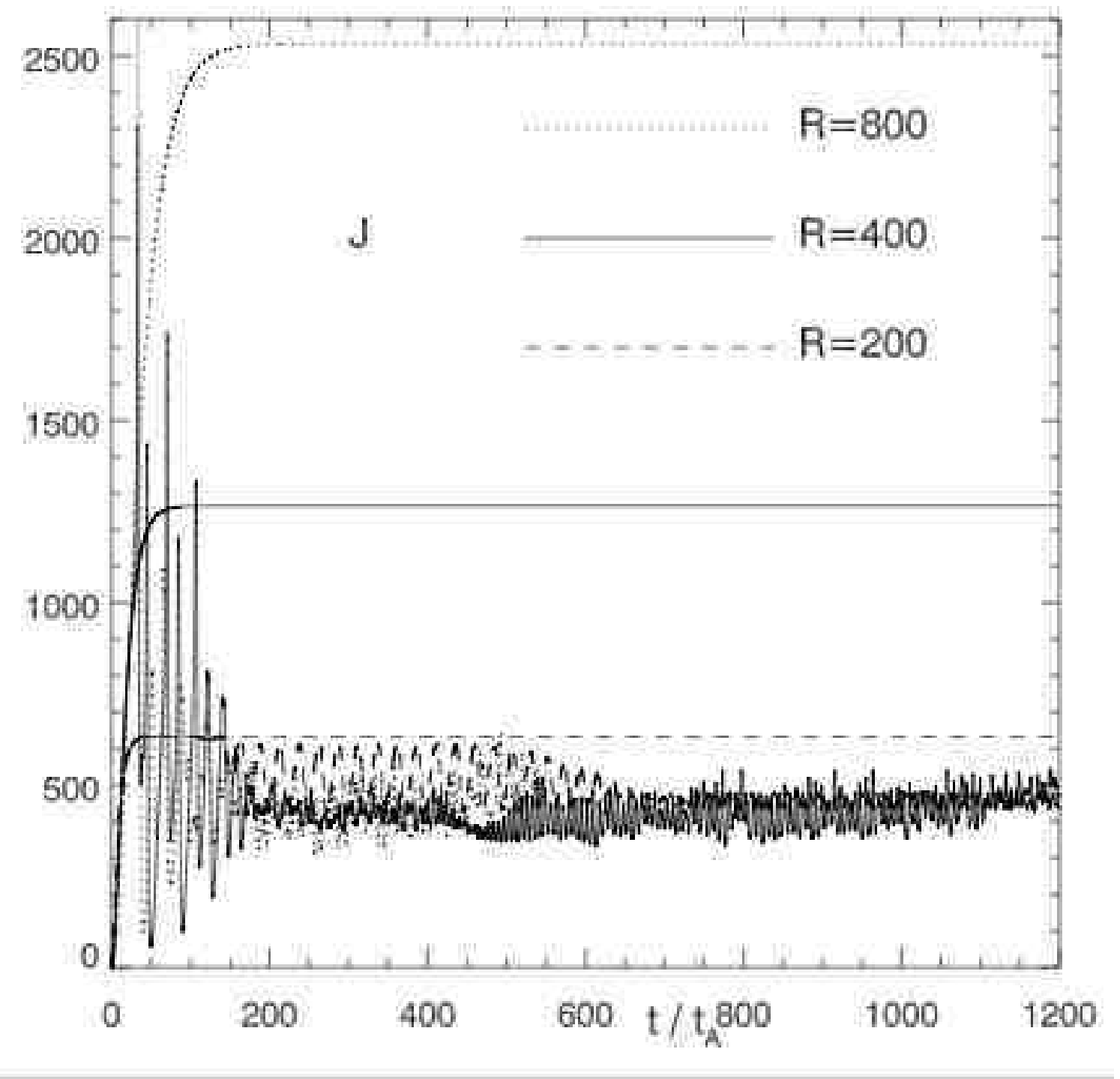}
\caption{Ohmic dissipation (and the corresponding linear saturation curve)
               for three simulations performed with
               different Reynolds numbers  $\mathcal R = 200, 400$ and $800$ . 
               The first peak of the simulation with $R=800$ reaches  the value 
               $\sim 5100$ (see Fig.~\ref{fig:ensjs}) which goes beyond the range of 
               the $y$ axis. In the fully nonlinear stage ohmic dissipations roughly
               overlap for the three simulations, reaffirming our hypothesis that 
               the dynamics is turbulent, so that beyond a threshold dissipation is 
               independent of the Reynolds number.
               \label{fig:jm}}
\end{center}
\end{figure}
carried out with 3 different Reynolds numbers, $\mathcal R = 200, 400$
and $800$. As the Figure clearly shows they overlap.
This supports our hypothesis that the dynamics are turbulent and that the 
dissipation is independent of the Reynolds number. In Fig.~\ref{fig:jturb} 
a close-up of 
\begin{figure}
\begin{center}
\includegraphics[width=0.8\textwidth]{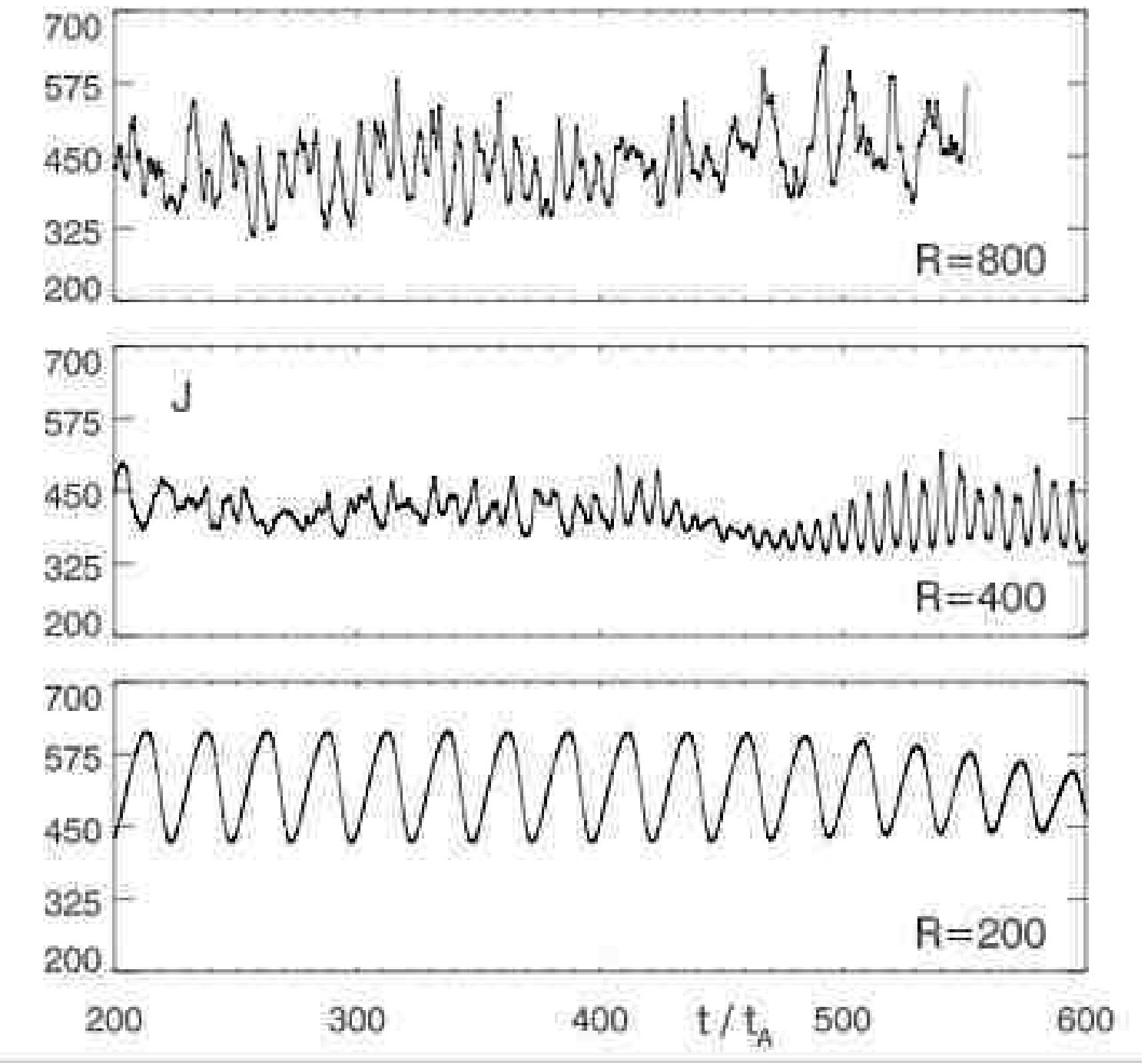}
\caption{Close-up of Ohmic dissipation as a function of time at
               different Reynolds numbers. A clear transition to turbulence
               is shown. In fact at higher Reynolds number  correspond  more
               fine structures becoming intermittent.
               \label{fig:jturb}}
\end{center}
\end{figure}
the dissipation functions with time at different Reynolds number is shown.
A clear transition to turbulence is present, 
the appearance of progressively 
more fine structure  as the Reynolds number increases.
Note that the dissipation is independent of the Reynolds number \emph{only
beyond a threshold} below which diffusion is important 
even at large scales, and consequently most of the energy is diffusively
lost from the field at large
scales before it can reach the small scales. In particular in the case 
of the sheared forcing considered in this simulation, if we decrease the 
Reynolds number below $\mathcal R =  50$ the diffusive time is so short
that no instability can grow, and the system follows the linear 
saturation curves, showing no sign at all of nonlinear dynamics.

\chapter{Nonlinear Analysis: Self-Organized Anisotropic Turbulence} \label{sec:nla}

As already mentioned in \S~\ref{sec:pi} one of the main original
results of this work is the \emph{connection between the heating of closed 
magnetic structures in the solar corona and anisotropic turbulence.} In an
environment threaded by a strong axial magnetic field, Alfv\'en wave packets
traveling in opposite directions are naturally present. In coronal loops they
are excited by photospheric motions on scales of the order of the cross-length
of a granule ($\sim 1000\, km$). Their  ``collisions'' give rise to a turbulent cascade 
which transfers energy from the large to the small scales, where it is finally dissipated, 
i.e.\ converted to heat and particle acceleration. Current sheets elongated in the axial
direction are a feature found in most of the previous numerical simulations
carried out to model a coronal loop. There have been many explanations about 
the formation of these structures. \emph{
In the framework of anisotropic turbulence, these structures spontaneously develop.}
In fact the cascade of energy mainly takes place in the orthogonal planes while it is 
strongly reduced in the axial direction, so the turbulent transfer of energy to smaller 
perpendicular scales results in sheet-like structures elongated in the direction of 
the main field.

In this section we analyze our system using the theory of anisotropic turbulence,
of which we have given a  review of the current state in  \S~\ref{par:bik}. As it usually 
happens in most of turbulence theory, the system studied is modeled
as a three-dimensional three-periodic one. Periodicity leads to the cancellation  
of all the flux terms at the boundary surfaces.
In a realistic model of a coronal loop, the axial direction cannot be modeled
as periodic, and in particular the energy flux originating from the photospheric
surfaces is just the source of energy  for the system, which of course cannot be neglected.
The inclusion of the energy flux terms at the boundaries gives to the system
the property to be \emph{self-organized}. In fact while the amplitude of the perturbations 
is usually imposed to be ``small'' on some physical ground, in a coronal loop 
it is the system itself that sets this amplitude, as described in the following sections.

\section{Self-Organization and Scalings} \label{sec:sos}

In a turbulent system energy is injected into the system at the rate $\epsilon_{in}$ 
at the scale $\ell_{in}$, it is subsequently scattered along the inertial range with
the transfer rate $\epsilon$, and finally dissipated at the dissipation scale
at the rate $\epsilon_d$. For balance all these fluxes must be equal
\be \label{eq:eqsf}
\epsilon_{in} = \epsilon = \epsilon_{d}
\ee
This equality still holds approximately when, as in our case, $\epsilon_{in}$
changes in time, because the more rapid dynamics at the small scales in the inertial
and dissipation ranges adjust the spectrum rapidly compared to the slower 
dynamics of the large scales.
In our case the value of $\epsilon_{in}$ is given by $S$~(\ref{eq:inr}), the integral of the 
Poynting flux at the boundaries, and the transfer rate $\epsilon$
is given by the anisotropic turbulence theory (see \S~\ref{par:bik}). Both 
$\epsilon_{in}$ and $\epsilon$ depend on the value of the internal fields,
so that the balance equality~(\ref{eq:eqsf}) fixes their value.

In our numerical simulations the injection scale is $\ell_{in} \sim \ell/4$,
that we approximate to $\ell$ in the following calculations.
Els\"asser variables $\bsy{z}^{\pm} = \bsy{u}_{\perp} \pm \bsy{b}_{\perp}$
are the fundamental variables in turbulence theory. The nonlinear terms
in the equations of reduced MHD~(\ref{eq:els1})-(\ref{eq:els4}) are
symmetric in the exchange of these two variables
\be
\left(  \bsy{z}^{\mp} \cdot \bnabla_{\perp} \right) \bsy{z}^{\pm}
\qquad \textrm{is symmetric for} \qquad \bsy{z}^{+} \Longleftrightarrow  \bsy{z}^{-}
\ee
while the linear terms simply describe a wave propagation in opposite directions
for the two fields $\bsy{z}^{\pm}$. Furthermore, in our case, the boundary 
conditions~(\ref{eq:bc0})-(\ref{eq:bcL}), that we rewrite for convenience, 
\begin{equation} \label{eq:bc0s}
\boldsymbol{z^{-}}  + \boldsymbol{z^{+}} = + 2 \, \boldsymbol{u}_{0} 
\quad \textrm{at} \ z=0 
\end{equation}
\begin{equation} \label{eq:bcLs}
\boldsymbol{z^{+}} + \boldsymbol{z^{-}} = + 2 \, \boldsymbol{u}_{L}  
\quad \textrm{at} \ z=L 
\end{equation}
are also symmetric in the exchange. In this way the rms values of $\bsy{z}^{\pm}$
are approximately equal at all scales, i.e.\ indicating with $\delta z_{\lambda}$
the rms value of the field $z$ at the scale $\lambda$
\be
\delta z^{+}_{\lambda} \sim \delta z^{-}_{\lambda} \sim \delta z_{\lambda}
\ee
This is equivalent to say that we expect the cross helicity
\be
H^C = \int_V \! \ud^{3}x\ \bsy{u}_{\perp} \cdot \bsy{b}_{\perp} = 
\frac{1}{4}\, \int_V \! \ud^{3}x\ \left( \left| \bsy{z}^{+} \right|^2 -
\left| \bsy{z}^{-} \right|^2 \right) \sim 0
\ee
to vanish. This is also confirmed by our linear analysis (see \S~\ref{sec:la}) which
has shown that the magnetic field is noticeably bigger than the velocity field,
and by our numerical simulations (see Figure~\ref{fig:atshc}).
As the velocity field is smaller than the magnetic field we can also
approximate in the Poynting flux integral~(\ref{eq:inr}) the magnetic field 
at the injection scale $\ell$ with the Els\"asser variable, i.e.\ 
$\delta b_{\ell} \sim \delta z_{\ell}$, so that identifying the injections
energy rate $\epsilon_{in}$ with S we have
\be \label{eq:pfi}
\epsilon_{in} = S \sim \ell^2\, v_{\mathcal A}\, u_{ph}\, \delta z_{\ell}
\ee
The energy injection rate is directly proportional to the rms of the Els\"asser 
fields at the injection scale $\ell$.
The velocity forcing at the boundaries injects energy with the rate~(\ref{eq:pfi}) 
at the scale $\ell_{in} = \ell / 4 \sim \ell << L$ smaller than the axial lenght $L$, 
developing a perturbation $ \delta z_{\ell} << v_{\mathcal A}$. 

The strongly reduced cascade along
the axial direction implies that the wave packets have size of the order
$L$ along this direction, while in the ortogonal plane a turbulent 
cascade with the characteristics described in \S~\ref{par:bik} takes place.
We now apply the theory of anisotropic turbulence (see \S~\ref{par:bik}) to have
the value of the transfer energy rate $\epsilon$. 

We start from the anisotropic
version of the Iroshnikov-Kraichnan theory described in 
equations~(\ref{eq:aik1})-(\ref{eq:aik2}) and characterized by the 
orthogonal spectrum $E_{k_{\perp}} \propto k_{\perp}^{-2}$.
Indicating with $\tau_{\lambda} = \lambda / \delta z_{\lambda}$
the eddy turn-over time and with $\tau_{\mathcal A} \sim L / v_{\mathcal A}$ 
the Alfv\'enic crossing time, the number of collisions for the fractional
perturbation to build up to order unity is
\be
N_{\lambda} \sim \left( \frac{\tau_{\lambda}}{\tau_{\mathcal A}} \right)^2 \gg 1
\ee
so that the energy transfer time is given by
\be
T_{\lambda} \sim N_{\lambda}\, \tau_{\mathcal A}  \sim \tau_\lambda^2 / \tau_{\mathcal A},
\ee
and the energy flux for unitary volume $\epsilon$ is given by
\begin{equation} \label{eq:scsf}
\epsilon \sim \frac{{\delta z_{\lambda}}^2}{T_{\lambda}} \sim
\frac{{\delta z_{\lambda}}^4}{\lambda^2} \frac{L}{v_{\mathcal A}}
\end{equation}
Usually the value of the spectral flux $\epsilon$ or equivalently of the perturbation
at the large scale are fixed to a ``small'' value on some physical ground.
In our model this value is \emph{self-consistently} established. 
Multiplying equation~(\ref{eq:scsf}) by the volume
$\ell^2 L$ in order to have the total flux, and considering the injection scale 
$\lambda = \ell$ we have
\be \label{eq:te}
\epsilon \sim  
\ell^2 L \, \frac{{\delta z_{\ell}}^4}{\ell^2} \frac{L}{v_{\mathcal A}}
\sim \frac{L^2}{v_{\mathcal A}} \, \delta z_{\ell}^4
\ee
This is the flux of energy that leaves the injection scale towards the small scales,
and that is supposed to have a constant value along all the inertial range. 
\emph{Equations~(\ref{eq:pfi}) and (\ref{eq:te}) show that the system is self-organized.} 
In fact both $\epsilon_{in}$ and $\epsilon$ depend by $\delta z_{\ell}$, the rms value of the 
Els\"asser variables at the injection scale $\ell$, so that the internal dynamics depends
by the injection of energy and the injection of energy depends by the internal
dynamics. As shown in
\begin{figure}[t]
\begin{center}
\includegraphics[width=0.5\textwidth]{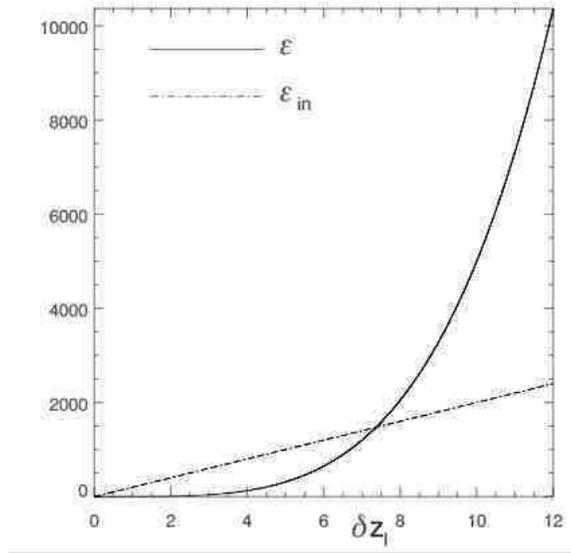}
\caption{Injection energy rate $\epsilon_{in}$~(\ref{eq:pfi}) in dashed line and 
transfer energy rate along the inertial range $\epsilon$~(\ref{eq:scsf}) in 
continuous line, as a function of the rms value of the Els\"asser variables 
$\delta z_{\ell}$ at the injection scale $\ell$. $\delta z_{\ell}$ can only
fluctuate around the equilibrium value~(\ref{eq:sbp}). The parameters
used are $\ell = 1$, $L=10$, $u_{ph}=1$ and $v_{\mathcal A} = 200$,
which have also been used to perform the numerical simulations.
 \label{fig:self1} }
\end{center}
\end{figure}
Figure~\ref{fig:self1} the two fluxes satisfy the equality condition~(\ref{eq:eqsf}) 
only for the value
\be \label{eq:sbp}
\delta z_{\ell}^{\ast} \sim \left( \frac{\ell v_{\mathcal A}}{L} \right)^{\frac{2}{3}} \, 
u_{ph}^{\frac{1}{3}}.
\ee
For a value of the perturbation smaller than~(\ref{eq:sbp}) the injection flux
is higher than the dissipation one, so that the pertubation $\delta z_{\ell}$ 
grows toward the equilibrium value. On the opposite for higher values 
than~(\ref{eq:sbp}) the dissipation flux is higher than the injection one, 
decreasing the value of the perturbation $\delta z_{\ell}$. The system is
dynamic and turbulent so that we will have fluctuations around this
equilibrium value. 

Substituting the equilibrium value~(\ref{eq:sbp}) in (\ref{eq:te}) or equivalenty
(\ref{eq:pfi}), we obtain the average spectral flux
\be \label{eq:hr}
\epsilon^{\ast} \sim \ell^2 \left( \frac{\ell}{L} \right)^{\frac{2}{3}}\,
v_{\mathcal A}^{\frac{5}{3}}
u_{ph}^{\frac{4}{3}}
\ee
As previously said (see (\ref{eq:eqsf})) the injection, transfer and dissipation rates
balance, so that equation~(\ref{eq:hr}) holds for all of them, in particular for the
dissipation rate $\epsilon_d$ that is also the \emph{heating rate}, i.e.\ the rate 
at which energy is converted to heat and particle acceleration at the dissipation
scale $\ell_d$.

As shown in \S~\ref{par:bik}, anisotropic turbulence is characterize by different
regimes to which correspond different slopes of the spectrum $E_{k_{\perp}}$.
The relative amplitude of the perturbation relatively to the axial Alfv\'enic velocity
$\delta z_{\ell} / v_{\mathcal A}$ indicates in which regime is the system.
In particular at smaller values of this ratio correspond higher slopes in the 
energy spectrum. In our case from equation~(\ref{eq:sbp}) we have that
\be \label{eq:rat}
\frac{\delta z_{\ell}^{\ast}}{v_{\mathcal A}} \sim 
\left( \frac{\ell}{L} \right)^{\frac{2}{3}}
\left( \frac{u_{ph}}{v_{\mathcal A}} \right)^{\frac{1}{3}}.
\ee
\emph{This means that changing the ratio $v_{\mathcal A} / u_{ph}$, the relative
amplitude of the perturbations changes, and we move among different
regimes of anisotropic turbulence}, in particular increasing the value of the
axial magnetic field $v_{\mathcal A}$ increaseas and the ratio~(\ref{eq:rat})
decreases.

Previous results can be generalized to the other scalings for anisotropic 
turbulence described in \S~\ref{par:bik}. The different scalings are characterized
by the different number of collisions $N_{\lambda}$ that a wave packet must
suffer for the perturbation to build up to order unity. Introducing the
parameter $\alpha$ we can write
\be \label{eq:atnc}
N_{\lambda} = \left( \frac{\tau_{\lambda}}{\tau_{\mathcal A}} \right)^{\alpha}
\ee
where $\tau_{\lambda} = \lambda / \delta z_{\lambda}$ and 
$\tau_{\mathcal A} = L / v_{\mathcal A}$ so that for the energy transfer time we have
\be \label{eq:btnc}
T_{\lambda} \sim N_{\lambda}\, \tau_{\mathcal A} \sim
\left( \frac{v_{\mathcal A}}{L} \right)^{\alpha -1} 
\left( \frac{\lambda}{\delta z_{\lambda}} \right)^{\alpha}
\ee
and finally for the transfer energy rate
\be \label{eq:sbe1}
\epsilon \sim \ell^2 L\, \frac{\delta z_{\lambda}^2}{T_{\lambda}} \sim 
\ell^2 L\, \left( \frac{L}{v_{\mathcal A}} \right)^{\alpha - 1} \, 
\frac{\delta z_{\lambda}^{\alpha + 2}}{\lambda^{\alpha}}.
\ee
Flux~(\ref{eq:sbe1}) is supposed to be constant along the inertial range, 
so that we can consider its value at the injection scale $\lambda = \ell$:
\be \label{eq:sbe2}
\epsilon \sim \ell^2 L\, \frac{\delta z_{\ell}^2}{T_{\ell}} \sim 
\frac{L^{\alpha}}{\ell^{\alpha - 2} \, v_{\mathcal A}^{\alpha - 1}} \, 
\delta z_{\ell}^{\alpha + 2}
\ee
\begin{figure}[!h]
      \centering
      %%---- start ----
      \subfloat[]{
               \label{fig:self:a}             %% label for subfigure
               \includegraphics[width=0.45\linewidth]{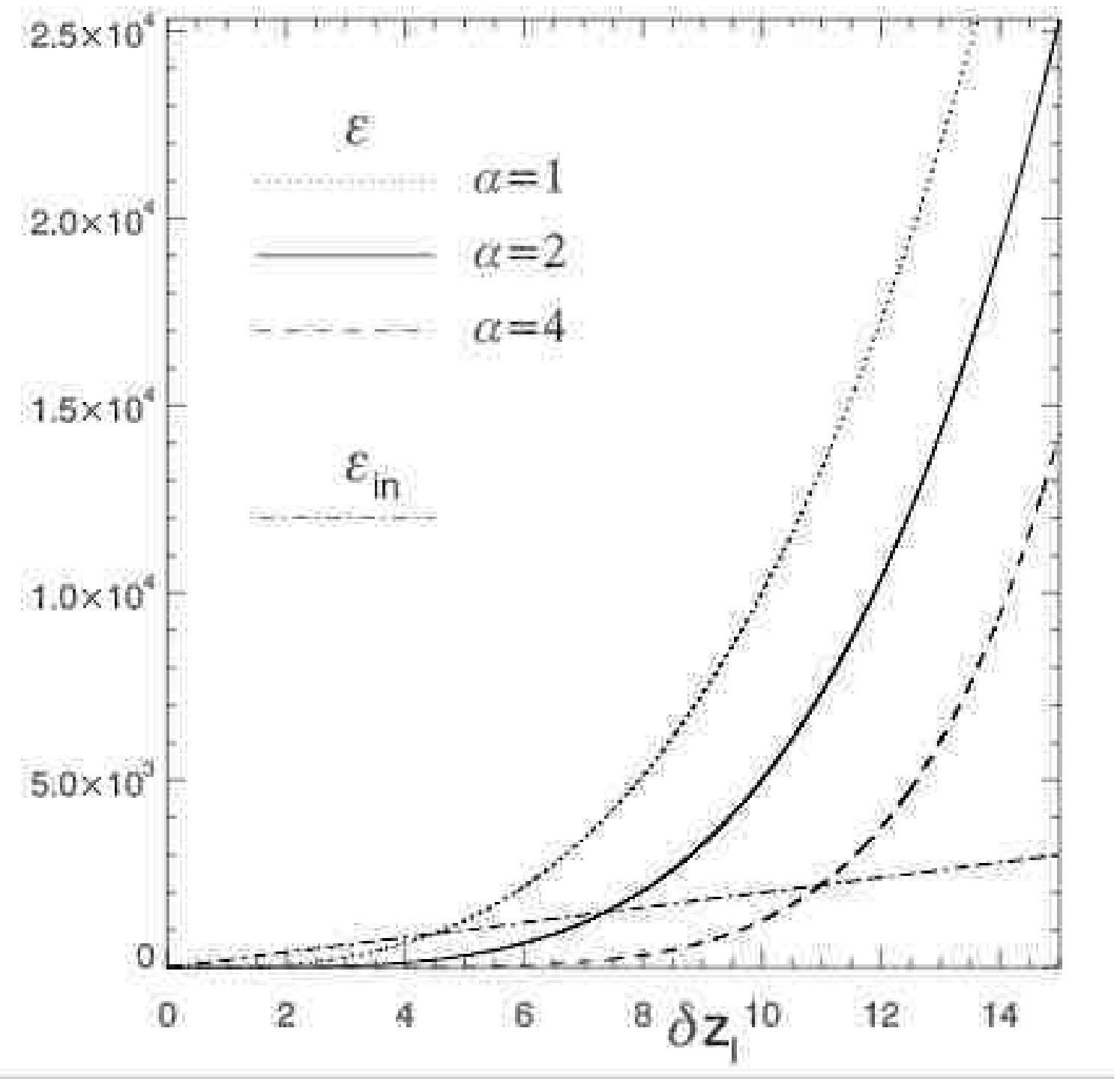}}
      \hspace{0.01\linewidth}
     %%---- start ----
      \subfloat[]{
               \label{fig:self:b}             %% label for subfigure
               \includegraphics[width=0.45\linewidth]{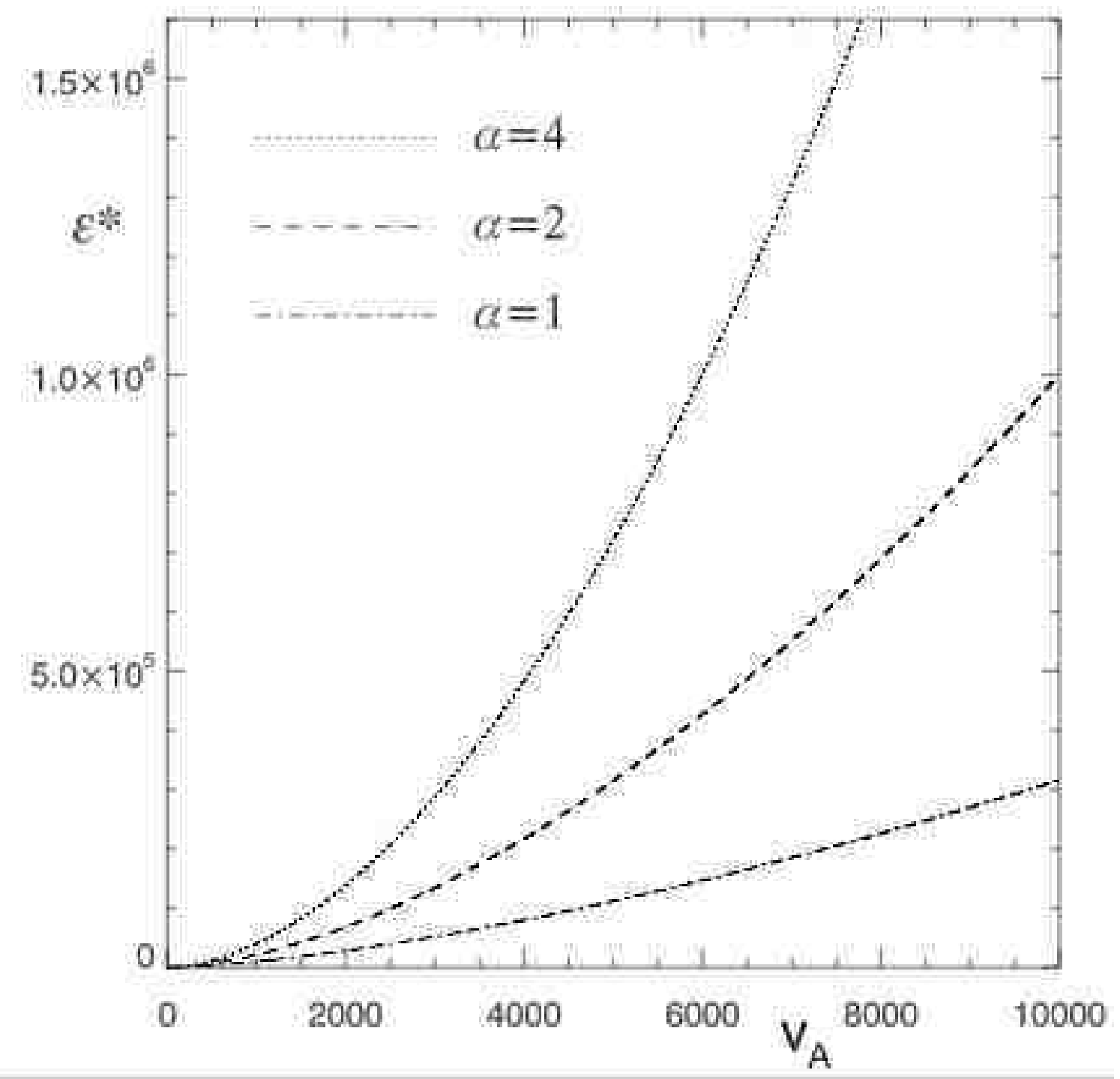}}
   \caption{(a) Injection energy rate $\epsilon_{in}$~(\ref{eq:pfi}) and 
   transfer energy rates along the inertial range $\epsilon$~(\ref{eq:sbe2})
   for different values of the parameter $\alpha$, as a function of the rms 
   value of the Els\"asser variables 
   $\delta z_{\ell}$ at the injection scale $\ell$. $\delta z_{\ell}$ can only
   fluctuate around the equilibrium value $\delta z_{\ell}^{\ast}$~(\ref{eq:sbpg}). 
   (b) Coronal heating functions $\epsilon^{\ast}$~(\ref{eq:chs}) as a function of 
   the axial Alfv\'enic velocity $v_{\mathcal A}$ for different values of the parameter
   $\alpha$, showing that to higher values of $\alpha$ (corresponding
   to weaker turbulent regimes) coronal heating rates are more efficient.
   The parameters used both in (a) and (b) are $\ell = 1$, $L=10$, $u_{ph}=1$ and 
   $v_{\mathcal A} = 200$.
        \label{fig:self}}             %% label for entire figure
\end{figure}
The injection energy rate is always given by~(\ref{eq:pfi}).
For $\alpha = 1$ we have the anisotropic Kolmogorov spectrum $k_{\perp}^{-5/3}$,
while for $\alpha = 2$ we have the anisotropic IK spectrum $k_{\perp}^{-2}$ and
for $\alpha = 4$ the SG94 spectrum based on 4-waves resonant interactions 
$k_{\perp}^{-7/3}$. In Figure~\ref{fig:self:a} energy fluxes 
$\epsilon$~(\ref{eq:sbe2}), for different values of the parameter $\alpha$, 
and $\epsilon_{in}$~(\ref{eq:pfi}) are plotted as a function of the perturbation $\delta z_{\ell}$.
There is always an equilibrium value where the two fluxes are equal, while for a higher
value of the perturbation the transfer rate $\epsilon$ is bigger than the injection rate 
$\epsilon_{in}$ and for a lower value of the perturbation the transfer rate is smaller
than the injection rate.
From (\ref{eq:sbe2}) and (\ref{eq:pfi}) we have that the equilibrium value for 
$\delta z_{\ell}$ is  given by
\be \label{eq:sbpg}
\delta z_{\ell}^{\ast} \sim \left( \frac{\ell v_{\mathcal A}}{L} \right)^{\frac{\alpha}{\alpha + 1}} \, 
u_{ph}^{\frac{1}{\alpha + 1}}.
\ee
and substituting this value in (\ref{eq:sbe2}) or (\ref{eq:pfi}) we obtain
the \emph{coronal heating scalings}
\be \label{eq:chs}
\epsilon^{\ast} \sim \ell^2 \left( \frac{\ell}{L} \right)^{\frac{\alpha}{\alpha+1}}\,
v_{\mathcal A}^{\frac{2\alpha +1}{\alpha+1}}
u_{ph}^{\frac{\alpha +2}{\alpha+1}}
\ee
which are plotted, for different values of the parameter $\alpha$, 
in Figure~\ref{fig:self:b}. To higher values of the parameter
$\alpha$ correspond higher heating rates.

The regime of anisotropic turbulence is determined by the ratio
$\delta z_{\ell} / v_{\mathcal A}$, which from~(\ref{eq:sbpg}) is given by
\be
\frac{\delta z_{\ell}^{\ast}}{v_{\mathcal A}} \sim 
\left( \frac{\ell}{L} \right)^{\frac{\alpha}{\alpha + 1}}
\left( \frac{u_{ph}}{v_{\mathcal A}} \right)^{\frac{1}{\alpha + 1}}.
\ee
As $\alpha \ge 1$, the property that to higher values of the Alfv\'enic velocity
$v_{\mathcal A}$ corresponds a smaller relative perturbations is preserved,
whatever the anisotropic turbulent regime is. This means that in general
none of the different scalings~(\ref{eq:chs}) corresponding to different
values of $\alpha$ is valid for the whole range of the possible values
of $v_{\mathcal A}$. In fact, starting form a low value of $v_{\mathcal A}$
strong anisotropic turbulence will develop, corresponding to a spectral
index $k_{\perp}^{-5/3}$ and a coronal heating rate~(\ref{eq:chs}) 
with $\alpha=1$ and $\epsilon \propto v_{\mathcal A}^{3/2}$. Increasing 
$v_{\mathcal A}$ the system transitions to an anisotropic IK regime
characterized by $\alpha = 2$, the spectral slope $k_{\perp}^{-2}$  and 
$\epsilon \propto v_{\mathcal A}^{5/3}$. Increasing still the value
of the Alfv\'enic velocity will cause the system to transition
in the regime characterized by $\alpha = 4$, the spectral 
slope $k_{\perp}^{-7/3}$  and 
$\epsilon \propto v_{\mathcal A}^{9/5}$. So that the scalings for the coronal
heating as a function of the Alfv\'enic velocity $v_{\mathcal A}$, is 
characterized by regions with increasing slopes in correspondence
of increasing values of $v_{\mathcal A}$, each region characterized
by the scaling~(\ref{eq:chs}) with an increasing value of the parameter~$\alpha$.

\section{Timescales}

From equations~(\ref{eq:atnc})-(\ref{eq:chs}) we can derive the values
of the eddy turn-over time $\tau_{\lambda} = \lambda / \delta z_{\lambda}$, 
the energy transfer rate $T_{\lambda}$~(\ref{eq:btnc}) and the number of 
collisions at the scale $\lambda$ $N_{\lambda}$~(\ref{eq:atnc}). After a 
few algebraic calculations we have
\be \label{eq:atts}
\tau_{\lambda} = 
\left( \tau_{\perp}\, \tau_{\mathcal A}^{\alpha} \right)^{\frac{1}{\alpha + 1}}\,
\left( \frac{\lambda}{\ell} \right)^{\frac{2}{\alpha + 2}}, \qquad
T_{\lambda} = 
\left( \tau_{\perp}^{\alpha}\, \tau_{\mathcal A} \right)^{\frac{1}{\alpha + 1}}\,
\left( \frac{\lambda}{\ell} \right)^{\frac{2\alpha}{\alpha + 2}}
\ee
and
\be \label{eq:atnc2}
N_{\lambda} = \left( \frac{\tau_{\lambda}}{\tau_{\mathcal A}} \right)^{\alpha} = 
\left( \frac{\tau_{\perp}}{\tau_{\mathcal A}} \right)^{\frac{\alpha}{\alpha + 1}}\,
\left( \frac{\lambda}{\ell} \right)^{\frac{2\alpha}{\alpha + 2}}
\ee
where $\tau_{\perp} = \ell/ u_{ph}$ is the crossing time used as characteristic 
time to render the reduced MHD equations dimensionless, and 
$\tau_{\mathcal A} = L / v_{\mathcal A}$ is the axial Alfv\'enic crossing time.
For our system 
\be
\tau_{\perp} \ll \tau_{\mathcal A}
\ee
so that $N_{\lambda} \gg 1$. The number of collisions has always the property
to decrease at small scales, leading to a stronger turbulent regime.
Except for $\alpha =1$ where correctly the energy transfer time and 
the eddy turn-over time are equal, for all the others regime ($\alpha \ge 2$)
\be
\frac{T_{\lambda}}{\tau_{\lambda}} = 
\left( \frac{\tau_{\perp}}{\tau_{\mathcal A}} \right)^{\frac{\alpha-1}{\alpha + 1}}\,
\left( \frac{\lambda}{\ell} \right)^{2\, \frac{\alpha-1}{\alpha + 2}} \gg 1
\ee
the energy transfer time is bigger than the eddy turn-over time, which is 
a characteristic of the Alfv\'en effect at the basis of the anisotropic turbulence theory.

An interesting quantity is the energy transfer time at the large scales $\lambda = \ell$
which can be identified with the nonlinear timescale $T_{NL}$, i.e.\
\be
T_{NL} = T_{\ell} = 
\left( \tau_{\perp}^{\alpha}\, \tau_{\mathcal A} \right)^{\frac{1}{\alpha + 1}}
\ee
which in dimensionless units is $T_{NL} = \tau_{\mathcal A}^{\frac{1}{\alpha + 1}}$.

The nonlinear time $T_{NL}$ is also the time at which the linear stage transitions
into the nonlinear stage. This results from the following calculations.
The linear stage dynamics is described in \S~\ref{sec:la}, and in particular
we use the solution~(\ref{eq:ts1})-(\ref{eq:ts6}) for the two-sided problem with 
velocity patterns $\bsy{u}^0$ and $\bsy{u}^L$, respectively in the boundary planes
$z=0$ and $z=L$. Averaging over time-scales bigger than the Alfv\'enic crossing time,
the magnetic field for this solution can be approximated as
\be
\bsy{b}_{\perp}  \left( x,y,z,t \right)  \sim \frac{t}{\tau_{\mathcal A}}\,
\left[ \bsy{u}^L \left( x,y \right) - \bsy{u}^0 \left( x,y \right) \right]  
\ee
Since during the linear stage the magnetic field has only large-scale components,
we can approximate
\be \label{eq:elsin}
\delta z_{\ell} \sim u_{ph}\, \frac{t}{\tau_{\mathcal A}}
\ee
and from~(\ref{eq:pfb}) we can write for the injection energy rate
\be
\epsilon_{in} \sim 
v_{\mathcal A} \int \! \ud a\,  \left( \bsy{u}^L - \bsy{u}^0 \right) \cdot \bsy b_{\perp}
\sim  
v_{\mathcal A} \int \! \ud a\,  \left| \bsy{u}^L - \bsy{u}^0 \right|^2 
\frac{t}{\tau_{\mathcal A}} \sim  
\ell^2\, v_{\mathcal A}\, u_{ph}^2\, \frac{t}{\tau_{\mathcal A}}
\ee
From this equation we can also note the important property that
during the linear stage the injection energy rate is always positive.
The energy flux $\epsilon$ along the inertial range is the rate at
which the energy is leaving the large scales towards the small
scales, and substituting the value~(\ref{eq:elsin}) for $\delta z_{\ell}$  
in~(\ref{eq:sbe2}) we have for the energy transfer rate
\begin{figure}[t]
\begin{center}
\includegraphics[width=0.6\textwidth]{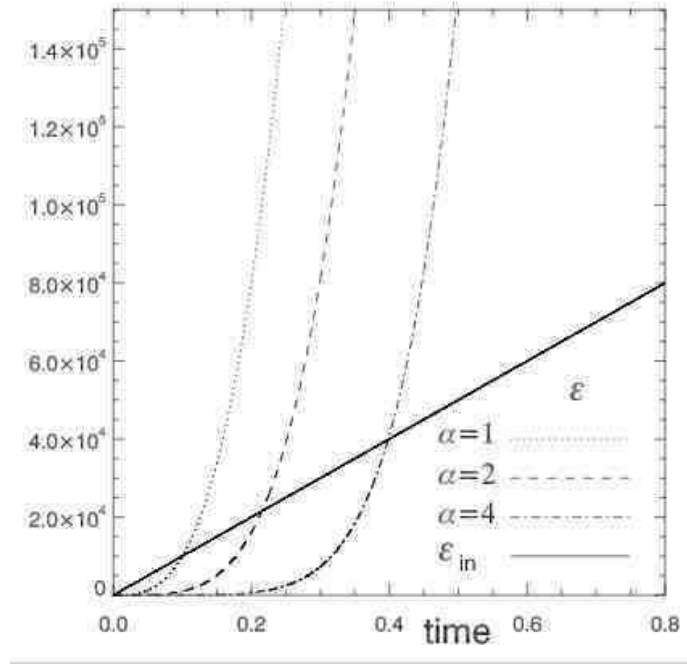}
\caption{Injection energy rate $\epsilon_{in}$~(\ref{eq:pfi}) in continuous 
line and 
transfer energy rate along the inertial range $\epsilon$~(\ref{eq:scsf}) in 
dashed lines during the linear stage, as a function of time.
The parameters
used are $\ell = 1$, $L=10$, $u_{ph}=1$ and $v_{\mathcal A} = 1000$,
to which corresponds a crossing Alfv\'en time 
$\tau_{\mathcal A} = L / v_{\mathcal A} = 10^{-2}$.
 \label{fig:selflin} }
\end{center}
\end{figure}
\be
\epsilon \sim
\frac{L^{\alpha}\, u_{ph}^{\alpha+2}  }{\ell^{\alpha - 2} \, v_{\mathcal A}^{\alpha - 1}} \, 
\left( \frac{t}{\tau_{\mathcal A}} \right)^{\alpha + 2}
\ee
Since $\alpha \ge 1$ the injection of energy is higher than the dissipation, until
time
\be
t = T_{NL} \sim 
\left( \tau_{\perp}^{\alpha}\, \tau_{\mathcal A} \right)^{ \frac{1}{\alpha + 1}}
\ee
where they become equal, and of course where the linear regime is no 
longer valid.

\chapter{Conclusions and Discussion}

In this thesis I have presented the results of the numerical and analytical investigations
of a reduced MHD realization of the Parker Scenario for coronal heating.
This is the 3D extension of earlier 2D investigations  by
Einaudi, Velli, Politano \& Pouquet~\cite{ein96}, Einaudi \& Velli~\cite{ein:1999},
Georgoulis, Velli \& Einaudi~\cite{georg98}. 

Parker~\cite{park79}, \cite{park83}, \cite{park86}, \cite{park88}
was the first to propose that the X-ray corona is heated by dissipation at the 
many small current sheets forming in a coronal loop as a consequence of the 
continuous shuffling and intermixing of the footpoints of the field
by the photospheric convection. Analogous to the linear analysis performed in 
\S~\ref{sec:la} the magnetic field results from a continuous mapping of the 
footpoints velocity pattern. He supposed that the magnetic field spontaneously
produces tangential discontinuities (current sheets) which become increasingly
severe with continued winding and interweaving, eventually producing
intense magnetic dissipation by magnetic reconnection.
This dissipation is characterized by bursts of rapid reconnection
that he named ``nanoflares''. 

This picture is, however,  ``static'' 
(as opposed to ``dynamic'') because the magnetic field-lines are assumed to
result from ``quasi-static'' displacements of the magnetic fields, and the current
sheets are a consequence of the displacement of two neighboring field lines
in opposite directions. The system does not show any dynamics,  the field lines
are passively bent by the photospheric motions, and the formation of the 
current sheets is almost ``geometric''.

The first numerical simulation of the Parker Scenario was been performed 
in 1989 by Mikic, Schnack \& Van Hoven~\cite{mik89}, who found that current sheets formed,
extended in the axial direction. Similar results were found later
by Longcope \& Sudan~\cite{long94} and Hendrix \& Van Hoven~\cite{hen96}
respectively in 1994 and 1996. The original ``static'' picture has 
remained the basis of all studies of the Parker Scenario. In particular,
Longcope \& Sudan~\cite{long94} derive a scaling law for coronal heating as if
this were due to magnetic reconnection of the large-scale field.
Hendrix \& Van Hoven~\cite{hen96} also found current sheets extended in the axial
direction and an inertial range but still thought that magnetic reconnection
is responsible for the turbulence, describing it as ``spontaneous
unstably driven MHD turbulence'' \cite{hen96}. All these simulations 
used  a relatively low resolution, i.e.\ a small Reynolds number,
and could not observe the full turbulent dynamics. 

In this work, using high-resolution simulations, we have conclusively shown 
that the transfer of energy from the large-scales, where energy is continuously
injected by photospheric motions, towards the small scales is due to 
\emph{weak anisotropic turbulence}. The small scale organizes into vortex-current
sheets (a common feature in MHD turbulence) that result from  the non-linear 
cascade. They may eventually  break up due
to tearing and magnetic reconnection, but are certainly not primarily generated
by a reconnection instability due to the large-scale field.
Magnetic reconnection of these cascade-produced current sheets 
certainly plays a role in the dissipation of energy, i.e.\ in the conversion of 
the energy flowing along the inertial range into heat and particle acceleration.
In any case the dissipation mechanism, which acts at the small scales, 
cannot be properly investigated with a fluid model because these are very
well known to break at small scales and/or high frequencies. Thus,
a kinetic investigation is appropriate.

In my view the difficulty in realizing that the dynamics of a coronal loop
is turbulent has been partially because  a theory of 
weak turbulence for an MHD system embedded in a strong axial field has
only recently been developed,
mainly due to Sridhar \& Goldreich (1994)~\cite{sg94} who  
studied this mechanism for the first time and find the $k_{\perp}^{-7/3}$ 
spectrum. The debate followed with the remarks of  
Montgomery \& Matthaeus (1995)~\cite{mm95},
while Goldreich \& Sridhar (1997)~\cite{gs97}   found  for the first 
time the $k_{\perp}^{-2}$ spectrum and propose
a multiple inertial range. Ng \& Bhattacharjee (1997)~\cite{bhatta97}, and
Galtier et al.\ (2000,2002)~\cite{gal00,gal02} then develop a kinetic model.

Weak turbulence has been barely investigated numerically.
Cho \& Vishniac (2000), Maron \& Goldreich (2001)  and
M\"uller \& Biskamp (2000) have performed three-dimensional 
simulations of strong MHD turbulence stretching the abilities
of the fastest supercomputers.  Simulations
of weak turbulence are  more difficult because each 
wavepacket must interact  many times with oppositely
directed ones before it cascades; by contrast, a wavepacket cascades 
in strong MHD turbulence in the time it takes to
cross a single oppositely directed wavepacket.

Our investigations show that:
\begin{itemize}

\item The dynamics are described by weak turbulence. Because the system is 
embedded in a strong axial field the cascade takes place in the 
orthogonal planes and is strongly inhibited in the axial direction.
Wave-packets are ``long-lived'' and they need to collide many times
before transferring energy to smaller scales.

\item The spectral slopes are in agreement with those derived from weak turbulence
theory. The energy spectra slopes change from $-2$ for 
$v_{\mathcal A} \sim 50$ to $\sim -3$ for $v_{\mathcal A} \sim 1000$.
The slope increase takes place while the rms of the orthogonal $\bsy{z}^{\pm}$
fields grows in absolute magnitude but its relative magnitude
$\bsy{z}^{\pm}/v_{\mathcal A}$ decreases, in accordance with the theory.

\item Small-scales structures do not homogeneously distribute in the planes
but organize in vortex-current sheets extended in the axial direction where
the energy flowing along the inertial range finally  dissipates.

\item In the framework of weak turbulence these extended structures 
are naturally formed because the cascade in the axial direction is 
strongly inhibited. Hence no boundary layer is present.
On the other hand in a real coronal loop a boundary layer, the transition
region, is present. This result implies that it has others physical origins.

\item Since the rate of injection of energy by photospheric motion depends 
not only on the photospheric velocity but also on the fields that develop into the 
computational box the system is self-organized. We have shown that
the rms amplitude of the magnetic field at the large scales, which would grow linearly
in time were it not for the non-linear dynamics, is fixed by the equality
of the injection energy rate $\epsilon_{in}$ and the rate at which energy
flows from the large scales towards the small scales, which is the transfer
energy rate along the inertial range $\epsilon$.

\item The weak turbulence and the self-organization of the system lead
to the new \emph{coronal heating scalings}~(\ref{eq:chs})
\be
\epsilon \sim \ell^2 \left( \frac{\ell}{L} \right)^{\frac{\alpha}{\alpha+1}}\,
v_{\mathcal A}^{\frac{2\alpha +1}{\alpha+1}}
u_{ph}^{\frac{\alpha +2}{\alpha+1}}
\ee
where $\epsilon$ is the rate of energy flowing from the injection scale towards
the small scales along the inertial range where finally dissipates.
We have shown in \S~\ref{sec:sos} that there is not a single scaling law, 
but that the dissipation rate depends on the regime of weak turbulence in which
the system relaxes, described in the previous equation by the 
parameter $\alpha$. We have also shown that to higher values
of the Alfv\'enic velocity $v_{\mathcal A}$ corresponds weaker
turbulent regimes characterized by a higher value of the parameter
$\alpha$ and thus leading to a more efficient heating rate.

\end{itemize}

The system we have studied is a rather schematic model of a coronal loop.
In particular we have not considered density or magnetic field variations
along the axis of the loop. The least realistic assumptions are the boundaries, 
at which the energy is injected and hence are 
critical to model. Our model should be considered to apply to
the extended central part of a loop, beyond the transition region,
where the hypothesis of homogeneity along the axial direction is justified.

The first step towards a more complex model is to consider
a more complex forcing velocity. A compressible code
gives the possibility to use more realistic photospheric motions,
including at the same time compressible effects into the dynamics.
The first compressible simulations using an incompressible
forcing velocity pattern (Dahlburg~\cite{dahlpvt}) confirm the picture 
shown in this thesis.
A future model for the numerical and theoretical 
investigation of a coronal loop dynamics  should consider 
the underlying regions
from the photosphere up to the transition region and of their mutual
influences.

\part{Slow Solar Wind}

\chapter{The Slow Solar Wind Acceleration}

During the first part of my Ph.D.\ I have completed a work about the acceleration
of the slow solar wind, which I had previously started for my ``Laurea'' thesis.
This work has led to the publication of a paper (Rappazzo et~al.~\cite{rapp05}) which
is reproduced in the Appendix. In the following a brief review is given.

Although the association between the slow solar wind and the streamer belt 
(e.g.\ Gosling et~al.~\cite{gosling}) and between the fast wind and the polar 
coronal holes is broadly recognized,
the mechanism which leads to the slow and fast acceleration is still a matter of debate.
Einaudi et~al.~\cite{ein:1999} developed an MHD model, with a current sheet 
embedded in a broader wake flow, that accounts for some of the typical features 
observed in the slow
component of the solar wind. Reconnection of the magnetic field occurs at the current
sheet and, in the non-linear regime, when the equilibrium magnetic field is substantially
modified, a Kelvin-Helmoltz instability develops, leading to the acceleration of density
enhanced magnetic islands.

The solar streamer belt is a structure consisting of a magnetic configuration centered
on the current sheet, which extends above the cusp of a coronal helmet. The region
underlying the cusp is made up of closed magnetic structures, with the cusp representing
the point where separatrices between closed and open field lines intersect. Further from
the Sun, at solar minimum, the streamer belt around the equator appears as a laminar
configuration consisting of a thick plasma sheet with a density about 1 order of
magnitude higher than the surrounding plasma, in which much narrower and complex
structures are embedded. As a first approximation, moving from the center of the
streamer in polar directions at radii greater than the radius of the cusp, the radial
component of the magnetic field increases from zero, having opposite values on the two
sides of the current sheet. As far as the flow distribution is concerned, the fast solar wind
originates from the unipolar regions outside the streamer belt, while the slowest flows are
located at the center of the sheet.

One of the most interesting findings of the LASCO instrument onboard the Solar and
Heliospheric Observatory (SOHO) spacecraft has been the observation of a continuous
outflow of material in the solar streamer belt. An analysis performed using a difference
image technique (Sheeley et~al.~\cite{sheeley:1997}, Wang et~al.~\cite{wang:1998}) 
has revealed the presence of
plasma density enhancements, called ÒblobsÓ, accelerating away from the Sun. These
plasmoids are seen to originate just beyond the cusps of helmet streamers as radially
elongated structures a few percent denser than the surrounding plasma sheet, of
approximately 1 Solar Radius ($R_{\odot}$) in length and $0.1~R_{\odot}$ in width. 
They are observed to
accelerate radially outward maintaining constant angular spans at a nearly constant
acceleration up to the velocity of 200-450 km/s, in the spatial region between about 5 and
$30~R_{\odot}$. It has been inferred that the blobs are ÒtracersÓ of the slow wind, being carried
out by the ambient plasma flow.

In my work~\cite{rapp05} I have included in the previous model
(Einaudi et~al.~\cite{ein:1999}) spherical geometry effects, taking into account either the radial
divergence of the magnetic field lines and the average expansion suffered by a parcel of
plasma propagating outward, using the Expanding Box Model (EBM), and the
diamagnetic force due to the overall magnetic field radial gradients, the so-called 
melon-seed force. I have found that the values of the acceleration and density contrasts 
can be in good agreement with LASCO observations, provided the spherical divergence of the
magnetic lines starts beyond a critical distance from the Sun and the initial stage of the
formation and acceleration of the plasmoid is due to the cartesian evolution of MHD
instabilities. This result provides a constraint on the topology of the magnetic field in the
coronal streamer, which observationally is unknown.

\blankpage
\chapter*{Appendix}
\addcontentsline{toc}{chapter}{Appendix}

In this appendix we have included the paper A.\ F.\ Rappazzo, M.\ Velli, G.\ Einaudi 
and R.\ B.\ Dahlburg  ``Diamagnetic and Expansion Effects on the
Observable Properties of the Slow Solar Wind in a Coronal Streamer'',
ApJ, 2005, volume 633, part 1, pages 474-488 \texttt{http://dx.doi.org/10.1086/431916}.

\blankpage

\nocite{*}

\bibliographystyle{plain}
\bibliography{phdthesis}

\end{document}